\documentclass[3p,sort&compress]{elsarticle}

\usepackage{fleqn}
\usepackage{pdflscape}
\usepackage{afterpage}
\usepackage[figuresright]{rotating}

\usepackage{graphicx}
\usepackage{amssymb,latexsym,amsthm,amsmath,dsfont,mathtools}
\usepackage{color}
\usepackage[utf8x]{inputenc}
\usepackage[T1]{fontenc}
\usepackage[polutonikogreek,english]{babel}
\usepackage{mathdots}
\usepackage{multirow}
\usepackage{hyperref}
\usepackage{colortbl}
\usepackage{longtable}
\usepackage{array}
\usepackage{caption}
\definecolor{mygray}{gray}{.9}

\usepackage{bm} 
\newcommand{\be}{\begin{equation}}
\newcommand{\ee}{\end{equation}}
\newcommand{\ba}{\begin{eqnarray}}
\newcommand{\ea}{\end{eqnarray}}
\def\bea{\begin{eqnarray}}
\def\eea{\end{eqnarray}}

\journal{Physics Reports}

\begin{document}

\begin{frontmatter}

\title{Statistical physics of vaccination} 

\author[ZW]{Zhen Wang}
\address[ZW]{Interdisciplinary Graduate School of Engineering Sciences, Kyushu University, Fukuoka, 816-8580, Japan}

\author[CTB]{Chris T. Bauch}
\address[CTB]{Department of Applied Mathematics, University of Waterloo, Waterloo, ON N2L 3G1, Canada}

\author[SB]{Samit Bhattacharyya}
\address[SB]{Department of Mathematics, School of Natural Sciences, Shiv Nadar University, India}

\author[AO2]{Alberto d'Onofrio}
\address[AO2]{International Prevention Research Institute, 95 Cours Lafayette, 69006 Lyon, France}

\author[PM]{Piero Manfredi}
\address[PM]{Department of Economics and Management, University of Pisa, Italy}

\author[MP,MP2]{\\Matja{\v z} Perc}
\address[MP]{Faculty of Natural Sciences and Mathematics, University of Maribor, Koro{\v s}ka cesta 160, SI-2000 Maribor, Slovenia}
\address[MP2]{Center for Applied Mathematics and Theoretical Physics, University of Maribor, Krekova 2, SI-2000 Maribor, Slovenia}

\author[NP]{Nicola Perra}
\address[NP]{Centre for Business Network Analysis, Greenwich University, Park Raw, SE10 9LS, London, United Kingdom}

\author[MS,MS2]{Marcel Salath{\'e}}
\address[MS]{School of Life Sciences, EPFL, 1015 Lausanne, Switzerland}
\address[MS2]{School of Computer and Communication Sciences, EPFL, 1015 Lausanne, Switzerland}

\author[DZ]{Dawei Zhao}
\address[DZ]{Shandong Provincial Key Laboratory of Computer Networks, Shandong Computer Science Center (National Supercomputer Center in Jinan), Jinan 250014, China}

\begin{abstract}
Historically, infectious diseases caused considerable damage to human societies, and they continue to do so today. To help reduce their impact, mathematical models of disease transmission have been studied to help understand disease dynamics and inform prevention strategies. Vaccination--one of the most important preventive measures of modern times--is of great interest both theoretically and empirically. And in contrast to traditional approaches, recent research increasingly explores the pivotal implications of individual behavior and heterogeneous contact patterns in populations.  Our report reviews the developmental arc of theoretical epidemiology with emphasis on vaccination, as it led from classical models assuming homogeneously mixing (mean-field) populations and ignoring human behavior, to recent models that account for behavioral feedback and/or population spatial/social structure. Many of the methods used originated in statistical physics, such as lattice and network models, and their associated analytical frameworks.  Similarly, the feedback loop between vaccinating behavior and disease propagation forms a coupled nonlinear system with analogs in physics.  We also review the new paradigm of digital epidemiology, wherein sources of digital data such as online social media are mined for high-resolution information on epidemiologically relevant individual behavior. Armed with the tools and concepts of statistical physics, and further assisted by new sources of digital data, models that capture nonlinear interactions between  behavior and disease dynamics offer a novel way of modeling real-world phenomena, and can help improve health outcomes. We conclude the review by discussing open problems in the field and promising directions for future research.

\end{abstract}


\end{frontmatter}


\section{Introduction}\label{intro}

\subsection{Background}

Infectious diseases have proven to be remarkably resilient foes of humanity.  Despite major achievements in other areas, such as discovering gravitational waves and landing humans on the Moon, we still have not eradicated most vaccine-preventable infectious diseases \cite{henderson2013lessons}.  Many of these diseases--measles, pertussis, influenza, and many others--have been burdening us for centuries, while new infectious diseases--such as Ebola, Zika virus, and SARS--continue to emerge from animal populations and make the jump to human populations, or spreading to new human populations due to climate change or other anthropogenic disturbances \cite{morens2004challenge}.

Sometimes, successful infection control is a matter of ensuring access to preventive technologies such as vaccines \cite{bonanni1999demographic}.  Barriers to access in lower-income countries remains an important reason why infectious disease burden remains high in many poor populations \cite{bonanni1999demographic}.  However, there are reasons to be cautiously optimistic that barriers to access will continue to fall, enabling more people to be vaccinated.  For instance, measles was completely eliminated from South America in the 1990s \cite{de1996measles}, prompting many health researchers and policy experts to conceive of global measles eradication as a potentially realistic goal for the first time.  Mathematical and statistical models, often based on methods and concepts borrowed from physics, have played a significant role in infection control by helping determine how to apply control measures in order to quickly and most effectively contain epidemics and helping predict and understand disease outbreak patterns, among their many other uses \cite{heesterbeek2015modeling}.

However, as our technologies advance and access to vaccines begin to become less important as a barrier, it is becoming clear that our own behavior may prevent local elimination or global eradication in many cases \cite{sturm2005parental}.  For instance, unfounded vaccine scares such as the ongoing measles-mumps-rubella vaccine autism scare which started in 1997, or the oral polio vaccine scare in northern Nigeria in 2003-2004 which arguably delayed global polio eradication by a decade, can enable a resurgence of disease in a population \cite{larson2011addressing}.  When the infection spreads to other populations in turn, local elimination or even global eradication can be threatened.  Similarly, traditional infection control methods such as isolation, quarantine, and contact tracing require adequate levels of support from the public.  When this is lacking, sometimes due to mismanagement of the situation by authorities or inadequate response of the international community, then infection control can fail, as was observed in the early stages of the 2014 West African Ebola outbreaks and in many other situations \cite{dhillon2015community}.

\subsection{Motivation and objective}

In summary, recent decades have seen the continual emergence of new infectious diseases together with growing examples where the technologies exist to eliminate long-established infections, but human behavior has prevented elimination.  Moreover, more data on human behavior are available than ever before, at very high levels of spatial and temporal resolution through sources such as internet search engine data, Bluetooth technologies, and social online media.

These trends may partially explain the surging interest in incorporating human behavior directly into mathematical models of infectious disease transmission, in order to understand the two-way interplay between the dynamics of human behavior and the dynamics of infectious diseases.  By accounting for human behavior, mathematical models may provide more insight, and better predictions, than models that neglect the enormous impact that behavior can have on infectious disease dynamics.  For brevity, we will refer to such models as `coupled behavior-disease models' through the remainder of this report, although models that only explore one direction of the coupling, from human dynamics to disease dynamics, or from disease dynamics to human dynamics, are also relevant to our report.

Much of this work relies upon methods and concepts from physics, either directly or indirectly.  This is neither new nor surprising.  For instance, exponential decay can describe radioactive decay of particles, or exponential growth can just as well describe the early growth of an invasive species.  Hence, physics in general, and statistics physics in particular, is highly relevant to the development of disease transmission models that take human behavior into account.

The two objectives of our report are (1) to give a state-of-the art overview of the literature on mathematical models of the impact of human behavior on disease dynamics (and \emph{vice versa}), with special emphasis on the connections to concepts and methods from physics; and (2) to provide an outlook on the future of the field, including open questions and promising directions for future research.  We attempted to include as much of the relevant literature as possible, but we were also cognizant of the need to limit the list of references to a realistic number.  Therefore we selected references that best captured the full breadth of research being undertaken in the field.

\subsection{Outline of the report}

In Section 2 of the report we provide a brief summary of basic ideas in epidemiology that are necessary to understand and develop mathematical models of infectious disease transmission.  Classic mean-field models of infectious disease transmission are then described in Section 3.  Subsequently we describe models of infectious disease transmission on networks and other spatially structured populations, including both simulation-based approaches (Section 4) and analytical approximations (Section 5) to network dynamics.  Sections 3-5 are concerned with models of infectious disease transmission that do not explicitly include human behavior, since understanding such models is a necessary precursor to understanding models that explicitly do include human behavior.  Section 6 provides a rationale for studying models of the interactions between human behavior and disease dynamics (some of which has already been described in the preceding subsections), while section 7 introduces some basic concepts and methods in the modelling of human behavior.  Section 8 provides an overview of mean-field models of coupled behavior-disease dynamics, while Section 9 provides an overview of such models in the context of networks and spatially-structured populations.  The next two sections are concerned with `traditional' sources of data on human behavior (Section 10) and digital sources of data on human behavior (Section 11) from which we can test and validate models of coupled behavior-disease dynamics.  The report concludes with a discussion in Section 12.

\section{Basic concepts in infectious disease epidemiology} \label{sec:basic_epi_concepts}

Infectious disease epidemiology is the study of contagious parasites of humans.  Parasitization is one of the oldest ecological relationships, and has been hypothesized to be a
major driving factor in the evolution of a very broad range of species (the `Red Queen hypothesis') \cite{van1973new}.  Indeed, infectious diseases are likely to have been with
humans for our entire existence.  The Bible and ancient Greek historical writing contain apparent references to epidemics and case isolation \cite{page1953thucydides,bennett1891diseases}.
This section describes the basic biology and ecology of infectious diseases, and how such diseases are controlled through interventions such as vaccines.

Parasites are traditionally divided into two main groups: microparasites and macroparasites. The distinction between them is their size and parasitic life cycle \cite{brown1987microparasites}. Macroparasites are larger forms of pathogens and include ticks, mites, helminthes and flukes \cite{anderson1979population}. Macroparasites grow in one host but reproduce by infective stages outside of the host organism \cite{morand2006micromammals}. They can be either external parasites (ectoparasitic) or internal parasites (endoparasitic) \cite{collinge2006disease}. Comparatively, microparasites are small, generally single-cell organisms, and are either viruses, bacteria or protozoans \cite{anderson1981population, swinton2002microparasite}. They generally exhibit short generation times and a high rate of direct reproduction within the host organism \cite{anderson1979population}. For simplifying assumptions in modeling, we focus on the body of literature pertaining to microparasites, where extensive long-term data and a solid understanding of disease transmission dynamics have contributed to the development of well-parameterized models \cite{keeling2008modeling}.

Infectious diseases (both micro- and macroparasitic) can be transmitted to susceptible hosts in a variety of ways. Routes of transmission are classified as direct and indirect transmission \cite{giesecke1994modern}. Direct transmission occurs when there is physical or close contact between an infected individual and a susceptible individual that permits transmission via droplets or aerosols \cite{kramer2010principles}. Although some exceptions exist, the majority of microparasitic diseases are directly transmitted and exhibit short survival periods outside of the host organism \cite{keeling2008modeling}.  Conversely, indirectly transmitted parasites are passed through vectors or are transmitted vehicularly via the environment \cite{kramer2010principles}. Most macroparasitic diseases are indirectly transmitted, spending part of their life cycle outside of their host organism \cite{keeling2008modeling}. Indirect transmission occurs via another organism through a vector or via an intermediate host \cite{nelson2014infectious}. A vector is an organism that does not directly cause disease, but that transmits infection of pathogens from one host to another \cite{porta2014dictionary}. Vectors can be mechanical or biological.  A mechanical vector transmits an infectious pathogen without it being altered while on the vector \cite{spickler2010emerging}.  In contrast, biological vectors harbor pathogens within their bodies and deliver pathogens to new hosts in an active manger, usually a bite \cite{goddard2009infectious}. Biological vectors are usually continuously infected with the disease and in some cases are a required part of the organisms life cycle \cite{spickler2010emerging}. Increases in populations and urbanization, habitat loss, and international travel and trade all contribute to the spread of vectors and indirect transmission of pathogens \cite{rogers1993vector}. Consequently, the mechanisms of transmission and transmissibility are key factors of pathogens that influence disease spread and what interventions, if any, are effective \cite{mayer2011social}. Timing of the onset of infectiousness relative to the onset of symptoms and duration of infectious periods are additional important factors in transmission dynamics \cite{mayer2011social}.

Integral to understanding infectious disease dynamics is distinguishing three time periods involved in these processes: the latent period, defined as the time from infection to the time when a host is able to transmit the pathogen to another host \cite{porta2014dictionary}; the incubation period, defined as the time from infection to the onset of disease \cite{merrill2013introduction}; and the infectious period, defined as the period from the end of the pre-infectious period until the time when a host is no longer able to transmit the infection to other hosts \cite{vynnycky2010introduction}. The distinction between the incubation period and the latent period of a disease is integral when considering the effect of control methods based on symptomatic surveillance in a population. If the latent period is shorter than the incubation period (i.e., infectiousness proceeds with the onset of symptoms), then interventions targeted toward symptomatic individuals are generally ineffective \cite{fraser2004factors}. Diseases whereby the onset of symptoms coincides with, or proceeds with infectiousness are more effective at being controlled using symptomatic surveillance methods \cite{nelson2014infectious}. Consequently, an important metric in the controllability of handling a disease outbreak requires the consideration of the proportion of transmission that takes place before the onset of symptoms occurs \cite{fraser2004factors}. Both incubation period and latent period are generally modeled following an exponential, log-normal, Weibull, or gamma distribution \cite{nelson2014infectious}.

Three important characteristics of pathogens that affect transmission of infectious diseases are infectivity, virulence and pathogenicity. In epidemiology, infectivity is the ability of a pathogen to cause infection in a susceptible host \cite{nelson2014infectious}. Infectivity can also vary between different strains of the same pathogen.  The route of the infection, and the age and innate resistance of the host are key factors which contribute to the infectivity of a disease \cite{thrusfield2013veterinary}. The gauge of infectivity is measured by the secondary attack rate (the proportion of exposed hosts who develop the disease \cite{thomas2001epidemiologic}). Infectivity has been shown to positively correlate with virulence \cite{cheng1988biologic,stewart2005empirical}. Virulence, although often used interchangeably with pathogenicity, refers to the severity of disease after infection occurs \cite{nelson2014infectious}. The capacity and vigor of the disease to cause severe and fatal cases of illness is virulence \cite{timmreck2002introduction}. Pathogenicity refers to the ability of a pathogen to induce disease \cite{nelson2014infectious}. Pathenogenicity and virulence are determined by the interaction between host and pathogen \cite{kramer2010principles}. Characteristics of infectious pathogens that affect virulence include their ability to replicate, invade organisms and damage the host \cite{kramer2010principles}. Host characteristics also affect the ability of a pathogen to cause disease, in that pathogeneticity and virulence depend on the resistance and immunity system of the host, in addition to genetic factors, age, gender, and physiological condition \cite{kramer2010principles}.

Herd immunity is a form of indirect protection from infectious disease that is brought about by the presence of immune individuals in a population \cite{fine1993herd}. With certain diseases that spread from host to host, the level of population immunity is critical in determining the potential of an epidemic from occurring, and therefore the risk of infection for a susceptible host in a population \cite{nelson2014infectious}. In that transmission occurs between an infected and susceptible host, if the number of immune individuals is large, then the likelihood that a susceptible individual being in contact with an infected individual is reduced. Immunity can be conferred through natural immunity, previous exposure to the disease, or through vaccination \cite{boslaugh2007encyclopedia}. The level of immunity required to attain herd immunity depends on the characteristics of the infectious disease. Generally a disease with high infectivity requires a higher proportion of immune individuals to achieve immunity than diseases with lower infectivity rates \cite{boslaugh2007encyclopedia}. Additional considerations are: seasonality, to what effect it has on population mixing and the viability of the pathogen; the number of susceptible individuals, their age and geographic distribution; and social habits of the population with respect to mixing and type of contacts made \cite{fox1971herd}. Mass vaccination to induce herd immunity is common practice and has proven successful in preventing the spread of many infectious diseases globally \cite{plotkin2006mass}. Opposition to vaccination has posed a challenge to herd immunity, allowing preventable diseases to persist in or return to communities that have inadequate vaccination rates \cite{gangarosa1998impact}.

An integral application in infectious disease modeling is the need for evaluating intervention strategies and their effects on new and re-emerging pathogens. The development and widespread use of antibiotics was an important advance in the treatment of infectious diseases \cite{mayer2011social}. Antibiotics are a type of antimicrobial used in the treatment and prevention of bacterial infection, by killing or inhibiting the growth of bacteria \cite{kohanski2010antibiotics}. Several antibiotics are effective against fungi and protozoans \cite{lorian2005antibiotics}, but are not effective against viruses \cite{gonzales2001excessive}. In many cases, their effectiveness and accessibility has led to their overuse, prompting bacteria to develop resistance \cite{seppala1997effect,blaser2011antibiotic,davies2010origins}. Increasing travel and migration have also contributed to the growth of antibiotic resistance problems \cite{zhang2006antibiotic}. The consequences of antibiotic-resistant infections may result in more prolonged illnesses, longer hospitalization, increased risk of death and higher health care costs than do infections with antibiotic-sensitive strains of the same species \cite{kramer2010principles}.

\begin{figure}
\begin{center}
\includegraphics[width=0.7\textwidth]{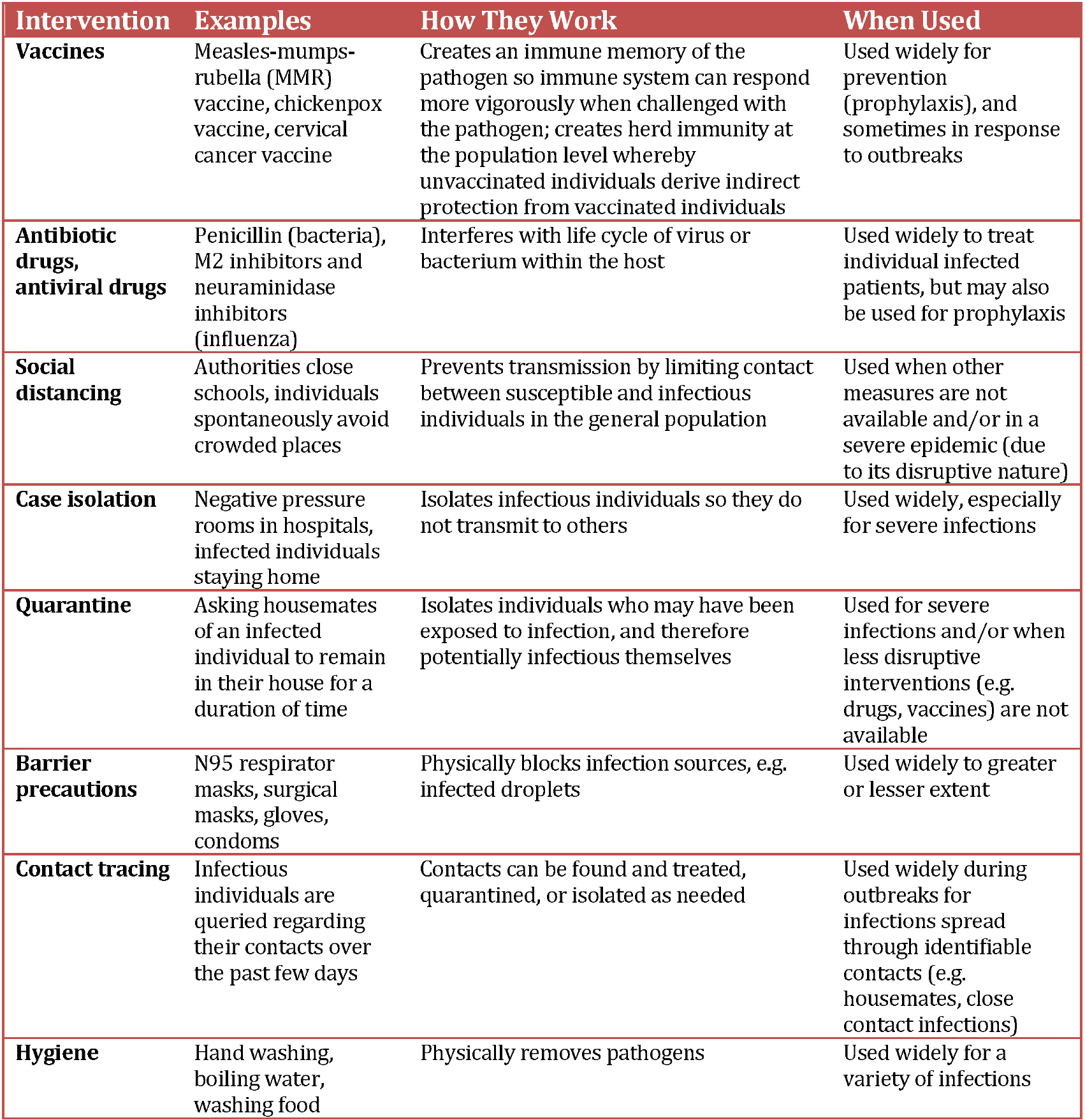}
\end{center}
\caption{Summary of commonly used interventions against infectious diseases.  We note that this list is not exhaustive, and interventions are often used in combination, such as contract tracing and isolation.}
\label{figure_interventions}
\end{figure}

A primary way to boost immunity in a population is through vaccination \cite{stephenson2003boosting}. Vaccines are any biologically derived substance that produces a protective immune response when administered to a susceptible host \cite{nelson2014infectious}. Vaccine effectiveness is the result of the vaccine's direct effect, which refers to its ability to protect individuals from infection, and its indirect effect, which is its ability to reduce the spread of infection in a population \cite{comstock1990vaccine,orenstein1988assessing}. Notably, no current vaccine is entirely effective. It is estimated that non-preventable primary vaccine failure rates range from 2 to 50 percent for licensed vaccines under ideal circumstances in clinical trials \cite{chen1996epidemiologic}. Vaccine effectiveness can be determined by comparing risk of the disease among vaccinated and nonvaccinated groups. Important variables to consider are that groups are comparable with respect to exposure, risk of infection, accessibility of the vaccine, and opportunity for diagnosis \cite{nelson2014infectious}. Vaccines, antibiotics, and other interventions are summarized in Fig.~\ref{figure_interventions}.

Social distancing is another strategy used to reduce the transmission of infectious diseases. Governments and public health agencies may promote targeted social distancing as a strategy, in the form of school or workplace closures. Social distancing also arises as individuals respond to news about disease spread, obtained from media reports, public announcements, rumours and individual experience \cite{blendon2004public}. Recognizing the importance of behavioral responses in populations has altered the way epidemiological modelers consider transmission \cite{gross2006epidemic,funk2009spread,zanette2008infection,yu2014early,xia2012sir,sun2016influence,sun2015effects}. Such models have shown that social distancing can be effective at reducing the attack rate of an epidemic \cite{glass2006targeted,poletti2009spontaneous} and that social distancing behavior is a plausible explanation for certain phenomena arising in real epidemics, such as multiple outbreaks or waves of infection \cite{poletti2009spontaneous, caley2008quantifying}.

Hygiene refers to environmental conditions and practices that contribute to maintaining health and prevent the spread of diseases \cite{cohen2000changing}. Aiello and Larson \cite{aiello2002evidence} examined the epidemiological evidence for causal relation between hygiene and infections and found that personal and environmental hygiene significantly reduced the spread of infections. In particular, the importance of hand hygiene in preventing infections has been widely recognized. Hands are a key step in the transmission of many infectious agents, and the importance of hand hygiene and interrupting hand contact with body fluids in preventing spread of infections is well documented \cite{pittet2000effectiveness,allegranzi2009role,fewtrell2005water}. Diseases related to hygiene in general disproportionately affect poorer members of society due primarily to accessibility of services and/or less polluted environments \cite{pruss2002estimating}.

Despite the complexities of the biology of infectious diseases on the cellular and organismal levels, it is often possible to mathematically model the transmission of infectious diseases
at the population level.  In fact, some of the earliest research in mathematical biology concerned the analysis of population-level transmission of infectious diseases using differential
equations \cite{kermack1927contribution}. The mathematical modelling of infectious disease transmission is the subject of the next section.

\section{Basic concepts and methods in epidemiological modeling and vaccination: compartmental (mean-field) models} \label{sec:compart-model}

Epidemiological modeling has nowadays become a global discipline, playing a key role in policy decisions about prevention and control of communicable diseases. With hindsight, much of this success is due to a few, extraordinarily fecund, deterministic models, first of all the two classical susceptible-infective-removed (SIR, see Table A1 in Appendix for this and other abbreviations used in our review) models for outbreaks and for endemic infections (and related extensions). These models represent the direct legacy of the intuitions of a few innovators who, at the beginning of the $1900s$, imported the mass-action law of statistical physics to describe infection transmission. This intuition, which represents the building block of the transmission process of a communicable disease, namely the social mixing between the individuals of a community, as collisions between particles in a certain medium, has allowed mathematical epidemiology to take-off as a scientific discipline. This section reviews the key concepts and results of basic deterministic mean-field models of theoretical epidemiology, departing from an historical overview of the founding role of the mass-action principle (section \ref{Mathepi}). Notably, the current \textit{behavioral epidemiology} revolution stems, one hundred years after the birth of mathematical epidemiology, from a twofold awareness, first that humans, unlike particles, are thinking agents that might depart from the mass-action rule especially when the threats from infections are serious, and second that we now have easier access to the tools that are necessary to finally model such departures. The rest of the section is organised as follows:  section \ref{Basic_concepts} introduces a basic compartmental modeling framework for infection transmission and the underlying key concepts; section \ref{BRN} reports a non-technical presentation of reproduction numbers, which are key summary parameters of transmission; the main results about models for infection with permanent immunity, namely the SIR and SEIR models for outbreaks and for endemic infections, follow in sections \ref{Epidemic_SIR} and \ref{Endemic_SIR}  respectively; the effects of mass vaccination for SIR infections are reported in section \ref{SIR_vaccination}, while the consequences of immunity losses follow in section \ref{SIS_SIRS_models}; the section ends with a discussion of a number of extensions of basic models (section \ref{metapop}), and of the epidemiological meanings and implications of the use of mean-field approximations (section \ref{stochastics_meanfield}).

\subsection{Mathematical  epidemiology: a story shaped by statistical physics} \label{Mathepi}
\begin{figure}
\begin{center}
\includegraphics[width=0.7\textwidth]{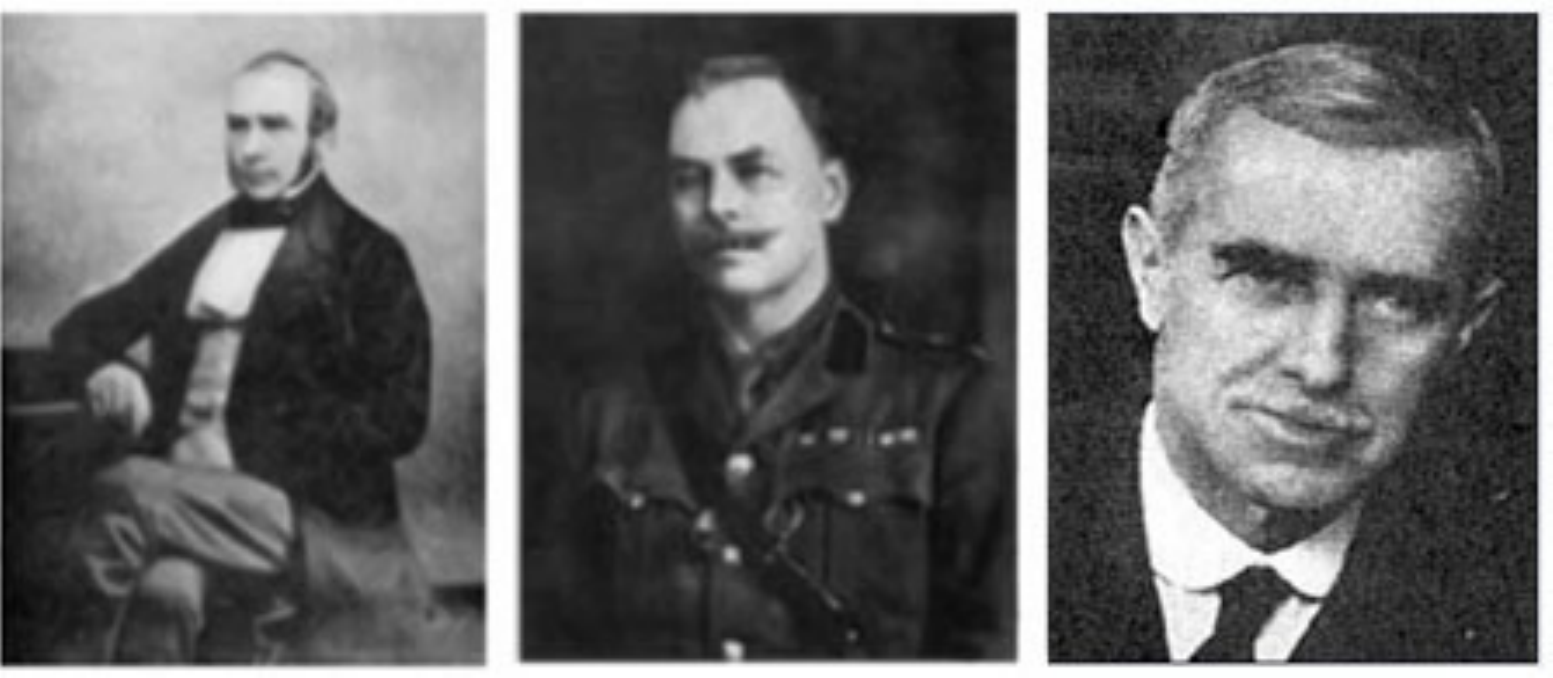}
\end{center}
\caption{Forerunners of mathematical epidemiology: John Snow (left), Sir Ronald Ross (centre), Anderson McKendrick (right).}%
\label{Forerunners}
\end{figure}

The first application of mathematics to epidemiology dates back to Daniel Bernoulli \cite{Daniel_Bernoulli} investigation of the gain in life expectancy that would be allowed by elimination of smallpox, then a major cause of humans' death. Though historically important and far-seeing - smallpox is so far the only infection ever eradicated in mankind history - Bernoulli's study cannot however be taken as the birthday of modern mathematical epidemiology, since it neglected the building block of the diffusion of a communicable disease, namely, the transmission process of an infective agent and its social spread in a population. The consideration of the transmission process took 150 further years i.e., following the nice historical reconstruction by Heesterbeek \cite{Heesterbeek_2005}, the lapse of time necessary to build a solid scientific background to the idea of contagion. It is indeed only in the 1870s that the conclusive confirmation by Koch and Pasteur that infectious diseases are caused by living organisms, definitely paved the way to the explanation of observed \textit{epidemic curves} (which today we call \textit{incidence curves}), giving credit to a number of central earlier intuitions. The main one was the John Snow (Fig. \ref{Forerunners}) hypothesis that epidemics come to an end when the availability of susceptible individuals - the  \textquotedblleft epidemic fuel\textquotedblright - goes down. Some decades later, while analyzing UK measles incidence data, William Hamer \cite{Hamer_1906} identified the biennial period of outbreaks in towns, and in particular explained the recurrence of measles epidemics by Snow's hypothesis, i.e. a new epidemics will start \textquotedblleft  when there is a sufficient accumulation of susceptible persons\textquotedblright. In addition he computed from data a relationship of the form \textquotedblleft  New cases at time $t+1$=(New cases at time $t$) $\times$ (susceptibles at time $t$)\textquotedblright. However, he failed to recognize this as a general mechanistic rule, nor did he draw any implications from it \cite{Heesterbeek_2005}. The key step finally turning mathematical epidemiology into a scientific discipline occurs when two fine innovators working independently but in strict touch, namely the 1908 Nobel Prize for medicine Sir Ronald Ross \cite{Ross_1916,Ross_1917} and Anderson McKendrick \cite{McKendrick_1912} (Fig. \ref{Forerunners}), formulated and analyzed the first modern epidemic models \cite{Heesterbeek_2005}. It was in particular McKendrick who, by importing from a stronger domain of knowledge, namely statistical physics, the \textit{mass-action} law of chemical reaction kinetics to describe infection transmission as the outcome of the social contact process between infective and susceptible individuals, sets-up the building block of modern epidemiology, thereby generating a true \textit{Gestalt} shift in scientific knowledge. McKendrick's ground-breaking intuition is summarized in his metaphor of \textquotedblleft collisions\textquotedblright, that he applies to an animal infection: \textquotedblleft The rate at which this epidemic will spread depends obviously on the number of infected animals, and also on the number of animals which remain to be infected, in other words the occurrence of a new infection depends on a collision between an infected and an uninfected animal.\textquotedblright \cite{McKendrick_1912}.

This observation makes McKendrick's seminal study a true - in contemporary language - transdisciplinary paper in statistical physics. He makes the counter-factual hypothesis that the population behaves as a collection of \textquotedblleft particles\textquotedblright  of  different types randomly moving in the environment, where each encounter (\textquotedblleft collision\textquotedblright) between a particle of \textit{susceptible}-type and one of \textit{infective}-type has a uniform probability that the $S$-type particle switches into $I$-type. Translating this into mathematical language for an infection without immunity in a stationary population of size $N$, letting $Y$ and $X$  respectively denote the numbers of infective and susceptible individuals ($X+Y=N$), and $k^{\prime}$ a proportionality constant, McKendrick derived the following ordinary differential equation (ODE) for the rate of epidemic spread
\begin{equation}
Y'=k^{\prime}Y(N-Y),
\end{equation}
as the first true model of modern epidemiology. Passing to the epidemiological fractions $S=X/N$, $I=Y/N$, $S+I=1 $ and defining $\hat{k}= k^{\prime} N$ one gets
\begin{equation}
I'=\hat{k}I(1-I).
\end{equation}
This simple logistic equation, having $I=1$ as a globally asymptotically stable (GAS) state reached by a sigmoid temporal pattern, implies that an infection without immunity spreading according to the mass action principle will eventually affect the entire population. McKendrick's \textit{full contagion principle} represents the first principle of theoretical epidemiology.

By pursuing further his mass-action formulation, McKendrick definitely launched mathematical epidemiology in a famous series of papers with Kermack \cite{KMK}, where they introduced the celebrated \textit{threshold theorem} for infections imparting permanent immunity, studied conditions for endemicity, and introduced and analyzed the first age-structure epidemiological models. The momentum impressed by Kermack and McKendrick work yielded, with a certain delay, an explosion of bio-mathematical studies, mostly devoted to the investigation of a number of variants of the key models introduced by the forerunners. For further historical details see \cite{Dietz_1985}.

There is a second major shift however, dating 1975-1980, when another group of innovators \cite{Dietz_1981,Hethcote_1984,AMay_1991} started and systematically using mathematical models to interpret epidemiological data and trends aiming to deep the perspectives for infection control. In this phase, dramatically accelerated first by the HIV/AIDS epidemics in the 1980s, and in the 2000s by the emergence of influenza pandemic threats, mathematical models of infectious diseases definitely left the boundary of bio-mathematics and became a central supporting tools for public health decisions, such as determining the duration of travel restrictions or of school closure during a pandemic event, or the fraction of newborn to be immunized for a vaccine preventable infection, as measles. It is this fine tuning of epidemiological models that eventually raised the key question of behavioral epidemiology: given that humans are not \textquotedblleft particles\textquotedblright, which is then the role they play for infection transmission and prevention? \cite{BDM}.

\subsection{Basic concepts: homogenous social mixing, infection incidence, force of infection} \label{Basic_concepts}
\begin{figure}
\centering \includegraphics[width=115mm,height=115mm]{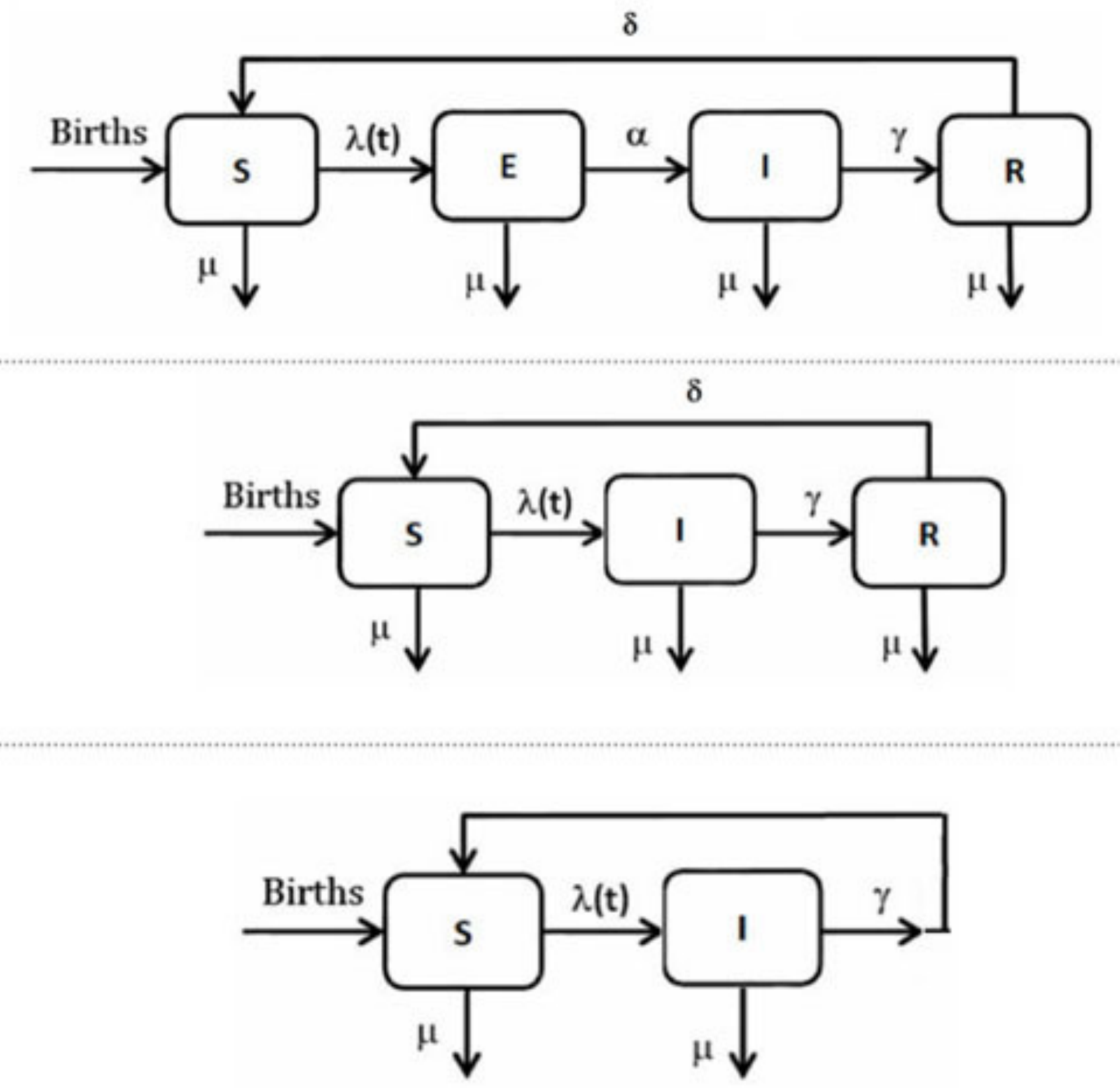}
\caption{Compartmental representation of the individual course for some important classes of infections: SEIRS model (top); SIRS model (middle); SIS model (bottom). Boxes identify epidemiological classes and arrows identify the processes governing transitions between classes, and in-out flows due to demographics. }%
\label{Fig_SEIRS_compartments}%
\end{figure}

Regardless of their nature (e.g., directly vs sexually transmitted), for most human infectious diseases the individual life-course of infection is conveniently represented by a sequence of transitions between a few epidemiological classes, or \textit{compartments}, triggered by specific processes and parameters \cite{Hethcote_2000,AMay_1991}. For new pathogens, such as a pandemic flu virus, all individuals are initially \textit{susceptible} to the infection (class $S$), which they acquire at a rate $\lambda$ entering the \textit{exposed} class ($E$), a latency or incubation phase where individuals are infected, i.e. the virus is replicating in their body, but not yet \textit{infective}, which they become at some rate $\alpha$. Individuals belonging to the infective class ($I$) will usually recover, at a rate $\gamma$, acquiring immunity and entering the \textit{removed} class ($R$). This Susceptible-Exposed-Infective-Removed course, \textit{SEIR} for brevity (Fig. \ref{Fig_SEIRS_compartments}), is shared for example by a number of vaccine-preventable infections of childhood yielding lifelong immunity, such as measles, mumps and rubella. For other, as pertussis and varicella, clinical reinfection through return to susceptibility (at some rate $\delta$) is possible, so that their actual course would be \textit{SEIRS}.


Based on this representation of the individual course, epidemiological models seek to describe the dynamics of infections at the population level, by associating to each compartment a dynamic variable denoting the number (or the proportion) of individuals in that class at each instant of time. Denoting by N the total population size, we let $X$ ($S=X/N$), $H$ ($E=H/N$), $Y$ ($I=Y/N$), $Z$ ($R=Z/N$) to denote the numbers (proportions) of individuals who respectively occupy the various classes at time t, with $N=X+H+Y+Z$, and $S+E+I+R=1$. The key process, namely infection spread, focuses on the number of transitions from compartment $S$ to compartment $E$ per unit of time, which is the absolute incidence ($U$) of new infections per time-units. In simple models the mass action formulation of incidence occurs by the concept of \textit{homogeneous} (or random) \textit{mixing} \cite{Hethcote_2000,AMay_1991}. This refers to an infection spreading by \textit {direct person-to-person contacts } in a closed homogenous social medium, the population, whose agents have an approximately identical social activity, summarized by the average number of contacts $C$ per unit time, the same probability $I=Y/N$ that a given contact occurs with an infective individual, and the same constant probability $\beta_{I}$ of infection transmission per single \textit{adequate} social contact (i.e., per single social link in networks jargon) between a susceptible and an infective individual. Therefore, the expected instantaneous per-capita rate $\lambda$ at which susceptible individuals acquire infection, called the \textit{force of infection} (FOI), can be written as: $\lambda=\beta_{I} CI$ or, setting $\beta=\beta_{I}C$, as: $\lambda=\beta I$. The corresponding absolute incidence $U$ is then given by
\begin{equation}
U=\lambda X=\beta (Y/N) X,
\end{equation}\label{True_mass_action_incidence}
which is the  \textit{standard mass action} formulation of the incidence and the FOI \cite{Hethcote_2000,DH_2000}. It depends on the complex parameter  $\beta$, representing the total \textit{transmission rate} per unit of time, which embeds distinct sub-processes, namely social behavior (mean number and type of contacts per unit of time) as well as the biological process of contagion. Note that, assuming the frequencies $S$ and $I$ as the "molar concentrations" of the "particles"  susceptible subjects and infectious, and the total population size $N$ as the "Avogadro's number" of the model in study, then the change per time unit of such "molar concentrations" is equal to $(U/N)$, which in turn is equal to $\beta S I$, i.e. the kinetic mass action law.

Also common in biomathematics is the \textit{bilinear incidence} $U=\widetilde{\beta} XY$ \cite{Capasso,Hethcote_2000}, which, however, is less supported empirically \cite{AMay_1991}, and moreover postulates that $C$ is linearly increasing in population size which seems to be reasonable only in a few special cases, and also in the quite artificial case where the numerical value of $\widetilde{\beta}$ scales as $1/N$. This means that one can write $U=\widetilde{\beta} XY$ $= \beta (Y/N) X$. In other words, in such cases the models, although formally built on the \textit{bilinear incidence}, in reality are ruled by the \textit{standard mass action} law for the contagion mechanism.

To complete the model formulation we make a few further hypotheses. First, we assume that all other epidemiological processes in Fig.  \ref{Fig_SEIRS_compartments}, namely, transfers $E\rightarrow I$, $I\rightarrow R$, $R\rightarrow S$, occur at constant rates, meaning that the duration of sojourn in the involved states is exponentially distributed. This means for example that $\gamma =1/D$, i.e. that the rate of recovery is the reciprocal of the average duration $D$ of the infective phase. Second, we assume that the population is large, so that the law of large numbers rules out large chance fluctuations, and closed to migration, i.e. it evolves over time due the natural balance between exits due to mortality (from all compartments), at a per-capita rate $m(N)$, and entries due to new births, occurring at a per-capita rate $b(N)$ into the susceptible class only. Last, we rule out complications such as infection-related mortality, or vertical transmission from mother to foeta during pregnancy. We can then translate the SEIRS diagram in Fig. \ref{Fig_SEIRS_compartments} into the following \textit{mean-field} system of ODEs

\begin{align}
X^{\prime}&= b(N)N + \delta Z - m(N)X -\beta X \frac{Y}{N}, \label{eqX}\\
H^{\prime}&= \beta X \frac{Y}{N} - \alpha H - m(N) H,  \label{eqH}\\
Y^{\prime}&= \alpha H  -\gamma Y - m(N) Y, \label{eqY}\\
Z^{\prime}&= \gamma Y  - m(N) Z - \delta Z, \label{eqZ}
\end{align}
\noindent where the population size obeys
\begin{equation}
N^{\prime} = (b(N)-m(N))N.
\end{equation}\label{eqN}
Given that the population evolution is known, one equation can be discarded, since e.g., $ Z = N(t) - X - H - Y $. Assuming that the population $N$ has a unique and globally stable equilibrium $N_e$ such that $ b(N_e) = m(N_e) = \mu $, and that the population is at its steady state $N_e$, yields
\begin{align}
S^{\prime}&= \mu( 1- S) + \delta R  -\beta S I, \label{eqS}\\
E^{\prime}&= \beta S I - (\alpha +\mu)E, \label{eqE}\\
I^{\prime}&= \alpha E -(\gamma +\mu)I,  \label{eqI}
\end{align}
\noindent
complemented by $ R = 1 - S - E -I $. Model of Eqs.~\ref{eqS}-\ref{eqI} is the so-called deterministic SEIRS model \cite{Hethcote_2000}.

By suitably restricting some parameters the SEIRS model can be specialized into various \textquotedblleft classical\textquotedblright models appeared in the biomathematical literature. For example, by assuming a negligibly short duration of the exposed phase (i.e. $\alpha >>1$) and applying a quasi-steady state approximation
\begin{equation}
E \approx \frac{\beta}{\alpha+\mu} SI
\end{equation}
yields
\begin{align}
S^{\prime}&= \mu( 1- S) + \delta R  -\beta S I, \label{eqSbis}\\
I^{\prime}&\approx\frac{\alpha}{\alpha+\mu} \beta S I -(\gamma +\mu)I  \approx \beta S I -(\gamma +\mu)I,  \label{eqIbis}
\end{align}
\noindent
i.e. the SIRS model (see (Fig. \ref{Fig_SEIRS_compartments}), to be complemented by the algebraic equation $R = 1- S-I$. Further assuming that the immune phase is permanent (i.e. setting $\delta =0$) yields the celebrated SIR model with vital dynamics
\begin{align}
S^{\prime}&= \mu( 1- S)   -\beta S I, \label{eqSsir}\\
I^{\prime}&= \beta S I -(\gamma +\mu)I.  \label{eqIsir}
\end{align}
\noindent
On the contrary, the assumption of absence of an immune phase (i.e. $\delta >>1$) brings to the SIS model (Fig. \ref{Fig_SEIRS_compartments})
\begin{align}
S^{\prime}&= \mu( 1- S)   -\beta S I +\gamma I, \label{eqSsis}\\
I^{\prime}&= \beta S I -(\gamma +\mu)I.  \label{eqIsis}
\end{align}
\noindent
Finally, by considering epidemic models for which the epidemiological time scales are fast compared to demographic ones, i.e. by setting in the above model $\mu \approx 0$, one obtains models for outbreaks, which include the celebrated SIR epidemic model by Kermack and McKendrick \cite{KMK}.

\subsection{Reproduction numbers of epidemic spread}
\label{BRN}
What happens when a new infectious agent, such as a pandemic virus, enters (e.g. via an infected traveler) a population of susceptible individuals? Will the virus start spreading and causing an \textit{outbreak}? And if it does, how fast and how many people will be eventually hit? Though detailed answers to such questions depend on the characteristics of the infection considered, a great deal of intuition can be obtained by a few summary parameters. The most well-known is  the \textit{basic reproduction number} (BRN) of infection, a dimensionless quantity universally denoted by $\mathcal{R}_0$, and intuitively defined as the expected number of secondary infection cases caused by a single \textit{typical} infective case during his entire period of infectivity in a wholly susceptible population \cite{DH_1990,AMay_1991, DH_2000,DHB_2012}. Under homogeneous social mixing $\mathcal{R}_0$ can be defined as
\begin{equation}
\mathcal{R}_0 = \beta_{C} C D=\beta D,
\end{equation}
where $D$ is the expected duration of the infective phase. In a large homogeneous population living a world without uncertainty, $\mathcal{R}_0$ represents, by definition, a threshold parameter stating whether the epidemics takes off, thereby invading the population ($\mathcal{R}_0 >1$) or dies out (otherwise). In case of invasion $\mathcal{R}_0$ also represents the exponential growth factor of infective cases counted per generations of infection. Let us consider for example an infection imparting immunity, as influenza, and take $\mathcal{R}_0 =2$, a figure close to the estimates for 1918 Spanish pandemic flu \cite{Ferguson_2005}. The initial case (the \textquotedblleft traveler\textquotedblright, representing the \textit{zero generation}) would cause $\mathcal{R}_0 =2$ cases in the first generation, which would in turn cause $\mathcal{R}_0 =2$ cases each one in the second generation, and so on, according to the geometric progression $2^n$ after $n$ generations. This exponential growth according to
\begin{equation}
U_n = \mathcal{R}_0^n
\end{equation}
\noindent holds for the initial \textit{invasion} phase, that is as far as the depletion of susceptible subjects is negliglible. The corresponding real time growth of infective cases will be $ U(t) \approx e^{r t} $. Assuming a stationary distance between any two consecutive generations of infection, what today we call the infection \textit{generation time} \cite{Wallinga_Lipsitch}, and that this equals $D$, the duration of the infective phase, leads to the fundamental relation between generational growth and real time growth: $ \mathcal{R}_0 = e^{r D}$, which enables to estimate $\mathcal{R}_0$ based on the observed rate of real time growth of cases and the infection generation time. Once epidemic spread has depleted the susceptible fraction $S$ to levels well below $100\%$, the definition of $\mathcal{R}_0$ does not apply anymore, and the outbreak evolution is described by the quantity (still assuming homogenous mixing)
\begin{equation}
\mathcal{R}_E(t) = \mathcal{R}_0 S(t)
\end{equation}
known as the \textit{effective reproduction number} (ERN), often denoted by $\mathcal{R}_E$. The ERN captures the fact that contacts with immune people do not contribute to the further growth of the epidemics. Therefore, still under the Spanish flu assumption $\mathcal{R}_0=2$, when the epidemics has immunized $25\%$ of the initial population, cases will temporary grow at the reduced generation speed $\mathcal{R}_E =1.5$, meaning that on average each infective case will now generate only 1.5 secondary cases. With further progressing of the epidemics, $\mathcal{R}_E$ further declines, and when the immune proportion reaches $50\%$, implying  $\mathcal{R}_E = 1$, the incidence curve reaches its maximum, after which the epidemics enters its declining phase, and eventually dies out.
%


Reproduction numbers allow to also understand some key effects of \textit{mass vaccination}. For sake of simplicity we consider a \textit{perfect} vaccine, i.e. one with $100\%$ take and ensuring lifelong protection against infection. Under the threat of an outbreak, vaccination by a perfect vaccine of a fraction $x$ of the population initially susceptible would decrease the transmission potential from $\mathcal{R}_0$ to $\mathcal{R}_V=\mathcal{R}_0 (1-x),$ the so-called \textit{vaccine reproduction number} (VRN). Therefore, choosing $x$ so as to bring the VRN below one, namely
\begin{equation}\label{thepc}
x>1-1/\mathcal{R}_0 =x_{c},
\end{equation}
would prevent the outbreak to occur. This argument extends to infections that prior to vaccination were persisting endemically. Assume that vaccination is administered to a constant proportion $x$ of newborn individuals. If vaccination is continued thoughout many consecutive birth cohorts, the susceptible fraction in the population will eventually follow $S \le (1-x)$, which in turn implies the inequality $\mathcal{R}_E =\mathcal{R}_0 (1-x)$. Therefore the condition $\mathcal{R}_0 (1-x)<1$, yielding again $x>x_{c}$, represents in this case a correct condition of infection elimination (Anderson and May 1991), since it implies that the ERN is definitely forced below $1$. The quantity $x_{c}$ is termed the \textit{critical vaccine coverage}. This result, indicating that a sufficiently large (but always smaller than $100\%$) vaccine uptake is capable to eliminate infection by breaking the transmission chain, is referred to as the \textit{herd immunity} principle. The critical coverage $x_{c}$ is increasing and concave in $\mathcal{R}_0$. Therefore, for measles, for which $\mathcal{R}_0$s was estimated to be about $15$ in England and Wales prior to vaccination, the corresponding critical coverage is $x_{c}=93.3\%$, requiring effective vaccination of nearly all newborns.

Both $\mathcal{R}_0$ and $\mathcal{R}_E$ actually are \textit{model-dependent} parameters. That is, any well posed mean-field epidemiological model as those presented in section~\ref{Basic_concepts} (i.e., SIS, SIR, SIRS, SEIR, SEIRS, etc) will have its own $\mathcal{R}_0$ and $\mathcal{R}_E$, depending more or less complexly on the model parametric structure, as illustrated in forthcoming subsections. However, fully general routes to compute RNs are available for very broad families of epidemiological models \cite{DH_1990,DH_2000,DHB_2012,VDD_Watmough}. The key underlying idea is that of \textit{next generation operator} (NGO), which is the positive operator associating to any given distribution of infective cases, representing a certain generation of infection, the number of cases in the subsequent generation. Properties of positive operators ensure that iteration of the NGO will eventually bring the exponential growth (per generation) of infective cases at a factor - the \textit{dominant eigenvalue} of the NGO - which is currently taken as the most general definition of RN in deterministic epidemiological models. Historical notes on the ubiquitous role of $\mathcal{R}_0$ in mathematical population dynamics are reported in \cite{Heesterbeek_2002}.

\subsection{SIR and SEIR models for epidemic outbreaks}
\label{Epidemic_SIR}
What does the framework of Eqs.~\ref{eqS}--\ref{eqI} further tells us about epidemic outbreaks? For example which is the outbreak's temporal profile?  When is the epidemic peak reached? Which is the fraction of the population eventually infected? Let us depart for simplicity from an SIR communicable infection (i.e. we disregard the $E$ class) that is introduced, by importation of a few infective cases, in a wholly susceptible population, that is with an initial susceptible fraction $S(0)\approx 1 $. As we are interested here in short outbreaks - such as seasonal influenza, whose outbreaks last a few months, as opposed to tuberculosis, whose historical onset required many decades - we also neglect the population change, setting $\mu=0$, which amounts to consider a \textit{fixed} population. This is a strong assumption making such a model purely transient. We are thus led to the classical Kermack and McKendrick's SIR epidemic model \cite{KMK}
\begin{align}
S^{\prime}&= -\beta S I, \label{Stra0}\\
I^{\prime}&= \beta S I - \gamma I, \label{Itra0}
\end{align}
where $\gamma = 1/D$.

As expected from the presence of a communicable disease, the susceptible fraction $S$ is decreasing, as reflected in Eq.~\ref{Stra0}. As for the dynamics of the infective fraction, or \textit{prevalence}, $I(t)$, it is useful to rewrite Eq.~\ref{Itra0} as
\begin{equation}\label{Irewr}
I^{\prime} = \gamma I \left( \mathcal{R}_0 S  - 1 \right),
\end{equation}
where $\mathcal{R}_0 = \beta/\gamma$ indeed represents the BRN specific to the above model, given that $\beta = \beta_I C$, and $\gamma=1/D$. $\mathcal{R}_0$ fully determines the behavior of the system. Indeed, since $S(t)<S(0)<1$, the differential inequality
$ I^{\prime} \le $ $\gamma I$ $ \left( \mathcal{R}_0 S(0)  - 1 \right) $ implies that if
\begin{equation}\label{R0le1}
\mathcal{R}_0  \le 1/S(0),
\end{equation}
then \textit{no outbreak occurs}. Note that the initial infective cases will cause a few other cases but the infection will not represent a threat since there is no increase in prevalence with respect to its initial level $I_0$. On the other hand, an outbreak is only possible if $I^{\prime}(0)>0$, i.e. if
\begin{equation}
\mathcal{R}_0>1/S(0) \label{R0gt1overS0}.
\end{equation}
Thus $\mathcal{R}_0$ is the key parameter which determines the \textit{threshold character} of the system, as first noted by Kermack and McKendrick \cite{KMK}. The result in Eq.~\ref{R0gt1overS0}, can be put in a number of informative ways. First, if the population is largely susceptible, i.e. it holds $S(0)\approx 1 $, the threshold value for model of Eqs.~\ref{Stra0}-\ref{Itra0} is $1$, as claimed in the previous subsection. In the presence of a substantial immune component (i.e. $S(0) < 1 $), as is typically the case for seasonal influenza, then the condition $\mathcal{R}_0>1/S(0)$ can be written as $\mathcal{R}_0 S(0)> 1$, i.e. $\mathcal{R}_E(0) > 1$, meaning that the occurrence of the outbreak requires an initial ERN greater than $1$ (see previous subsection). A further noteworthy reading of condition in Eq.~\ref{R0gt1overS0} is Kermack and McKendrick's original one \cite{KMK}: $S(0) > 1/\mathcal{R}_0$, which, in line with John Snow's seminal intuition, states that an outbreak requires a large initial susceptible fraction.

The temporal dynamics of the outbreak under Eq.~\ref{R0gt1overS0} shows, by linearization of Eq.~\ref{Irewr}, an initial \textit{exponential} growth in prevalence according to $I(t) \approx $ $I(0) Exp( \gamma (\mathcal{R}_0 S(0)  - 1)t ) $, which will be faster the bigger the difference between the initial value of the ERN and its threshold. This initial exponential growth is followed by a non-linear deceleration due to the decline in the susceptible pool $S$, and consequently in the ERN $\mathcal{R}_(t) = \mathcal{R}_0 S(t)$. Moreover, at the time $t_x$ when $ S(t_x) = 1/\mathcal{R}_0 $
it is $ I^{\prime}(t_x)=0 $, while for all $t>t_x$ it is $ I^{\prime}(t)<0 $, given that $S(t) < S(t_x)$ due to the decrease of $S(t)$. In words, the infective  prevalence attains its maximum when the susceptible fraction $S$ becomes equal to the inverse of the BRN, after which the epidemics enter its declining phase until it eventually dies out.

The maximum of the prevalence can be determined by the curve $I(S)$, which is obtained from Eqs.~\ref{Stra0}-\ref{Itra0} by taking
\begin{equation}\label{dIdS}
\frac{dI}{dS}= \frac{1}{\mathcal{R}_0 S}-1,
\end{equation}
whose solution in $ 0< S \le S(0) $ (remind that $S(t)$ is decreasing) is
\begin{equation}\label{IdiS}
I(S) = I(0)+S(0)-S + \frac{1}{\mathcal{R}_0}Log\left(\frac{S}{S(0)}\right).
\end{equation}
This implies that the extremum of $I$ is
\begin{equation}\label{Imax}
I_{max}= I(0)+S(0) -\frac{log(1+ \mathcal{R}_0 S(0))}{\mathcal{R}_0}.
\end{equation}
Therefore, given a wholly susceptible initial population ($S(0)\approx 1$), the maximum prevalence is increasing in $\mathcal{R}_0$. For example, under  the Spanish flu hypothesis $I_{max} \approx 45\%$, while for $\mathcal{R}_0 \approx 15$, as commonly observed for measles, then  $I_{max} \approx 80\%$.

Moreover, the cumulative fraction hit by the infection at time $t$ is given by
\begin{equation}
J(t) = S(0)-S(t) = I(t)-I(0)+R(t)-R(0).
\end{equation}
Therefore the fraction eventually infected at the end of the outbreak, the so-called final \textit{attack rate} ($AR$), is given by
\begin{equation}
AR = J(\infty) \approx 1- S(\infty).
\end{equation}

Setting in Eq.~\ref{IdiS} $I=0$, $S(0)+I(0)=1$, $S_{\infty} \approx 1-AR$ yields the following trascendental equation for the $AR$
\begin{equation}\label{R_infinity_equation}
AR = 1 - exp(-\mathcal{R}_0 AR ).
\end{equation}
The AR is therefore an increasing and concave function of $\mathcal{R}_0$. For example, under the Spanish flu hypothesis ($\mathcal{R}_0 \approx 2$) the predicted AR is close to $80\%$. More important, Eq.~\ref{R_infinity_equation} shows that for infections imparting immunity, the outbreak will never hit the entire population (though the AR may reach levels only slightly lower than $100\%$ for measles-like values of $\mathcal{R}_0$). Also this \textit{incomplete contagion} principle was first noted by Kermack and McKendrick \cite{KMK}. Starting from Eq.~\ref{IdiS}, the case where an initial pool of immune subjects is present ($R_0>0$) - as in the case of seasonal influenza outbreaks - is straightforward.

\begin{figure}
\begin{center}
\includegraphics[width=0.9\textwidth]{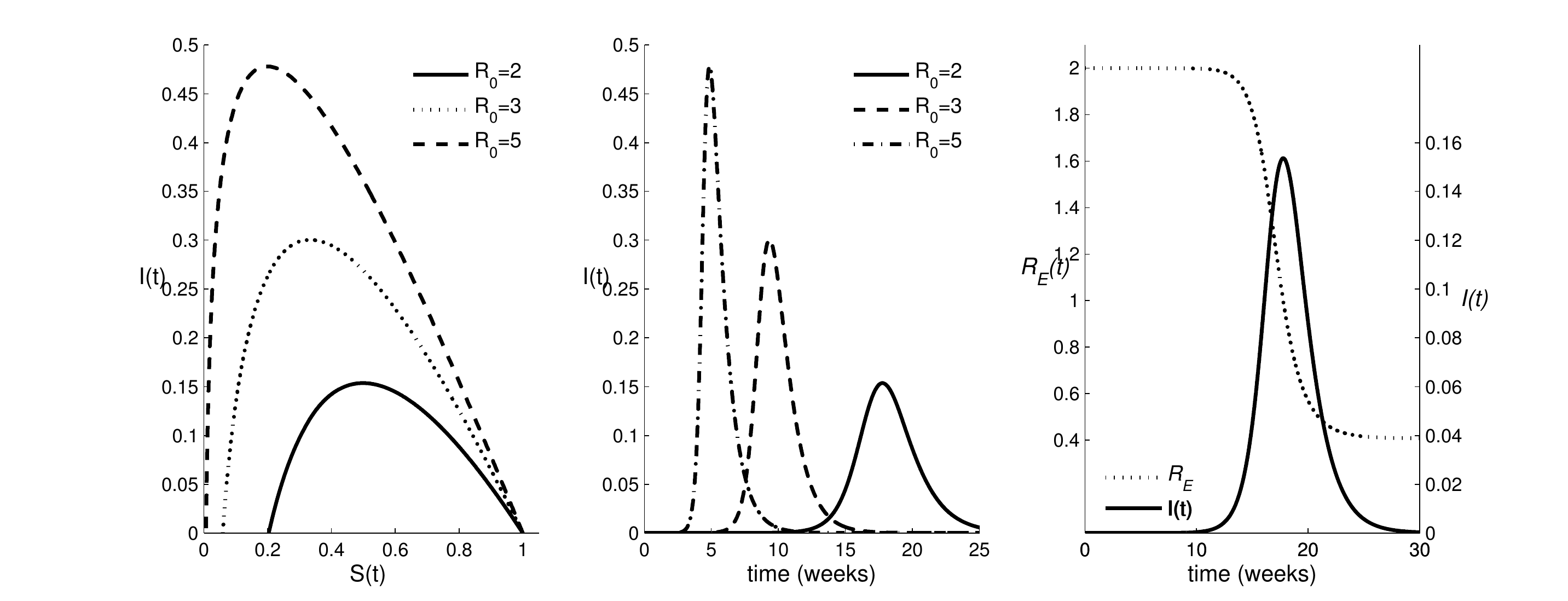}
\end{center}
\caption{Dynamics of the epidemic SIR model for different values of the BRN: $\mathcal{R}_0=2,3,5$. Left panel: dynamics in the $S,I$ phase plane, based on Eq.~\ref{IdiS}. Central panel: time course of the infective prevalence $I(t)$. Right panel: time course of the infective fraction $I(t)$ and the ERN for $\mathcal{R}_0 =2$. The duration of the infective phase $D=1/ \gamma$ is set to 3 days}%
\label{Figure_epidemic_SIR}%
\end{figure}

The behavior of Eqs.~\ref{Stra0}-\ref{Itra0} above the invasion threshold ($\mathcal{R}_0 > 1$) (Fig. \ref{Figure_epidemic_SIR}) clarifies the nonlinear tuning of $\mathcal{R}_0$ on both the intensity (left panel, based on Eq.~\ref{IdiS}) and the time-course of the outbreak (central panel). In particular the decline of the ERN following the decline in the susceptible pool during the outpbreak modulates the changes in the form of the infective prevalence (right panel of Fig. \ref{Figure_epidemic_SIR}).

Public health actions aimed at reducing the AR, or at slowing down the outbreak course,  e.g. by vaccination and/or by reducing transmission, can be easily accomodated. For example, the effects of an awareness campaigns aimed at favouring behavioral changes affecting the contact rate, could be modeled by means of a time-varying decreasing transmission rate $\beta(t)$.

\subsubsection{The SEIR model for outbreaks}
The more realistic case where a time-lag between exposure and onset of infectivity exists due to the presence of an exposed, non-infective, phase, can be tackled by the epidemic SEIR model
\begin{align}
S^{\prime}&= -\beta S I, \label{StraSEIR}\\
E^{\prime}&= \beta S I -\alpha E, \label{EtraSEIR}\\
I^{\prime}&= \alpha E - \gamma I, \label{ItraSEIR}
\end{align}
As for Eqs.~\ref{Stra0}-\ref{Itra0}, it holds $ S(t) < S(0)$. The joint dynamics of $(E,I)$ is, instead, more complex. However, observe that the subsystem $(E,I)$ is cooperative \cite{donofrioMBS02} since
\begin{align}
\frac{\partial E^{\prime}}{\partial I}&= \beta S >0,\\
\frac{\partial I^{\prime}}{\partial E}&= \alpha >0.
\end{align}
Thus the dynamics of $(E(t),I(t))$ is such that for all $t$ it holds $ (E(t),I(t))<(U(t),W(t)),$ where

\begin{align}
U^{\prime}&= \beta S(0) W -\alpha U, \label{UtraSEIR}\\
W^{\prime}&= \alpha U - \gamma W, \label{WtraSEIR}
\end{align}
(with initial conditions $U(0)= E(0)$ and $W(0)= I(0)$), whose characteristic polynomial is
$ \lambda^2 +$ $ (\alpha+\gamma)\lambda $ $+ \alpha (\gamma -\beta S(0)) $. Therefore, if $ \mathcal{R}_0 <1/S(0)$ the infection dies out also for SEIR model. Note also that the system of Eqs.~\ref{UtraSEIR}-\ref{WtraSEIR} is the one (approximately) governing the initial phase of the epidemics, when $S(t)\approx S(0)$. Thus, if $ \mathcal{R}_0 > 1/S(0)$ then there is an initial phase of exponential growth. Finally, for those $t>0$ such that
$S(t)$ remains bigger than $1/\mathcal{R}_0$, it is $ (E,I)$ $ > (U_1,W_1)$, where
\begin{align}
U_1^{\prime}&= \beta \left(\epsilon + \frac{1}{\mathcal{R}_0}\right) W_1 -\alpha U_1, \label{UtraSEIRbis1}\\
W_1^{\prime}&= \alpha U_1 - \gamma W_1, \label{WtraSEIRbis1}
\end{align}
(with initial conditions $U_1(0)= E(0)$ and $W_1(0)= I(0)$), which is unstable (characteristic polynomial: $ \lambda^2 +$ $ (\alpha+\gamma)\lambda$ $ - \alpha \beta \epsilon$). In other words, as long as $S(t)$ remains bigger than $1/\mathcal{R}_0$, it follows that $(E,I)$ is bigger than an increasing vector.

Summarizing, in vectorial sense, the BRN of the SIR transient model governs also the dynamics of the SEIR transient model. However, note that a vectorial increasing system, i.e. characterized by eigenvalues that have all positive real part,
may have some of its components that can be transitorially decreasing. For example, if $E(0)= 0$ then initially $I(t)$ is decreasing:
$ I^{\prime}(0) = -\gamma I(0)<0$.

Finally, let us consider the case of a latent phase of average duration $1/\alpha$ much longer than the infective phase: $ \alpha << \gamma $ and such that $E(0) =0$, and that there is initially a single infectious subject $I(0) =1/N$, so that $S(0) = 1-1/N \approx 1$. In such a case, in the time interval $[0,D]$ there is no new infectious subjects $ I(t) =$ $ constant =$ $ 1/N$. It follows from
\begin{equation}
J = N\int_{0}^{D}\beta S(t) I(t)dt \approx \beta D = \mathcal{R}_0,
\end{equation}
where the number $J$ of secondary infections caused by one infectious subject during his entire infectious period in a wholly susceptible population is equal to the BRN.

\subsection{Emergence and features of endemicity: the SIR and SEIR models with vital dynamics.} \label{Endemic_SIR}
\begin{figure}
\begin{center}
\includegraphics[width=0.7\textwidth]{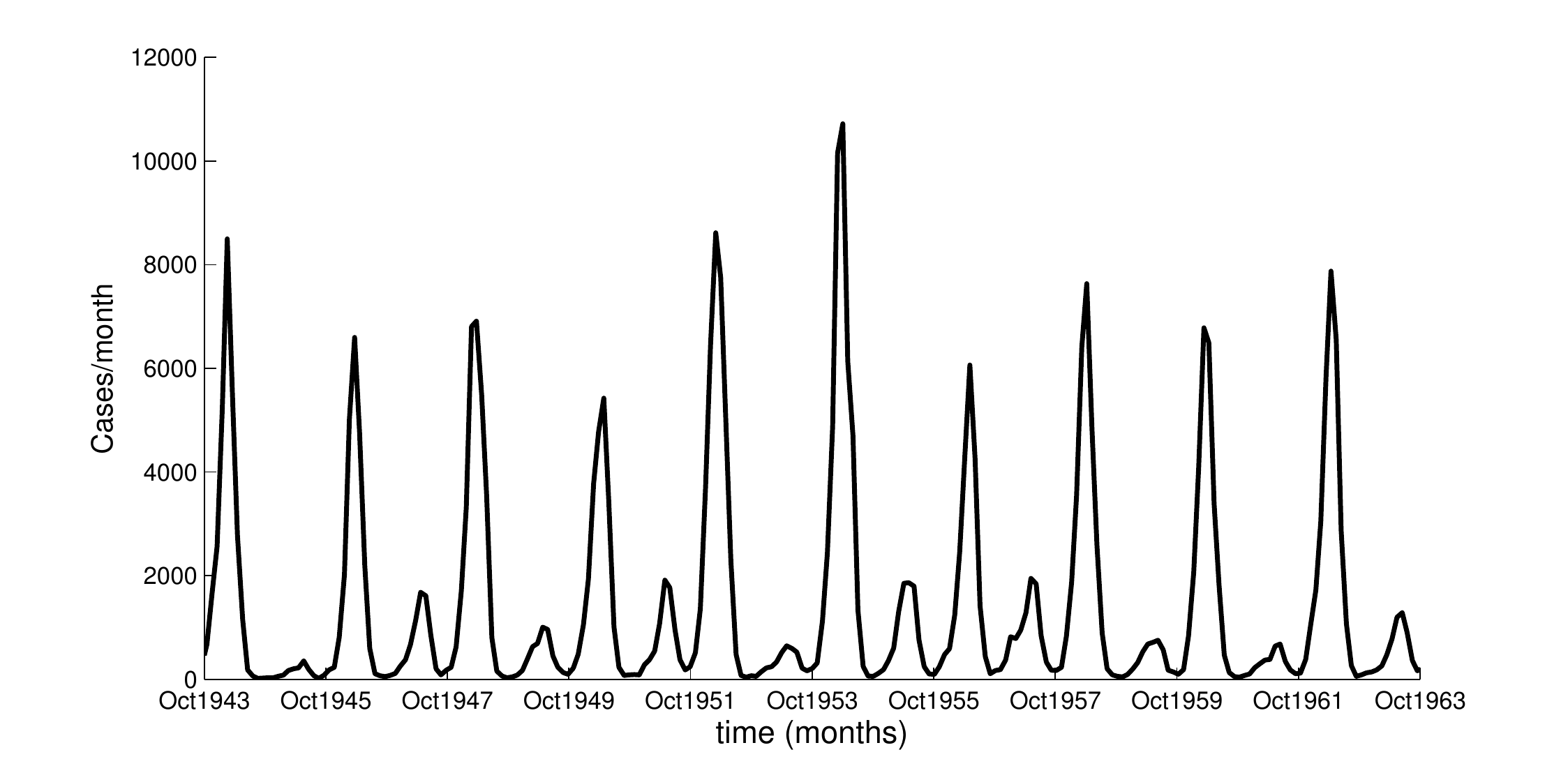}
\end{center}
\caption{Biennal pre-vaccination oscillations of measles in New York, 1943-1963}%
\label{Measles_New_York}%
\end{figure}

The SIR and SEIR epidemic models unavoidably predict the extinction of the infection due to the depletion of the susceptible population. This means that further factors are necessary to sustain \textit {endemicity}, that is the ability of a communicable disease to persist in a population in the long-term, with its peculiar patterns, such as the biennial pre-vaccination oscillations of measles in large UK and US cities (Fig. \ref{Measles_New_York}).

The first candidate among such factors is the birth process which allows to rebuild the susceptible pool after an outbreak. The simplest reference model is therefore the SIR model with \textit{stationary vital dynamics} \cite{AMay_1991,Hethcote_2000}, which is obtained from Eqs.~\ref{eqS}-\ref{eqE} by disregarding the exposed phase:
\begin{align}
S^{\prime}&= \mu( 1- S) -\beta S I, \label{SIR_vitaldyn_eqS}\\
I^{\prime}&= \beta I S -(\gamma +\mu)I.  \label{SIR_vitaldyn_eqI}
\end{align}
Note that system of Eqs.~\ref{SIR_vitaldyn_eqS}-\ref{SIR_vitaldyn_eqI} has a disease-free equilibrium $DFE = (1,0)$. The BRN specific to this model is given by
\begin{equation}
\mathcal{R}_0 = \frac{\beta}{\gamma +\mu}.
\end{equation}
\noindent
Indeed, from $ I^{\prime}=$ $ (\gamma +\mu)I( \mathcal{R}_0 S -1)<$ $ (\gamma +\mu)I(\mathcal{R}_0  (1-I) -1)  $ it follows that if $\mathcal{R}_0 \le 1$ then $I(t) \rightarrow 0$ and $S(t)\rightarrow 1$, i.e. DFE is globally attractive. Moreover, linearizing Eq.~\ref{SIR_vitaldyn_eqI} around the DFE gives, letting $i$ to denote the linearised prevalence,
$i^{\prime} =$ $ (\gamma +\mu)( \mathcal{R}_0  -1)i $, implying that if $$ \mathcal{R}_0>1 $$ then the DFE is unstable and an introduction of infective subjects $i(0)<<1$ in a wholly susceptible population induces an outbreak, initially growing as
$ i(t) \approx $ $i(0) Exp( (\gamma +\mu)( \mathcal{R}_0  -1)t)$. Moreover, for $\mathcal{R}_0>1$ an equilibrium with positive infective prevalence, termed an \textit{endemic equilibrium}, appears
\begin{equation}\label{Se}
EE= (S_e, I_e) = \left( \frac{1}{\mathcal{R}_0}, \frac{\mu}{\mu+\gamma}\left(1-\frac{1}{\mathcal{R}_0}\right) \right).
\end{equation}
Note that in the case where $\gamma >>\mu$, common for childhood infections, it is $I_e < (\mu/\gamma)<<1$. The EE is globally attractive, as it follows by applying the Poincar{\'e}-Bendixon theorem with multiplying factor $1/I$
\begin{equation}
div\left( \frac{1}{I}(S^{\prime}, I^{\prime}) \right)= -\frac{\mu}{I}-\beta.
\end{equation}

To sum up, technically speaking, $\mathcal{R}_0>1$ induces a \textit{transcritical bifurcation} at $\mathcal{R}_0 = 1$, such that below threshold only the DFE exists and is GAS, while above threshold the endemic state appears, which inherits the GAS, and promotes an infective prevalence which is strictly increasing in $\mathcal{R}_0>1$. In words, if $\mathcal{R}_0 \le 1$ then the infection is not able to self-sustain and dies out, whereas if $\mathcal{R}_0>1$ the infection will invade the population and remain endemic. The behavior of the endemic state is studied by linearizing around the EE and computing the associated characteristic polynomial
\begin{equation}
\lambda^2 + \mu \mathcal{R}_0 \lambda + \mu (\mu+\gamma)(\mathcal{R}_0 - 1)=0.
\end{equation}

\noindent The associated discriminant is $\Delta= $ $\mu^2 \mathcal{R}_0^2 -$  $4 \mu (\mu+\gamma)(\mathcal{R}_0-1) \approx$  $- 4 \mu (\mu+\gamma)(\mathcal{R}_0-1)<0 $, implying dumping oscillations of prevalence about its endemic level with natural (pseudo-) period:
\begin{equation}\label{T_natural}
T = \frac{2 \pi}{\sqrt{\mu(\mu+\gamma)(\mathcal{R}_0-1)}}. \end{equation}
\noindent In the case of measles, for which $1/\gamma \approx 1  week$, and estimates of $\mathcal{R}_0$ in the region of $15$ \cite{AMay_1991} were common in western countries in the pre-vaccination epoch, by taking a life expectancy at birth of $1/\mu =75  years$, one finds $T \approx 2.01 years$, therefore reproducing the measles inter-epidemic period observed in the UK and the US cities before the advent of vaccination.

\begin{figure}
\begin{center}
\includegraphics[width=0.6\textwidth]{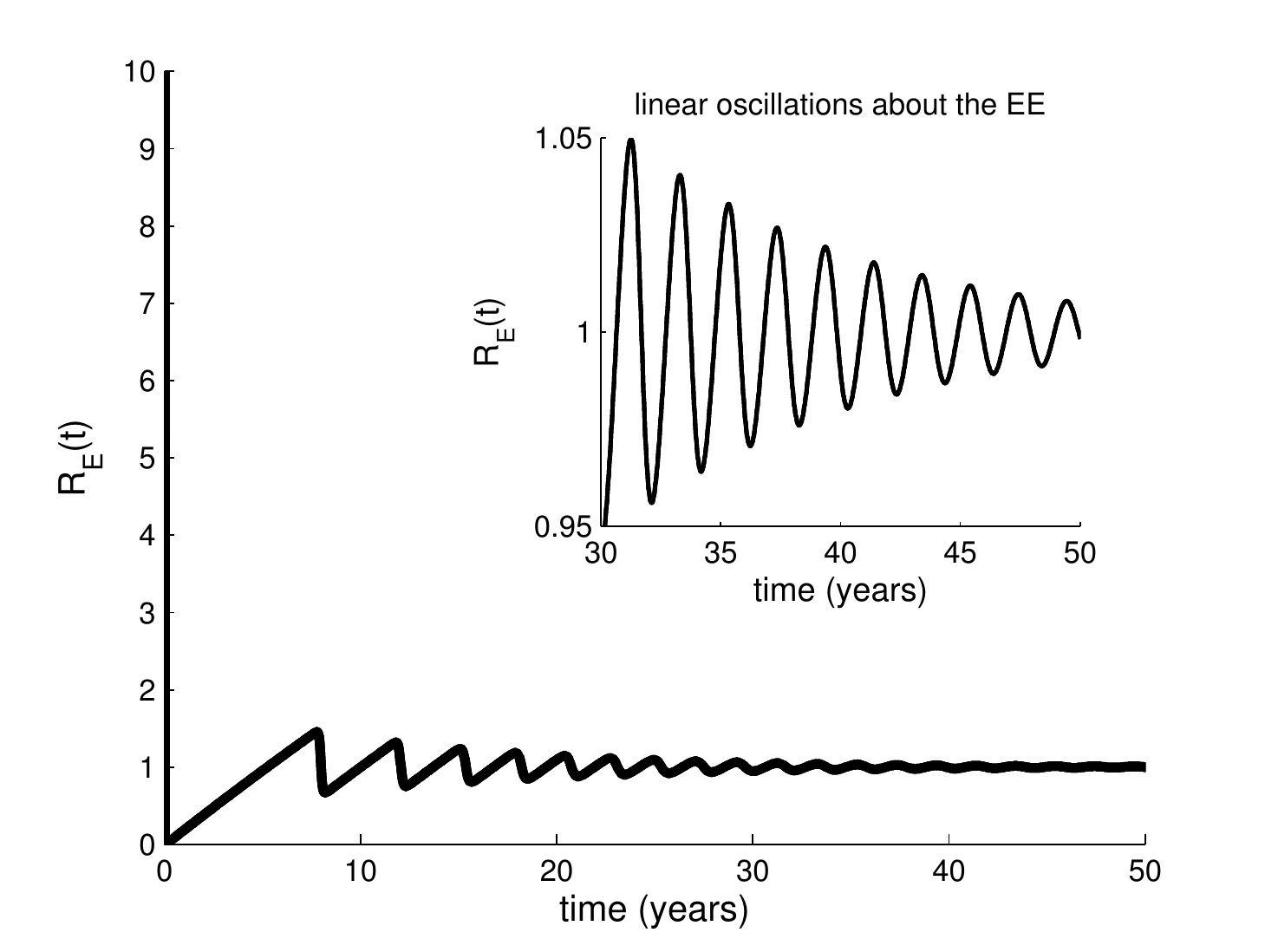}
\end{center}
\caption{The SIR model with vital dynamics: pre-vaccination dynamics of the effective reproduction number ($\mathcal{R}_E(t)$), with convergence to the endemic state. The pattern of linear oscillations about the $EE$ is reported in the subgraph, showing the biennial inter-epidemic period. The demographic and epidemiological parameters are: $\mu = (1/75)year^{-1}$, $\gamma = 52year^{-1}$, $\mathcal{R}_0 =15$.}
\label{SIR_VD_prevacc_dynamics}
\end{figure}

The global dynamics of the system, particularly the interplay between infection and demographics, can be unfolded by a time-scale argument, departing from the introduction of the infection in a wholly susceptible population. Assume for simplicity that $\mathcal{R}_0$ is greater than $1$ and large, and that epidemiological and demographic time scales are well-separated i.e. $\gamma >> \mu$. In this case there is an initial phase where model dynamics is essentially described by model of Eqs.~\ref{Stra0}-\ref{Itra0} for outbreaks, yielding a large initial epidemics, which exhausts most of the susceptible population, bringing the ERN (and prevalence) to very low levels, as clear by the thick line overlapped with the vertical axis in Fig. \ref{SIR_VD_prevacc_dynamics}. This implies that thereafter the $S$-equation takes the approximate form $S^{\prime} \approx \mu$, which has a twofold effect: it promotes a phase of rebuilding of the susceptible pool due to the new births on the one hand, while on the other hand it slows down the decline in prevalence, which avoids the infection to simply die out as in Eqs.~\ref{Stra0}-\ref{Itra0}. The growth in susceptible population will eventually end by a new, but less severe, epidemics, which will in turn be followed by a new phase of replenishment of the susceptible pool, and so on \cite{AMay_1991} until the endemic equilibrium is achieved (Fig. \ref{SIR_VD_prevacc_dynamics}). The global dynamics then consists in continued alternation between epidemic phases, followed by epochs of demographic replenishment, which is how we know the \textit{natural history} of many infectious diseases. The biennal period of quasi-linear oscillations about the EE is emphasised in the sub-graph of Fig. \ref{SIR_VD_prevacc_dynamics}.

\subsubsection{Periodic contact rates}
The simple endemic SIR model with constant transmission rate $\beta$ and stationary population dynamics plays a fundamental role as trigger of the studies on vaccine-preventable infections of childhood, since it well reproduces, by minimal ingredients, their main qualitative and quantitative features such as the endemic oscillation and the observed inter-epidemic period \cite{AMay_1991,Hethcote_2000}. However, the model fails to capture a number of further characteristics, ranging from the sustained biennial pre-vaccination oscillations of measles in US and UK cities (see again Fig. \ref{Measles_New_York}), to the more complicate dynamics observed in other contexts, both prior and after mass vaccination \cite{Olsen,Earn_chaos_2000, Dalziel_2016}. Motivated by the fact that both the weather and certain social processes, namely school-recruitment following the yearly school calendar, have one year periodicity, an important strain of the literature has focused on $1-year$ periodic $\beta(t)$ in endemic SIR models. If the corresponding period $T_\beta$ is around 1 year, then nonlinear resonance may appear, inducing phenomena such as biennal and multi-year periodicity, and chaos \cite{Olsen,gross,aron,SchwartzSmith,Smith, bacaer,rebelo,Dalziel_2016} that partially explain a number of observed epidemiological time-series.

The effect of periodic forcing in the contact rate on the stability of the disease-free equilibrium, is found by defining
\begin{equation}\label{Rovar}
\mathcal{R}_0(t)=\frac{\beta(t)}{\mu+\gamma},
\end{equation}

\noindent
and operating as in the case of constant BRN, i.e. by differential inequality and linearization. Letting $\langle \mathcal{R}_0(t)\rangle$ to denote the average of $\mathcal{R}_0(t)$ over $[0,T_\beta]$, one easily gets that: 1) $ \langle \mathcal{R}_0(t)\rangle \le 1$ implies the global attractiveness of DFE; 2) $\langle \mathcal{R}_0(t)\rangle > 1 $ implies the instability of the DFE. It is important to note that in the SIR model the stability and instability of DFE are independent of the specific shape of $\beta(t)$: they only depend on its average value. However, this independence of the behavior on the entire shape of $\beta(t)$ is a peculiarity of the SIR and of some other epidemic models, but it cannot be generalized.

The inclusion of a latency period in the SIR endemic model leads to the SEIR endemic model
\begin{align}
S^{\prime}&= \mu( 1- S) -\beta S I, \label{eqS1}\\
E^{\prime}&= \beta I S -(\gamma +\alpha)E,  \label{eqE1}\\
I^{\prime}&= \alpha E -(\gamma +\mu)I,  \label{eqI1}
\end{align}
whose BRN is given by
\begin{equation}\label{RoSEIR}
\mathcal{R}_0^{SEIR} = \frac{\alpha}{\alpha+\mu}\frac{\beta}{\mu+\gamma}.
\end{equation}

\noindent Note that if $\mu<<\alpha$ then $\mathcal{R}_0^{SEIR} \approx \mathcal{R}_0^{SIR}$. It holds that i) if  $ \mathcal{R}_0^{SEIR}\le 1 $ then the DFE is the unique equilibrium and it is GAS; ii) if $ \mathcal{R}_0^{SEIR} > 1 $ then there is a unique EE, which is GAS. The latter property has been demonstrated by Li and Muldowney \cite{limu} as an application of their 3D extension of Poincar{\'e}-Bendixon theorem. As for the SIR case, nonlinear resonances are possible for a $1-year$ periodic contact rate under parameter constellations characteristic of childhood infections.

Interestingly, in the SEIR model the deterministic periodic fluctuations of the contact rate have nontrivial effects also on the stability of the DFE. As regards the global stability of the DFE, also here the cooperativity of subsystem $(E,I)$ implies that for all $t>0$ $ (E(t),I(t))<$ $(U(t),W(t)), $ where
\begin{align}
U^{\prime}&= \beta(t) W -\alpha U \label{UtraSEIRperiodic},\\
W^{\prime}&= \alpha U - \gamma W \label{WtraSEIRperiodic},
\end{align}
with initial conditions $U(0)= E(0)$ and $W(0)= I(0)$. It follows that if both the eigenvalues of the Floquet's matrix $F$ associated to Eqs.~\ref{UtraSEIRperiodic}-\ref{WtraSEIRperiodic} are interior to the unit circle in the complex plane, then
\begin{equation}
(U(t),W(t)) \rightarrow (0,0) \Rightarrow (E(t),I(t)) \rightarrow (0,0)
\end{equation}
in turn implying that $ S(t) \rightarrow 1$. As regards the instability of DFE, by linearizing the SEIR model around the DFE one gets a linear time-varying system equivalent to Eqs.~\ref{UtraSEIRperiodic}-\ref{WtraSEIRperiodic}. Thus if the eigenvalues of the matrix $F$ are outside the unit circle of the complex plane, then the DFE is unstable.

It is of interest to note that although $\mathcal{R}_0^{SEIR}(t) \approx \mathcal{R}_0^{SIR}(t)$, for the SEIR endemic model the stability of the DFE is deeply different from the SIR case. Indeed, in the SIR model the behavior of the DFE is uniquely determined by the average of the transmission rate $\beta(t)$, while in the SEIR case the stability of the DFE depends on the "whole shape" of $\beta(t)$.

The computation of Floquet's matrices is usually nontrivial. However, matrix $F$ can be easily computed in the noteworthy case where $\beta(t)$ is switching between two constant values, which mimicks well the realistic school-holiday scenario where students have a large number of contacts during the school period, which remarkably reduces during holidays \cite{Earn_chaos_2000,keeling}. Setting
\begin{equation}
\beta(t) =
  \begin{cases}
    \beta_1,       & \quad \text{if } t \in [0,T],\\
    \beta_2,  & \quad \text{if } t \in (T,1],
  \end{cases}
\end{equation}
the linear time-varying system of period one year is equivalent to the following two systems
\begin{equation}
X_1^{\prime}=M_1 X_1 ; 0\le t \le T,
\end{equation}
\begin{equation}
X_2^{\prime}=M_2 X_2 ; T < t \le 1,
\end{equation}
where for $i=1,2$
\begin{equation}
M_{i} = \begin{bmatrix}
  -\alpha & \beta_i \\
  \alpha & -\gamma
 \end{bmatrix}.
\end{equation}
As a consequence $F$ is the product of two matrix exponentials:
\begin{equation}
F= Exp(M_2(1-T))Exp(M_1 T).
\end{equation}
The generalization to more complex patterns of alternance between school and holidays can also easily handled.

In the general case, if one eliminates the first state variable in the above equivalent linear systems, one obtains the following \textit{family} of Newton's equation\cite{donofrioMBS02}
\begin{equation}\label{Newton}
w^{\prime\prime} + (\gamma+\alpha)w^{\prime} + (\gamma -\alpha \beta(t))w=0.
\end{equation}
Further, by setting $ y(t) = w(t) Exp((\gamma+\alpha)t/2) $ \cite{minorsky}, one yields the Hill family of equations
\begin{equation}\label{Newton2}
y^{\prime\prime} + \left(-\frac{(\gamma+\alpha)^2}{4}+\gamma -\alpha \beta(t)\right)y=0.
\end{equation}
The families of Eqs.~\ref{Newton}-\ref{Newton2} are ubiquitous and widely studied in various fields of mathematical physics \cite{Arscott64,mcl,minorsky,farkas,cesari,arnold,landau}. For example, in mechanics they model the periodic perturbations in the elastic force of damped or undamped oscillators respectively, which are at the base of several theories, such as that of parametric resonance \cite{cesari,arnold,landau}; while in applied electromagnetism Eq.~\ref{Newton2} models the frequency modulation\cite{mcl}. Analytic solutions for $t \in [0,1]$ are known for a number of particular forms of $\beta(t)$. For example, if $\beta(t)$ is linearly increasing, the solutions of Eq.~\ref{Newton2} in $ t \in [n T, n T+1)$ are Airy's functions, whereas if the contact rate is sinusoidally varying around an average value, the solutions of Eq.~\ref{Newton2} are Mathieu's functions.

\subsubsection{Nonlinear contact rates as a phenomenological route to behavior change}
A limitation of mass-action-based SIR and SEIR models with and without vital dynamics, and of other epidemic models, is that they assume that the behavioral component is either constant (resulting in a constant $\beta$) or influenced by exogenous phenomena such as the school/work calendar (resulting in a periodic time-varying $\beta(t)$ with one-year period). Capasso and Serio were first in pointing out this shortcoming \cite{Capasso_Serio}, and in stressing that agents might respond to the threat posed by the outbreak by changing their social behavior (i.e. reducing their average number of contacts $C$), and/or by adopting protections against transmission. They modelled these effects by a \textit{prevalence-dependent} transmission rate of the form $\beta(I)$, $\beta'(I)<0$, showing that behavior change may deeply modify the quantitative dynamics of an outbreak \cite{Capasso_Serio}. In a more general study on the SIR endemic model \cite{noibeta}, it was stressed that prevalence-dependent behavior is influenced by a range of information variables, some of which can be made available to agents with time-lags. The inclusion of delayed prevalence-dependent behavior in the SIR endemic model can destabilize the endemic equilibrium by triggering sustained oscillations, and, in turn, these endogenous oscillations may interact with the one-year periodicity of the contact rate, resulting in chaotic fluctuations \cite{noibeta}.
Finally, another seminal paper is Hadeler and Chavez \cite{hadeler1995core}, which models the spread of a vaccine/education preventable sexually transmitted disease in a population split in two groups: an inactive group with $\beta =0$ and a \textit{core} active group where $\beta >0$. The model included a unidirectional prevalence-depending flux of subjects from the inactive to the \textit{core} group, with tranfer rate $r(I)$ decreasing with $I$.

\subsection{Mass vaccination and herd immunity for vaccine preventable infections: the SIR model with vaccination at birth}
\label{SIR_vaccination}
In modern communities new vaccines, targeting different infections and different populations, appear at a continuously increasing rate, making the mosaic of characteristics of the different available vaccines quite articulated. Nonetheless, much of the main dynamic effects of mass vaccination in controlling communicable diseases can be understood by the following simple variation of the endemic SIR model of Eqs.~\ref{SIR_vitaldyn_eqS}-\ref{SIR_vitaldyn_eqI}
\begin{align}
S^{\prime}&= \mu( 1- x ) - \mu S -\beta S I, \label{SIR_vitaldyn_vacc_eqS}\\
I^{\prime}&= \beta I S -(\gamma +\mu)I,  \label{SIR_vitaldyn_vacc_eqI}\\
V^{\prime}&= \mu x - \mu V.  \label{SIR_vitaldyn_vacc_eqV}
\end{align}
\noindent
Model of Eqs.~\ref{SIR_vitaldyn_vacc_eqS}-\ref{SIR_vitaldyn_vacc_eqI}-\ref{SIR_vitaldyn_vacc_eqV} (of course $R= 1- S-I-V$) considers vaccination of a time-invariant, randomly chosen, proportion  $x$ of newborn susceptible individuals per unit of time by a \textit{perfect vaccine}, i.e. one with $100\%$ take and lifelong duration (see section \ref{BRN}). Under a perfect vaccine the fraction successfully vaccinated coincides with the  \textit{vaccine coverage}, i.e. the fraction vaccinated. Vaccinated individuals are fully protected against infection (and ensuing disease) and therefore behave, in practice, as removed individuals. This does not need to be the case if, for example, the vaccine has waning immunity, or leaves some residual susceptibility - as is the case e.g. for the varicella vaccine - in which case further compartments are required.

After discarding the $V$ equation, the analysis of Eqs.~\ref{SIR_vitaldyn_vacc_eqS}-\ref{SIR_vitaldyn_vacc_eqI} goes along exactly the same lines as Eqs.~\ref{SIR_vitaldyn_eqS}-\ref{SIR_vitaldyn_eqI}. Since we are interested in infections that are endemic prior to vaccination, we assume $ \mathcal{R}_0 >1 $, where $ \mathcal{R}_0$ still represents the BRN. The key threshold quantity is now represented by the VRN
\begin{equation}\label{VRN}
\mathcal{R}_V = \frac{\beta}{\gamma +\mu} (1-x).
\end{equation}
It can then be easily shown that Eqs.~\ref{SIR_vitaldyn_vacc_eqS}-\ref{SIR_vitaldyn_vacc_eqI} have the $ DFE = (1-x,0) $, which is globally attractive if $\mathcal{R}_V \le 1 $, and unstable for $ \mathcal{R}_V>1 $. For $\mathcal{R}_V>1$ the model also has the globally attractive (by the same Poincar{\'e}-Bendixson argument) endemic state
\begin{equation}\label{EE_SIR_vacc}
EE_V= (S_e, I_e) = \left( \frac{1}{\mathcal{R}_0}, \frac{\mu}{\mu+\gamma}\left(1-x-\frac{1}{\mathcal{R}_0}\right) \right).
\end{equation}
This endemic state is also characterised by damped oscillations, with pseudo-period
\begin{equation}
T_V = \frac{2 \pi}{\sqrt{\mu(\mu+\gamma)(\mathcal{R}_0 (1-x) -1)}} . \end{equation}\label{T_SIR_vacc}

Though the mathematical analysis is the same as for the basic endemic SIR model, there is a number of substantive implications of vaccination. Indeed, the threshold for the VRN implies the already met coverage threshold: $ x_{c}= 1-1/\mathcal{R}_0$ (i.e. formula \ref{thepc}), so that if $x\ge x_{c}$ then the infection is eliminated (technically, the introduction of mass vaccination above threshold causes the EE to disappear, leaving only the DFE, which becomes GAS).
In the opposite scenario, i.e. if the coverage is insufficient to achieve elimination ($px\le x_{c}$), the infection will continue to persist endemically, in the new long-term regime described by Eqs.~\ref{EE_SIR_vacc}. This \textit{post-vaccination} regime  is characterized by the same susceptible proportion  $(1/ \mathcal{R}_0)$ as in the pre-vaccination history, but by a different distribution of immune individuals between natural and vaccine-induced, and especially by a smaller prevalence and an increased inter-epidemic period with respect to the case where no vaccination campaign is enacted. The increase in $T$ straightforwardly follows from the reduced replenishment of the susceptible class due to vaccination. For example, considering again a measles-type parametrization ($\mathcal{R}_0 =15$, $1/\gamma =1 $ week, $\mu=(1/75)/year$), yielding a pre-vaccination inter-epidemic period of about $2$ years, a mass vaccination program with $x=0.9$ would increase the inter-epidemic period to a post-vaccination figure $T_V$ in excess of $10$ years.

\begin{figure}
\begin{center}
\includegraphics[width=0.6\textwidth]{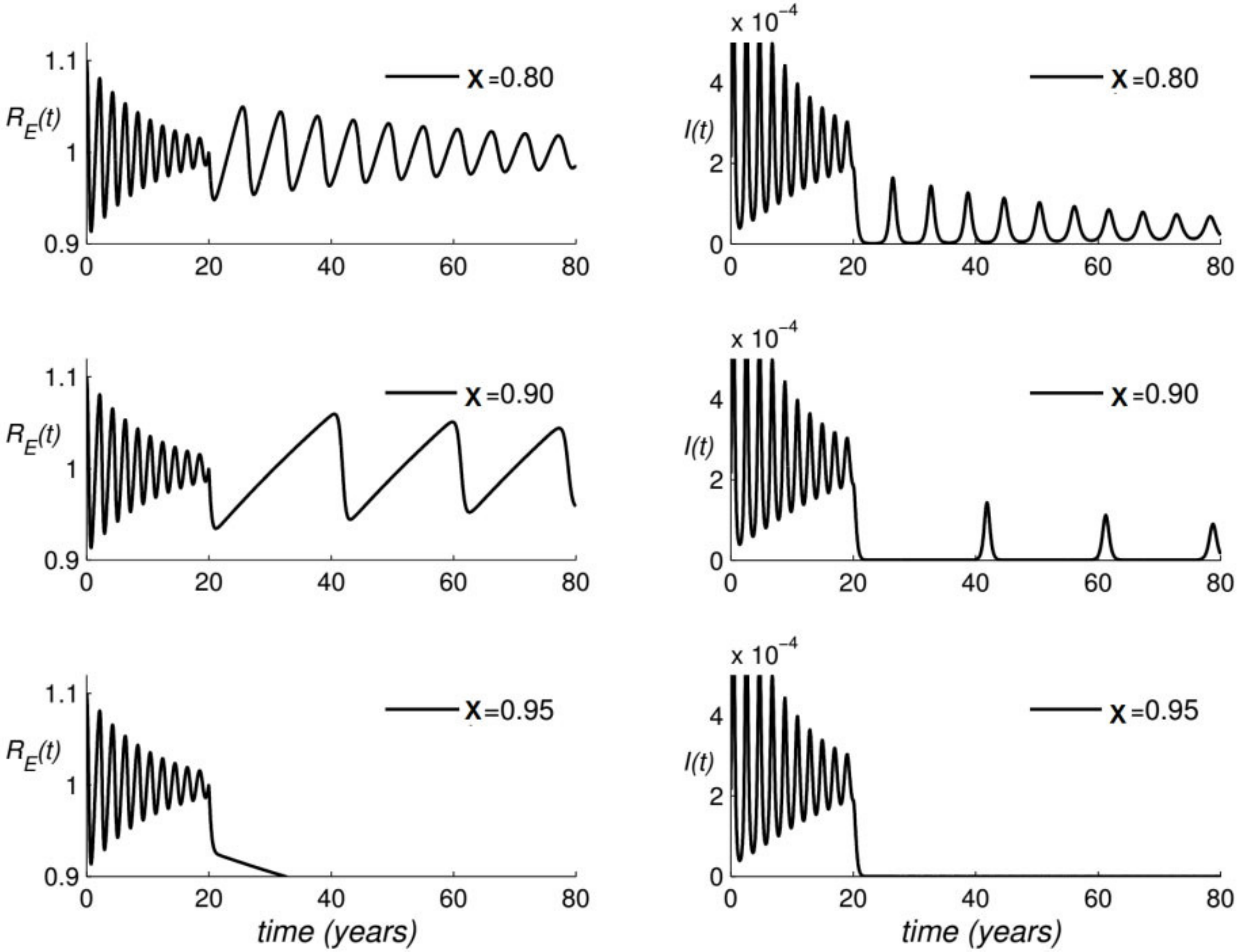}
\end{center}
\caption{The SIR model with vital dynamics: pre- and post- vaccination dynamics of the effective reproduction number $\mathcal{R}_E(t)$ (left panels) and the infective prevalence $I(t)$ (righ panels) under different programs of vaccination at birth initiated at time $t=20 years$ with coverage of $x=0.80$ (upper panels), $x=0.90$ (central panels) and $x=0.95$ (right panels) respectively. The demographic and epidemiological parameters are: $\mu = (1/75)year^{-1}$, $\gamma = 52 year^{-1}$, $\mathcal{R}_0 =15$.}%
\label{SIR_VD_pre_postvacc_dynamics}%
\end{figure}

The dynamic implications of the introduction of a mass vaccination for a measles-like infection are illustrated in Fig. \ref{SIR_VD_pre_postvacc_dynamics} for different vaccination programs (respective coverages: $x=0.80$, $x=0.90$ and $x=0.95$) initiated at time $t=20 years$ under the hypothesis that infection spread for $t<20$ was evolving in its pre-vaccination regime, characterized by an almost biennial inter-epidemic period. As expected for $x=0.95$ the infection is eliminated, since $x>x_{c}$. In the two remaining cases the infection remains endemic with a much lower prevalence but with an amazing increase in the inter-epidemic period. For example, for $x=0.90$ the inter-epidemic period initially increases to a level in excess of $20 years$ before gradually contracting and converging in the long-term to the above predicted figure of $T_V \approx 10 years$. The inflation in the inter-epidemic period has far reaching consequences since it implies that unvaccinated individuals who did not acquire infection early in life might do so at somewhat later ages, where they might be exposed at a much larger risk of serious disease from the infection \cite{AMay_1991,Hethcote_2000}. The increase in the average age at infection in the post-vaccination regime is the consequence in the decline in the individual risk of infection, as clear from the decline in the endemic FOI $\beta I_e$, which shifts from its pre-vaccination level of $\mu (\mathcal{R}_0 - 1)$ to $\mu (\mathcal{R}_0 (1-x) - 1)$. These phenomena are of importance, besides measles, for essentially all vaccine preventable infection, particularly for rubella \cite{AMay_1991,Hethcote_2000}.

\subsection{Return to susceptibility: SIS and SIRS models}
\label{SIS_SIRS_models}
If infection-acquired immunity is not lifelong, recovered subjects will eventually become again susceptible. A first possible scenario is the one where infective subjects directly return to susceptibility, yielding the so-called SIS model (Fig.  \ref{Fig_SEIRS_compartments})

\begin{align}
S^{\prime} &= \mu (1-S) -\beta(t) I S + \gamma I,  \label{SIS-S}\\
I^{\prime} &= \beta(t) I S  - \mu I- \gamma I.  \label{SIS-I}
\end{align}
Since $S+I=1$, one gets the scalar model
\begin{equation}\label{SIS}
I^{\prime} = (\mu+\gamma) I ( \mathcal{R}_0(t) (1- I) - 1),
\end{equation}
where
\begin{equation}
\mathcal{R}_0(t) = \frac{\beta(t)}{\mu+\gamma}
\end{equation}
and we still take $\beta(t)$ as a 1-year periodic function. Unsurprisingly, if $ \langle \mathcal{R}_0(t) \rangle  \le 1 $ then the DFE $I = 0$ is globally attractive, whereas if $ \langle \mathcal{R}_0(t) \rangle > 1 $ the DFE is unstable, and $I(t)$ tends to a one-year periodic solution, i.e. the model predicts that the disease remains endemic via a series of annual epidemic outbreaks. For a constant transmission rate $\beta$ Eq.~\ref{SIS} is logistic, therefore implying a constant, globally attractive, endemic state:
\begin{equation}
I_{EE} = 1 - \frac{1}{\mathcal{R}_0},
\end{equation}
\noindent
where the susceptible fraction is still given by the inverse of $\mathcal{R}_0 $. Note that the establishment of a globally attractive EE also holds in absence of vital dynamics (i.e. for $\mu =0$, suggesting that the loss of immunity \textit{per se} is a further source of endemicity through rebuilding of the susceptible pool, as first noted by \cite{KMK}.

A second more general scenario occurs when return to susceptibility is buffered by a long sojourn time in the recovered class, yielding the SIRS endemic model (Fig.  \ref{Fig_SEIRS_compartments})

\begin{align}
S^{\prime}&= \mu( 1- S) -\beta S I +\delta R, \label{eqSsirs}\\
I^{\prime}&= \beta I S -\gamma I - \mu I, \label{eqIsirs}\\
R^{\prime}&= \gamma I - \mu R -\delta R, \label{eqRsirs}
\end{align}
where $1/\delta$ is the average length of the immunity conferred by the disease. From $S=1-I-R$ one gets
\begin{align}
I^{\prime}&= (\mu+\gamma) I ( \mathcal{R}_0 (t)(1-I-R) -1), \label{eqIsirs1}\\
R^{\prime}&= \gamma I - (\mu + \delta) R, \label{eqRsirs1}
\end{align}
where
\begin{equation}\label{tvBRN}  \mathcal{R}_0(t) = \frac{\beta(t)}{\mu+\gamma}. \end{equation}
Again, the infection ultimate fate, namely extinction vs endemicity, depends on whether
$ \langle \mathcal{R}_0(t)\rangle$ is smaller or bigger than one. Under constant $\beta$ , the threshold condition $ \mathcal{R}_0 >1 $ allows the existence of an endemic state:
\begin{equation}\label{valSee} S_{EE} = \frac{1}{\mathcal{R}_0}, I_{EE} = \left(1- \frac{1}{\mathcal{R}_0}\right)\frac{\mu+\delta}{\mu+\delta+\gamma},\end{equation}
which can be shown to be globally attractive by applying the same Dulac-Bendixon test used for the SIR endemic model.

Mass vaccination by a perfect vaccine can be introduced in SIS and SIRS models along the same route followed for the SIR model, with analogous results.
However, for infections with only temporary, or without, immunity, the hypothesis of a perfect vaccine is mostly unattainable. If the vaccine is \textit{imperfect}, several scenarios are possible i.e. vaccinated individuals might either (i) not produce any immune response, therefore remaining fully susceptible, or (ii) produce a weaker immune response. This weaker immune response might occur in the form of a temporary immunity which however typically wanes faster than natural one, eventually returning the individual to susceptibility, or in the form of reduced susceptibility, putting the individual at risk of \textit{breakthrough} infection, at some reduced force of infection $\lambda = \sigma \beta I$ with $\sigma<1$. For example, all these possibilities are known to occur for the current varicella vaccine \cite{Varicella_vaccine_imperfect}.
This opens a more complicate scenario which makes infection elimination more difficult. Indeed, as it can be demonstrated even for simple SIS models with vaccination at birth, there is a range $(x_{c}^{0},x_{c}^{*})$ of the vaccinated proportion $p$ where a bistability phenomenon arises, i.e. where the disease-free equilibrium is locally stable but it coexists with two endemic equilibria: one unstable and one locally stable. Therefore, infection elimination requires that $x>x_{c}^{*}$, whereas if $x<x_{c}^{0}$ then there is only one endemic steady state, coexisting with the DFE which is unstable. This whole scenario is termed as \text{backward bifurcation} \cite{Zaleta_Velasco_Hernandez}.

\subsection{Remarks on the wise use of the mean-field principle in theoretical epidemiology}
\label{stochastics_meanfield}
Thanks to the use of a few important deterministic models, theoretical epidemiology has become a key supporting tool to public health decision makers. These models, historically derived from classical physical chemistry and other deterministic applications of statistical mechanics, are, however, only approximations of the underlying real-world stochastic processes, obtained from the mean-field limits of stochastic birth and death processes analogous to those adopted in chemical kinetics \cite{gardiner}.

In this subsection we therefore concisely remind some well-known tools of nonequilibrium statistical physics \cite{gardiner, vank} with the aim to recall how deterministic epidemic models arise as limits on the underlying stochastic processes on the one hand, and their shortcomings, on the other hand.

For example, let us consider an epidemic (i.e. no vital dynamics) SIR infection process, with state variables $(X(t),Y(t),Z(t) = N -X(t)-Y(t))$. In this process there are two types of events: the \textit{infection}, responsible for the $S \rightarrow I$ transition with infinitesimal probability
\begin{equation}\label{ev1}
 Prob\left( (X(t+dt),Y(t+dt)) = (X(t)-1,Y(t)+1 )\right) = \beta \frac{Y(t)}{N}X(t)dt,
\end{equation}
and \textit{recovery}, responsible for the $I \rightarrow R$ transition with infinitesimal probability
\begin{equation}\label{ev2}
 Prob\left( (X(t+dt),Y(t+dt)) = (X(t),Y(t)-1 )\right) = \gamma Y(t)dt.
\end{equation}
For the SIS model (still without vital dynamics), since $X(t) = N -Y(t)$, the $S \rightarrow I$ transition has probability $\beta (Y/N)(N-Y)$.

The behavior of the realizations of these epidemic stochastic processes can be simulated exactly by means of the Gillespie algorithm \cite{gill}, whereas the probability density function of its state variables is described by a multivariate Master equation (ME) \cite{gardiner,vank}, which is bivariate for the epidemic SIR process, and univariate for the SIS process. For example, for the epidemic SIS model the ME reads
\begin{equation}\label{cmeSIS}
\frac{d}{dt}P(t,Y) = \beta \frac{Y-1}{N}\left(N-(Y-1)\right)P(t,Y-1)-\beta \frac{Y}{N}\left(N-Y\right)P (t,Y)+\gamma(Y+1)P(t,Y+1)-\gamma Y P(t,Y).
\end{equation}
Taking the average $\langle Y(t) \rangle = \sum Y P(t,Y)$ from Eq.~\ref{cmeSIS}, we get
\begin{equation}\label{meanY}
\frac{d}{dt}\langle Y(t) \rangle = (\beta -\gamma)\langle Y(t) \rangle - \beta \frac{1}{N}\langle Y^2(t) \rangle.
\end{equation}
Essentially the mean field hypothesis applied in theoretical population models stands, for large populations, in neglecting the correlations between the state variables and assuming that the average of the product of two state variables is approximately equal to the product of the averages \cite{mckanenewt}, for example in Eq.~\ref{meanY} above $ \langle Y^2(t) \rangle \approx (\langle Y(t) \rangle)^2 $.

However, due to the approximate nature of deterministic epidemic model and their impact of public health sciences, it is important to go beyond the mere application of the mean-field limit, and to understand under which conditions deterministic differential equations (whose state variables are continuous function of time) can safely be used in place of the originary "exact" stochastic formulation of the epidemic process (whose state variables are time-continuous stochastic processes taking integer non-negative values). Moreover, it is also important to have a feeling of the behavior of the system at least in the regime of small stochastic deviations from the deterministic behavior.

If the total population size is small, birth-death stochastic epidemic models must be used. On the other hand, when the size of the state variables becomes large (which is the case intuitively leading to the deterministic approximation), it is hard to simulate the process due to the crowding of events. In such a case, however, one can use a system-size approximations~\cite{gardiner,vank}. This methodology allows approximating, respectively, the solution of a multivariate  ME with a multivariate Fokker-Plank equation (FPE) \cite{gardiner} (this approximate equation is sometime called Chemical FPE~\cite{gardiner}). In turn, from the obtained FPE one can derive a corresponding system of Stochastic Differential equations (SDEs) that approximate the dynamics of the state variables vectors. In physical chemistry it has been proposed a reverse equivalent approach, the Chemical Langevin equation (CLE)\cite{gill2000}, where first the approximating SDE system is obtained from the definition of the model, and only then the FPE is written. Namely, a spatially homogeneous system of particles/subjects with state variable $u$ characterized by $M$ "events" having probability $a_j(u)dt$ $j=1,\dots,M$ can be approximated by the following  system of SDEs \cite{gill2000}
\begin{equation}\label{cle}
u_i^{\prime} = \sum_j \nu_{i,j}(a_j(u) + \sqrt{a_j(u)} \xi_j(t) ),
\end{equation}
where the $\xi_j(t)$ are $M$ uncorrelated white noises, and $\nu$ is the \textit{stechiometric} matrix where $\nu_{i,j} \in \{0,-1,+1\}$ depending on the changes induced by the $j$-th event into the $i$-th state variable. Note that here the original state vector, whose components were integer non-negative numbers, is approximated by a real vector.

For example, in the case of the SIR epidemic model, the CLE reads
\begin{align}
X^{\prime} &= -\beta \frac{Y}{N}X - \xi_1(t) \sqrt{\frac{Y}{N}X}, \label{CLE-X} \\
Y^{\prime} &= \beta \frac{Y}{N}X -\gamma Y + \xi_1(t) \sqrt{\frac{Y}{N}X}-\xi_2(t)\sqrt{\gamma Y}.\label{CLE-Y}
\end{align}
Since $N$ is constant, we can pass to the fractions $(S,I)$
\begin{align}
S^{\prime} &= -\beta SI - \xi_1(t) \frac{1}{\sqrt{N}}\sqrt{\beta SI}, \label{CLE-S} \\
I^{\prime} &= \beta SI -\gamma I + \xi_1(t) \frac{1}{\sqrt{N}}\sqrt{\beta SI}-\xi_2(t) \frac{1}{\sqrt{N}}\sqrt{\gamma I}.\label{CLE-I}
\end{align}
As a consequence, in the limit $N\rightarrow +\infty$ we obtain the deterministic model.

It is interesting to note that the use of the force of infection FOI=$\beta_1 Y$ leads, after passing to fractions, to a stochastic term $\xi_1(t) \sqrt{\beta_1 SI}$ which is not scaling as $1/\sqrt{N}$, and thus it does not disappear in the limit $N\rightarrow +\infty$. Moreover, note that an hypotetical model where the contact rate is decreasing to $0$ with the population size $N$, would yield the following trivial uncoupled model of the type $S^{\prime} = 0$, $I^{\prime}=-\gamma I$.

As stressed by a few authors among which Gillespie \cite{gill2000}, the accuracy of the approximation given by Eq.~\ref{cle} or by those derived by the system-size expansion \cite{gardiner,vank}, depends on the fulfillment of specific conditions that are \textit{time-varying} since they are state-variable dependent. We can roughly summarize these conditions as follows: for all times
$ a_j(u) dt $ must be sufficiently larger than $\sqrt{a_j(u) dt } $. This in turn implies that  all state variables must be sufficiently large. Thus the above mentioned approximations  require caution, since \textit{a rigori} they are invalid even when only one of the state variables becomes small.

For the specific case of the SIR model, note that in the initial epidemic phase if one sets $X(t)\approx X(0) \approx N$ the ensuing stochastic process for $Y(t)$ is linear and its mean is given by $E[Y] = Y(0) Exp((\beta S(0)-\gamma)t) $ as in the approximation of the deterministic SIR epidemic model. The corresponding variance is $Var[Y] = Y(0) (e^{(\beta S(0) - \gamma)t} -1 )/(\beta S(0) - \gamma)$. Of course, this approximation is valid for a relatively short time-interval. However, as Bartlett noted in his pioneering work on measles \cite{Bartlett_CCS}, the most critical difference between deterministic and stochastic models arises in the intervals between epidemics (for example during the undamped oscillations caused by periodic changes in the contact rate) when the prevalence can become so small that extinction of the infective population almost certainly occurs in the stochastic formulation \textit{also in presence of a large population} while deterministic models (e.g. the periodically forced SIR and SEIR models with vital dynamics) simply predict that system trajectories will be positive for all times. This is of relevance in presence of vaccination because stochasticity may cooperate with the further reduction in prevalence caused by vaccination in inducing infection elimination \cite{pej}.

The study of stochasticity-induced infection extinction led Bartlett to introduce \cite{Bartlett_CCS} the key concept of \textit{critical community size} (CCS). This is the (heuristically defined) threshold over which, in absence of vaccination, the infection is likely to persist, and below which extinction is very likely. Over the CCS it is therefore safe to "read" the output of a deterministic model as an average behavior. For example pre-vaccination measles data indicate a CCS in the region of $250 000$- $400 000$ \cite{KG,Bartlett_CCS}. The CCS is only weakly associated to the BRN, while it is strongly dependent on the duration of the infectious and latent periods. As for the latter, it was noted  that simple stochastic models based on exponentially distributed durations, tend to severely overestimate the CCS \cite{KG}, while instead adequate predictions can be obtained by resorting to more realistic, e.g. bell-shaped, distributions  of the latent and infectious periods \cite{KG}. On the other hand, the effect of space on the CCS is extremely complex \cite{pej}.

Stochastic epidemic models are characterized by a range of interesting phenomena not shown in their deterministic counterparts. The simplest is obviously stochastic extinction after importation, i.e. the fact that an infectious individual introduced in a fully susceptible population has a probability of recovering before being able to infect anyone (given by $1/\mathcal{R}_0$ ($1/\mathcal{R}_0^k$) \cite{pej}), in which case no epidemics will occur, unlike the deterministic model. Other stochastic effects are more subtle. For example, in the SIR model with vital dynamics, the stochastic fluctuations of the total population $N(t)$ around its deterministic equilibrium  (which can be approximated by the CLE: $ N^{\prime} = \xi_b(t) \sqrt{\mu N} - \xi_d(t) \sqrt{\mu N}$), may transform the damped oscillations around the endemic equilibrium of the corresponding deterministic model in sustained oscillations\cite{Kuske}. This phenomenon, termed \textit{stochastic amplification}, is an instance of the more general phenomenon of \textit{coherence resonance} \cite{Kuske}. The theoretical possibility of strochastically inducing the disease eradication with a limited amount of vaccines via periodic delivery of vaccinations so that to induce resonances, and the related role of seasonal fluctuations of the contact rates, are invstigated in \cite{Meerson} by means of methodologies of the theory of fluctuation-induced population extinction.   Further details on the stochastic theory of epidemics can be found in the complementary books \cite{tb,lja}.

\subsection{Space and beyond: mean-field metapopulations}
\label{metapop}
Everytime a new epidemic focus appears, the first concern of public health is to assess whether this isolated focus will remain localized or will spatially spread. This leads to a key missing dimension in the previous simple models, namely \textit{space}. The most straightforward way of modelling the geographical spread of
infections is to add spatial diffusion to the basic SIR and SEIR model \cite{Capasso,murray2,bailey}, leading to equations that are analogous to the reaction-diffusion equations of theoretical biochemistry \cite{murray1,murray2}. This approach, which is proved useful to understand some basic aspects of spatial epidemic behavior by using tools of reaction-diffusion theory, has obvious shortcomings. The first one is that human beings have patterns of movement dramatically different from that of particles moving under Fickian diffusion \cite{Gonzalez_Barabasi,Balcan_Vespignani}. This is a major complication leading to complex integro-differential equations where the diffusion part interacts with the reaction part, i.e. transmission \cite{mendez}. Moreover, human movement often includes rapid commuting between areas with highly clustered population, such as cities \cite{pej,pej2002}.

To cope with the above problems, population ecology and mathematical epidemiology have increasingly used the \textit{metapopulation}, or \textit{multi-patch} approach, where a large populations is split in a finite number of \textit{patches}, for example cities, where: i) each city obeys its own deterministic \textquotedblleft local \textquotedblright  epidemic model, e.g. SIR-type; ii) local dynamics are linked through population fluxes,  e.g. due to commuting for work. Commuting plays a fundamental role in the geograpical spread of infectious diseases in contemporary societies \cite{pej2002,pej}, by favouring the geographical spread and persistence of infections through the interlink between patches. The multi-patch approach was used in the pioneering work by Baroyan and Ravchev, who linked a large number of SIR epidemic models to simulate the spread of flu in URSS cities \cite{bailey}. Multi-patch epidemiological models have since then developed fast both on the theoretical and the applied side, e.g. for predicting the world-wide spread of pandemic influenza \cite{Vespignani_flu_2011,wang2012estimating,nah2016predicting,nah2016estimating}.

As for the mathematical viewpoint, multipatch epidemic models show the same spatial phenomena we may observe in traditional reaction-diffusion models, namely: i) emergent phenomena such as the onset of spatial heterogeneities by means of spatial instability of spatially homogeneous equilibrium points (Turing-like instabilites) or even of limit cycles \cite{lloydjmb,lloydmbs}; ii) the onset of traveling waves and solitons, representing the spatial displacement of epidemic outbreaks \cite{pej}; iii) the emergence of various spatio-temporal asynchronous/synchronous complex patterns \cite{pej,lloydeco}. To these classes of phenomena, one has to add those induced by parametric heterogeneities between patches \cite{pej}, which might mirror heterogeneities in the epidemiology (e.g. in social contacts), public health interventions (e.g. in vaccine uptake), demographics (e.g. population density and age-structure), as well as environmental (e.g. climate, temperature, etc).
As regards the onset of Turing-like patterns is concerned, let us consider the following system \cite{lloydjmb} for an infection with $m$ epidemiological states ($m=4$ for a SEIR system) spreading among $n$ interacting patches. If patches are isolated, the local spread is modeled by $n$ $m$-dimensional ODE systems of the form $u_j^{\prime} = f(u_j) $. In presence of linear population flows between patches one has the following interlinked dynamical systems
\begin{equation}\label{interpop}
u_j^{\prime} = f(u_j) + \sum_{i=1}^{n} c_{i,j} M u_i,
\end{equation}
where $u_i \in R^m$ are the state variables associated to patch $i$, $ U = (u_1,\dots,u_n) $ is the $m*n$ matrix of all state variables, $M$ is a $m*m$ diagonal matrix reporting the different mobility coefficients of individuals in the different epidemiological compartments (e.g. the mobility of infectious subjects might be reduced with respect to that of susceptible and removed \cite{bailey}), $\mathcal{C}$ is the $n*n$ \textit{connection} matrix whose elements are the transmission rates between individuals of different patches.

A \textit{spatially homogeneous} solution of Eq.~\ref{interpop} is one such that in all the patches the state vector is the same $ u_j = s(t) $, with $s(t) \in R^m$. Linearizing around the spatially homogeneous solution and applying some similarity transformation one gets the following $n$ independent systems in $C^m$ \cite{lloydjmb}
\begin{equation}\label{indep}
\psi^{\prime} = \left(  Df(s(t)) + \lambda_h M \right)\psi, h =1,\dots,n,
\end{equation}
where $\lambda_h$ are the eigenvalues of the connection matrix $C$. Eq.~\ref{indep} is the discrete equivalent of the linearization equation arising in the study of the Turing bifurcation \cite{murray2}. Notably, this methodology can also be applied in the case of nonlinear population fluxes \cite{lloydmbs}.

Among multi-patch models, a fairly well-studied class is represented by the following \textit{multi-group, cross-coupled} SEIR model (\cite{lloydmbs} and refs therein) in a constant population
\begin{align}
S_i^{\prime} &= \mu(1-S_i) - S_i \sum_{j=1}^{n} \beta_{ij}I_j,  \label{Scrosspop}\\
E_i^{\prime} &= S_i \sum_{j=1}^{n} \beta_{ij}I_j - (\mu + \alpha) E_i, \label{Ecrosspop}\\
I_i^{\prime} &= \alpha I_i - (\mu + \gamma)I_i, \label{Icrosspop}\\
R_i&= 1- S_i - E_i -I_i. \label{Rcrosspop}
\end{align}

\noindent Unlike Eq.~\ref{interpop} where all contacts are local, through inter-patch movements, in Eq.~\ref{Scrosspop} fluxes between patches are disregarded on the rationale that movements from own patch to other patches are short-lasting. The core of the description stands in the \textit{spatial transmission matrix} $\beta_{ij}$ assigning the per-capita rate at which a susceptible from patch $i$ is infected by an infective individual belonging to patch $j$ regardless of the patch where the at-risk contact occurred \cite{lloyd_may_1996,lloydmbs,pej}. For Eq.~\ref{Scrosspop} fairly general results are available, incuding a decoupling equivalent to those of Eq.~\ref{indep} \cite{lloydmbs}. In particular, Eq.~\ref{Scrosspop} was used to study the conditios under which the local (i.e., in each patch) oscillations of SIR and SEIR models become synchronised~\cite{lloyd_may_1996,lloydmbs}, as observed for measles in UK cities in the pre-vaccination era \cite{Earn_chaos_2000}. Notably, synchronization of unforced SIR and SEIR models arises at very low levels of patches connection, while seasonal forcing, besides inducing complicate behavior, allows maintenance of some phase differences in local dynamics~\cite{lloyd_may_1996}, and yields more plausible prevalence levels at the inter-epidemic intervals compared to the simple SIR and SEIR models.

Introduction of vaccination can have a complex impact on the spatio-temporal dynamics of multi-patch systems \cite{pej}. For example metapopulation models can explain the observed de-correlation observed in the time-series of UK major cities after the introduction of measles vaccination \cite{Bolker_Grenfell_pnas96,Earn_chaos_2000}.

\subsubsection{Large scale meta-population networks}
In very recent times, the increasing availability of data on both the spread  of infectious diseases and human movements is allowing to finely model the dynamics of outbreaks on larger and larger scales, even world-wide \cite{Vespignani_flu_2011}. This has led to a number of investigations on the role of the topology of the networks of inter-patches contacts as resulting from population mobility \cite{vittoriaalex}. These analyses have shown that though the condition $\mathcal{R}_0>1 $ still guarantees that the epidemics is not self-extinguishing and that there are patches here and there in the network where the infection is circulating, in order to have global macroscopic spread a second threshold has to be overcome \cite{vittoriaalex,vittoriaalexjtb}. This second threshold, larger than the first one, depends both on population mobility and on the topology of network. Not suprisingly, if the network is heterogenous, i.e. its degree distribution $P(k)$ (which summarizes the probabiity for a node of the underlying graph of having $k$ links to other nodes of the networks) has a heavy tail, global epidemic spread is easier. However, for very large heterogeneous network structure the second threshold is close to the classical one, quite independent of population mobility. This can have profound implications as it implies that measures aimed at limiting international mobility to reduce infection spread might have little impact.

\subsubsection{Metapopulation models beyond space: social and age-specific heterogeneities}
Multi-group and metapopulation models are one of the most active and effective fields of investigation in the theoretical epidemiology of infectious diseases, as  they \textit{latu sensu} represent a convenient framework for modeling a much wider class of epidemiological phenomena than geographical spread only. This has been done by straightforwardly extending the interpretation of patches to different types of internally homogeneous groups, such as social groups, age groups, etc, still relying on appropriate contact or transmission matrice, having elements $\beta_{i,j}$ reflecting heterogeneous contact patterns between social or age groups. Though the first works in the field were following a more mathematical vein, i.e. the need to remove the homogeneous mixing hypothesis, the public health modelling revolution described in section \ref{Mathepi} is much the legacy of this intuition. Indeed, the multigroup SIS model analogous to the SEIR model of Eqs.~\ref{Scrosspop}-\ref{Rcrosspop} \cite{lloydmbs}, first introduced by Lajmanovich and Yorke \cite{laj} to model the spread of an infection among societal groups with different social/sexual behavior, and later extended to model gonorrhea transmission in the US~\cite{Hethcote_1984}, has represented the departure point for the explosion of studies on the modeling of sexually transmitted infections that was subsequently triggered by the onset of the HIV/AIDS epidemics in the early eighties. Noteworthy extensions of Ref.~\cite{laj} to heterogeneous networks of contacts are available in  \cite{pastor2001epidemic,olinky,don}.

On the other hand, in the field of common childhood infections, such as measles and pertussis, the large availability of age-specific infection data triggered the development of the \textquotedblleft standard \textquotedblright  age-structured model \cite{Dietz_1981,AMay_1991,Hethcote_2000}. The \textquotedblleft standard \textquotedblright  model is nowadays extensively used to investigate the nonlinear effects of mass vaccination programs, including the complicate patterns of increase in the average age at infection following the reduction in transmission, and the related perverse increase in the number of cases of serious disease, as it may be the case for rubella, measles and varicella \cite{Dietz_1981,AMay_1991,Hethcote_2000}. The key ingredient of the \textquotedblleft standard \textquotedblright  model is represented by the age-specific transmission matrix of elements $\beta_{i,j}$, representing the per-capita transmission rates in contacts between individuals of different age groups. In a first phase these matrices were computed by resorting to simplifying hypotheses about contacts patterns between age groups, aimed to reduce the number of unknowns in the $\beta_{ij}$ matrix from $n^2$ to $n$, thereby making them identifiable from standard age-specific epidemiological data. These approaches include the so-called Who Acquires Infection From Whom (WAIFW) matrices (\cite{AMay_1991} and refs therein), and the \textit{proportionate} and \textit{preferred} mixing hypotheses (\cite{Hethcote_2000} and refs therein). Recently, considerable advances were made thanks to studies aimed to directly observe population contact patterns \cite{Mossong,wallinga,wang2013how,wang2013impact}, or to reconstruct them from official data \cite{Zagheni,Fumanelli}.

Apart the anagraphical age of subjects, two other chrono-biological variables impact on the in-population dynamics of infectious diseases. The first is the so called \textit{age since infection} \cite{thieme2003mathematics}, which is the time elapsed from the infection, a key factor modulating both the contact rate of infected subjects, the disease duration and the disease-related risk of death. Quite remarkably, this variable was first introduced in mathematical epidemiology by Kermack and McKendrik in their 1927 paper \cite{kermack1927contribution}, but later ignored for decades \cite{thieme2003mathematics}. The second is the \textit{age since vaccination} \cite{noi4,iannelli2005strain}, which is the time elapsed since the vaccine inoculation, and which modulates both the vaccine-induced immunity \cite{iannelli2005strain} and the occurrence of vaccine-related side effects \cite{noi4}. Note that if any of these temporal variables is modeled as a continuous variable, then the resulting model, due to the impact of these variables on the contact rate and other parameters, is not differential multi-patch but integro-differential, as in the seminal work \cite{kermack1927contribution}.

Finally, a number of recent re-assessments of the main future challenges for multi-group and metapopulation studies can be found in \cite{gianpi,Mick_Roberts,Metcalf,Many}.
\section{Basic concepts and methods in (non-behavioral) epidemiological modeling and vaccination: network models}\label{concep-net}

A real population is mix of heterogeneous individuals and the heterogeneity exists in several aspects, especially in contacts. The foundations of epidemiology and early epidemiological models were based on population wide random-mixing, which was well described in the last section. However, in practice, each individual has a finite set of contacts or individuals whom they can pass the infection. The ensemble of all such contacts forms a \textit{network}. Structural pattern of this contact network among individuals allows models to compute the epidemic dynamics at the population scale from the individual-level process of spreading infections.

In the recent years, overwhelming data explosion in human sciences shows complex and heterogeneous connectivity patterns in a wide range of biological and social systems. Network theory provides a natural gateway to understand all these complex patterns. Ideas from network theory have also inspired research in many fields to study interactions between elements and their dynamics in areas like computer science, system biology, social sciences, economic and statistical physics. However, epidemiological study has also used potentiality of network theory like any other scientific disciplines. The complex properties of networks have a profound impact on studying equilibrium and non-equilibrium phenomena in epidemic processes. Dated back to mid-eighties, the researchers have shown that understanding the spreading process of an infection have considerable overlap with the studies of network properties [151, 152]. The group of individuals and their connections in a population naturally defines a network in which disease spreads from one node (individual) to another distantly related node. So, insights from topological structure of the underlying network provides useful information that helps predicting spreading process including growth of infection at the first stage and subsequently its distribution over the entire contact network.

Although it has long been acknowledged that network theory is the key ingredient of epidemic modeling, recent abundance of data exemplifies the huge complexity in the spreading and persistence of infection in a population and thus calls for detailed theoretical understanding of the interplay between epidemic processes and networks. Previous works have shown that most real-world networks exhibit dynamic self-organization and are statistically heterogeneous -- typically all marks of complex systems \cite{albert2002statistical, baronchelli2013networks, boccaletti2006complex, caldarelli2007scale, cohen2010complex, costa2007characterization, dorogovtsev2002evolution, newman2010networks, newman2003structure}. A regular lattice rarely represents the real-world networks of relevance for epidemic spreading. In a real population, there are few individuals that may act as hubs (or `super-spreader'), whereas the majority of the population have very few interactions. Both social and infrastructure networks are organized in communities of tightly interconnected nodes of individuals \cite{girvan2002community}. Although randomness in the connection process of nodes is always present, organizing principles and correlations in the connectivity patterns define network structures that are deeply affecting the evolution and behavior of epidemic and contagion processes \cite{pastor2001epidemic, pastor2001aepidemic, moreno2002epidemic, kretzschmar1996measures}. Furthermore, the complex features of networks often find their signature in statistical distributions which are generally heavy-tailed, skewed, and varying over several orders of magnitude \cite{watts1998collective}. The evidence of large-scale fluctuations, clustering and communities characterizes the connectivity patterns of real-world systems; and this has prompted the need for mathematical approaches capable to deal with the inherent complexity of networks.

However, the connection between infectious disease epidemiology and the network theory goes far beyond the general discussion. Understanding network topology may help identifying disease transmission routs, which may help designing strategies to control the disease. For instance, `contact tracing’ is a very powerful public health strategy during first phases of the spreading process. It relies on underlying transmission routes to control further spread without even any information on epidemiology of the infectious disease [167]. Thus studying network theory and how it can be applied to relate the infection spreading process may be a vital tool for understanding the disease and, designing control strategies. With a discussion of basic network metrics and topologies, here we review the development of current research at the interface of epidemiological studies and network theory including immunization to control the disease.

\subsection{Networks: types and topologies}
\subsubsection{Definitions and Notations }
Graph theory \cite{ore1962theory} can be used to mathematically formalize networks \cite{newman2003structure}. A graph is a collection of points, called \textit{vertices} (or \textit{nodes}), and a set of connections, called \textit{edges} (or \textit{links}). Edges indicate link between vertices, representing the presence of an interaction or relations between those vertices. Interaction can be bidirectional which defines \textit{undirected} networks, or it can be directional which defines  \textit{directed} networks or \textit{digraphs}. We represent a graph of size $N$ (say, with $N$ vertices, and $L$ links) using $N \times N$ adjacency matrix $A$, with elements $a_{ij} = 1$ if an edge is connecting nodes $i$ and $j$ and zero otherwise. $A$ is symmetric matrix in undirected graphs, and asymmetric in directed graphs.\\

\noindent$\bullet$  \textbf{Degree and Degree distribution:} The \textit{degree} (or  \textit{connectivity}) $k_i$ of vertex $i$ in an undirected network is the number of edges emanating from $i$, i.e. $k_i = \sum_j a_{ij}$. In the case of directed networks, we have both \textit{in-degree}, $k^{\mbox{in}}$, and \textit{out-degree}, $k^{\mbox{out}}$, as the number of edges that end in $i$ or start from $i$, respectively. The total degree, however, in directed network is defined by $k_i = k^{\mbox{in}}+k^{\mbox{out}}$. The \textit{degree distribution} $P(k)$ of a network is the probability that a randomly chosen vertex has degree $k$. In a finite network, it denotes the fraction of vertices with degree exactly equal to $k$. In case of directed networks, there are instead two different distributions, the \textit{out-degree} $P^{\mbox{out}} (k^{\mbox{out}})$ and the \textit{in-degree} $P^{\mbox{in}} (k^{\mbox{in}})$ distributions, though in-degree and out-degree of a given vertex might not be independent. It is also useful to consider the \textit{moments} of the degree distribution, $<k^n> = \sum_k k^n P(k)$. The first moment $<k> = 2L/N$ is called the \textit{average degree}, provides information about the density of the network. The degree distribution defines classes of networks with similar statistical features. A network is called \textit{sparse} if its number of links $L$ grows at most linearly with the network size $N$; otherwise, it is called \textit{dense}. In directed networks, since every edge contributes to one node in-degree and other node out-degree we have that $<k^{\mbox{in}}> = <k^{\mbox{out}}>$.\\

\noindent$\bullet$  \textbf{Connectedness, Shortest path length, and Diameter:}
A path $P_{i,n}$ is a sequence of different edges ${(i_j,i_{j+1})}, j = 0,...,n-1$, connecting vertices $i_0$ and $i_n$; the number of edges traversed, $n$, is also called the length of the path. A graph is \textit{connected} if there exists a path connecting any two vertices in the graph. A component $\mathcal{C}$ of a graph is defined as a connected subgraph. The \textit{shortest path} $l_{ij}$ between two nodes $i$ and $j$ is defined as the length of the shortest path (though may not be unique) joining $i$ and $j$, and the \textit{average shortest path length} $<l>$ is the average of the value of $l_{ij}$ over all pairs of vertices in the network. The \textit{diameter} of a network is the maximum value of all the pairwise shortest path lengths.\\

\noindent$\bullet$  \textbf{Degree correlations:} The \textit{degree correlation} between two vertices is defined as the conditional probability $P(k'|k)$ that an edge departing from a vertex of degree $k$ is connected to a vertex of degree $k'$ \cite{pastor2001epidemic}.  A network is called \textit{uncorrelated} if this conditional probability is independent of the originating vertex $k$. In this case, $P(k'|k)$ can be simply estimated as the ratio between the number of edges pointing to vertices of degree $k'$, $k'P(k')N/2$, and the total number of edges, $L = <k>N/2$, to yield $P^{\mbox{un}}(k'|k) = k'P(k')/<k>$.
The most popular measure of degree correlation is \textit{Pearson correlation coefficient} $r$. Proposed by Mark Newman \cite{Newman_mixing,newman2002assortative}, the degree correlation coefficient is defined as
\begin{equation}
r = \sum_{_{jk}}\frac{jk(e_{jk}-q_jq_k)}{\sigma^2},
\end{equation}\label{degree-correlation-eq}
where \[\sigma^2 = \sum_kk^2q_k-\bigg[\sum_k kq_k\bigg]^2.\]
$e_{ij}$ is the degree correlation matrix that represents the probability of finding node with degrees $i$ and $j$ at the two ends of a randomly selected link, and $q_k$ is the probability that there is a degree-$k$ node at the end of the randomly selected link. The degree correlation varies between $-1\leq r \leq1$. For instance, $r > 0$  the network is assortative, for $r = 0$ the network is neutral and for $r < 0$ the network is disassortative. In reality, social networks are assortative, whereas technological and biological networks are disassortative \cite{newman2002assortative}. For example, for the scientific collaboration network we obtain $r = 0.13$, in line with its assortative nature; for the protein interaction network $r = - 0.04$, supporting its disassortative nature and for the power grid we have $r = 0$. Another related measure of correlations is the \textit{average degree} of the nearest neighbors of vertices of degree $k$, denoted by $\bar{k}_{nn}(k)$ which is formally defined as \cite{pastor2001epidemic} \[\bar{k}_{nn}(k) = \sum_{k'} k' P(k'|k).\]\\

\noindent$\bullet$  \textbf{Centrality and Betweenness:} The concept of \textit{centrality} encodes the relative importance of a node inside a network. Together with the \textit{degree} and the \textit{closeness of a node} (defined as the inverse of the average distance from all other nodes), another standard measure of node centrality is \textit{node betweenness}, which is obtained by counting the number of geodesics going through it. More precisely, the betweenness $b_i$ of a node $i$, sometimes referred to as $load$, is defined as
\begin{equation}
b_i = \sum_{j,k\in N, j\neq k} \frac{n_{jk}(i)}{n_{jk}},
\end{equation}
where $n_{jk}$ is the number of shortest paths connecting $j$ and $k$, while $n_{jk}(i)$ is the number of shortest paths connecting $j$ and $k$ and passing through $i$. Betweenness distributions have been investigated in \cite{molloy1995critical, molloy1998size, newman2001random, newman2002random, adamic2001search, goh2001universal, barthelemy2004betweenness, guimera2005worldwide}. Betweenness-betweenness correlations and betweenness-degree correlations have been studied respectively in \cite{ravasz2003hierarchical} and in \cite{vazquez2002large}. The concept of betweenness can also be extended to edges. The \textit{edge betweenness} is defined as the number of shortest paths between pairs of nodes that run through that edge \cite{newman2004finding}.\\

\noindent$\bullet$  \textbf{Clustering:} \textit{Clustering} refers to transitivity property of the network, i.e. the relative propensity of two nodes to be connected, provided that they share a common neighbor. The \textit{clustering coefficient} $C$ is defined as the ratio between the number of loops of length three in the network (i.e. triangles), and the number of connected triples (three nodes connected by two edges). An alternative definition is also available by Watts and Strogatz \cite{Strogatz2001}. A local measure $c_i$ of clustering is defined as the ratio between the actual number of edges among the neighbors of a vertex $i$, $e_i$, and its maximum possible value, measuring thus directly the probability that two neighbors of vertex $i$ are also neighbors of each other. The \textit{mean clustering} of the network $<c>$ is defined as the average of $c_i$ over all vertices in the network. The clustering spectrum $\bar{c}(k)$ is defined as the average clustering coefficient of the vertices of degree $k$ \cite{ravasz2003hierarchical, vazquez2002large}, satisfying $<c> = \sum_k P(k)\bar{c}(k)$.\\

\noindent$\bullet$  \textbf{Community Structure:} A \textit{community} or \textit{cluster} (or \textit{cohesive subgroup}) of a given network of vertices $N$ and set of edges $L$ is a subgraph $G'(N', L')$, whose nodes are tightly connected, i.e. cohesive. Since the structural cohesion of the nodes of $G'$ can be quantified in several different ways, there are different formal definitions of community structures.

A related terminology of community structure of a network is \textit{clique}. A \textit{clique} is a maximal complete subgraph of three or more nodes all of which are adjacent to each other, and such that no other nodes exist adjacent to all of them. This definition can be extended by weakening the requirement of adjacency into a requirement of reachability: a \textit{n-clique} is a maximal subgraph in which the largest geodesic distance between any two nodes is no greater than $n$. A detail description is given in \cite{boccaletti2006complex, fortunato2010community}.

\subsubsection{Type of networks}
Based on the qualitative and quantitative features of the degree distribution, the networks can be classified into a series of important types of networks, which are also useful for the epidemiological applications.\\

\noindent$\bullet$ \textbf{Random graph:} The simplest network model is the classical random graph model proposed by Erd{\"o}s and R{\'e}nyi (also named ER graph) \cite{erdHos1959random, gilbert1959random, solomonoff1951connectivity,erdHos1959random}. A random graph model $G_p(N)$ is constructed from a set of $N$ nodes in which each one of the $N(N-1)/2$ possible links is present with probability $p$. The degree distribution is given by a binomial form, which, in the limit of constant average degree (i.e. $p = <k>/N$) and large $N$ can be approximated by a Poisson distribution $P(k) = e^{-<k>} {<k>}/{{k!}}$ . The clustering coefficient is given by $<c> = p$, and the average shortest path length is $<l> \sim \log N/ \log<k>$ \cite{dorogovtsev2013evolution}. This model is therefore adequate in the case of networks governed only by stochasticity, although $G_p(N)$ tends to a regular graph for large N and constant $p$. Extensions of the basic model to allow for other degree distributions lead to the class of models known as \textit{generalized random graphs}, \textit{random graphs with arbitrary degree distributions} and the \textit{configuration model} \cite{newman2003properties}.\\

\noindent$\bullet$ \textbf{Small-world (SW) network:} A more sophisticated and tractable model of a network with small proximity and high clustering coefficient is the \textit{small-world network model} proposed by Watts and Strogatz (i.e. WS model) \cite{watts1998collective}. This network model is based on rewiring procedure of an ordered lattice.  The initial configuration starts with a ring of $N$ vertices, each one of which symmetrically connected to its $2m$ nearest neighbors. This represents the model with a structure which has large clustering coefficient and large average shortest path length. Starting from it, a fraction $p$ of edges in the network are rewired, by visiting all $m$ clock-wise edges of each vertex and reconnecting them, with probability $p$, to a randomly chosen node. There is another version of the model proposed by Monasson \cite{monasson1999diffusion}, where a fraction $p$ of edges are added between randomly chosen pairs of vertices. The overall effect of the rewiring processes is to add long-range shortcuts, that, even for a small value of $p \sim N-1$, greatly reduce the average shortest path length, while preserving a large clustering for not very large values of $p$. This model, although better suited for social networks with high clustering coefficient, has a degree distribution and centrality measures decaying exponentially fast away from the average value. The small-world model thus generates homogeneous networks where the average of each metric is a typical value shared, with little variations, by all nodes of the network.\\

\noindent$\bullet$ \textbf{Scale-free (SF) network - static and dynamic:} In contrast to random graph, and small-world network models, real-world networks are structured in a hierarchy of nodes with a few nodes having very large connectivity (the hub nodes), while the vast majority of nodes have much smaller degrees. In precise, these heterogeneous networks exhibit heavy-tailed degree distributions often approximated by a power-law behavior of the form $P(k) \sim k^{-\zeta}$ , which implies a non-negligible probability of finding vertices with very large degree. The degree exponent $\zeta$ of many real-world networks takes a value between 2 and 3. In such cases, these network models are called \textit{scale-free}. Several variations of the classical random graph models have been proposed in order to generate networks with a power-law degree distribution. For example, \textit{configuration model} \cite{bender1978asymptotic, molloy1995critical, aiello2000random}, \textit{fitness model} by Caldarelli et al. \cite{caldarelli2002scale}, \textit{threshold graph model} by Masuda et al. \cite{masuda2004global} and \textit{gradient network models} by Toroczkai and Bassler \cite{toroczkai2004network}, where directed graphs induced by local gradients of a scalar field distributed on the nodes.

So far, we have not discussed evolution process of these scale-free networks, i.e. nodes and links do not change over time. There are many examples of real networks in which the structural changes are ruled by the dynamical evolution of the system. Most well-known and useful evolving network model is \textit{Barab{\'a}si-Albert model} \cite{barabasi2002evolution}. The Barab{\'a}si-Albert (BA) network is a model of network growth inspired to the formation of the World Wide Web and is based on two basic ingredients: i) \textit{growth} (of nodes) and ii) \textit{preferential attachment} (connection of vertices). A description of the growth precess and attachment function is discussed in \cite{boccaletti2014structure}. Variations in preferential attachment function produced several other evolving network models like \textit{Dorogovtsev-Mendes-Samukhin} (DMS) model (linear preferential attachment), \textit{Krapivsky} model (nonlinear attachment probability), etc.\cite{dorogovtsev2000structure}. The Barab{\'a}si-Albert (BA) model has been generalized further by the same authors, and also by others including certain characteristics of underlying growth process of nodes \cite{boccaletti2014structure}.\\

\noindent$\bullet$ \textbf{Weighted networks:} In the real world settings, many networks exhibit varied intensity of connections. For example, social network, where existence of strong and weak ties between individuals are seen \cite{marchiori2000harmony, latora2001efficient, krause2003compartments, latora2003economic}, metabolic reaction pathways, predator-prey interactions in food webs \cite{polis1998ecology, mccann1998weak}, different capabilities of transmitting electric signals in neural networks \cite{sporns2000connectivity}, unequal traffic on the Internet \cite{barrat2004architecture} or of the passengers in airline networks \cite{li2004statistical}. These types of complex systems can be explained with model networks in which each link carries a numerical value measuring the strength of the connection. Considering weights of the connections puts another dimension of complexity in the topology of weighted network. Similarly, we can also have networks with weighted nodes. Detail descriptions of the measuring and modeling weighted networks has been discussed in~\cite{boccaletti2006complex}.

There are also other type of networks, which are important to study real world phenomenon. These networks include \textit{spatial network} (whose nodes occupy a precise position in two or three-dimensional Euclidean space, and whose edges are real physical connections) \cite{li2014spatial,li2011dimension,sun2016pattern,wang2011evolution}, adaptive networks \cite{gross2008adaptive}, temporal or time-varying networks \cite{holme2011temporal,nicosia2014measuring,zhang2012towards} (for more details, please also see Section~\ref{section:digital_epidemiology}).

\begin{figure}[th]
\centering
\includegraphics[width=122mm,height=73mm]{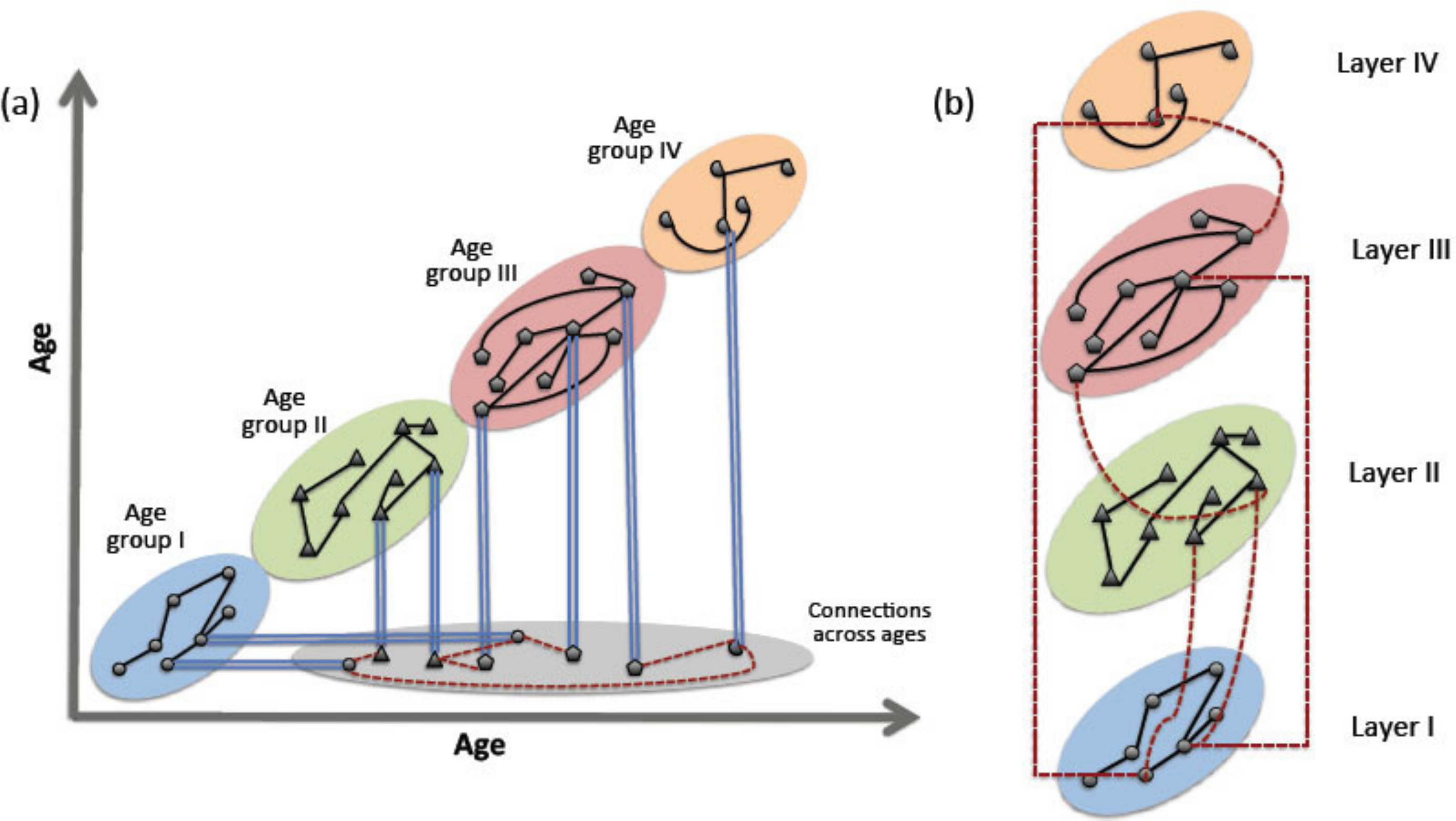}
\caption{Schematic illustration of age-dependent contact networks into a multilayer system. Network  with different color shades in (a) indicates network of individuals in the same age group, where network in gray region denotes connections across different age groups. Blue lines point to the same individuals from colored regions to gray region. (b) The same network can be conceived as multilayer networks.}
\label{AgeMultilayer}
\end{figure}

All the previous networks are single-layer. But there are many real world complex systems that can be conceived as multiple layers of such single networks. Most complex systems from physical, social, engineering, information and biological sciences include multiple subsystems and layers of connectivity, and they are often open, value-laden, directed, multilevel, multicomponent, reconfigurable systems of systems, and placed within unstable and changing environments. An example of multilayer networks conceived from age-dependent contact network is shown in Fig.~\ref{AgeMultilayer}. The recent development of \textit{multilayer networks} in network theory provides radical views of understanding these multilayer systems. For more information about single-layer and multilayer networks, please see the comparison in  Table~\ref{multi}. Descriptions and examples of more different type spatial and multilayer networks are given in \cite{boccaletti2014structure}.

\begin{sidewaystable}%
\caption{Definitions, Measure and Expressions of network properties of single-layer and multilayer networks}
\begin{center}
\begin{tabular}{|c|c|c|}
\hline
&&\\
Definition & Single-layer networks & Multilayer networks\\
&&\\
\hline
&&\\
Notation& $\mathcal{G} = (\mathcal{N, L})$& $\mathcal{M} = (\mathcal{G, C})$\\

&&layers: $\mathcal{G} = \{\mathcal{G_{\alpha}}: \alpha \in \{1,\ldots, M\}\}$, $\mathcal{G_{\alpha} = (X_{\alpha},E_{\alpha}})$\\

&&links between layers: $\mathcal{C} = \{\mathcal{E_{\alpha\varphi} \subset X_{\alpha} \times X_{\varphi}}; \alpha \in \{1,\ldots, M\}, \alpha \neq \varphi\}$\footnote{$\alpha, \varphi$ denote layers in multilayer networks}\\
&&\\
\hline
&&\\
Degree \& Centrality &degree: $k_i = \sum_{j \in \mathcal{N}} a_{ij}$& degree: $\mathbf{k_i} = (k_i^{[1]}, \ldots k_i^{[M]})$, $k_i^{[\alpha]} = \sum_ja_{ij}^\alpha$\\
&betweenness: $b_i = \sum_{j,k\in N, j\neq k} \frac{n_{jk}(i)}{n_{jk}}$& overlapping degree: $o_i = \sum^M_{\alpha = 1} k_i^{[\alpha]} $\\

&degree correlation coefficient $c_i = \frac{1}{\lambda_I}\sum_{i\in \mathcal{N}}a_{ij}c_j$&degree correlation coefficient $c_i = (c_i^{[1]}, \ldots c_i^{[M]})$\\
&&\\
\hline
&&\\
Clustering coefficients&$c_i = \frac{2e_i}{k_i(k_i-1)}$&$C_\mathcal{M}(i) = \frac{2\sum_{\alpha=1}^M|\bar E_\alpha(i)|}{\sum_{\alpha=1}^M|\mathcal{N}_\alpha(i)|(|\mathcal{N}_\alpha(i)| -1)}$\\
&&\\
\hline
&&\\
Characteristic path length&$\frac{1}{N(N-1)}\sum_{i,j\in \mathcal{N}, i\neq j}d_{ij}$&$\frac{1}{N(N-1)}\sum_{u,v\in \mathcal{X_M}, u\neq v}d_{uv}$\\
&&\\
\hline
\end{tabular}
\end{center}
\label{multi}
\end{sidewaystable}%

\begin{table}
\caption{Epidemic thresholds for \textbf{SI, SIS} and \textbf{SIR} models over Erd{\"o}s and R{\'e}nyi (ER) and Scale-Free (SF) networks}\footnotesize
\begin{center}
\begin{tabular}{|c|c|c|c|c|}
\hline
&&&\\
Model Type&Model expression& Characteristic time ($\tau_c$) & Spreading rate ($\lambda_c$)\\
&&&\\
\hline
&&&\\
SI&$\frac{di_k}{dt}=\beta_I [1- i_k(t)]k\theta_k(t)$&SF $(\zeta\geq 3)$: $\frac{<k>}{\beta_I(<k^2> - <k>)}$&SF $(\zeta > 3)$: 0 \\
&&SF $(\zeta < 3)$: 0&SF $(\zeta \leq 3)$: 0\\
&&ER:~~~$\frac{1}{\beta_I <k>}$&\\
&&&\\
\hline
&&&\\
SIS&$\frac{di_k}{dt}=\beta_I [1- i_k(t)]k\theta_k(t)-\gamma i_k(t)$&SF: $\frac{<k>}{\beta_I <k^2> - \gamma <k>)}$&$\frac{<k>}{<k^2>}$\\
&&ER:~~~$\frac{1}{\beta_I(<k>+1) - \gamma}$&$\frac{1}{<k>+1}$\\
&&&\\
\hline
&&&\\
SIR&$\frac{di_k}{dt}=\beta_I s_k(t)\theta_k(t)-\gamma i_k(t)$&$\frac{<k>}{\beta_I<k^2> - (\gamma+\beta_I)<k>)}$&$\frac{<k>}{<k^2>-<k>}$\\
&$s_k(t) = 1-i_k(t) - r_k(t)$&&\\
&&&\\
\hline
\end{tabular}\\
\end{center}
\scriptsize $i_k$: infected nodes with degree $k$.\\
\scriptsize $s_k$: susceptible nodes with degree $k$.\\
\scriptsize $r_k$: recovered or immune nodes with degree $k$. \\
\scriptsize $\theta_k$: fraction of the neighbors of node $k$ that are infected.\\
\scriptsize $\beta_I$: transmission rate.\\
\scriptsize $\gamma$: recovery rate.\\
\scriptsize $\zeta$: degree exponent SF network.\\
\label{DiseaseThresh}
\end{table}%

\subsection{Epidemic spread over networks}
Most infectious diseases such as influenza, SARS, HIV spread through human populations by physical contact between infective and susceptible individuals. The pattern of these disease-causing contacts forms a network, which has substantial influence on success of the spreading process. The traditional epidemic models such as classic \textbf{SIS} or \textbf{SIR}  epidemic models do not explicitly incorporate the topological structure of the underlying contract network that facilitates the spread of a pathogen. Instead they assume that any individual can come into contact with any other individual (homogenous mixing hypothesis) and so all individuals have comparable number of contacts. It also assumes that probability of disease transmission is same across all individuals in the network. Both of these assumptions are limited in most of the diseases: in reality an individual can transmit a pathogen only to those it comes into contact with and also transmission occurs in varied probabilities. Hence pathogens spread can be assumed on complex contact networks.

Spreading epidemics on complex networks has been studied by many researchers including May and Lloyd \cite{may2001infection}. Most detail and systematical study is done by Newman \cite{newman2002spread}. Considering an SIR model with uniform recovery rate $\gamma$, i.e. where infected nodes become removed at rate $\gamma$ after infection, and infection rate $\beta_I$. Then the transmissibility $T$ is defined as the probability that the infection will be transmitted from an infected node to a connected susceptible neighbor before recovery takes place. For continuous-time dynamics, the transmissibility computed as  \cite{newman2002spread}:
\begin{equation}
T = 1-\lim_{\delta t\rightarrow 0}(1-\beta_I\delta t)^{\gamma/\delta t} = 1-e^{-\gamma\beta_I}.
\end{equation}

In general, $\beta_I$ and $\gamma$ will vary between individuals, and assume initially that these two quantities are independent and identically distributed (iid) random variables chosen from some appropriate distributions $P(\beta_I)$ and $P(\gamma)$. Hence, transmission of the disease between two individuals is simply the average $T$  (i.e.  homogeneous case) of $T_{ij}$ over the distributions $P(\beta_I)$ and $P(\gamma)$, which is
\begin{equation}
T =<T_{ij}>= 1-\int^{\infty}_0 P(\beta_I)P(\gamma) e^{-\gamma\beta_I} d\beta_I d\gamma.
\end{equation}
Computing threshold transmissibility for spreading the disease in a special non-degenerate case of SIR model has been discussed in \cite{boguna2002epidemic}.

Now, starting from a single infective individual, disease spreads across the contact networks and a outbreak occurs. The ultimate size of the outbreak would be precisely the size of the cluster of vertices that can be reached from the initial vertex by traversing only occupied edges (i.e., edge in the graph across which the disease is transmitted, say with average probability $T$). This is precisely a similar model  equivalent to a bond percolation model with bond occupation probability $T$ on the graph representing the community. The connection between the spread of disease and percolation was in fact one of the original motivations for the percolation model itself. Below in the next subsection, we discuss how disease spread can be realized as the percolation problem  on random graphs with arbitrary degree distributions discussing size of outbreaks, presence of an epidemic, transmissibility thresholds, etc.

\subsubsection{Percolation theory and spreading process}
Like mean-field method (mentioned in section 3), which is a useful tool to analyze the transmission dynamics in homogeneous mixing population, percolation theory also plays a significant role in prediction of epidemic or vaccination dynamics, such as emerging threshold and phase transition, in network population. One convenient universal approach to deal percolation problem with bond occupation probability $T$ on complex networks is considering \textit{generating functions} \cite{newman2002spread}. Generating functions and the problem of percolation in networks have been discussed in many earlier studies \cite{callaway2000network, cohen2000resilience, molloy1995critical, parshani2010epidemic}. We define generating functions $G_0(u)$ and $G_{1}(u)$ to describe a network as follows:
\begin{equation}
G_0(u) = \sum_{k=0}^\infty P(k)u^k,
\end{equation}
\begin{equation}
G_1(u) = \frac{G'_0(u)}{G'_0(1)},
\end{equation}
where $P(k)$ is the degree distribution of random graph $G$ with $N$ vertices. We can calculate back the degree distribution $P(k)$ and also all the higher moments
using $G_0(u)$ with
\begin{equation}
<k^n> = \sum_{k}k^nP(k) = \bigg[\bigg(u\frac{d}{du}\bigg)^nG_0(u)\bigg]_{u=1}.
\end{equation}
$G_{1}$ calculates the distribution of the degrees of vertices reached by following a randomly chosen edge. A modified set of generating functions is required to solve percolation problem with bond probability $T$: $G_0(u, T)$ and $G_{1}(u, T)$, where $G_i(u, T) = G_i(1+(u-1)T)$ \cite{van2012epidemics, van2009virus}. $G_i(u,T),~i=0,1$ represents the distributions of the number of \textit{occupied} edges attached to a vertex, as a function of the transmissibility $T$.

Outbreak size of a disease on a network precisely depends on the distribution of sizes of clusters of vertices together by occupied edges in the corresponding percolation model. Using the generation functions $G_0(u, T)$ and $G_{1}(u, T)$, we can find the mean outbreak size $<s>$ in closed form as
\begin{equation}
<s> = 1+\frac{G_0'(1;T)}{1-G_1'(1;T)}.
\end{equation}
So, given the values of generating function $G_i$, we can completely evaluate this expression to get the mean outbreak size for any value of $T$ and degree distribution. The outbreak size critically depends on the relation $TG'_1(1,T) = 1$. This point indicates the onset of an epidemic; it is the point at which the typical outbreak ceases to be confined to a finite number of individuals, and expands to fill significant fraction of the graph. This defines the critical transmissibility $T_c$, at which the critical transition occurs\\
\begin{equation}
T_c = \frac{1}{G_1'(1)} = \frac{\sum_kkP(k)}{\sum_kk(k-1)P(k)}= \frac{<k>}{<k^2> - <k>}.
\end{equation}
For $T>T_c$, we have an epidemic, or ``giant component'' in the language of percolation. This equation is valid for degree of uncorrelated networks which have no loops in the clusters. However, in the case of heterogeneous networks with degree distribution $P(k)\sim k^{-\zeta}$, it was found that the percolation threshold tends to zero for  $\zeta<3$ in the limit of an infinite network size $N\rightarrow \infty$ (\cite{cohen2000resilience}). That is, an epidemic can spread in this network regardless of how small the infection probability and how quick is the recovery process. In general, the expansion of the generating functions around the nonzero solution yields the scaling behavior of the order parameter in the vicinity of the critical point: $P_G(T) \sim (T-T_c)^{{\beta_I}_{c}}$, where the critical exponent ${\beta_I}_{c}$ assumes following values
\begin{equation}
{\beta_I}_{c} = \left\{
\begin{array}{lll}
1/(3-\zeta)&\mbox{for} &\zeta<3,\\
1/(\zeta-3)&\mbox{for} &3< \zeta \leq 4,\\
1& \mbox{for}& \zeta \geq 4.
\end{array}
\right.
\end{equation}
For the case $\zeta = 3$, a stretched exponential form $P_G(T) \sim e^{1/p}$ is expected. For more details about disease thresholds, we provide a brief summary in Table ~\ref{DiseaseThresh}.

There are other approaches of modeling spreading process of infectious disease over networks such as \textit{Degree-based mean-field theory} (DBMF) and \textit{Individual-based mean-field theory} (IBMF). DBMF considers the dynamics of node of degree $k$ to be infected or remain susceptible, and assume that all the nodes with degree $k$ are statistically equivalent. The value of the epidemic threshold can be obtained by means of a linear stability analysis \cite{boguna2002epidemic, pastor2001epidemic}. It was the first approach to the study of the SIS model in complex networks \cite{pastor2001aepidemic}. Global stability of the disease-free state in such model, the role of seasonal oscillations and the coexistence of an endemic equilibrium state via multistability have been investigated in  \cite{don}. In contrast, the state of the system in the disease model (such as SIS) in the IBMF framework  is fully defined by a set of Bernoulli random variables $X_i(t) \in {0, 1}: X_i(t) = 0$ for a healthy, susceptible node and $X_i(t) = 1$ for an infected node; altogether it defines a $2^N$ Markov chain for a network with $N$ nodes \cite{van2012epidemics, van2009virus}, which specifies exactly the time evolution of the disease model. However, the Markov chain approach complicates analytical calculations of the model. A discussion on both DBMF and IBMF can be found in \cite{pastor2015epidemic}.

\subsubsection{Disease spreading in multilayer networks}
In many situations, disease spread can be modeled over \textit{interconnected networks} instead of a single-layer network (or called \textit{monoplex} network). For example, spreading of an infection amongst groups of communities, or even groups of individuals of different age profiles \ref{AgeMultilayer}. Studying the diffusion of pathogens over interconnected network or \textit{multilayer networks} has got very recent attention in epidemiology. Multilayer networks are made of multiple layers, where each layer is a monoplex network, where the same nodes can appear in multiple layers and nodes on different layers can be connected to each other. One important aspect of spreading infection across multilayer networks is that it can also spread from one layer to another. Generally, there are three possibilities for spreading process \cite{salehi2015spreading, min2014layer}: \textit{same-node, inter-layer}, when infection switches layer but remains on the same node, e.g., when an infected individual travels from one city to another city; \textit{other-node inter-layer}, when infection  continues spreading to another node in another layer, e.g., spreading of infection between individuals in different age profiles through direct physical contact. In third type, \textit{other-node intra-layer}, when infection spreads across the same community. In the context of interacting spreading processes in multilayer networks, two types of thresholds have recently been introduced, called \textit{survival threshold} measuring if infection will survive and \textit{absolute-dominance threshold} indicating whether it can completely remove another competing process \cite{salehi2015spreading}.

Most of the works on modeling the dynamics of diffusion over multilayer networks have used epidemic models such as SIS, SIR \cite{min2014layer, magnani2011ml, wang2013effect, lee2014multiplex, berlingerio2013multidimensional, wasserman1994social, gao2014single, boccaletti2014structure, dickison2012epidemics, zhao2014multiple,wang2014epidemic}. In a similar approach with single-layer networks, the dynamics of epidemic spreading according to the SIS and SIR models over multilayer networks are described as a two- or three-state process, respectively. Thereafter, many extensions have been applied to SIR and SIS models. One of the most important extensions, Goldenberg et al. \cite{goldenberg2001talk} proposed a discrete-time version of the SIR model called Independent Cascade Model (ICM), where time proceeds in discrete time steps. In this model, each infected node $u$ at time $t$ can infect each of its neighbors. If the infection succeeds, then neighbor $v$ will become infected at step $t + 1$. ICM is often used in the literature on influence spreading. In \cite{salehi2015spreading}, the authors extended this model to analyze the dynamics of multiple cascades over multiplex networks.

Unlike a monoplex network, the infection may diffuse over inter- and intra-layer connections at different speeds over multiple networks, meaning that we may have different infection rates (i.e., transmissibilities) across the links of each layer and also the links between the layers. Therefore, most of the works on spreading processes over multilayer networks \cite{dickison2012epidemics, wang2011effects, qian2012diffusion, zhao2014multiple, marceau2011modeling, buono2014epidemics, funk2010interacting, wang2013effect, sanz2014dynamics} have extended epidemic-like models by considering different infection rates dependent on the types of the layers. A recent contribution in this context proposed a generalized epidemic mean-field (GEMF) framework, which was capable of modeling epidemic-like spreading processes with more complex states in multiplex network layers (compared to two or three states in the SIS and SIR models).

\subsubsection{Simulation-based models}
Simulation-based modelling, like analytical models described above, has similar appeal in epidemic theory \cite{eames2002modeling, eubank2004modelling, meyers2005network, read2003disease, wallinga1999perspective, watts1998collective, chao2010flute, merler2010role}. There are many good examples of simulation based network models of infectious diseases such as sexually transmitted diseases (STD) that highlight the importance of network structure (e.g., role of core groups or interconnected individuals with a large number of contacts) in disease transmission and persistence \cite{garnett1996sexually, ghani1997role, morris1997concurrent, potterat1999network, klovdahl2001networks, rothenberg2001net, mcelroy2003network, szendroi2004polynomial, doherty2005determinants}.

There are several simulation-based works that attempt to model airborne disease outbreaks \cite{halloran2002containing, cohen1997social, wallinga1999perspective}. Such models consider large population, like of order $\sim 10^6$ with certain specified social connections among individuals. The networks used are generated by computer simulation to conform to several observed demographic and social characteristics such as given age distribution, household size, public crowding. Variable activities in the entire day are also assigned to the individuals such as working place, public gathering, or even spending time at home. Different activities reflect the change in contact pattern in different time of the day. Children are assigned to school, day-care centre or play group where they interact (and therefore form connections) with other children. The simulation model puts emphasis particularly on transmission within households and family groups, which are probably the main routes of transmission for this disease. There are models that attempt similar tasks, using census data to determine interaction patterns estimating demographic and disease parameters \cite{eubank2004modelling}. Despite the number of approximations involved, the inherent stochasticity of such microsimulations allows a direct estimation of the variability between epidemics. For example, Meyers et al \cite{meyers2005network} generates a network model to characterize the spread of SARS in Vancouver, British Columbia. They have used census data of Vancouver and using computer simulations generated a plausible contact network that reflects an urban setting. They had chosen $N = 1000$ households at random from the Vancouver household size distribution, which yields approximately 2600 people. Household members were given ages according to the census data of Vancouver. Kids were then assigned to schools according to school and class size distributions to occupations according to (un)employment data, to hospitals as patients and caregivers according to hospital employment and bed data, and to other public places. Within each location then random connections were created among individuals with variable probabilities at households, hospitals and schools, workplaces, and other public places. They have also chosen the parameters of the Poisson and power law networks so that all three networks share the same epidemic thresholds. Conventional epidemiological models shows clear threshold effects, that is, the disease is expected to die out when transmission probability falls below a critical threshold level. Olinky and Stone \cite{olinky2004unexpected}  have shown that epidemic propagation depends equally on the infection process and also on the network topology. Their recent analyses of infection processes on highly heterogeneous networks (e.g., scale-free networks) concludes that diseases spread and persist even for vanishingly small transmission probabilities.

However, network-based simulation models have certain pitfalls as well. As mentioned by Keeling and Eames \cite{keeling2005networks}, these models are limited to ascertain the sensitivity of the epidemiological results to the details of the network structure. For example, the study by Halloran et al \cite{halloran2002containing} may be limited to answer questions like whether the network is representative of an average American community, whether variation between communities will bias the results if large population sizes are considered and whether rare but epidemiologically important contact structures are missing from the network. It is difficult to answer such questions or gain an intuitive understanding of network structure if our experience is limited to simulations of sampled networks. However, a range of idealized networks and analytical tools have been developed that can reveal the elements of network structure, which are important determinants of epidemic dynamics.

\subsection{Vaccinations over networks}
Designing optimal vaccination strategies to prevent and control infection has been widely addressed in the epidemic modeling of infectious diseases. Vaccination over networks can be modeled as a site percolation problem \cite{albert2000error}. Each vaccinated node can be regarded as a site which is removed from the network. The goal of the vaccination process thus is to reduce the transmissibility and pass the percolation threshold, which leads to minimization of the number of infected individuals in the network. Thus implementing the SIR model and vaccination can be considered as a site - bond percolation model, and the vaccination is considered successful if the network is below the percolation threshold \cite{albert2000error,madar2004immunization}. However though, the epidemic threshold decreases with the standard deviation of the connectivity distribution in the heterogeneous networks \cite{anderson1992infectious, pastor2001epidemic, pastor2001aepidemic}. This feature is paradoxically amplified in SF networks that have diverging connectivity fluctuations. In fact, epidemic processes in SF networks do not possess, in the limit of an infinite network, an epidemic threshold below which diseases cannot produce a major epidemic outbreak or the inset of an endemic state. SF networks are, therefore, prone to the spreading and the persistence of infections, whatever virulence the infective agent might possess. In view of this, it becomes a major task to public health policy makers designing optimal vaccination strategies oriented to minimize the risk of epidemic outbreaks on networks such as SF networks. However, population with heterogeneous connected networks may take different paths to extinction and thus control of infection. A recent paper by Hindes and Schwartz (2016) discussed such issues on heterogeneous random networks under various configuration \cite{hindes2016epidemic}.

The simplest hypothetical way to experiment the effect of vaccination over networks is the random introduction of immune nodes in the network population, in order to get a \textit{uniform vaccination} density \cite{anderson1992infectious, heesterbeek2000mathematical}. Immune nodes cannot become infected and, thus, do not transmit the infection to their neighbors. In this case, for a fixed spreading rate $\beta_I$, the relevant control parameter is the immunity $x$, defined as the fraction of vaccinated nodes present in the network. At the mean-field level, the presence of uniform vaccination will effectively reduce the spreading rate $\beta_I$ by a factor $(1-x)$; i.e. the probability of finding and infecting a susceptible and non-immune node. In homogeneous networks, such as the Watts-Strogatz (WS) model, it is easy to show that in the case of a constant $\beta_I$, the critical vaccination rate $x_c$ is given by
\begin{equation}
x_c = 1-\frac{{\beta_I}_c}{\beta_I}.
\end{equation}
Thus, the critical vaccination that achieves eradication is related to the spreading rate and the epidemic threshold of the infection, which implies that the critical vaccination allows the complete eradication of the disease over the network.

However, in an heterogeneous population with a uniform vaccination scheme, it is necessary to vaccinate a fraction of the population larger than the estimate given by a simple (homogeneous) assumption \cite{anderson1992infectious}. In this case, it can be proved that optimal vaccination programs can eradicate the disease by  vaccinating a smaller number of individuals. A straightforward way to reintroduce an intrinsic vaccination threshold in heterogeneous networks consists in using different fractions of vaccinated individuals according to their connectivity. Let us define $x_k$ as the fraction of immune individuals with a given connectivity $k$ and averaging $x_k$ over the various connectivity classes, we have the critical vaccination threshold as
\begin{equation}
x_c = \sum_{k>\beta_I^{-1}}\bigg(1-\frac{1}{\beta_I k}\bigg)P(k).
\end{equation}

Scale-free networks can be considered as a limiting case of heterogeneous systems and it is natural to look for specifically devised vaccination strategies. This however, does not work on SF network. Because of the absence of any epidemic threshold in SF networks, it becomes impossible to find any critical vaccination rate that ensures the eradication. In scale-free (SF) networks, the pervious equation takes the form
\begin{equation}
1-x_c = \frac{1}{\beta_I} \frac{<k>}{<k^2>}.
\end{equation}
This shows that only a complete vaccination $(x_c = 1)$ of the network works to eradicate the disease over scale free network with $<k^2> \rightarrow \infty$.

While uniform or proportional vaccination fails to protect contagion over scale-free networks due to its peculiar nature, it has been seen that scale-free networks are strongly affected by selective damage. For example, if a few of the most connected nodes are removed, the network suffers a dramatic reduction of its ability to carry information \cite{albert2000error, callaway2000network, cohen2000resilience}. With this idea, one can design a \textit{targeted vaccination} scheme to progressively make immune the most highly connected nodes, i.e., the ones more likely to spread the disease. In scale-free networks, it produces an significant increase of the network tolerance to infections at the price of a tiny fraction of immune individuals \cite{barabasi2009scale,pastor2002immunization,madar2004immunization}.

Suppose, a fraction $x$ of all nodes with connectivity $k>k_c$ are vaccinated in a network. Then $x = \sum_{k>k_c}P(k)$. At the same time, this also implies that all links emanating from vaccinated individuals can be considered as if they were removed. So, if we assume that the fraction of links is effectively removed from the network, the new connectivity distribution after the vaccination of a fraction $x$ of the most connected node is
\begin{equation}
P_x(k) = \sum_{q>k}^{k_c}P(q){q\choose{k}}(1-p)^kp^{q-k}.
\end{equation}
This gives, calculating the first two moments, the critical fraction of vaccination needed to eradicate the infection given by the relation:
\begin{equation}
\frac{<k^2>_{x_c}}{<k>_{x_c}}\equiv\frac{<k^2>_c}{<k>_c}[1-p(x_c)]+p(x_c) = \beta_I^{-1},
\end{equation}
where $<k>_c = \sum_{k_{min}}^{k_c}kP(k)$, $<k^2>_c = \sum_{k_{min}}^{k_c}k^2P(k)$. An explicit calculation for Barab{\'a}si-Albert network model, the approximate solution for the vaccination threshold is given by
\begin{equation}
x_c \sim \exp(-2/k_{min}\beta_I).
\end{equation}
This clearly indicates that the targeted vaccination program is extremely convenient in scale-free networks where the critical vaccination is exponentially small in a wide range of spreading rates $\beta_I$. Also in this case, the present result can be generalized for scale-free networks with arbitrary connectivity exponent $\zeta$.

A comparative analysis of targeted vaccination over SW and BA models exhibit contrastive results \cite{pastor2002immunization}. In the case of the SW networks, the behavior of the prevalence as a function of $x$ is equivalent in the uniform and targeted vaccination procedures. On the contrary, in the case of the BA networkss, we observe a drastic variation in the prevalence behavior, though it is sensitive to random choice of immune nodes. This however, confirms that targeted strategies do not have a particular efficiency in systems with limited heterogeneity, but are highly sensitive to the network with higher degree of heterogeneity such as scale-free networks. Targeted vaccination of a small fraction of the most connected nodes in scale-free networks shows significant impact on spread process of infection.

Another target-based strategy is \textit{acquaintance vaccination} that tries to target all of the most highly connected nodes for vaccination \cite{pastor2002immunization, gallos2007improving,madar2004immunization, liu2009common, cohen2003efficient}, but this strategy requires no knowledge of the node degrees or any other global knowledge. In this approach a random group of nodes are chosen and then a random set of their neighbors are selected for vaccination. The most highly connected nodes are far more likely to be in this group of neighbors. So immunizing this group results in targeting the most highly connected nodes, and it requires far less information about the network \cite{cohen2003efficient, christakis2010social, krieger2003focus}.  Another variant of this strategy again calls for the random selection of nodes but instead asks for one of their neighbors with a higher degree, or at least more than a given threshold degree and immunizes them \cite{gallos2007improving}. These degree based strategies consistently require fewer nodes to be vaccinated and as such improve a network's chances against epidemic attacks \cite{cohen2003efficient,madar2004immunization}.

This however, is just a brief summary about vaccination on networks.  For more details about mathematical and physics framework, please see next section.


\section{Non-behavioral epidemiological vaccination on networks}\label{sec:vac-net}

Due to the potential threat posed by the spreading of infectious disease, ample research has been devoted to the mitigation and prevention of epidemics. To date, one of the most popular and effective methods is network vaccination \cite{pastor2002immunization,barrat2008dynamical,pastor2015epidemic,barabasi2016network,bornholdt2003handbook,gao2011network,liu2016biologically}, where certain nodes in network are effectively immunized by a perfect vaccine, and are thus no longer able to transmit the disease to their neighbors \cite{cornforth2011erratic}. More precisely, given a fixed effevctive spreading rate $\beta_I/\gamma$ ($\beta_I$ and $\gamma$ are the infection rate and recovery rate respectively. Without loss of generality, $\gamma=1$ is fixed since it only affects the definition of the time scale of epidemic spreading.), the relevant control parameter is the immunity rate $x$, which is simply defined as the fraction of vaccinated nodes in a network. The challenge is how to arrive at a minimal value of $x$, which is the so-called vaccination threshold $x_c$, to reduce the spreading rate of the disease $\beta_I$ under the critical threshold ${\beta_I}_c$, such that the disease dies out completely \cite{wang2003epidemic,chakrabarti2008epidemic,restrepo2006characterizing}.

This section is devoted to the systematic review of such non-behavioral vaccination techniques, where networks are effectively immunized by means of a targeted vaccination of certain nodes. We will review research done on single-layer and multilayer networks, on static and adaptive networks, and theoretical as well as empirical results.

\subsection{Uniform (or random) vaccination}
The simplest vaccination strategy that one can conceive, the implementation of which requires no preparation or information at all, is uniform or random vaccination~\cite{anderson1992infectious,pastor2002immunization}. Here a fraction $x$ of nodes are randomly selected and then vaccinated. Accordingly, only the remaining $1-x$ fraction of nodes contributes to the spreading of the
disease. The effective number of neighbors that each
susceptible node possesses thus decreases from its degree $k$ to $k (1-x)$. If one considers the standard SIS model as an example, then after vaccination the evolution equation of the probability $\rho^I_k(t)$ that a node with degree $k$ is infected at time
$t$ is
\begin{equation}\label{}
\frac{d\rho^I_k(t)}{dt}=-\rho^I_k(t)+\beta_I k (1-x)[1-\rho^I_k(t)]\sum_{k'}P(k'|k)\rho^I_{k'}(t),
\end{equation}
where the first term defines the rate at which infected nodes of degree $k$ recover and become susceptible again, and in the second term $P(k'|k)$ refers to the conditional probability that an infected node with degree $k'$ points to the node with degree $k$.

It can be observed that in the mean-field model, where individuals interact randomly with one another, the uniform vaccination is equivalent to the simple decrease of the spreading rate of the disease from $\beta_I$ to $\beta_I (1-x)$. Since eradicating an epidemic by uniform vaccination requires
$\beta_I$ to be below its critical epidemic threshold ${\beta_I}_c$, we obtain \begin{equation}\label{x_c}
\beta_I(1-x_c)={\beta_I}_c,
\end{equation}
where $x_c$ is the critical vaccination threshold, above which the density of infected individuals in the stationary state is zero.

However, in predominantly homogeneous networks, such as random or small-world networks, ${\beta_I}_c=1/(\langle k\rangle+1)$ for the SIS model and ${\beta_I}_c=1/\langle k\rangle$ for the SIR model \cite{barrat2008dynamical,pastor2001epidemic}. Based on Eq.~\ref{x_c}, the vaccination threshold is
\begin{equation}\label{}
 x_c=1-\frac{1}{\beta_I}\frac{1}{\langle k\rangle+1}
\end{equation}
for the SIS model, and
\begin{equation}\label{}
 x_c=1-\frac{1}{\beta_I}\frac{1}{\langle k\rangle}
\end{equation}
for the SIR model. Thus, if the uniform vaccination level is larger than $x_c$, homogeneous
networks will be completely protected and no large epidemic outbreaks will be possible.

Conversely, the situation is much different for strongly heterogeneous networks, such as scale-free networks, where the epidemic threshold vanishes \cite{cohen2000resilience,cohen2001breakdown}. Here uniform vaccination will fail. In particular, the absence of an epidemic threshold $({\beta_I}_c=0)$ in the thermodynamic limit implies that no matter how one rescales
${\beta_I}\rightarrow {\beta_I} (1-x)$, the epidemic will not be stopped unless $x=1$.

Based on the fact that in heterogenous networks ${\beta_I}_c=\langle k\rangle/\langle k^2\rangle$ for the SIS model and ${\beta_I}_c=\langle k\rangle/(\langle k^2\rangle-\langle k\rangle)$ for the SIR model \cite{pastor2001epidemic,lloyd2001viruses,moreno2002epidemic}, and by using
Eq.~\ref{x_c}, the threshold of uniform vaccination is given by
\begin{equation}\label{}
 x_c=1-\frac{1}{{\beta_I}}\frac{\langle k\rangle}{\langle k^2\rangle}
\end{equation}
for the SIS model and
\begin{equation}\label{}
 x_c=1-\frac{1}{{\beta_I}}\frac{\langle k\rangle}{\langle k^2\rangle-\langle k\rangle}
\end{equation}
for the SIR model.

In a scale-free network with degree exponent $\zeta <3$, we have $\langle
k^2\rangle\rightarrow \infty$, which indicates $x_c\rightarrow 1$ for both the SIS and the SIR model \cite{pastor2002immunization,pastor2001epidemic}. In other words, a complete vaccination
of the network is required to stop an epidemic. As emphasized before, uniform
vaccination is largely inefficient for disease eradication in heterogenous
networks \cite{callaway2000network,cohen2000resilience}. This theoretical prediction is also consistent with the finding that
many infectious diseases require that 80\%-100\% of the
population be vaccinated to prevent an outbreak. For example, measles require 95\% of the population be vaccinated \cite{anderson1992infectious}, while for digital viruses the strategies
relying on uniform vaccination call for practically 100\% of the
computers to install the appropriate antivirus software
\cite{pastor2001epidemic,jeong2000large,cohen2000resilience}.

\subsection{Targeted vaccination}

\subsubsection{Degree-based targeted vaccination: theoretical prediction of the vaccination threshold}

The problems encountered with uniform vaccination are rooted in the
vanishing epidemic threshold. Therefore, to successfully eradicate a
disease in a heterogenous network, we must increase the epidemic
threshold. This requires us to modify the underlying contact network
to reduce the degree variance $\langle k^2\rangle$. Evidently, large-degree nodes
are known to be responsible for the large variance of heterogenous
networks. Thus, if we vaccinate only the nodes whose degree exceeds a
preselected threshold value $k_c$, it should be easy to decrease the variance and increase the
epidemic threshold \cite{pastor2002immunization,dezsHo2002halting,barabasi2016network}. To this aim, Pastor-Satorras and Vespignani have proposed a degree-based targeted vaccination strategy in which the most highly connected
nodes are progressively immunized, given that these are the ones that are most likely to spread the disease \cite{pastor2002immunization,chen2008finding,borgatti2005centrality,cohen2003efficient,dezsHo2002halting,eames2009epidemic,gallos2007improving,gao2011network,miller2007effective,schneider2011suppressing,vidondo2012finding,wang2009imperfect}.

The vaccination of a fraction $x$ of large-degree nodes can be regarded as the removal of nodes whose degree is larger than a certain value $k_c$. Accordingly, $x$ thus can be implicitly defined as
\begin{equation}\label{x}
 x=\sum_{k=k_c+1}^\infty P(k).
\end{equation}
Moreover, the average $\langle k\rangle_{c}$ and the second moment $\langle k^2\rangle_{c}$ of the degree distribution with maximum degree $k_c$ of the network are
\begin{equation}\label{k}
\langle k\rangle_{c}=\sum_{k=k_{min}}^{k_c}kP(k),
\end{equation}
and
\begin{equation}\label{k2}
\langle k^2\rangle_{c}=\sum_{k=k_{min}}^{k_c}k^2P(k),
\end{equation}
where $k_{min}$ denotes the minimal degree of the node in a network.

However, the removal of large-degree nodes will modify the degree distribution of the remaining nodes, owing to the deletion of the links between removed and remaining nodes. The probability that the link connected to the removed nodes is also removed equals
the probability that the link points to a node whose degree is larger than $k_c$. We thus have\\
\begin{equation}\label{f}
 f=\sum_{k=k_c+1}^\infty \frac{kP(k)}{\langle k\rangle},
\end{equation}
\\
and the degree distribution of the resulting network becomes~\cite{pastor2002immunization,cohen2001breakdown}\\
\begin{equation}\label{}
P'(k')=\sum_{k=k'}^\infty\left (\begin{array}{l} k\\k'\end{array}\right ) f^{k-k'}(1-f)^{k'}P(k).
\end{equation}
\\
Finally, the average degree $\langle k'\rangle$ and the second moment $\langle k'^2\rangle$ of the resulting network are \cite{pastor2002immunization,cohen2001breakdown}\\
\begin{equation}\label{k'}
\langle k'\rangle=\sum_{k'=k_{min}}^{k_c}k'P'(k')=(1-f)\langle k\rangle_{c},
\end{equation}
\\
and\\
\begin{equation}\label{k'2}
\langle k'^2\rangle=\sum_{k'=k_{min}}^{k_c}k'^2P'(k')=(1-f)^2\langle k^2\rangle_{c}+f(1-f)\langle k\rangle_{c},
\end{equation}
\\
respectively.

For the SIS model, the epidemic threshold of the resulting network can be calculated as\\
\begin{equation}\label{}
{\beta_I}'_c=\frac{\langle k'\rangle}{\langle k'^2\rangle}=\frac{(1-f)\langle k\rangle_{c}}{(1-f)^2\langle k^2\rangle_{c}+f(1-f)\langle k\rangle_{c}}.
\end{equation}
\\
Assuming $2<\zeta<3$ and combining Eqs.~\ref{k}-\ref{f}, we obtain~\cite{barabasi2016network}\\
\begin{equation}\label{}
{\beta_I}'_c=\bigg[\frac{3-\zeta}{\zeta-2}k_c^{3-\zeta}k_{min}^{\zeta-2}-\frac{3-\zeta}{\zeta-2}k_c^{5-2\zeta}k_{min}^{2\zeta-4}+k_c^{2-\zeta}k_{min}^{\zeta-2}\bigg]^{-1}.
\end{equation}
\\
A similar approach to the SIR model yields\\
\begin{equation}\label{}
{\beta_I}'_c=\bigg[\frac{3-\zeta}{\zeta-2}k_c^{3-\zeta}k_{min}^{\zeta-2}-\frac{3-\zeta}{\zeta-2}k_c^{5-2\zeta}k_{min}^{2\zeta-4}+k_c^{2-\zeta}k_{min}^{\zeta-2}-1\bigg]^{-1}.
\end{equation}
\\
In both the SIS and the SIR model, if $k_c \gg k_{min}$, we have \cite{barabasi2016network}\\
\begin{equation}\label{}
{\beta_I}_c\approx
\frac{\zeta-2}{3-\zeta}{k_{min}^{2-\zeta}}{k_c^{\zeta-3}}.
\end{equation}
\\
For here it follows that if more large-degree nodes are vaccinated, and thus $k_c$ becomes smaller, the epidemic threshold will become larger. By vaccinating an appropriate fraction of large-degree nodes, we enable ${\beta_I}_c$ to increase beyond the spreading rate ${\beta_I}$ of the disease. Accordingly, the network is then protected completely.

In order to assess the efficiency of degree-based targeted vaccination, it is useful to derive the explicit expression for the vaccination threshold in an uncorrelated network with a power-law degree distribution $P(k)=ck^{-\zeta}$ and $c \approx (\zeta-1)/k_{min}^{-\zeta +1}$, where $\zeta=3$ \cite{pastor2002immunization}. In this case, we obtain
$x=k_{min}^2k_c^{-2}$, $f=x^{1/2}$, $\langle k\rangle_{c}=2k_{min}$, and $\langle k^2\rangle_{c}=2k_{min}^2\textit{\textit{ln}}(x^{-1/2})$ based on Eqs.~\ref{x}-\ref{f}. The critical vaccination threshold $x_c$ needed to eradicate the disease is then
\begin{equation}\label{b}
\frac{\langle k'\rangle}{\langle k'^2\rangle}={\beta_I}.
\end{equation}
By inserting above values into Eq.~\ref{b} and combining Eqs.~\ref{k'}-\ref{k'2},
we obtain the approximate solution for the vaccination threshold in the case of targeted vaccination \cite{pastor2002immunization}
\begin{equation}\label{}
x_c\sim \emph{\emph{exp}}(-2/k_{min}{\beta_I}).
\end{equation}
This equation indicates that the degree-based targeted vaccination strategy can be very effective, giving a critical vaccination threshold that is exponentially small for a wide range of spreading rates.

\begin{figure}
\centering
\includegraphics[scale=0.5,trim=50 0 50 0]{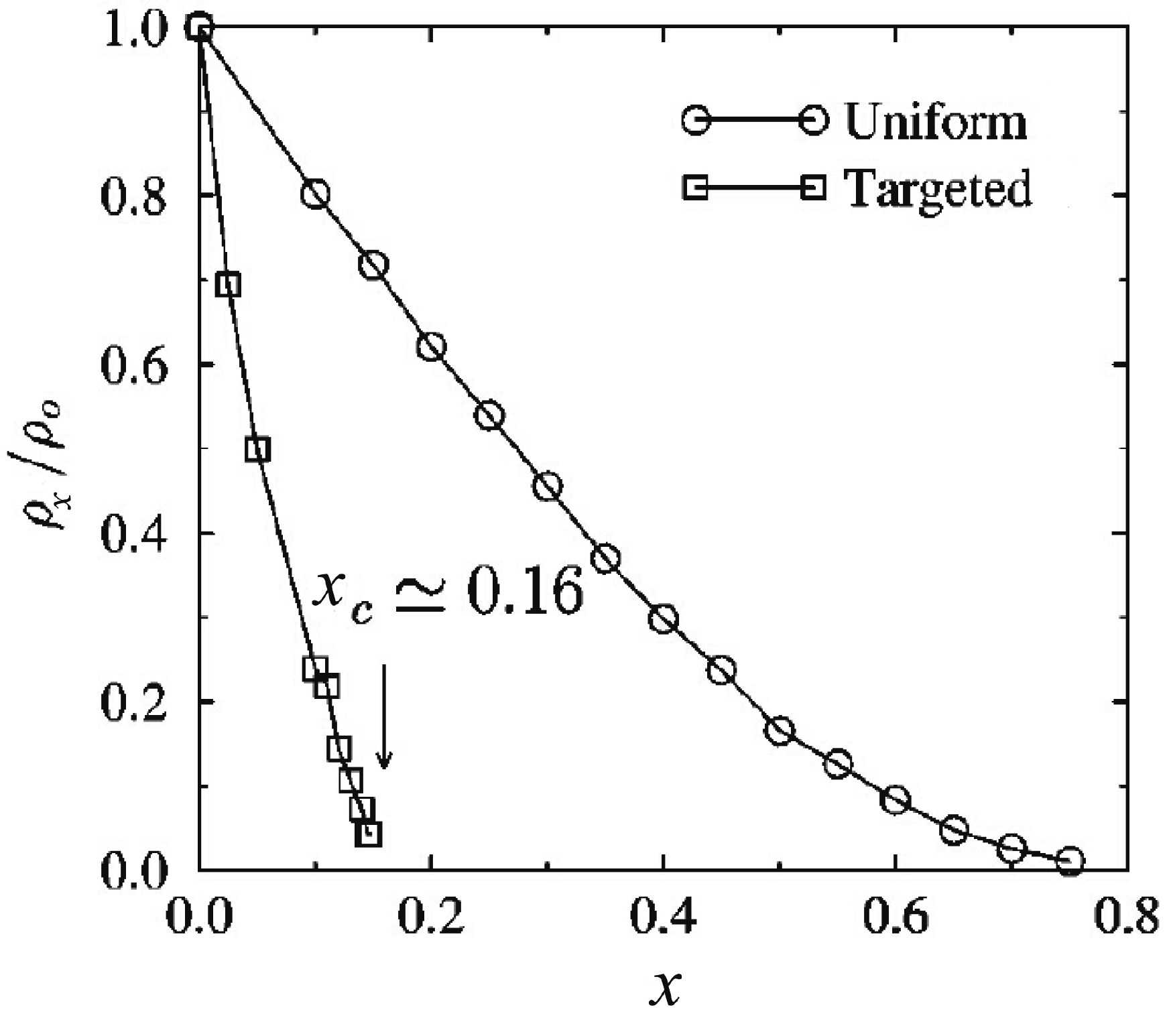}
\caption{Reduced prevalence of disease $\rho_x/\rho_0$ for the SIS model on a scale-free network, as obtained with random (circles) and targeted (squares) vaccination at a fixed spreading rate ${\beta_I} = 0.25$, respectively. A linear extrapolation from the largest values of $x$  yields an estimation of the threshold $x_c\simeq  0.16$ for targeted vaccination.
Source: Adapted with permission from Ref.~\cite{pastor2002immunization}. Copyrighted by the American Physical Society.}\label{threshold}
\end{figure}

This theoretical prediction can be tested numerically with simulations of the SIS model on a scale-free network. Results presented in Fig.~\ref{threshold} compare the efficiency of targeted vaccination and random vaccination, respectively. It can be observed that for random vaccination the disease prevalence decays very slowly and vanishes only in the $x\rightarrow 1$ limit. Conversely, for targeted vaccination there is a very sharp drop towards
the vaccination threshold (indicated by the arrow), above which the
system is infection-free. A linear regression from the largest
value of $x$ yields an approximate vaccination threshold $x_c
\simeq 0.16$ in this case, which proves that scale-free networks are
very sensitive to degree-based targeted vaccination even if a very small
fraction of large-degree nodes is selected for vaccination

\subsubsection{Centrality-based targeted vaccination: definitions and simulation-based models}

The procedure of degree-based targeted vaccination is equivalent to altering the structure of the network. Namely, by vaccinating the large-degree nodes, we fragment the contact network, thus making it more difficult for the disease to reach the nodes in the other components. Based on this principle, it is possible to devise alternative vaccination strategies which take advantage of the heterogenous connectivity patterns in order to achieve a high level of tolerance towards infections~\cite{pastor2002immunization}. Accordingly, other centrality indexes, such as betweenness, closeness, eigenvalue, or PageRank, can be used to design targeted vaccination strategies~\cite{friedl2010critical,chen2008finding,hebert2013global,latora2007measure,borgatti2005centrality,schneider2011suppressing,ventresca2013evaluation,salathe2010dynamics,schneider2011suppressing,schneider2012inverse,bonacich1987power,restrepo2006characterizing,miller2007effective,page1999pagerank,yang2016immunization,yang2011control,du2016identifying}. In what follows, we will briefly survey these centrality indexes and their applications in targeted vaccination.

\emph{Betweenness centrality}.  Betweenness centrality is defined as
the number of shortest paths between pairs of nodes that pass through a given node~\cite{freeman1978centrality,christley2005infection,holme2002attack}. More precisely, if $\sigma_{vw}$ is the total number of shortest paths from node $v$ to node
$w$ and $\sigma_{vw}(i)$ is the number of those shortest paths that pass through the node $i$, the betweenness of $i$ is given by $b_i=\sum_{v\neq w\neq i}\frac{\sigma_{vw}(i)}{\sigma_{vw}}$.

Since nodes with large betweenness centrality are often bridges between different communities in the network \cite{yu2010finding,salathe2010dynamics}, the vaccination of these nodes has a good potential to break up a large network into smaller parts.
Hence, if an epidemic starts in a particular component of the network, it cannot infect nodes in
other components. Additionally, betweenness centrality is an
effective measure to identify high risk individuals, as it determines the
volume of flow passing through each node \cite{borgatti2005centrality}. That is also one of the main reasons why largest betweenness-based vaccination is considered to be the most
effective targeted vaccination algorithm \cite{shams2014using}.

\emph{Random-walk centrality}. Betweenness is, in some sense, a
measure of the influence a node has over the spread of the epidemic
through the network. By counting shortest paths, betweenness-based vaccination simply assumes that epidemics spread only along those shortest paths, which, however, is inconsistent with most empirical observations. To relax this hypothesis, Newman proposed  the random-walk centrality that involves the contributions from essentially all the paths between nodes, though it still gives more weight to short paths \cite{newman2005measure}. The random-walk centrality of a node $i$ is a measure based on random walks, counting how often the node $i$ is traversed by a random walk
between any pair of nodes $s$ and $v$
\begin{equation}\label{}
r_i=\sum_{s<v}I_{sv}(i),
\end{equation}
where $I_{sv}(i)=\frac{1}{2}\sum_jA_{ij}|T_{is}-T_{iv}-T_{js}+T_{jv}|$
for $i\neq s, v$, and $T_{is}$ is the element in the voltage matrix. This measure is particularly useful for finding high-centrality vertices that do not happen to lie on geodesic paths or on the paths formed by maximum-flow cut-sets, which are still important for the epidemic spreading \cite{newman2005measure}. Salath\'e and Jones have proven that vaccination strategies based on random-walk
centrality can result in the lowest number of infected cases if the
vaccination coverage is low \cite{salathe2010dynamics}.

\emph{Closeness centrality}. Closeness centrality is based on the
assumption that nodes with a short distance to other nodes can spread
the disease very effectively across the network \cite{borgatti2005centrality,freeman1978centrality}. It is defined as
\begin{equation}\label{}
c_i=\frac{1}{\sum_{j\neq i}l_{ij}},
\end{equation}
where $l_{ij}$ is the distance between nodes $i$ and $j$. This measure
gives a large centrality to nodes which have small shortest path
distance to the other nodes. Therefore, immunizing largest-closeness nodes not only vaccinates high risk individuals, but also postpones epidemic spreading through the network \cite{christley2005infection,freeman1978centrality}.

\emph{Eigenvector centrality}. Degree centrality awards one
``centrality point'' for every network neighbor a node has. In reality, however,
all the neighbors are not equivalent. Under many circumstances, a node's
importance in one network increases via building connections to other
nodes that are themselves important, which is the concept behind
eigenvector centrality \cite{shams2014using,bonacich1972factoring,miller2007effective,bonacich2007some}. Instead of awarding just one point for each neighbor, eigenvector centrality gives each node
a score proportional to the sum of the score of its neighbors. That
is, the eigenvector centrality $e_i$ of a node $i$ is proportional
to the sum of the eigenvector centrality of the nodes it is
connected to. It is defined as
\begin{equation}\label{}
e_i=\Lambda^{-1}\sum_j a_{ij}e_j,
\end{equation}
where $a_{ij}$ denotes the adjacency matrix of the network, $\Lambda$ is the largest eigenvalue and $e_i$ is the $i^{th}$
component of the eigenvector associated with $\Lambda$ of the
network.

The impact of eigenvector centrality on epidemic control has been
evaluated by \cite{bonacich1972factoring,tomovski2012simple}. Tomovski and Kocarev used nonlinear system stability analysis of the SIS model and proved that the
epidemic threshold equals to the reciprocal value of the largest
eigenvalue of the network adjacency matrix \cite{tomovski2012simple}, that is
\begin{equation}\label{}
{\beta_I}_c=\frac{1}{\Lambda}.
\end{equation}
Therefore, the risk of infection through networks can be reduced
by decreasing $\Lambda$. Along the same lines, Masuda defined $I_i$ as the decrement
of $\Lambda$ owing to the removal of node $i$ \cite{masuda2009immunization}
\begin{equation}\label{}
I_i=-\frac{\Delta \Lambda}{\Lambda}\approx\frac{e_i^2}{\sum_j e_j^2}.
\end{equation}
From here it is clear that the vaccination of a node with a large $e_i$ ($I_i$) will
significantly increase the epidemic threshold, and thus the epidemic will be effectively
suppressed.

\emph{PageRank centrality}. The PageRank centrality \cite{page1999pagerank},
introduced by Google for webpage ranking, is given by
\begin{equation}\label{}
p_i=c\sum_{j\in B_i}\frac{p_j}{N_j}+(1-c)1/N,
\end{equation}
where $i$ is a web page, $B_i$ represents the set of pages that point to $i$,
and $N_j$ denotes the number of pages that $j$ points to. Moreover, $c$ is a random jumping factor which is introduced by assuming that the web surfer
will browse the web pages along the links with probability $c$, and leave the current page and open a random page with probability $1-c$. Similar to eigenvector centrality, PageRank supposes that the importance of a node is determined by both the quantity and
the quality of the nodes that point to it. In terms of epidemic spreading, nodes with large PageRank centrality are more likely to be infected
or infect others \cite{miller2007effective}. Their
vaccination can thus eliminate a large number of potentially very effective disease transmission routes. Since nodes with large PageRank centrality have many low-degree neighbors, largest PageRank vaccination will immunize influential nodes whose vaccination choice can strongly protect their neighbors \cite{miller2007effective}.

\begin{figure}[!htb]
\centering
\includegraphics[scale=0.4,trim=50 0 50 0]{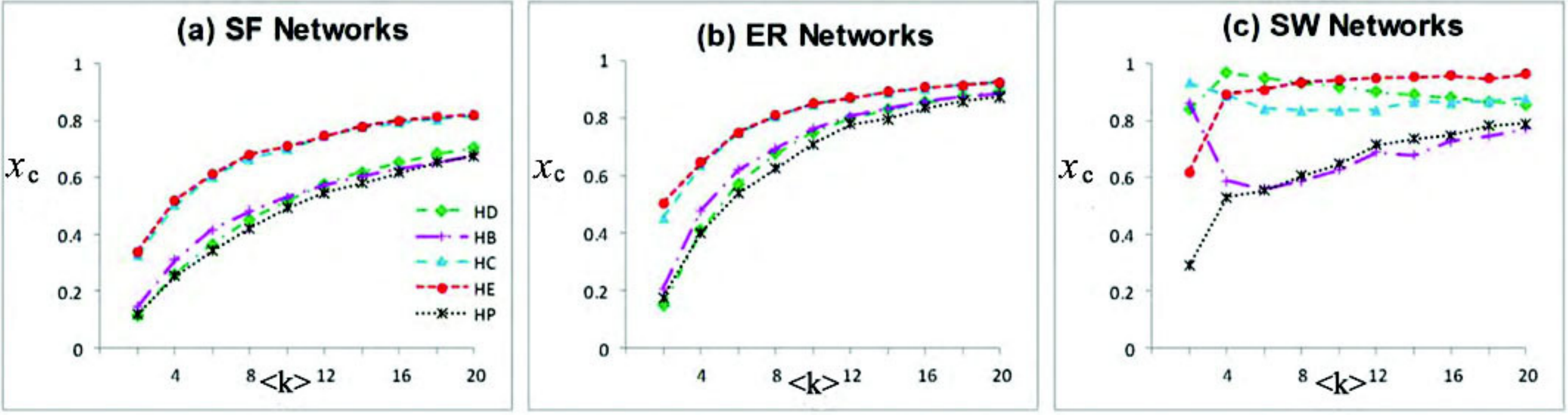}
\caption{The critical threshold of vaccination $x_c$, as a function of the average degree $\langle k\rangle$ on different networks. The legend refers to the following vaccination strategies: targeted vaccination based on degree (HD), betweenness (HB), closeness (HC), eigenvalue (HE), and PageRank (HP). Source: Reprinted figure from Ref.~\cite{shams2014using}}
\label{gcvsk}\end{figure}

To obtain a better understanding of all these proposed vaccination schemes, Shams et al. compared the efficiency of different centrality indexes in terms of the vaccination threshold $x_c$, and plotted it as a function of the average degree $\langle k\rangle$ for different network models \cite{shams2014using}. As shown in Fig.~\ref{gcvsk}, it can be observed that HP (for the meaning of the abbreviation please refer to the caption of Fig.~\ref{gcvsk}) delivers the best performance on all the networks, HB is second-best, while HE and HC deliver the worst performance. Despite of the weak performance of HD in small-world networks, the approach performs well on scale-free and ER random networks.

Moreover, we point out that targeted vaccination techniques can be improved by adaptive strategies, in which the centrality is recalculated for the network of unvaccinated nodes at each step of the vaccination process \cite{holme2002attack}. For example, degree-based vaccination with dynamic re-ranking entails an immediate update of node degrees, and then the nodes with the largest number of unvaccinated neighbors are vaccinated at each time step \cite{miller2007effective}. When a node has been vaccinated, the adaptive betweenness centrality is recalculated for the remaining network composed of unvaccinated nodes~\cite{holme2002attack,schneider2011suppressing,schneider2012inverse,miller2007effective}. Although such adaptive approaches are more efficient from the theoretical viewpoint, they usually require more information, and the complexity of the procedure might make a practical implementation difficult for all but very small populations \cite{schneider2012inverse}.

\subsection{Vaccination without global knowledge}

Although targeted strategies are in principle very effective, an important drawback in terms of practical applications in the real world is that they require complete knowledge of the network structure. Only so it is possible to identify and then vaccinate the most influential nodes as determined by the employed metric. However, the complete structure of the network is seldom, if ever, known, and often it is also not well-defined. In social networks, for example, the number of links an individual has depends strongly on the criteria based on which the network is constructed. To overcome the lack of global knowledge of the network structure, several different strategies have been proposed that require information only about the local structure of the network. The efficiency of these vaccination strategies can nevertheless be very similar to that of targeted vaccination, as reviewed in what follows.

\subsubsection{Acquaintance vaccination: theoretical prediction of the vaccination threshold}

The so-called acquaintance vaccination was first proposed by Cohen et al.~\cite{cohen2003efficient,madar2004immunization}, in which a fraction $p$ of nodes is
selected at random, with the constraint that each node is required to have at least one connection to another individual in the network. Subsequently, this neighbor, rather than the originally selected node, is vaccinated. The strategy thus requires no knowledge of the degree of nodes or other global knowledge about the network.

More precisely, the probability that a node with degree $k$ is selected for vaccination is $kP(k)/(N\langle
k\rangle)$. This quantifies the known fact that randomly selected
acquaintances have, on average, a higher degree than randomly
selected nodes \cite{feld1991your,newman2002assortative}. Based on percolation theory, Cohen et al. developed a theoretical framework to determine the critical fraction $p_c$ and vaccination threshold $x_c$. They followed a possible branch in the course of the epidemic, starting from a random link of the spanning
cluster. That is, they studied the possible spreading of the
epidemic by considering nodes that are not vaccinated, and are therefore susceptible to the epidemic and may become infected \cite{cohen2003efficient}. If $n_l(k)$ denotes the number of nodes with degree $k$ in some layer $l$ (hop-distance from the starting point), the number $n_{l+1}(k')$ of nodes with degree $k'$ that are susceptible in layer $l+1$ is given by\\
\begin{equation}\label{l+1}
n_{l+1}(k')=\sum_k n_l(k)(k-1)p(k'|k,s_k)p(s_{k'}|k',k,s_k),
\end{equation}
\\
where $k-1$ means that the node with degree $k$ in layer $l$ has $k-1$ new neighbors in layer $l+1$ (excluding the one through which we arrived). Moreover, $p(k'|k,s_k)$ is the probability of reaching a node with degree $k'$ by following a link from a susceptible node of degree $k$, while $s_k$ is the probability that a node with degree $k$ is susceptible. Lastly, $p(s_{k'}|k',k,s_k)$ is the probability that the reached node is also susceptible.

Based on Bayes' rule, it holds that\\
\begin{equation}\label{}
p(k'|k,s_k)=\frac{p(s_{k}|k,k')p(k'|k)}{p(s_{k}|k)}.
\end{equation}
\\
Furthermore, assuming that the network is uncorrelated, we have\\
\begin{equation}\label{}
\phi(k')=p(k'|k)=k'P(k')/\langle
k\rangle,
\end{equation}
\\
which is independent of $k$. The probability that the acquaintance is not selected in one particular vaccination attempt by a random node of degree $k$, is $1-1/(Nk)$, and in all $Np$ attempts, it reads\\
\begin{equation}\label{}
v_p(k)=(1-\frac{1}{Nk})^{Np}\approx e^{-p/k}.
\end{equation}
\\
However, if the neighbor's degree is not known, the average probability becomes $v_p=\langle v_p(k)\rangle$. Then the probability that a node with degree $k'$ is susceptible (not vaccinated) is\\
\begin{equation}\label{}
p(s_{k'}|k')=\langle v_p(k)\rangle^{k'},
\end{equation}
\\
if no other information on its neighbors exists. But when the degree of one neighbor is known to be $k'$, we have\\
\begin{equation}\label{}
p(s_{k}|k,k')=e^{-p/k'}\times\langle e^{-p/k}\rangle^{k-1}.
\end{equation}
\\
Since the fact that a neighbor with known
degree is vaccinated does not provide any further information about a node's probability of vaccination, it follows $p(s_{k}|k,k')=p(s_{k}|k,k',s_{k'})$.

Substituting above results in Eq.~\ref{l+1}, we obtain\\
\begin{equation}\label{l1}
n_{l+1}(k')=n_{l}(k')\sum_k\phi(k)(k-1)v_p^{k-2}e^{-2p/k}.
\end{equation}
\\
If the sum in the above equation is larger than 1, the branching
process will continue forever, which is referred to as the percolating phase. On the contrary, if it is smaller than 1, the vaccination is subcritical and the epidemic will be contained. Thus, we obtain a relation for $p_c$, which is\\
\begin{equation}\label{}
\sum_k\frac {P(k)k(k-1)}{\langle k\rangle}v_{p_c}^{k-2}e^{-2p_c/k}=1,
\end{equation}
\\
and the  vaccination threshold is easily obtained
from the fraction of nodes which are not susceptible, namely\\
\begin{equation}\label{}
x_c=1-\sum_kP(k)p(s_{k}|k)=1-\sum_kP(k)v_{p_c}^k.
\end{equation}
\\

\begin{figure}
\centering
\includegraphics[scale=0.4,trim=50 0 50 0]{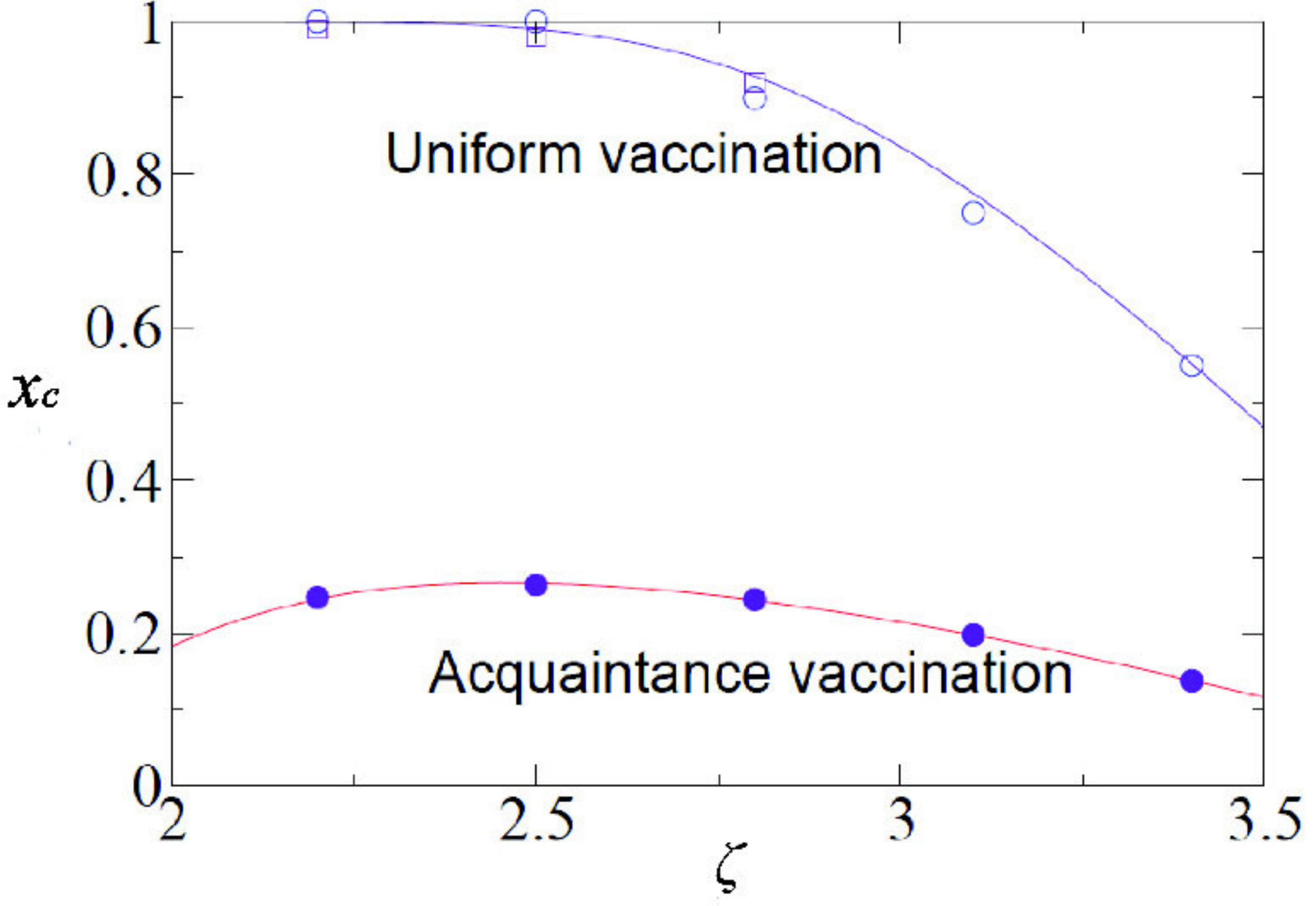}
\caption{The critical vaccination threshold $x_c$ as a function of the degree exponent $\zeta$ of the contact network on which the disease spreads via the SIS model. The curves
correspond to two distinct vaccination strategies: uniform vaccination (top curve) and acquaintance vaccination (lower curve).
The continuous lines represent the analytical results while the
symbols represent simulation data for $N = 10^6$ and $k_{min} = 1$. Source: Adapted with permission from Ref.~\cite{cohen2003efficient}. Copyrighted by the American Physical Society.}\label{acquaintance vaccination}
\end{figure}

Results presented in Fig.~\ref{acquaintance vaccination} show the critical threshold that is required to eradicate a disease in a scale-free network. It can be observed that acquaintance
vaccination is more efficient than random vaccination for both high and low $\zeta$ values, corresponding to homogeneous and heterogenous networks, respectively.

\subsubsection{Other strategies: definitions and simulation-based models}

In addition to acquaintance vaccination reviewed above, similarly efficient strategies can be devised based on local information about the structure of the network. An example is the so-called random walk vaccination strategy \cite{noh2004random}, in which a random walker diffuses through the network, and every node that it visits is vaccinated until a given fraction of the population is immune~\cite{ke2006immunization}. Given that a random walker visits a node with degree $k_i$ with the probability proportional to $k_i$, this strategy could lead to the same effectiveness as achieved by acquaintance vaccination \cite{pastor2015epidemic}. Another alternative to acquaintance vaccination is common acquaintance vaccination, where only the common neighbors of the randomly chosen nodes are selected for vaccination~\cite{liu2009common}. This may alleviate inefficiencies related to the fact that, of course, only very few of the randomly selected nodes and their neighbors will be the hubs of the network. Indeed, common acquaintance vaccination proves to be somewhat more effective than the simpler acquaintance vaccination.

The acquaintance vaccination strategy can also be improved by taking into account additional information about the local structure of the network. For example, if each node has the local knowledge about the degree of its nearest neighbors, and selects the neighbor with the degree to be vaccinated, then the efficiency in comparison to acquaintance vaccination is significantly improved \cite{holme2004efficient}. Similarly, the random walk vaccination strategy can also be improved via a bias favoring the exploration of large-degree nodes during the random walk process \cite{ke2006immunization,lee2009centrality}.

Along the same line, if even more local information is available, for example if nodes have the knowledge about the degree of their neighbors within short paths of length $D$, then the so-called $D$-steps vaccination strategy can be employed~\cite{gomez2006immunization,echenique2005distance}. In particular, for every node $i$, simply look for the largest-degree node within the distance $D$ and vaccinate it. In case several nodes within the distance have the same largest degree, then choose one of them uniformly at random.

\begin{figure}
\centering
\includegraphics[scale=0.5,trim=50 0 50 0]{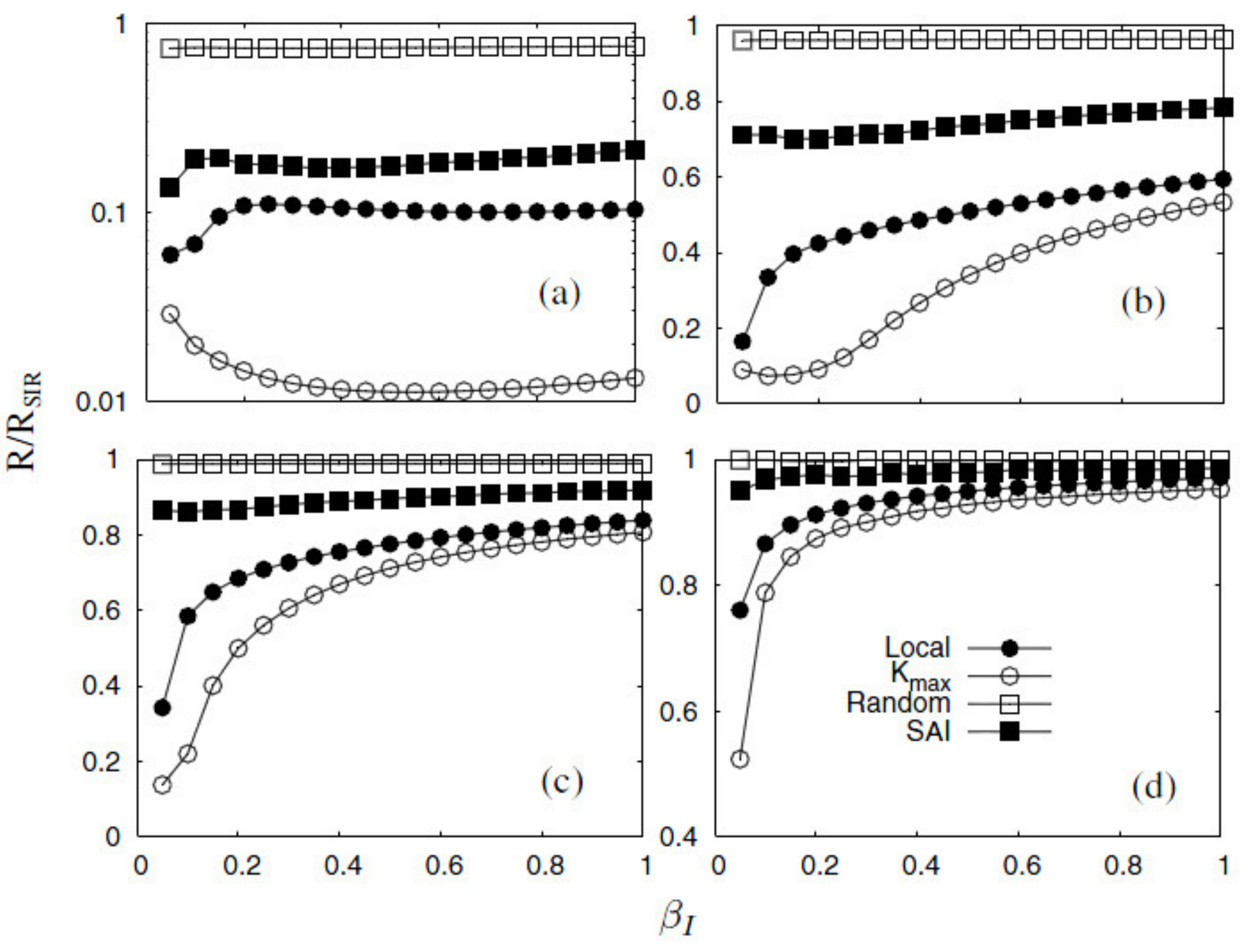}
\caption{Comparison of different vaccination strategies for the Internet router map. $R/R_{SIR}$ represents the ratio between the epidemic incidence ($R$) of the four proposed vaccination strategies and the epidemic incidence $(R_{SIR})$ without vaccination. The legend refers to the following vaccination strategies: $D$-steps vaccination (Local), targeted vaccination
($K_{max}$), random vaccination (Random) and single acquaintance vaccination (SAI). In each case, 1\% of the non-vaccinated nodes were initially infected at random. The distance considered in the local algorithm is: (a) $D = 1$; (b) $D = 2$; (c) $D = 3$; (d) $D = 5$. Source: Adapted with permission from Ref.~\cite{gomez2006immunization}. With permission of Springer.}
\label{dstep}\end{figure}

Based on these considerations, in \cite{gomez2006immunization} detailed numerical simulations of four different vaccination schemes, including $D$-steps vaccination, degree-based targeted vaccination, random vaccination, and acquaintance vaccination, were performed and tested in terms of their efficiency. As shown in Fig.~\ref{dstep}, targeted vaccination expectedly produces the best results, and this irrespective of the interaction topology. Also understandably, the performance of random vaccination is not affected by the structure of the network. Turning to local algorithms, it is found that the vaccination scheme based on the covering algorithm performs
better than acquaintance vaccination, even for small
values of $D$. In fact, the $D$-steps vaccination is only outperformed by the targeted procedure and lies between the most efficient and the acquaintance vaccination scheme for all the values of $D$ and ${\beta_I}$. Moreover, from a practical point of view, the covering strategy could be a good policy since it balances the degree of
local knowledge and the efficiency of the vaccination.  As most empirical network topologies are neither completely known nor completely unknown, this method thus allows us to fine-tune the value of $D$ on a case-by-case basis, i.e., according to the degree of
local knowledge of the network.

\begin{figure}
\centering
\includegraphics[scale=0.45,trim=50 0 50 0]{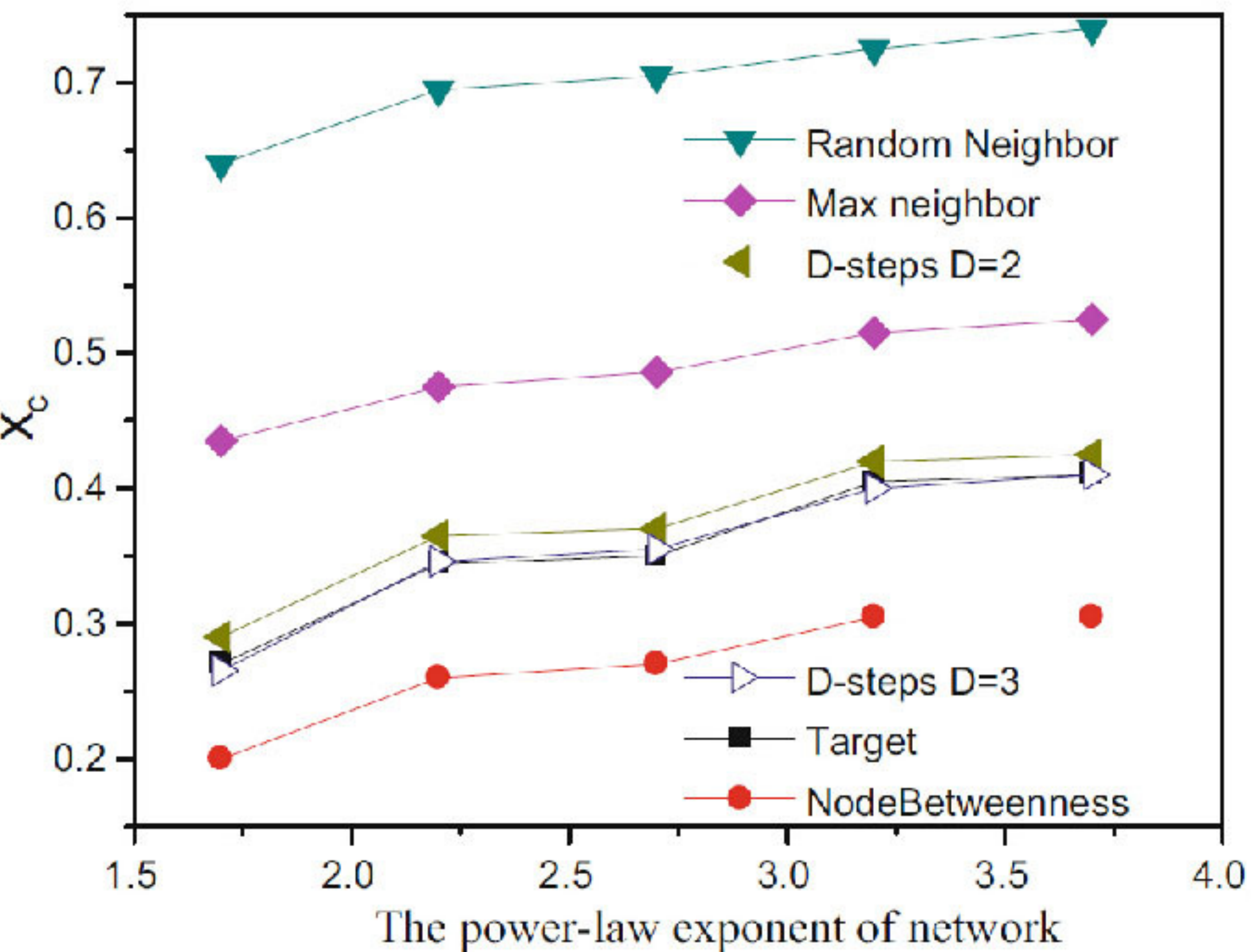}
\caption{The critical vaccination threshold $x_c$ as a function of
the power-law exponent of the degree distribution of the network. The legend refers to the following vaccination strategies: acquaintance vaccination (Random Neighbor and Max neighbor), $D$-steps vaccination  (D-steps D=2 and D-steps D=3), node degree-based targeted vaccination
(Target) and node betweenness-based targeted vaccination (NodeBetweenness). Source: Adapted with permission from Ref.~\cite{gao2011network}. With permission of Springer.}
\label{gcvspower}\end{figure}

In addition, Gao et al. evaluated the relationship between the threshold $x_c$ of the aforementioned vaccination strategies and the power-law exponent of the degree distribution of the network, which is shown in Fig.~\ref{gcvspower}~\cite{gao2011network}. A low level of $x_c$ with a small variation indicates that the immunization strategy is robust. Results presented in Fig.~\ref{gcvspower} also show that the robustness of the $D$-steps vaccination strategy is close to that of targeted vaccination.

Vaccination plays an obviously important role in the study of epidemic transmission and prevention. Based on the above-reviewed research, we can conclude that the efficiency of vaccination depends significantly on the amount of information we have about the structure of the network, and whether this information concerns the global or the local structure. In the rare case information about the global structure is available, targeted vaccination works best. If only information about the local structure is available, then the $D$-steps vaccination seems to provide the optimal compromise between efficiency and flexibility. Table~\ref{table-vaccination} gives a systematic summary of the reviewed vaccination strategies, for a quick overview.

\begin{sidewaystable}\newcommand{\tabincell}[2]{\begin{tabular}{@{}#1@{}}#2\end{tabular}}\centering{
\caption{Summary of non-behavioral epidemiological vaccination strategies, with details concerning vaccinated nodes, network models, and the required knowledge about the structure of the network.}\footnotesize \label{table-vaccination}
\begin{tabular}{|l|l|l|l|}
\hline
Vaccination strategy & Vaccinated nodes & Applied network models & Required knowledge of topology\\\hline
uniform (or random) vaccination & random chosen nodes & \tabincell{l}{ homogeneous networks \\ less heterogeneous networks} & no any need of network information \\\hline
targeted vaccination &  nodes with high centrality\footnote{Centrality includes degree centrality, betweenness centrality, random-walk centrality, closeness centrality, eigenvector centrality and PageRank centrality.} & \tabincell{l}{heterogeneous networks \\networks with special centrality} & global knowledge of network topology\\\hline
acquaintance vaccination\footnote{Except for simple acquaintance vaccination, it also includes common acquaintance vaccination, random walk vaccination, D-steps vaccination strategy.} & random neighbor of random chosen node\footnote{Correspondingly, it also involves common neighbor of random chosen nodes, nodes visited by random walker, largest-degree node within the distance $D$ of random chosen node.} & \tabincell{l}{ homogeneous networks \\ heterogeneous networks} & local knowledge of network topology\\
\hline
\end{tabular}}
\end{sidewaystable}

\subsection{Vaccination on other types of networks}

Up to now, we have focused mainly on research concerning vaccination on traditional network models. This research provides vital insight in terms of the efficiency of vaccination under different circumstances, yet empirical networks usually have particular topology properties that require special care. This includes taking into account community structure~\cite{girvan2002community,newman2004finding,palla2005uncovering,newman2006modularity,liu2005epidemic,wu2008community}, changes in the network structure over time~\cite{gross2006epidemic,gross2008adaptive,gross2009adaptive,shaw2008fluctuating,marceau2010adaptive,shaw2010enhanced,gao2013modeling}, as well as multilayer properties that are typical for a large plethora of real-world networks~\cite{boccaletti2014structure,kivela2014multilayer,wang2016suppressing,wang2014asymmetrically,liu2015impacts,wang2014epidemic,gomez2013diffusion,cozzo2013contact,zhao2014multiple,zhao2014immunization,salehi2015spreading,zhao2016roin,gao2016competing,zhao2015finding}. In what follows, we will review vaccination programs that take into account the particular aspects of these properties on various types of networks.

\subsubsection{Vaccination on community networks}

Community structure is ubiquitous in a variety of real complex systems, such as human contact networks and social networks. The hallmark of this fact is that the connections among members
of the same community are much more common than connections among members of different communities \cite{girvan2002community,newman2004finding,palla2005uncovering,newman2006modularity,liu2005epidemic,wu2008community}. For networks with community structure, the weak ties that
connect a pair of nodes belonging to different communities, usually named the bridge nodes, therefore typically provide pathways for information and disease
to propagate from one community to the other. These nodes thus play a more important role in the spreading of disease than the nodes with no or fewer inter-community links \cite{onnela2007structure,zhao2010weak}. However, their importance
is not necessarily reflected by their degree centrality. In this regard, identifying
the bridge nodes in community networks is crucial in preventing
epidemic outbreaks~\cite{masuda2009immunization,gong2013efficient}. Identification of community bridges can be partially achieved by betweenness centrality
or random-walk centrality, since both locate nodes on the basis of
important paths of epidemic spreading. On a similar note, Chen et al. proposed a vaccination strategy based on an equal graph partitioning algorithm
to identify the minimum group that separates a network into several
clusters of approximately equal size \cite{chen2008finding}. The role of the
minimum separator group is actually similar to that of the bridge nodes
connecting different communities.

\begin{figure}
\centering
\includegraphics[scale=0.4,trim=50 0 50 0]{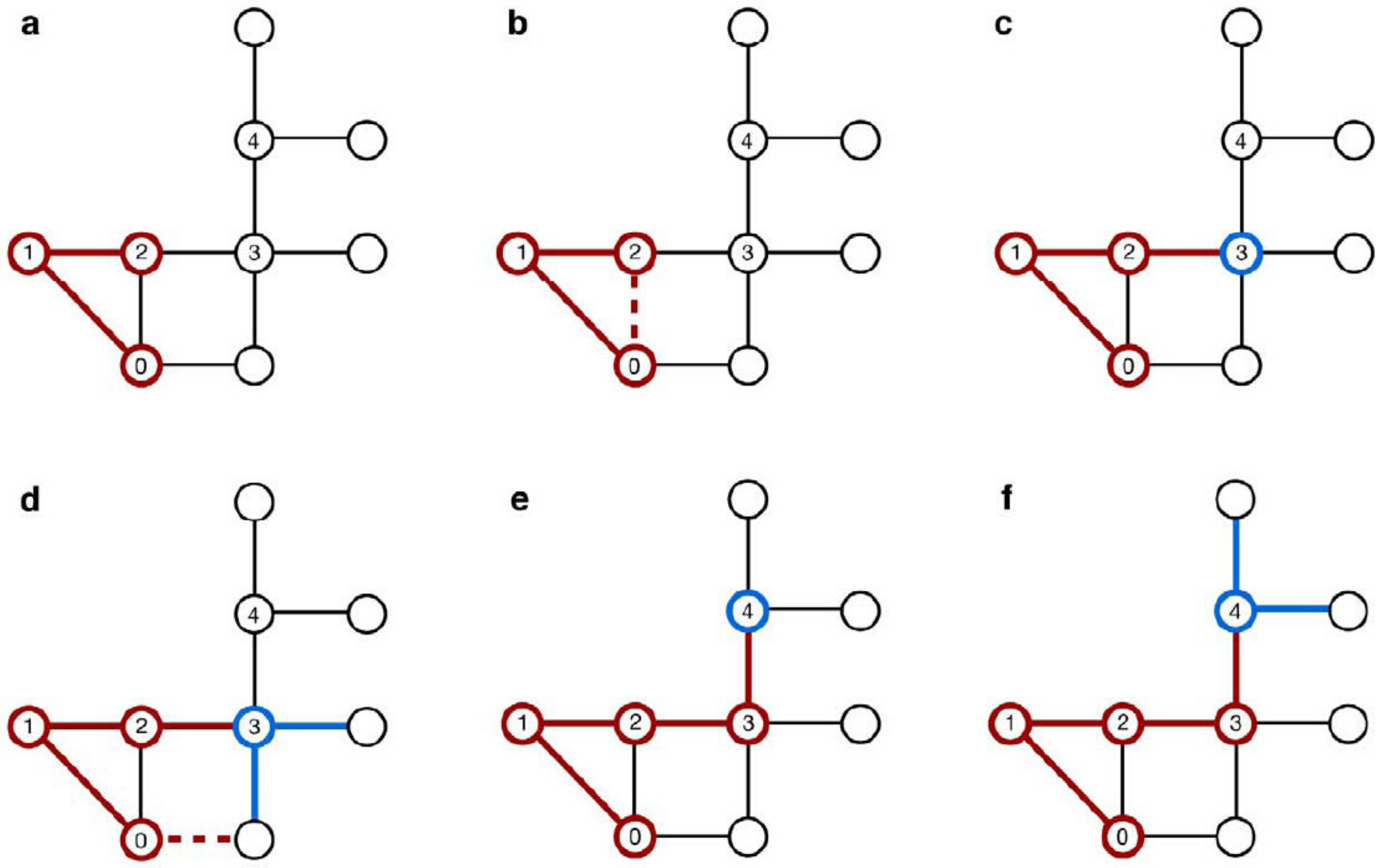}
\caption{A schematic presentation of the community bridge finder (CBF) algorithm. (a) A
random walker follows the path starting from $v_0$ to $v_1$ and $v_2$,
at which point it starts checking the connections from $v_2$ to $v_0$
and $v_1$. (b) Since there is more than one connection ($v_2-v_1$
and $v_2-v_0$), the walker turns to $v_3$. (c) Except for the obvious
$v_3-v_2$, there are no connections from $v_3$ to any of the
previously visited nodes, so $v_2$ is a potential target. (d) The
algorithm then picks two random neighbors of $v_3$ to check
connections with previously visited nodes and finds one (to $v_0$).
(e) Hence, $v_2$ is dismissed as a potential target, and the random
walker comes to $v_4$. Again, $v_4$ does not connect to any
previously visited node (except to $v_3$), and thus
$v_3$ is identified as a potential target.(f) Again, two random
neighboring nodes are picked to check connections with previously
visited nodes. Since no back connections can be found, $v_3$ is
identified as a target and vaccinated. Source: Reprinted figure from Ref. \cite{salathe2010dynamics}}\label{CBF}
\end{figure}

Salath{\'e} et al.~\cite{salathe2010dynamics} subsequently proposed a dedicated community bridge finder (CBF) algorithm that searches for the bridge nodes, thereby requiring only local structural
information about the structure of the network. Their research showed that such a method is more efficient than vaccination strategies targeting different kinds of hubs. The algorithm is based on the self-avoiding walk, which starts from a randomly chosen node
$v_0$, and then follows the procedure outlined in Fig.~\ref{CBF}. After $t$ steps ($t\geq 2$), the set of all the visited nodes is denoted by $\{v_{t'}\}$ for
$t'=0,1,\ldots,t$. Thus, $v_t$ is the node where the walker is located after $t$ steps, and $v_{t-1}$ is the node visited at step
$t-1$. The first process is to examine whether the node $v_t$ has a
link or several links to the nodes in the set $\{v_{t'}\}$, other
than the link between $v_t$ and $v_{t-1}$. If yes, the
self-avoiding walk proceeds towards step $t+1$; otherwise $v_{t-1}$
will be considered as a possible target of the bridge node. To determine
whether the node $v_{t-1}$ is one bridge node, two nodes are randomly
chosen among all the possible nodes that the walker could visit in
step $t+1$, i.e. two neighbors of the node $v_t$ are randomly chosen (except for the node $v_{t-1}$ due to the self-avoiding restriction of the walk). If there exists a path from any of the two chosen nodes back to any node in the set $\{v_{t'}\}$, then the node $v_{t-1}$ will not be a bridge node and the walker moves to node $v_{t+1}$. If there exists no path back to the set $\{v_{t'}\}$ from both chosen nodes, $v_{t-1}$ is
regarded as a bridge node that connects two communities and is vaccinated. Then a new self-avoiding walk starts and the above procedure is repeated until the desired vaccination ratio is attained. An important idea behind the CBF algorithm is actually that a community is formed by a circle of
close friends. Thus, when two randomly chosen neighbors of
$v_t$ cannot trace back to the community that $v_{t-1}$ belongs
to, the link between $v_{t-1}$ and $v_{t}$ is likely to be a bridge
between two communities, and the node $v_{t-1}$ is hence a bridge node.

It has been found that community networks typically exhibit a heterogeneous distribution
in the number of weak ties originating from the bridge nodes \cite{guimera2005functional,guimera2005worldwide}. Inspired by this finding, Gong et al. turned their attention to a particular kind of bridge nodes, named bridge hubs, which connect a community to many other communities, i.e., bridge hubs have a large number of weak ties~\cite{gong2013efficient}. They proposed a vaccination strategy named bridge-hub detector (BHD) to identify the bridge hubs. This algorithm extends the self-avoiding searching scheme to examining the overlap and the existence of links from all the neighbors of the last node back to the union of the friendship circles of all the nodes in the trail of the walk. A pair of nodes, a bridge node and a bridge hub, are
searched for vaccination via a self-avoiding walk. Compared with other local vaccination strategies such as acquaintance vaccination and CBF, this strategy is proven to be
more effective in preventing the epidemic~\cite{gong2013efficient}.

Also of note, Masuda recently proposed a vaccination strategy based on the eigenvector centrality for community-structured networks~\cite{masuda2009immunization}. In his work, a module based strategy technique is employed for measuring the contribution of each node to the weighted community network in order to preferentially vaccinate nodes that bridge important communities. The number of links that these nodes share gives the weight of
a link between any two communities. The influence of a node is thus related to the importance of the community together with its connectivity to other communities.

\subsubsection{Vaccination on adaptive networks}

From the epidemiological viewpoint, when an infectious disease occurs in a population, individuals are likely to adopt some self-protection measures to protect themselves from the disease. For example, susceptible people may break their contacts with infected partners, which can significantly alter the structure of the contact network, thus influencing the pathway of the epidemic spreading \cite{gross2006epidemic,shaw2008fluctuating,marceau2010adaptive,shaw2010enhanced,yang2012efficient,ruan2012epidemic,zhao2013efficient,zhou2012epidemic}. Both SIS and
SIR models have been studied on adaptive networks in which susceptible nodes rewire their links adaptively from infected neighbors towards other non-infected nodes~\cite{gross2006epidemic,shaw2008fluctuating}.
Such adaptation typically increases the epidemic threshold and
reduces the number of infectious cases, thus contributing favorably to the containment of epidemic spreading.

\begin{figure}
\centering
\includegraphics[scale=0.42,trim=50 0 50 0]{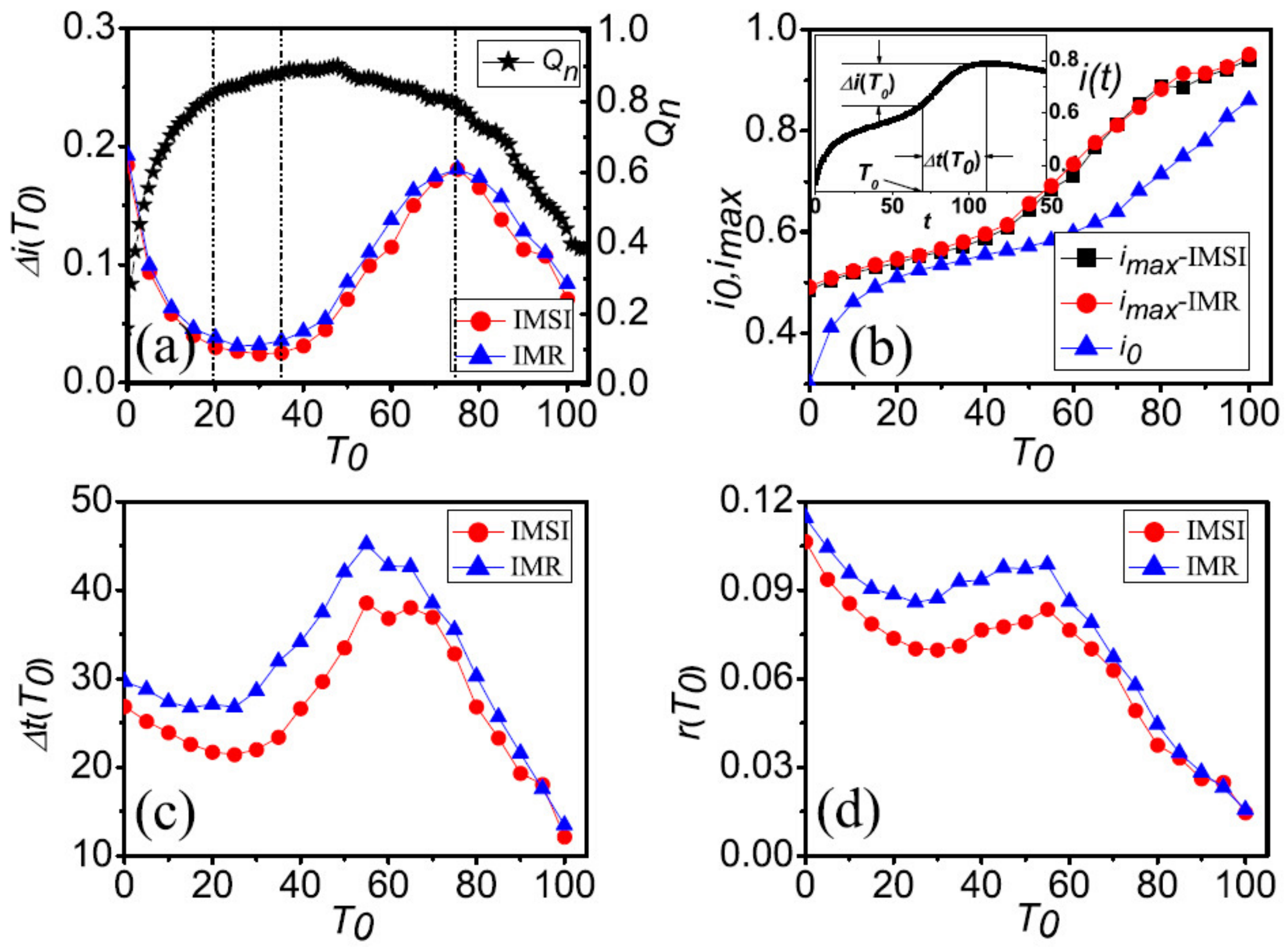}
\caption{The effect of IMR strategy and IMSI strategy versus the
starting time $T_0$ of quarantine for (a) $\Delta i_0(T_0)$, (b) $i_0$,
$i_{max}$, (c) $\Delta t(T_0)$ and (d) $r(T_0)$. Source: Reprinted figure from Ref.~\cite{yang2012efficient}}
\label{IMR}
\end{figure}

Moreover on the subject, Shaw et al. have studied vaccine control of disease spreading on an adaptive
network to model the specifics of disease avoidance behavior \cite{shaw2010enhanced}. Control
was implemented by adding a Poisson-distributed vaccination to
the susceptible individuals. Their research showed that vaccine control is much more effective
in adaptive networks than static networks due to the feedback
loop between the adaptive rewiring and the administration of the vaccine. On the other hand, Yang et al. have focused on the transient process rather than
the steady state, and found that strong community structure ($S$ nodes
community and $I$ nodes community, connected by $SI$ links) is
induced by the rewiring mechanism in the early stages of the epidemic
spreading \cite{yang2012efficient}. Subsequently, they proposed and examined various vaccination strategies that build on this fact, targeting in particular such $SI$ links. In particular, the vaccination of randomly selected $S$ bridge nodes from $SI$ links (IMSI) and the vaccination of randomly selected $I$ bridge nodes
from $SI$ links (ISSI) were considered. To evaluate the efficiency of IMSI and ISSI strategies, the vaccination of randomly selected $S$ nodes from the network (IMR) and the vaccination of randomly
selected $I$ nodes from the network (ISR) were also employed.

\begin{figure}
\centering
\includegraphics[scale=0.42,trim=50 0 50 0]{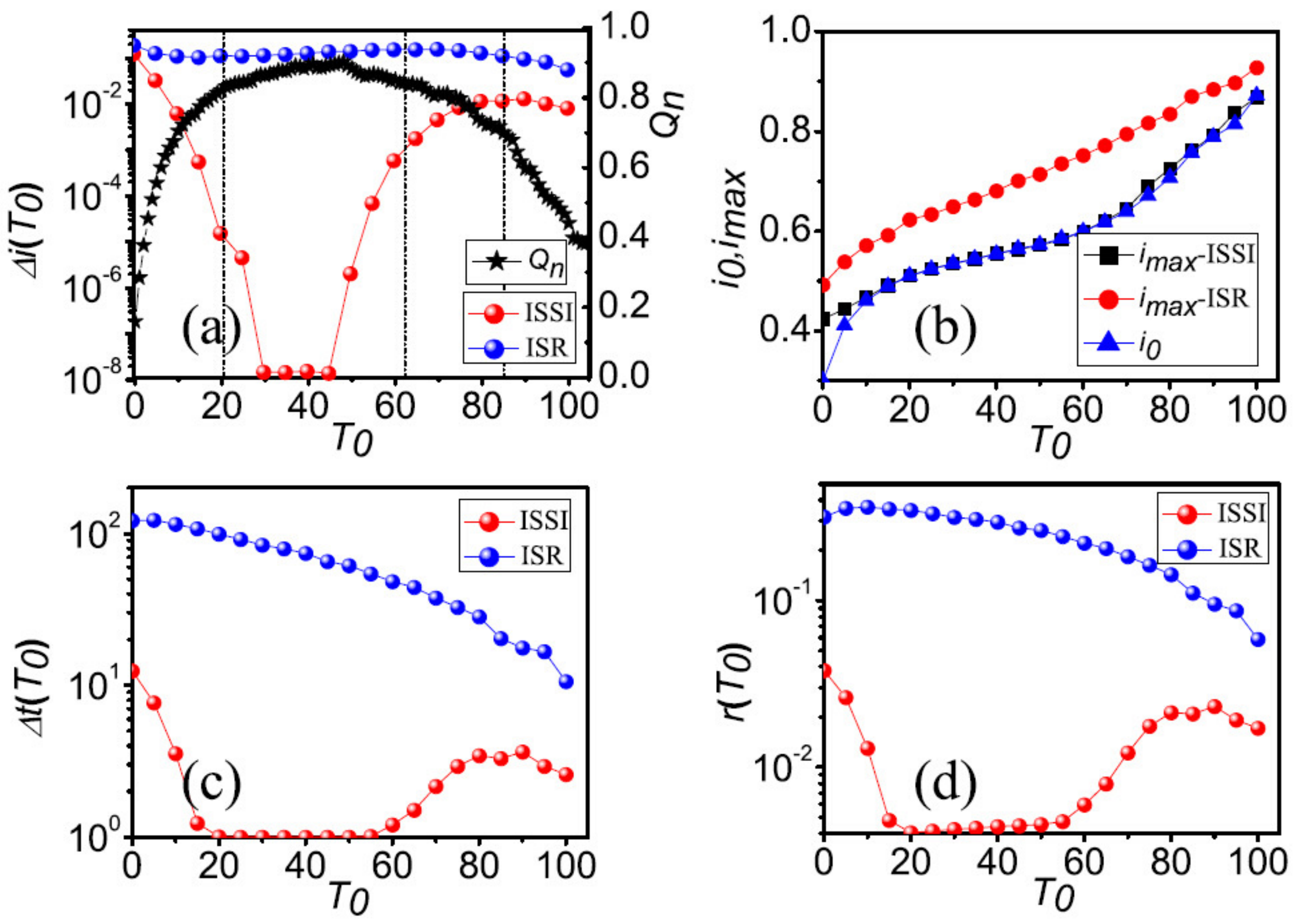}
\caption{The effect of ISR strategy and ISSI strategy versus the starting time $T_0$ of quarantine for (a) $\Delta i_0(T_0)$, (b) $i_0$, $i_{max}$, (c) $\Delta t(T_0)$ and (d) $r(T_0)$. Source: Reprinted figure from Ref.~\cite{yang2012efficient}}
\label{ISR}
\end{figure}

In terms of their results, in Figs.~\ref{IMR} and \ref{ISR}, $T_0$ denotes the starting time of
vaccination, $\Delta i_0(T_0)$ is the difference between the maximal
density of infected individuals $i_{max}(T_0)$ that can be reached after
vaccination and the density of infected individuals $i_0(T_0)$ at time $T_0$, i.e. $\Delta i_0(T_0)=i_{max}(T_0)-i_0(T_0)$. $\Delta t(T_0)$ represents the time interval of this process, and $r(T_0)$ denotes the total percentage of vaccinated nodes in $\Delta t(T_0)$.
In Fig.~\ref{IMR}, the impact of the IMR strategy is almost equal to
that of the IMSI strategy. Because most $S$ nodes are located on $SI$ links
when an epidemic prevails in an adaptive network, randomly vaccinating $S$
nodes can give rise to similar effects as vaccinating
$S$ nodes on $SI$ links. Nevertheless, for ISR and ISSI,
quarantining $I$ nodes on $SI$ links is significantly more effective than quarantining $I$ nodes randomly, since the former can
efficiently cut the pathways along which the disease could be invading the $S$ cluster
(Fig.~\ref{ISR}). Because of this fact, the ISSI strategy is more efficient in controlling the epidemic and has a larger optimal region. More importantly, the consideration of these different vaccination strategies reveals a counterintuitive conclusion: ``the
earlier the better'' does not necessarily hold for the implementation of vaccination measures. Optimal results can be expected more realistically if smart vaccination strategies are implemented when the community structure emerges after the initial rewiring of links that sets in due to an epidemic.

\subsubsection{Vaccination on multiplex networks}

\begin{figure}
\centering
\includegraphics[scale=0.5,trim=50 0 50 0]{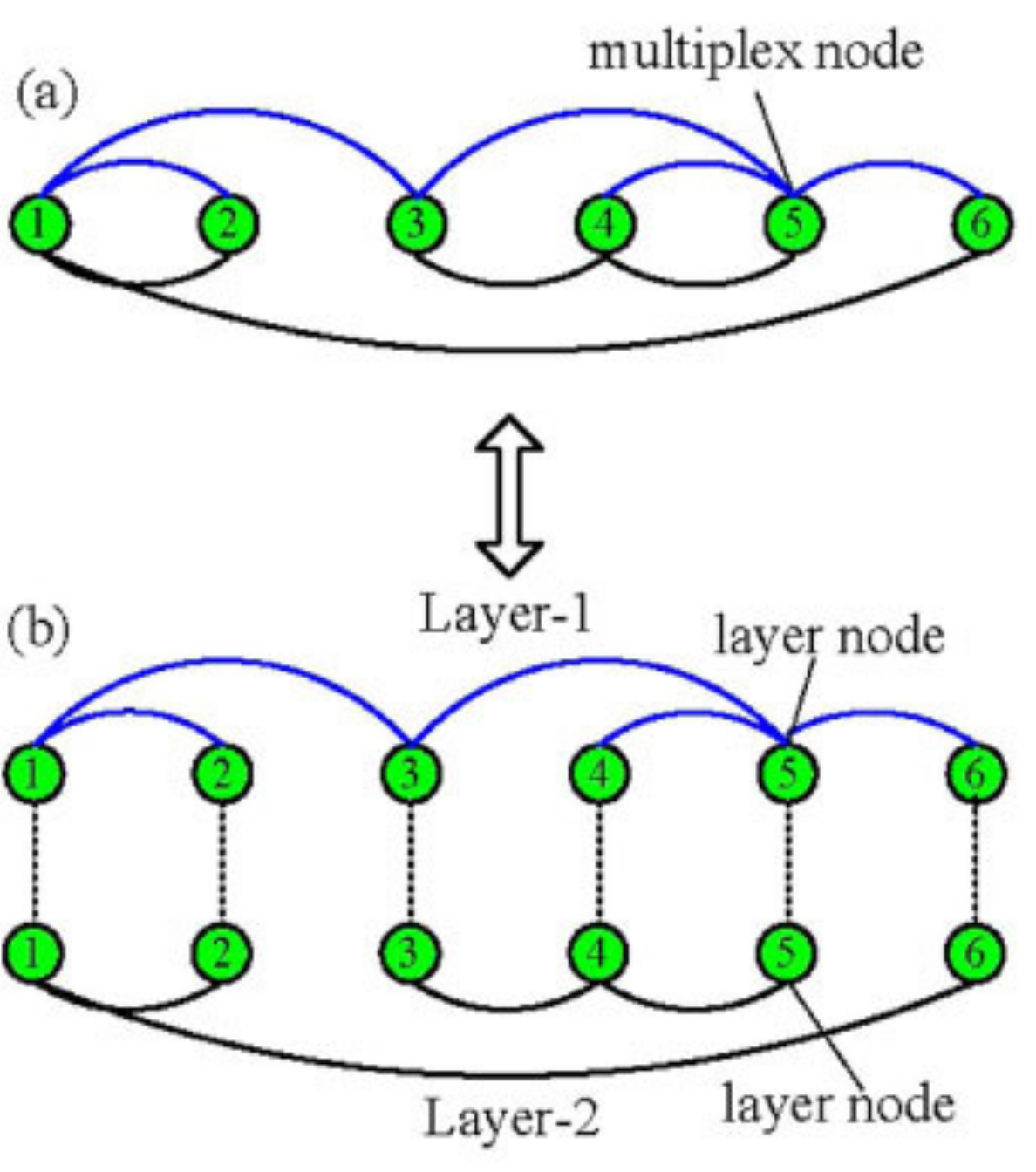}
\caption{Schematic illustration of a multiplex configuration. (a) In a multiplex network, six nodes are connected through two different types of links, i.e., blue and black links. In this configuration, each node simultaneously belongs to two layers, and it is thus termed as a multiplex node. (b) Multiplex networks are composed of several, for example two, layers, each of which is composed by the same type of links. The term layer node indicates that the node is connected only through the one type of link in the corresponding layer. See main text for details with regards to the importance of this classification for vaccination strategies. Source: Reprinted figure from Ref.~\cite{zhao2016robustness}.}
\label{multi-layer}
\end{figure}

Multiplex networks, as a typical kind of multilayer network structure, can be regarded as the combination of several network layers, which contain the same nodes (or share at least some fraction of nodes) yet different intra-layer connections \cite{boccaletti2014structure,kivela2014multilayer,wang2016suppressing,wang2014asymmetrically,liu2015impacts,wang2014epidemic,peng2010models,gomez2013diffusion,cozzo2013contact,zhao2014multiple,zhao2014immunization,salehi2015spreading,zhao2016robustness,du2016analysis}. For each node, we term its counterpart in every layer as a `replica'. Along the above definition, may real-world social and engineering systems, such as online social networks \cite{min2014layer,hu2014conditions,zagenczyk2015multiplex,ma2015social,xie2014construction,shen2015novel,guo2014varibale,centola2010spread}, airport traffic networks \cite{hosseini2016review,du2016analysis,du2016physics}, biological metabolic networks \cite{de2016mapping,sporns2016modular}, and scientific collaboration and citation networks \cite{battiston2015emergence}, can be put into the framework of multiplex networks. Multiplex networks naturally shed new light into the research of epidemiology (see Refs. \cite{boccaletti2014structure,salehi2015spreading} for a comprehensive understanding). For example, the well known acquired immune deficiency syndrome (AIDS) usually propagates via three types of ways, namely sexual activity, blood, and breast milk. Different transmission routes could be mapped into different network layers which contain the same individuals yet different network topologies and dynamic properties \cite{zhao2014multiple}. Moreover, coupling methods from different layers will greatly affect the onset of disease and its outbreak threshold \cite{zhao2014multiple}. In what follows, we will focus on recent research concerning vaccination in multiplex networks.

To distinguish the node of a multiplex network and its replica in each network layer, Zhao et al. proposed the terminology of the multiplex node and the layer node~\cite{zhao2014immunization}. The former refers to the node which has neighbors in each layer via different intra-layer connections, like node 3 of Fig.~\ref{multi-layer}(a), whose neighbors are nodes $1$, $5$ and $4$ via blue and black links. While the latter is the partial case of a multiplex node and just involves the local connections of a node in one given layer. For example, if we only consider blue (black) links in Fig.~\ref{multi-layer}(b), node $3$ is the layer node of Layer-1 (Layer-2). Correspondingly, vaccination on multiplex networks could be classified into multiplex node-based vaccination and layer node-based vaccination. Multiplex node-based vaccination means that all the replicas of one node simultaneously possess immunity in all the layers, while layer node-based vaccination provides protection solely to nodes in a given network layer.

\begin{table}
\newcommand{\tabincell}[2]{\begin{tabular}{@{}#1@{}}#2\end{tabular}}
\caption{Vaccinated probability of layer nodes and multiplex nodes for different vaccination strategies}\label{Vaccinated probability}
\begin{tabular}{|c|c|c|}
  \hline
vaccination type \& probability & layer node-based vaccination &  multiplex node-based vaccination\\
  \hline
random vaccination & $x(k_{j_i})=x_i $, for $\forall j$&  $x(\overrightarrow{k_j})=x$, for $\forall j$\\
  \hline
targeted vaccination & $x(k_{j_i})=\left\{\begin{array}{ll} 1,\ \ \emph{\emph{if}}\ k_{j_i}>k_{ci}\\
f_i,\ \ \emph{\emph{if}}\ k_{j_i}=k_{ci}\\ 0,\ \ \emph{\emph{if}}\
k_{j_i}<k_{ci}\end{array}\right.$  &  \tabincell{l}{$x(\overrightarrow{k_j})=\left\{\begin{array}{ll} 1,\ \ \emph{\emph{if}}\ K_j>K_c\\
f,\ \ \emph{\emph{if}}\ K_j=K_c\\ 0,\ \ \emph{\emph{if}}\
K_j<K_c\end{array}\right.$ \\ where $K_j=\sum\limits_{i=1}^m{\beta_I}_i{k_j}_i$} \\
  \hline
acquaintance vaccination & \tabincell{l}{$x(k_{j_i})=1-$\\ $(1-q_i\sum\limits_{\overrightarrow{k'_j}}
\frac{k'_{j_i}p(\overrightarrow{k'_j})}{z_i}
\frac{1}{k'_{j_i}})^{k_{j_i}}$} & \tabincell{l}{$x(\overrightarrow{k_j})=1-$\\ $\prod_{i=1,\cdots,m}(1-\frac{q}{m}\sum\limits_{\overrightarrow{k'_j}}
\frac{k'_{j_i}p(\overrightarrow{k'_j})}{z_i}
\frac{1}{k'_{j_i}})^{k_{j_i}}$}\\
  \hline
\end{tabular}\\
   \scriptsize $\overrightarrow{k_j}=({k_j}_1,{k_j}_2,...,{k_j}_m)$: the degree expression of multiplex node $j$; ${k_j}_i$ refers to its degree in layer $i$.\\
   \scriptsize $x(k_{j_i})$: the vaccinated probability of layer node $j$ in layer $i$ with degree $k_{j_i}$.\\
   \scriptsize $x(\overrightarrow{k_j})$: the vaccinated probability of multiplex node $j$. \\
   \scriptsize ${\beta_I}_i$: the infectious rate of the disease in layer $i$.\\
   \scriptsize $k_{ci}$: the cutoff degree for vaccination of layer node in layer $i$.\\
   \scriptsize $K_j$: the spreading degree of multiplex node $j$.\\
   \scriptsize $K_c$: the cutoff spreading degree for vaccination of multiplex node.\\
   \scriptsize $z_i$: the average degree of layer $i$.
\end{table}

Similar to the vaccination strategies on single-layer networks, Refs.~ \cite{zhao2014immunization,wang2015immunity,buono2015immunization,zuzek2015epidemic} proposed the corresponding vaccination strategies (including random vaccination, targeted vaccination, and acquaintance vaccination) on multiplex networks. Here, however, the aim of vaccination is specified to layer nodes and multiplex nodes. These vaccination strategies are determined by the vaccinated probability of layer nodes and multiplex nodes, which are summarized in Table~\ref{Vaccinated probability}

\emph{Multiplex node-based vaccination} refers to the case where a fraction
of multiplex nodes is vaccinated. If $x(\overrightarrow{k_j})$ is defined as the vaccinated probability
of multiplex node $j$, the generating function of the joint
degree distribution under multiplex node-based vaccination is
given by\\
\begin{equation}
G_0(\overrightarrow{u})=\sum\limits_{\overrightarrow{k_j}}
p(\overrightarrow{k_j})(1-x(\overrightarrow{k_j}))\prod\limits_{i=1}^{m}u_i^{{k_j}_i}.
\end{equation}
\\
The generating function of the remaining joint degree distribution by
following a randomly chosen link of layer $i$ is defined as\\
\begin{equation}
G_1^{(i)}(\overrightarrow{u})=\frac{1}{z_i}\frac{\partial}{\partial
u_i}G_0(\overrightarrow{u}).
\label{u-1}
\end{equation}
\\
Then, the probability $\upsilon_i$ that a multiplex node
connects to the infected cluster by following a random link of layer $i$ is given by the coupled self-consistency equation\\
\begin{equation}
\upsilon_i=G_1^{(i)}(\overrightarrow{1})-G_1^{(i)}(\overrightarrow{1-{\beta_I}
\upsilon}),
\label{u}\end{equation}
\\
where $\overrightarrow{1-{\beta_I} \upsilon}=(1-{\beta_I}_1 \upsilon_1,...,1-{\beta_I}_m
\upsilon_m)$. It is worth mentioning that the cluster of multiplex networks is defined as a set of connected multiplex nodes. A pair of multiplex nodes is regarded as having a connection if there exists at least one type of link between them.

Thus, the existence of an epidemic regime under multiplex node-based
vaccination just requires the largest eigenvalue $\Lambda$ of the
Jacobian matrix of Eq.\ref{u} at (0,0,...,0) to be larger than unity
\cite{min2014network}. Consequently, the critical vaccination threshold  $x^{M}_c$ will
be the value of immunity $x^{M}=\sum p(\overrightarrow{k_j})x(\overrightarrow{k_j})$ satisfying $\Lambda=1$ in Eq. \ref{u}.

\emph{Layer node-based vaccination}, in contrast, aims to vaccinate layer nodes. To calculate the critical vaccination threshold, we just need to replace $G_0(\overrightarrow{u})$ by\\
\begin{equation}
G'_0(\overrightarrow{u})=\sum\limits_{\overrightarrow{k_j}}
p(\overrightarrow{k_j})\prod\limits_{i=1}^{m}(x_i({k_j}_i)+(1-x_i({k_j}_i))u_i^{{k_j}_i}),
\end{equation}
\\
in Eq.~\ref{u-1}, where $x(k_{j_i})$ is the vaccinated probability of layer node $j$ in layer $i$ with degree $k_{j_i}$. Since the vaccination of layer nodes is performed independently in each layer, the critical vaccination threshold becomes a vector $(x^{L}_1,...,x^{L}_m)_c$ which can be obtained when the immunity $x^{L}_i=\sum\limits_{{k_j}_i}p_i({k_j}_i)x_i({k_j}_i)$ is satisfying $\Lambda=1$ in Eq.~\ref{u}.

According to the above theoretical analysis, computer simulations have revealed that both multiplex node-based random vaccination and layer node-based random vaccination strategies are more effective in controlling disease spreading on multiplex ER networks~\cite{zhao2014immunization}. On the contrary, multiplex node-based targeted vaccination and layer node-based targeted vaccination strategies provide better protection on multiplex SF networks. As on single-layer networks, random vaccination strategies work well only in largely homogeneous networks, while targeted vaccination strategies are required for useful results on heterogeneous networks.

With regards to the efficiency of acquaintance vaccination in multiplex networks, we note that it is closely related to the correlation between network layers~\cite{wang2015immunity}. Interestingly, as the correlation coefficient increases, the vaccination threshold decreases for multiplex node-based acquaintance vaccination, but slowly increases under the layer node-based framework. Moreover, Buono et al. considered vaccination of only one layer and studied its effect on all the other layers while disregarding degree-degree correlations among them~\cite{buono2015immunization}. Though, relative to the case of no vaccination, the size of the epidemic is drastically reduced in the layer where the vaccination strategy is applied, the targeted strategy is still not as efficient as in single-layer networks. Thus, the selection of the vaccination strategy has a major effect on the layer where it is employed, but does not efficiently protect the individuals on other layers. Similar conclusions have also been reported in Refs.~\cite{vaidya2016modeling,zuzek2015epidemic}, thus corroborating the need to reconsider single-layer vaccination strategies when applied to multilayer networks.

\section{Rationale for studying behavior-disease dynamics} \label{sec:rationale}

As demonstrated by the foregoing sections of this review, epidemiological models that do not capture behavioral dynamics have been very successful over the past century, in terms of yielding rich theoretical opportunities for research as well as in terms of providing insights into epidemiological mechanisms and how to improve infection control. However, the impact of human behavior on disease dynamics is ubiquitous, and vaccinating behavior is no exception.

This section is concerned with explaining the rationale for studying and modeling coupled behavior-disease dynamics for vaccines, especially on networks.  We start with a brief
summary of vaccine use in public health, discuss why vaccinating behavior may become an increasingly important part of disease control in coming decades, and then discuss
how models can help us to better understand and predict behavior-disease dynamics.

The twentieth century saw enormous progress in public health, and especially in terms of preventing and treating infectious diseases.  For instance,
medical use of antibiotics began during World War II, and antibiotics have since changed how bacterial infections are treated and saved many lives.  Improvements
in sanitation and hygiene are likewise credited with preventing many cases of cholera,  typhoid fever, yellow fever, dysentery, tuberculosis, and malaria by providing
cleaner drinking water and living conditions and improved hygienic practice \cite{fink2011effect}.  The use of vaccines, although already available for smallpox since the eighteenth century,
expanded considerably in the twentieth century as  vaccines were invented for more infectious diseases, and were distributed to more people.  Safe and effective vaccines
now exist for many of the most common infectious diseases that impose significant morbidity and mortality upon populations, such as smallpox, measles, pertussis,
influenza, hepatitis A and B, chickenpox, diphtheria, and others \cite{bonanni1999demographic}.  Global smallpox eradication was achieved in large part due to the success of ring vaccination \cite{fenner1988smallpox}.

The use of vaccines has been estimated to save millions of children's lives per year in the 1990s alone \cite{bonanni1999demographic}.  In the 2000s, special immunization activities (SIAs)--ambitious large-scale immunization campaigns in the world's poorest regions and especially in sub-Saharan Africa--reduced the number of deaths due to measles from an estimated 766,000 per year in 2000 to an estimated 164,000 per year
in 2008 \cite{dabbagh2009global}.  Global eradication of measles in the coming decades is becoming seen as a real possibility for the first time in history \cite{levin2011global}.
Thus, although challenges still remain, it is becoming significantly easier to get vaccines administered to the people who need them.  There are, however, important
exceptions.  For instance, the challenges of administering vaccines in war zones have contributed to a large outbreak of polio in Syria as recently as 2015
\cite{cousins2015syrian}.  At the same time in Pakistan, violence against WHO vaccinators has stymied efforts to combat polio \cite{ganapathiraju2015endgame}.
Despite this, global polio eradication is coming closer \cite{cochi2014global}.

While some countries with resource limitations struggle to contain vaccine-preventable infectious diseases, other comparatively wealthy countries are also struggling
with infectious disease control, albeit for different reasons.  In countries such as the United Kingdom, Switzerland, and Germany, measles circulation in the 2000s and
2010s has been relatively widespread compared to previous decades, and has often re-established endemicity in some populations due to parental vaccine refusal or vaccine hesitancy on account
of unfounded concerns about the link between the measles-mumps-rubella (MMR) vaccine and autism \cite{de2001seroepidemiology, brown2010factors, larson2014understanding}.
Similarly, polio was on the cusp of global eradication and had been limited to northern Nigeria when an oral polio vaccine scare in 2003 caused resurgence of the disease and
spread of polio to countries like Syria and Pakistan that had previously eliminated the infection \cite{serpell2006parental}. These are only two examples where  population resistance
to vaccines due to unfounded concerns has led to reduced vaccine coverage and resurgent infections, even in resource-rich countries where it is easy for children to
get vaccinated, if their parents want it.

\begin{figure}
\begin{center}
\includegraphics[width=0.6\textwidth]{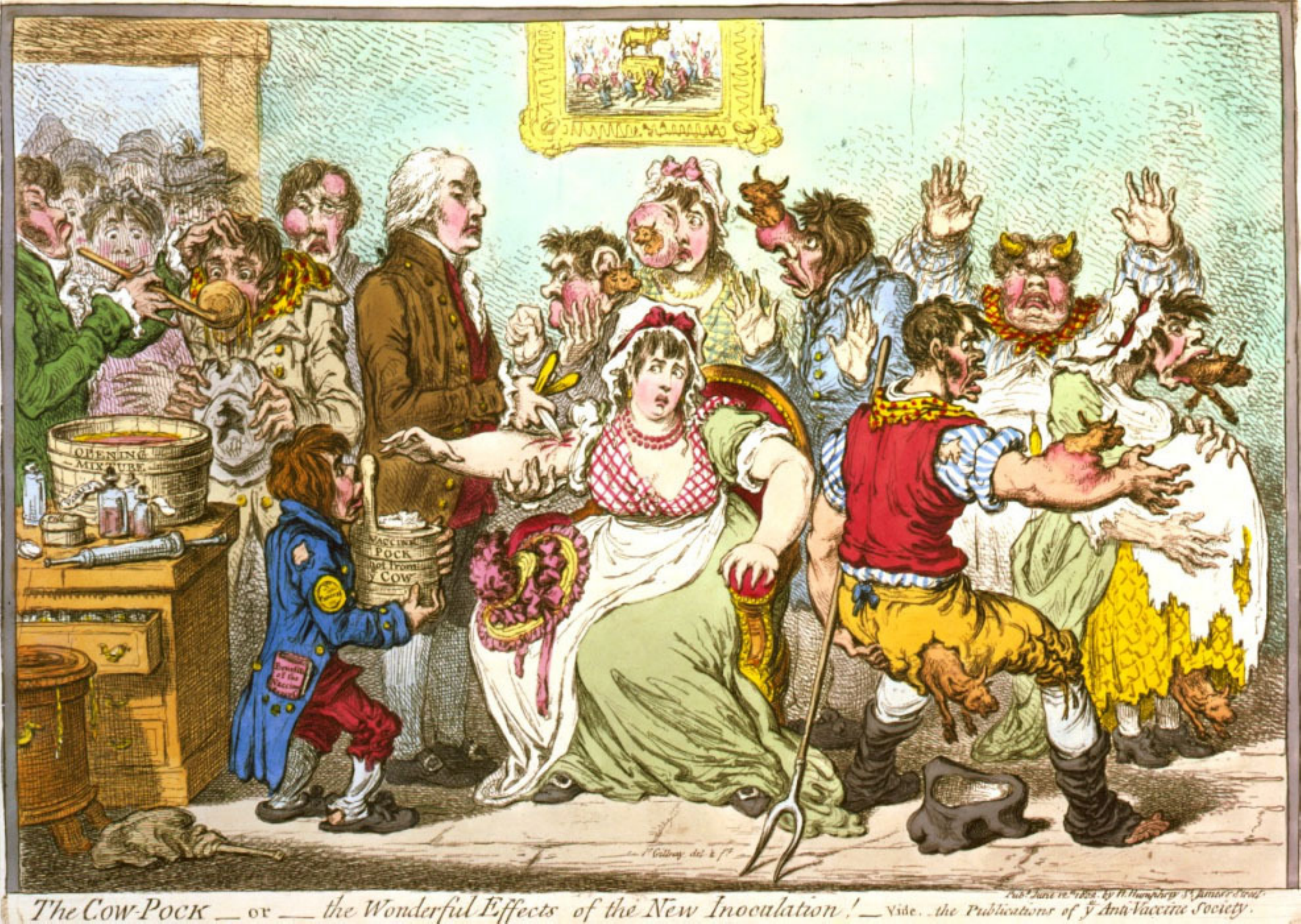}
\end{center}
\caption{Satirical cartoon by James Gillray capturing public fears of the smallpox vaccine in the 18th century, the first ``vaccine scare".  Edward Jenner stands in the middle, administering the smallpox vaccine.  Vaccinated persons grow cow parts out of various parts of their bodies (the smallpox vaccine of the time was based on the virus that causes cowpox, hence the fear among some that the vaccine would turn humans into cow-like hybrids).  A painting of the Worshippers of the Golden Calf hangs in the background. Source: Wikimedia Commons.}
\label{smallpox_cartoon}
\end{figure}

Logistic, financial, technological, and administrative barriers to vaccination are likely to continue receding in the coming decades, enabling it to become easier to improve
vaccine coverage everywhere in the world.  As this happens, we speculate that vaccine refusal and vaccine hesitancy such as recently observed in the case of MMR vaccine
will become an important barrier--and perhaps even the most important barrier--to global eradication of vaccine-preventable infections.  Many factors contribute
to the decision to become vaccinated \cite{chapman1999predictors,sturm2005parental,brown2010factors}.  Also, vaccine refusal is not a new phenomenon, and even the
first vaccine to be invented--the smallpox vaccine--was subject to considerable resistance (see Fig.~\ref{smallpox_cartoon}). However, growing herd immunity generated
by vaccine programs themselves can bring about the preconditions for vaccine refusal.

Vaccine refusal is often construed in the context of ``free-riding'': individuals who do not vaccinate are consciously free-riding
on the herd immunity provided by those who do vaccinate.  Debate on the usefulness of coupled behavior-disease models often centers on
whether individuals consciously free-ride when they make vaccinating decisions or not.  However, it is not necessary for individuals to consciously free-ride in order for vaccine
refusal to be linked to growing herd immunity, or for behavior-disease models to be useful (although it is certainly possible that some individuals will consciously free-ride when
making vaccinating decisions \cite{ibuka2014free}).  It is well-known that perceived infection risk and past exposure to infections are determinants of whether or not individuals
will choose to be vaccinated \cite{chapman1999predictors}.  In particular, a high perceived infection risk and/or a past history of infection are positively associated with
decisions to accept vaccination.  Therefore, lack of recent infection, such as caused by vaccine-generated herd immunity, can decrease the motivation to become vaccinated.
Parents who have never seen a case of measles infection leading to hospitalization will tend to under-estimate the risk and/or probability of measles infection.  Similarly,
lack of a recent outbreak can cause carelessness or de-prioritization on the part of physicians, and therefore lower vaccine uptake \cite{swennen2001analysis}.
Whether or not non-vaccinators consciously free-ride is secondary in some respects. What matters is that vaccine-generated herd immunity brings
about lack of recent infections in a population, which can in turn cause vaccinating to become less of a priority, as well as creating fertile conditions for false rumours
about vaccine risk to spread.  Vaccine programs are victims of their own success.

This is also echoed in observed behavior for pertussis and MMR vaccines.  When declining vaccine coverage caused in partly by vaccine refusal leads to an outbreak, vaccine coverage responds by increasing during and after the outbreak \cite{goldstein1996effect, bauch2012evolutionary}.  This vividly demonstrates the coupled nature
of disease dynamics and vaccinating choices, completely aside from whether we need to characterize this behavior as free-riding, or whether non-vaccinating
individuals are aware that is what they are doing.  Individuals may not answer truthfully when surveyed about their motives for becoming vaccinated or not becoming
vaccinated, or more importantly they may answer very differently in surveys conducted during an outbreak, than during a time of elimination. However,
when we observe vaccinating behavior responding to changes in disease prevalence, as in the Disneyland, California measles outbreak of 2014 \cite{CDPH2016} and other examples,
we see the proof in the behavioral pudding that vaccinating behavior and disease dynamics are strongly linked to one another, particularly near the elimination threshold.

The Disneyland, California measles outbreak and similar outbreaks illustrate the important role played by social factors in determining individual vaccinating choices.
The influences of peer groups and medical professional opinions on vaccine decision-making are well-documented \cite{sturm2005parental,allen2010parental}.
Social contact networks can be mapped out using Bluetooth and other technologies (see Section \ref{section:digital_epidemiology} for more details) \cite{salathe2010high}.  Moreover,
social influence has often been described as a contagious process in itself, as ideas and behaviors spread through networks \cite{christakis2007spread, campbell2013complex}.
Therefore, social processes can be very naturally modelled by defining processes on networks representing social contacts, and the methods of statistical physics concerned
with the properties of networks could therefore be very useful \cite{bansal2007individual}.

Because vaccine-generated herd immunity tends to create complacency toward vaccine programs in the way we have described above, we expect vaccine hesitancy and
outright vaccine refusal or vaccine ``scares" to become more common as eradication and elimination thresholds are approached for more vaccine-preventable infectious diseases.  Moreover,
we have seen how the significant role of social context in influencing individual decision-making is very naturally represented using the network paradigm.  The study
of coupled behavior-disease dynamics on networks could therefore help us address this potentially growing problem \cite{wang2015coupled}.  In particular, mathematical models, often based on
methods borrowed from physics, can be a useful way of better understanding and predicting these coupled dynamics.

Human behavior is doubtless more difficult to mathematically model than the motion of a single billiard ball, for instance, or many of the slightly more complex systems studied
by physicists.  However, at some levels of organization (in particular, the aggregate population level) it is often possible to develop simple models that have some predictive power.
For instance, mathematical models incorporating behavior-disease interactions have been fitted to time series data of infectious disease prevalence, and model selection approaches
such as Aikaike Information Criterion (AIC) have shown that models that include behavior can explain the data more effectively compared to models that neglect behavior, with little or no parsimony penalty
\cite{bauch2012evolutionary,he2013inferring,oraby2014influence}.  Such models have also shown retrospective predictive power \cite{bauch2012evolutionary}. Other model fitting and data
analysis exercises without mechanistic representations of behavior have likewise shown the importance of accounting for behavior when explaining observed epidemic
patterns \cite{chowell2006transmission, bootsma2007effect}. Similarly, the results of experimental games and surveys have been useful
in the parameterization of coupled behavior-disease models \cite{galvani2007long,shim2012influence}.  The advent of online social media
promises another source of data on both individual-level and population-level human behavior and this will be explored in greater depth in a later section.

Hence, there appears to be a strong public health and scientific rationale for studying behavior-disease interactions. The existing research shows that behavior and
disease are strongly coupled to another.  This coupling is particularly important near the elimination threshold, hence vaccine refusal may become an increasingly
important barrier to achieving elimination and eradication, as logistic barriers recede into the distance.  Moreover, models show promise to help us better understand
coupled behavior-disease dynamics, and comparing models with data have often shown that  behavior is important for understanding observed epidemic patterns and
that behavioral models may have predictive power.  Finally, these models can show interesting dynamics that are similar to those studied in physics--such as phase
transitions, oscillatory solutions, and spatio-temporal patterns--and therefore similar methodologies to those used in physics may be helpful in their study.  In the next
section, we will introduce some of the basic concepts of behavioral modeling that have been applied to study coupled behavior-disease interactions.

\section{Basic concepts in behavioral modeling} \label{sec:behaviormodeling}

Much of the recent research on modelling the coupled dynamics of vaccinating behavior and disease dynamics has been conducted by investigators with
training in the natural sciences, including physics, applied mathematics and epidemiology.  These investigators have borrowed concepts from the social
sciences in order to find behavioral models to couple with the existing infection transmission models they are more familiar with.  These behavioral models
can be broken down into three main categories: phenomenological models, game theoretical models, and psychological models.  However, we note
that the distinctions between these categories are not always well defined, and the delineations depends to some extent on subjective opinions.
We also note that our goal here is to characterize different ways of doing behavioral modelling, rather than different ways of modelling coupled
behavior-disease interactions \emph{per se} (see Refs.~\cite{funk2010modelling, perra2011towards} for classifications of coupled behavior-disease models).

\subsection{Phenomenological models}

Phenomenological approaches to modeling behavior mathematically describe the observed effects without positing mechanisms behind the observed
behavior.  For instance, arguably the first model to capture the impact of adaptive human behavior on disease transmission used the mean-field
ordinary differential equations \cite{capasso1978generalization}
\begin{eqnarray}
\frac{dS}{dt} & = & -g(I)S, \nonumber \\
\frac{dI}{dt} & = & g(I) S, - \gamma I \\
\frac{dR}{dt} & = & \gamma I, \nonumber
\end{eqnarray}
where $S$ is the number of susceptible individuals, $I$ is the number of infectious individuals, $R$ is the number of recovered individuals, $\gamma$
is the per capita recovery rate, and $g(I)$ is the force of infection.  The form of $g(I)$ was chosen to reflect the phenomenological impact of unspecified psychological effects.  One of the functional forms explored in
Ref.~\cite{capasso1978generalization} included
\begin{equation}
g(I) = \frac{\beta I}{1 + \beta \delta I},
\end{equation}
which saturates at high prevalence of infection, $I$.  Such saturation of the transmission rate at high levels of prevalence could occur if individuals reacted
to high prevalence by reducing their contact rate through hand-washing or other means, which is a plausible assumption for a sufficiently dangerous infectious
disease.  We describe this model as phenomenological because, although it is motivated by psychological factors, the equations only describe the impact of
such psychological factors on the transmission rate without actually modelling specific psychological processes such as individual cognition or social learning.

Similarly, as noted in the previous section, a rapidly growing body of research uses the contagion metaphor to describe the transmission of ideas and behaviors through
social networks \cite{christakis2007spread,campbell2013complex}.  We classify these approaches as phenomenological because they describe the effect of
behaviors and ideas being transmitted between individuals, without positing the mechanisms that might bring about this transfer.  Contagion of ideas is
a metaphor, and caution must be exercised when applying metaphors to complex social phenomena.  Vaccine scare and others panics undoubtedly share common
features with contagious processes, and using this metaphor can simplify and facilitate analysis of social processes.  Moreover, the contagion metaphor
comes naturally to investigators experienced in modelling contagious diseases.  However, the downside of such metaphor is that it runs the risk of being
too facile for some applications, by ignoring important subtleties of real-world social behavior \cite{alshamsi2015beyond}.  For instance, some social science
literature distinguishes between descriptive and injunctive social norms \cite{cialdini1990focus}.   Descriptive social norms are those which describe what other
individuals do, whereas injunctive social norms describe what other individuals approve or disapprove \cite{cialdini1990focus}.  It is moreover possible to distinguish
between social learning and social norms.  Social learning theory emphasizes the social aspects of cognitive processes: individuals do not learn many behaviors
by cogitating in isolation from others, but rather they learn through a more efficient process of imitating others  \cite{bandura1963social}.  An individual may use
social learning to choose among descriptive or injunctive social norms that seem to be most acceptable, or most successful.  It is not immediately clear how
the contagion metaphor applies to these nuanced distinctions of how we learn ideas and behaviors from others.  Indeed, empirical research seems to indicate
that the contagion metaphor (unsurprisingly) is not universally applicable \cite{alshamsi2015beyond}. Moreover, many of the usual criticisms of memetics also
apply to using the contagion metaphor for the spread of ideas \cite{atran2001trouble}.

However, in many contexts, these limitations of the contagion metaphor may not matter.  Science works by adopting the simplest explanation that captures the
observed behavior (Occam's Razor) and there are some contexts in which the contagion metaphor may be perfectly adequate to explain the observations.
This is often the value of phenomenological models over mechanistic models.  For instance, to prove that behavior influences the spread of infectious diseases,
it is sufficient to construct a phenomenological model that captures the impact of behavior and to show that it is superior to models that do not capture
behavior \cite{bauch2012evolutionary, he2013inferring}.  And, precisely because phenomenological models are often simpler in structure, they may make
it easier to understand how simple mechanisms such as information spread through networks can give rise to complex emergent phenomena, such as
control of infection spread in a spatially structured population \cite{funk2009ZhenPNAS}.   In other contexts, mechanistic models may be desirable, and
in the next two subsections we discuss two broad categories of mechanistic models: game theoretical models and  psychological models.

\subsection{Game theoretical models}

\begin{figure}
\begin{center}
\includegraphics[width=0.25\textwidth, angle = 270]{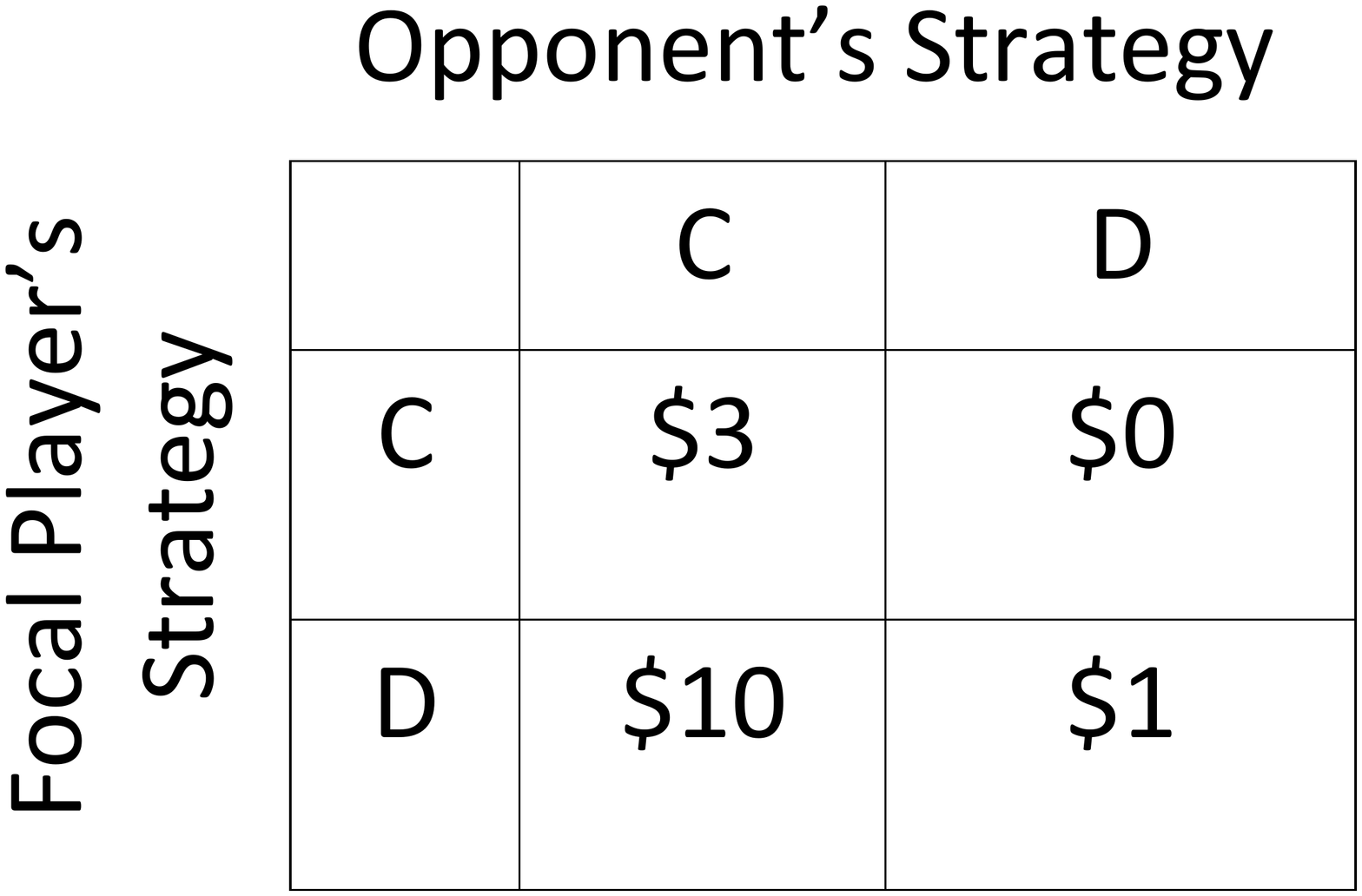}
\end{center}
\caption{Payoff matrix describing the Prisoner's Dilemma, a classic example from game theory. Players must choose whether to Cooperate (C) or Defect (D) against their opponent, with the payoffs for taking these two different courses of action also dependent on what their opponent chooses. The money values are the payoff to the focal player, and the payoffs for the opponent are symmetrical hence payoffs for the opponent are the same, if they are treated as the focal player instead.}
\label{prisoners_dilemma}
\end{figure}

Game theory is the formalization of strategic interactions between individuals in a group of two or more individuals \cite{von2007theory}. Game theory is
well exemplified by the Prisoner's Dilemma game (so called because it was originally explained as a decision process made by two prisoners being
interrogated in separate jail cells who must decide whether to confess their crime, or stick to their story of being innocent).  Each player can choose to
either cooperate ('C') or defect ('D'), and receives payoffs depending on what s/he chooses, and what her/his opponent chooses (hence, the strategic
interaction of the game).  If you, (the focal player) cooperates and your opponent also cooperates, then you receive \$3 for instance (Fig.~\ref{prisoners_dilemma}).  But if you cooperate and your opponent defects, you get no money.  On the other hand, if you defect and your
opponent cooperates, you get a huge \$10 payoff, but if your opponent also chooses to defect, you only get \$1.  Your opponent has the
same payoff outcomes, and therefore the payoff matrix is called symmetric.

What should you--the focal player--do in this situation?  If your opponent cooperates, then it is optimal for you to defect since you get \$10 instead
of \$3.  If your opponent defects, then it is also optimal for you to defect, since you get \$1 instead of \$0.  Therefore, regardless of what your
opponent does, you should choose `D', defect.  Your opponent is thinking the same way, so s/he also chooses to defect.  Therefore you both
defect and you get a payoff of \$1 each.  This is a shame, since if you had both cooperated, you would each get \$3, but the selfishly optimizing
logic of maximizing personal gain in anticipation of what your opponent will do prevents this socially (Pareto) optimal outcome.  The Defect-Defect
outcome is known as a Nash equilibrium, since neither player has an incentive to unilaterally change his or her strategy away from Defect and
therefore we expect the Nash equilibrium strategy choices to be stable.

The Prisoner's Dilemma has been used to illustrate the clash between what is optimal for the individual and what is optimal for the group in
a variety of contexts, including vaccination \cite{may2000enhanced, bauch2003ZhenPNAS, bauch2004ZhenPNAS}. In the case of high levels of vaccine coverage,
non-vaccinators are the Defectors, who benefit from herd immunity generated by vaccinators without paying any real or perceived cost of
vaccination, while vaccinators are the Cooperators.  However, this analogy only applies at high levels of vaccine coverage, since at low
levels of vaccine coverage it is optimal for individuals to accept at least some vaccination.

Game theoretical models of interactions between vaccinating behavior and disease dynamics typically define strategies such as Vaccinator
or Non-vaccinator, define payoffs for those strategies, and attempt to solve for a Nash equilibrium \cite{brito1991externalities,bauch2004ZhenPNAS,heal2005vaccination}
or the analogue from evolutionary biology, the evolutionarily stable state (ESS) \cite{smith1982evolution}.  The approaches may also attempt to
determine whether the Nash equilibrium is convergently stable, which indicates whether players will evolve, over time, toward the Nash equilibrium.
However, classical game theory may not be well suited to describing real-world vaccinating behavior. It assumes that individuals are selfishly rational
optimizers, but in fact, effects like social learning, bounded rationality, and imperfect information are very important in vaccinating decisions
\cite{sturm2005parental, allen2010parental}, and this can impact predicted behavior \cite{bauch2012evolutionary}.

Research has explored the statistical physics of games such as the Prisoner's Dilemma or the Snowdrift Game played out on lattices \cite{nowak1992evolutionary,szabo1998evolutionary,hauert2004spatial,xia2015dynamic}, subsequently expanded to networks \cite{szabo2007evolutionary,tanimoto2007dilemma,dengentropy,deng2015Generalized,deng2014belief,deng2016novel,huang2015understanding,huang2015cooperative}.
Likewise, game theoretical concepts have been applied to vaccination decisions on networks
\cite{perisic2009social,fu2011ZhenPRSB, mbah2012impact, wells2013policy}.
However, the resemblance of this work with `game theory' gradually fades, as investigators use increasingly computational approaches
in their work (and less rigorous proof of Nash equilibria) as a result of including more realistic cognitive processes.  Such cognitive processes
better reflect the importance of social influences and social norms, and the reality of imperfect information which makes individuals use `rules
of thumb' to make decisions, rather than assuming individuals can predict the future.  In the end, the only resemblance to `game theory'
may be the use of utility functions where individuals make decisions (either non-strategic, explicitly strategic, or implicitly strategic) based on how
much utility (a measure of preference) they will receive for different actions such as vaccinating or not vaccinating
\cite{perisic2009social,fu2011ZhenPRSB,mbah2012impact, wells2013policy}.  Hence, these more complex
game theoretically-inspired models blur the distinction between `game theoretical' models and `psychological' models,
the latter of which is the focus of the next subsection.

\subsection{Psychological models}\label{PsychologicalModels}

Some models of the interaction between vaccinating choices and disease dynamics invoke psychological theory in order to formulate
mechanisms underlying vaccinating decisions.  An example of a model that uses social and psychological assumptions is the following SIR model modified
to include mechanistic modelling of vaccinating behavior according to an evolutionary game theoretical approach \cite{bauch2005ZhenPRSB,reluga2006evolving}:
\begin{eqnarray}
\frac{dS}{dt} & = &  \mu (1 - x) - \beta S I - \mu, \nonumber \\
\frac{dI}{dt} & = &  \beta S I - \gamma I - \mu S,  \\
\frac{dx}{dt} & = &  \hbar x (1 - x) (-1 + \omega I), \nonumber
\end{eqnarray}
where $S$ is the proportion of susceptible individuals, $I$ is the proportion of infectious individuals, $x$ is the proportion of individuals who adopt the vaccinator
strategy, $\mu$ is the per capita birth/death rate, $\beta$ is the transmission rate, $\gamma$ is the recovery rate, $\hbar$ is a rescaled parameter
that essentially captures the rate of social learning, and $\omega$ is a rescaled parameter that essentially captures the relative perceived risk of
the disease compared to the vaccine.  In this model, individuals are not perfectly rational optimizers because they only switch strategies
through social learning ($\hbar$), after they have interacted with individuals playing a different strategy and moreover, they make the
`rule of thumb' assumption that their probability of eventually being infected depends on the current proportion of infectious persons, $I$,
rather than assuming that they could solve the SIR equations explicitly and figure out their chances of becoming eventually infected. This approach
can be straightforwardly extended to spatial settings, where individuals imitate successful strategies of their network neighbours \cite{fu2011ZhenPRSB}. A similar approach has also been adopted in models of coupled human-environment systems \cite{innes2013impact,barlow2014modelling}. 

Other approaches incorporate specific theories from psychology and social psychology in order to formulate mechanisms in coupled behavior-disease
models.  For example, random walk subjective expected utility (SEU) theory posits that individual decision-making can be modelled as a biased random
walk in a one-dimensional space of possible states with two decision outcomes on either end of the space.  Individuals move toward one decision or the other depending
on the probability of certain outcomes of their decisions, and the subjective utility associated with those outcomes, as formulated through a decision
tree \cite{busemeyer1993decision}.  Once a certain threshold has been crossed close to one end of the one dimensional space or the other, the individual
has finally decided on a course of action, and takes it. This has been applied to vaccinating decisions for influenza on a contact network, to explore
how decisions about influenza vaccination and decisions about contact precautions interact with one another through their mutual coupled influence on
disease dynamics \cite{andrews2015disease}.

Another relevant theory from psychology is prospect theory, which mathematically formalizes the phenomenon whereby humans tend to over-estimate
the probabilities of very rare events.  Since vaccine adverse events and disease complications are rare events, the relevant of prospect theory to vaccinating
decisions is obvious.  Prospect theory has been applied in mean-field models of coupled vaccinating behavior and disease dynamics for paediatric infectious
diseases, and has been found to significantly impact model predictions \cite{oraby2015bounded}.

A third example is the application of Dempster-Shafter theory, which is a general statistical framework (closely related to Bayesian statistics) for dealing
with decision-making under uncertainty \cite{shafer1976mathematical}.  This theory has been incorporated into models of coupled dynamics of vaccinating
decisions and disease dynamics, where investigators have used the theory to explore how awareness of disease-related events and vaccine-related events
spreads through a group of socially connected individuals \cite{xia2014belief}.

\section{Behavior-vaccination dynamics in well-mixed (mean-field) populations}
\label{sec:beh-vac-mean}

The dynamics and control of vaccine-preventable infectious diseases under voluntary vaccination can be quite complex, because of the role of human decisions. These are in turn based on information and \textit{rumors} gathered on: i) the \textit{perceived} risks of getting the disease; ii) its spread; iii) the \textit{perceived} risks of being affected by vaccine-related side effects. This complexity becomes apparent in the modeling phase, due to the intrinsic difficulty of quantitatively  representing human behavior, and is mirrored by the features of the resulting mean-field models, whose solutions are radically different from those of the basic model with mandatory vaccination illustrated in Section \ref{sec:compart-model}. Here we review the recent literature on the dynamic implications of the introduction of human behavior in mean-field SIR-like models for vaccine preventable infections under voluntary vaccination. The presentation broadly follows the categorization proposed in  \cite{Funk_review2010} (also refer to Section~\ref{sec:behaviormodeling} for classification), contrasting \textit{phenomenological}, or \textit{behavior-implicit} models of vaccinating behavior, with \textit{behavior-explicit} ones, such as \textit{psychological} or \textit{game-theoretical} models. However, this categorization is not always effective from the modeling viewpoint since, under certain conditions, the various approaches can become mathematically indistinguishable. More mathematically focused reviews can be found in \cite{ourbook,Funk_review2010,Bauch_Galvani_2013}.

\subsection{\textit{Phenomenological} SIR models with information-dependent vaccine uptake}
\label{InfoDependentCoverageNoiTPB2007}

A general phenomenological framework is represented by the SIR model with \textit{information-dependent} vaccination introduced in \cite{noi1,noi2,noi3,noi8}. The underlying rationale is that the possibly complex behavioral rules parents follow when deciding whether to vaccinate or not their children will often  translate into simple relationships between the individual probability to vaccinate, i.e. the \textit{vaccine demand}, and the gathered information on infection spread. These models therefore postulate relatively simple feedback rules whereby trends of infection (and of vaccine adverse effects, VAE) can affect the current vaccine uptake, eventually feeding-back on infection dynamics.

These ideas led to the following modified SIR model with vaccination by a perfect vaccine \cite{noi1} (note that for basic demographic and epidemiological quantities we use the notations adopted in Section \ref{sec:compart-model})
\begin{align}
S^{\prime}&=\mu\left(  1-x(M)\right)  -\mu S-\beta(t)SI, \label{General_SI_S}\\
I^{\prime}&=I(\beta\left(  t\right)  S-(\mu+\gamma))  \label{General_SI_I},
\end{align}
where the vaccine demand $x(.)$ is an increasing function of a new variable, $M$, which summarizes the information available on current and/or past disease trend, used e.g. to formulate expectations about future risks. Here we considered uniquely vaccination of newborns, however models considering  'delayed' vaccination strategies targeted also to later ages, i.e. of the type
\begin{equation}
S^{\prime} = \mu(1-S) -\psi(M) S - \beta(t) SI,
\end{equation}
where $\phi(M)$ is an increasing function of $M$, have a qualitative behavior that is similar to the one of the model \ref{General_SI_S}-\ref{General_SI_I} \cite{noi1}.

Focusing on the perceived risk of disease as the driving force of vaccination decision, $M$ might be any continuous and increasing function $g$ of the \textit{current} infection prevalence or incidence. For linear $g$ it would hold $M=h(t)\beta(t)SI$ (with $h(t)>0$) or $M=k(t)I$ (with $k(t)>0$). Namely, the latter expression defines the perceived risk of serious disease as the product of the perceived risk of infection, proportional to infection prevalence $I$, times a constant perceived risk of serious disease given infection.

More realistically $M$ also depends on past values of state variables, due e.g. to time-delays in the reporting/acquisition of information, and mainly to the fact that parents might also include their experience of past disease trends, etc. In this case, $M$ can be modeled as follows
\begin{equation}
M(t)=\int\limits_{-\infty}^{t}g(I(\tau))K(t-\tau)d\tau,
\label{eqM}%
\end{equation}
where the \textit{delaying kernel} $K$ \cite{Mac} is a probability density function. Besides the trivial case $K(t)=\delta(t)$ (yielding back the unlagged case), the literature has focused on kernels allowing reduction to ordinary differential equations\cite{noi1,noi2,noi3,noi8, Reluga1}, in particular on the \textit{exponentially fading memory} kernel $K(t)=a\exp(-at)$, which describe an exponential decay of the "`memory"' of agents, with characteristic memory lenght $T=1/a$ \cite{Mac}. Another kernel investigated in \cite{noi8} is the \textit{acquisition-fading} kernel  $K(t) = a_1a_2(a_2-a_1)^{-1} (e^{-a_1 t}-e^{-a_2 t} )$. This unimodal kernel is null at $t=0$, and it mimics not only the exponential decay of memory, but also an initial phase of information acquisition.

Finally, function $x$ can be written as \cite{noi1,noi3,noi8}:
\begin{equation}
x(M)=x_{0}+x_{1}(M)\ \ , 0<x_{0}<1,\label{pM}%
\end{equation}
where $x_{0}$ is a non-null baseline constant coverage, mirroring awareness of disease severity also in absence of spread, and $x_{1}(M)$ is a sufficiently regular increasing function, with $x(M)\le 1$.

For the sake of simplicity, from now on we shall assume that the information index $M$ only depends on (current or past values of) $I$, the so called \textit{prevalence-dependent} case \cite{Funk_review2010,Geoffard97}.  In the current case this assumption implies that epochs of increasing prevalence would rapidly promote an increase in vaccine uptake in children (and vice-versa).

Two main substantive questions arise about model of Eqs.~\ref{General_SI_S}-\ref{pM}, namely how prevalence-dependent vaccination affects infection control on the one hand, and, on the other hand, how it might affect the dynamical pattern of infection and vaccination, e.g. by triggering oscillations. From now on we always assume that in absence of vaccination the infection is endemic, i.e. $\mathcal{R}_0>1$.

\subsubsection{Implications of information-dependent vaccination: elimination \textquotedblleft mission impossible\textquotedblright}
The basic SIR model with constant mandatory vaccination $x$ at birth shows a simple threshold behavior governed by the \textit{vaccine reproduction number} (VRN) $\mathcal{R}_{V}=(1-x)\mathcal{R}_0$ such that if $\mathcal{R}_{V}\le 1$ (i.e. $x\ge x_{c}$) then the infection can be eliminated, otherwise the disease remains endemic. This simple result no longer holds under prevalence-dependent vaccination  which makes elimination impossible. This is a general fact holding regardless of the form of the information index $M$, of the memory kernel $K(.)$ and of the nature of the transmission rate $\beta(t)$. This may be inferred by investigating the local as well as the global stability of the disease-free equilibrium (DFE): $DFE=(1-x_{0},0,0)$ of model in Eqs.~\ref{General_SI_S}-\ref{pM}, yielding the following global eradication condition
\begin{equation}
B=\frac{1-x_{0}}{\mu+\gamma}\frac{1}{\theta}\int\limits_{0}^{\theta}\beta
(u)du \le 1\label{C_GAS_DFE},%
\end{equation}
(where $B$  represents the average VRN when the vaccine uptake is set to the baseline coverage $x_{0}$)  while if $B>1$ the DFE is unstable. For constant $\beta$ the condition $B\le1$ is equivalent to the condition $x_{0}>x_c$, in other words elimination can only be achieved if the baseline vaccine uptake lies above the elimination threshold. Actually, it is a documented fact that western countries have been able to maintain very high coverages - i.e. above the estimated elimination threshold - against e.g. polio and diphtheria for decades. This however has been possible under situations where these  vaccinations were essentially mandatory. Whether similar results could be maintained under fully voluntary vaccination regimes seems to be quite remote, given that scenarios where prevalence is very small would also make the perceived reward of vaccination to vanish, so that people would start escaping from vaccination.

\subsubsection{Memory-triggered oscillations}
\label{delay_oscillations}
Given that infection elimination is ruled out, the question turns into the types of dynamics that might be triggered by prevalence-dependent vaccination.
For  $\mathcal{R}_{0}(1-x_{0})>1$ and constant $\beta$, the system of Eqs.~ \ref{General_SI_S}-\ref{pM} has a unique endemic equilibrium
$EE=(S_{e},I_{e},M_{e})$.

In the unlagged case the system becomes 2-dimensional and it is easy to prove that the endemic state is globally asymptotically stable (GAS), so that only damped oscillations are possible \cite{noi1}. Instead, in the delayed case, stable oscillations - via Hopf bifurcation of the endemic state - already appear under the simplest pattern of delay, namely the exponentially fading memory. In particular, assuming as
key parameter the inverse of the average memory/delay lenght $a=1/T$, and taking for simplicity $g(I)=kI$, it holds that \cite{noi1}  if and only if%
\begin{equation}
\left(  \beta I_{e}+\mu\right)  ^{2}-\beta I_{e}\mu kx_{1}^{\prime}%
(M_{e})+2\left(  \beta I_{e}+\mu\right)  \sqrt{\beta I_{e}(\gamma+\mu
)}<0,\label{Cond_Hopf}%
\end{equation}
there is a range $[  a_{1},a_{2}]$ for $a$ such that: i) for $a\in(a_{1},a_{2})$ Yabucovich oscillations \cite{yf3} occur via Hopf bifurcations at $a_1$ and $a_2$; ii) for
$a\notin\left[  a_{1},a_{2}\right]$ the endemic equilibrium EE is locally stable. The global stability of EE, suggested by numerical sumulations, is nontrivial and it has been investigated in \cite{mbsbbdonlac} by means of the Li-Muldowney geometric theorem, which extends the Dulac-Bendixon theorem.

Note that Eq.~\ref{Cond_Hopf} shows that sustained oscillations can appear only if $x_{1}^{\prime}(M_{e})$ is sufficiently steep, that is only if the behavioral response of vaccine uptake to changes in the perceived risks is sufficiently intense. In this case oscillations occur in an appropriate intermediate window of the average delay $1/a$. These results continue to hold for more realistic patterns of delay, such as acquisition-fading and Erlangian kernels \cite{noi1,noi2,noi3,noi4,noi8}, for generic increasing $g(I)$ functions \cite{noi3}, in presence of disease-related mortality and a non-constant population \cite{noi3}, and when the incidence of VAE is taken as the key determinant of the vaccine demand \cite{noi4}. Finally, as shown in \cite{buonomo2013modeling}, oscillations in behavior-depending vaccination models are also observed in absence of memory mechanisms but in presence of disease-intrinsic delays, such as in the SEIR model (the latent period being, of course, equivalent to an exponentially distributed delay in the onset of infectivity).

\subsubsection{Dynamic pattern for a measles-like infection}
\label{Dynamics_pheno_model_delay}
We illustrate some of the above described dynamic oscillatory patterns, under the parameter values adopted in section \ref{SIR_vaccination}. As for behavioral rules we choose $g(S,I)=I$, and a Michaelis-Menten-type \cite{murray1} coverage function $ x_{1}(M)=(1-x_{0}-\varepsilon) D M/(D M+1)$ where $D>0$ is a shape parameter tuning the reactivity of $x_{1}$, and $1-x_{0}-\varepsilon$ represents the \textit{ maximal coverage } arising for large $D M$ values. Behavioral parameters are chosen in the oscillatory region, with the average delay $T=1/a$ set to 4 months, $x_{0}=0.75$, $\varepsilon=0.01$ and $D=15\times 10^4$. The selected values of $\varepsilon$ and $D$ allow, in principle, the vaccine uptake to reach values close to $100\%$ during epochs of high perceived risk.

\begin{figure}[t]
\centering \includegraphics[width=0.7\textwidth] {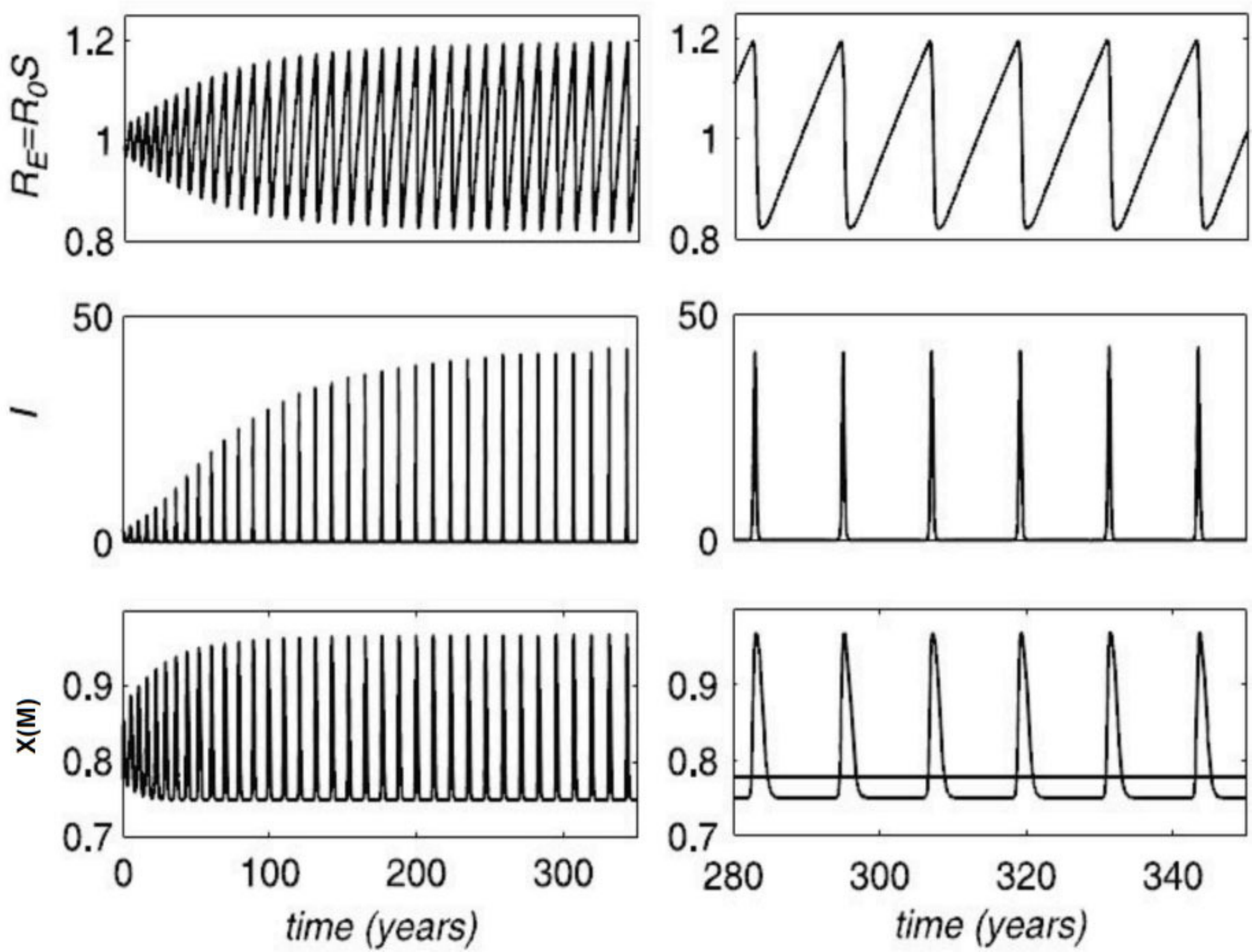}
\caption{Dynamics of model in Eqs.~\ref{General_SI_S}-\ref{pM} for $g(S,I)=I$ under an exponentially fading kernel ($T=4$ months) and a Michaelis-Menten-type prevalence-dependent component of vaccine uptake ($D=500$). Transient (left-hand side) and long-term (righ-hand side) paths of: (top) effective reproduction number $R_E=R_0 S$; (middle) infective prevalence $I$ (normalized to its equilibrium value); and (bottom) overall vaccine uptake $x(M)$ (compared to its time average given by the flat line in the right-bottom graph). $R_0$ is equal to $15$. Source: Reprinted figure from Ref.~\cite{noi8} } 
\label{TPB}       
\end{figure}

The model is initialized at time $t=0$ at the endemic state determined by a constant coverage set at the baseline level $x_0$.  Fig.~\ref{TPB} reports the transient (left-hand side) and long-term (righ-hand side) time paths of the ERN $R_E(t)=\mathcal{R}_0 S(t)$ (top), of the infective prevalence (medium), and of the overall vaccine uptake $x(M)=x_0+x_{1}(M)$ (bottom), jointly with its time average. State variables converge to a stable limit cycle in about 150 years, with a long-term inter-epidemic period around $12$ years, i.e. about 2.5 times the pseudo-period of the SIR model with constant coverage $x_{0}=0.75$. The most striking fact is that though the vaccine uptake $x(M)$ reaches levels as high as $96\%$ during epochs of high perceived risk, this is totally insufficient for elimination, as witnessed by the average long-term coverage, which remains below $80\%$, i.e. well below the critical coverage $x_{c}\approx 0.93$.
This oscillatory nature appears to be intrinsic to SIR systems with delayed prevalence-dependent vaccine demand: epochs of increasing prevalence will yield, after a certain period of time, an increase in the \textit{demand} for vaccines, which will in turn reduce infection prevalence, therefore eventually feeding-back on the vaccine uptake and so on, as first noted in \cite{Geoffard97}.

%
%

\subsubsection{Further extensions and remarks on phenomenological models}
The information-dependent framework is a flexible one from several standpoints. As mentioned above, a model analogous to the one presented here has been applied to investigate the effects of VAE \cite{noi4} which, in the current western situation where incidence of most infections is persistently low, appear to represent the most important component of the vaccine demand. In particular, in \cite{noi4} the key determinant of the vaccine demand was the information on the incidence of VAE. Role of VAE as vaccination determinants will be considered more in detail in the \textit{behavior-explicit} models of next subsection. However, many other applications are possible. For example the framework in~\cite{noi1} was also used to consider the possibility  that parents adopt the strategy of delaying the age at vaccination of their children, later extended in a more general behavior-explicit setting in \cite{BhattaBauch2010}.

More important, the phenomenological model of Eqs.~\ref{General_SI_S}-\ref{eqM} represents a general form in that, as we have remarked, also \textit{behavior-explicit} (e.g., game theoretic) formulations for vaccine uptake eventually collapse into the same structure, once the vaccine uptake function resulting from the adopted formulation is set within the framework of a dynamic epidemiological model \cite{Reluga1, Geoffard97}. For example, game-theoretic approaches (discussed in section \ref{game-theoretic models}) often yield step-wise \textit{best response} behavioral functions. The simplest example for the vaccination problem would be \textit{not to vaccinate} with probability one when the risk of infection is below a certain threshold, \textit{to vaccinate} above that threshold, and to vaccinate with a certain probability when the risk is exactly at threshold. However, a number of reasons (e.g., bounded rationality) suggest to approximate the best response function by a corresponding \textit{smoothed responses} \cite{Xu_Cressmann_2016}. The shape of smooth response function is typically analogous to that of function $x_{1}(.)$ used here.

\subsection{\textit{Behavior-explicit} psychological models of vaccinating behavior}
\label{ImitationBasedVaccination} 
Most available psychological models of vaccinating behavior in previous sections relied on what we will term \textit{Imitation Game Dynamics} (IGD), an important class of models developed within \textit{Evolutionary Game Dynamics Theory} (EGDT)\cite{nowak:2006a,nowak:2006b,Hofbauer98}, which is a powerful approach to modeling behavior by setting ideas from game theory into a population dynamics framework \cite{nowak:2006a,nowak:2006b,Hofbauer98}. \textit{Imitation} represent a main avenue through which the different strategies enacted by players can spread within a population, through mutual learning between individuals \cite{Bauch2}. The first application of IGD to investigate vaccinating behavior and its feedback on infection dynamics and control was developed in a seminal paper by Bauch \cite{Bauch2}, later extended \cite{noi5} by a more general formulation which will prove useful to link the class of imitation-based models with the phenomenological models of previous section.

\subsubsection{Imitation dynamics}
IGD is usually presented using the framework of game theory, both in its economic and in bioevolutionary standpoints. However, we note that at its very core there is again a mass-action diffusion process based on statistical physics \cite{Hofbauer98}. Let us consider a binary decision problem, i.e. one for which only two, mutually exclusive, strategies are played: strategy  $V$ and $N$ (these might for example be \textit{to Vaccinate}, and \textit{Not to Vaccinate}). Let us also assume that individuals can switch from the strategy they are currently playing to the alternative one by \textit{learning and imitating} after social encounters with subjects that are playing the other strategy. Considering a population of constant size, and indicating with $y_V$ and $y_N$ the fractions of the $N$ and $V$ sub-populations ($y_V +y_N=1$), leads to the following mean-field \textit{double infection} model:\\
\begin{align}
y_V^{\prime} =& - K_{VN}y_N y_V + K_{NV}y_N y_V \label{sV},\\
y_N^{\prime} =& - K_{NV}y_N y_V + K_{VN}y_N y_V \label{sN},
\end{align}
\noindent where the coefficients $K_{NV}, K_{VN}$ have a clear "epidemiological" interpretations as the \textit{strategy-specific} transmission rates
following social contacts with individuals playing the other strategy. Following EGDT \cite{Hofbauer98}, these coefficients are proportional to the perceived benefit, or \textit{payoff}, arising from strategy switching.
Since $y_N = 1 - y_V $, it follows
\begin{equation}\label{igdeq}y_V^{\prime} = (K_{NV}- K_{VN})y_V(1-y_V).   \end{equation}
Note that if $K_{NV}- K_{VN}$ is strictly positive (negative) in all the state space, strategy $V$ ($N$) will eventually prevail, by spreading to the entire population with a logistic-type dynamics, as is intuitive given that the quantity $K_{NV}-K_{VN}>0$ must be proportional to the perceived \textit{net payoff gain} when switching from strategy $N$ to $V$. More interesting, dynamic situations arise when the net payoff $K_{NV}-K_{VN}$ has non-constant sign, which happens when it is made dependent on other state variables of the system, as is the case for vaccination games \cite{Bauch2,noi5,BhattaBauch2010,BauchBhattaPLOSCB2012, oraby2014influence, OrabyBauch2014_bounded_rationality}.\\

\subsubsection{Basic SIR models with imitation-based vaccinating behavior}
\label{VAE_prevalence_models}
Unlike previous subsection where the vaccine uptake at birth $x$ was taken as a "static" function of $I$ (and/or of $S$), we assume that $x$ is a state variable determined by the switches of parents between strategy $N$ (not to vaccinate) and $V$ (vaccinate) according to the above described IGD. Therefore, $x(t)$ actually represents the proportion of parents favorable to vaccination at time $t$, which is taken as a proxy for the actual emerging vaccine uptake. Setting $y_V = x,$ yields the following family of mean-field SIR models with vaccination \cite{noi5,noibeta}
\begin{align}
S^{\prime}  &  =\mu(1-x)-\mu S-\beta SI\label{sirvs},\\
I^{\prime}  &  =\beta SI-(\mu+\gamma)I\label{sirvi},\\
x^{\prime}  &  =\kappa_{1} \Delta E x(1-x) \label{sirvp},%
\end{align}
where the term $K_{NV}-K_{VN}$ has been represented by the product between the scale coefficient $\kappa_{1}$, tuning the speed of the imitation process, and the net expected \textit{payoff gain} of vaccination, $\Delta E$. The latter can in turn be defined as the difference $e_{V}-e_{N}$ between the \textit{vaccinator payoff} and the \textit{non vaccinator payoff}, or alternatively as $C_{V}-C_{N}$ where $C_{i}=-e_{i}$ represents the corresponding \textit{cost}.

Note preliminarly that irrespective of the specific form of $\Delta E(t)$, the family of models in Eqs.~\ref{sirvs}-\ref{sirvp}, always has the following three equilibria: (i) a disease-free state with no vaccinators $A=(1,0,0)$, which is unstable due to the above assumption that $\mathcal{R}_0>1$; (ii) a \textit {pure-vaccinator} disease-free equilibrium $B=(0,0,1)$ where everyone is vaccinated; (iii) the pre-vaccination endemic equilibrium $C=\left(
S_{SIR}, I_{SIR},x_{SIR})=(\mathcal{R}_0^{-1},\mu(1-\mathcal{R}_{0}^{-1})/(\mu+\gamma),0\right)$. On the other hand, the stability of $B$ and $C$ as well as the existence of further equilibria (induced by the behavioral component) depend on the chosen form of the perceived payoff gain $\Delta E(t)$.

\subsubsection{The baseline SIR model with imitation-based vaccinating behavior}
\label{Bauch_2005}
The seminal study \cite{Bauch2}, motivated by need to understand the causes and dynamics of \textit{vaccine scares}, such as the pertussis or the MMR vaccine scares,
rests on two main hypotheses, namely: (a) a prevalence-dependent cost of not vaccinating (see Fig. 21), of the form $C_N=r_{I}m I(t)$, where $m I(t)$ is the current perceived risk of infection (taken as an estimate of the actual FOI $\lambda=\beta I$), and $r_{I}>0$ is the perceived (conditional) risk of serious disease given infection; and (b) a constant cost of vaccinating  $C_V = r_{V}>0$, representing the perceived risk of suffering a VAE per single vaccination.  Hence
\begin{equation}
\Delta E(t)=r_{I}mI(t)-r_{V}=r_{V}\left( \vartheta I(t)-1\right),
\end{equation}
where $\vartheta=mr_{I}/r_{V}$ is a measure of the
\textit {relative cost} of disease with respect to that of vaccination. Based on these hypotheses, a number of substantive results were shown \cite{Bauch2}. First, the pure-vaccinator equilibrium $B$ is unstable, reflecting the fact that in a wholly vaccinated population, prevalence is equal to zero and therefore the payoff of vaccination is strictly negative ($\Delta E=-r_{V}$), removing incentive to vaccinate, so that vaccination will start declining. Second, there is a \textit{partial vaccinator} endemic state $D$, that is an endemic state with positive vaccine uptake  $D=\left(\mathcal{R}_0^{-1},\vartheta^{-1},\widehat{x}\right)$, where
$\widehat{x}= \left(1+\gamma/\mu\right)\left(I_{SIR}-\vartheta^{-1}  \right).$ The equilibrium coverage $\widehat{x}$ is meaningful only for values of the behavioral parameters ensuring that $I_D<I_{SIR}$, i.e. such that the corresponding endemic prevalence is lower than the prevalence observed in absence of any vaccination. This in turn requires  $\vartheta > I_{SIR}^{-1}$, meaning that $D$ exists only for sufficiently high levels of the relative cost of disease. Finally, at $\vartheta=\vartheta_{0}=I_{SIR}^{-1}$  there is a transcritical bifurcation between $C$ and $D$. In words, given that endemic infection for which at some stage a vaccine becomes available, then: (i) if the relative cost of the infection is low (i.e. the perceived cost of vaccination is high with respect to infection) then the vaccine would not be adopted due to lack of incentive; (ii) if the relative cost of infection is high then the vaccine would spread in the population, therefore destabilizing the pre-vaccination endemic state $C$, and sustaining the appearance of the new, locally stable, endemic state $D$ with positive vaccination. In turn, the stability of $D$ depends on $\kappa_{1} (\vartheta-\vartheta_{0})$, showing in particular that appropriately large values of the imitation coefficient can yield sustained oscillations by Hopf bifurcation of $D$, indicating that fast social learning is a source of oscillations.

Previous results add robust evidence, based on a behavior explicit model, to those reported in section \ref{InfoDependentCoverageNoiTPB2007}, indicating that prevalence-dependent vaccinating behavior represents a serious threat for programs aimed to infection elimination.

\begin{figure}[t]
\centering \includegraphics[width=0.7\textwidth] {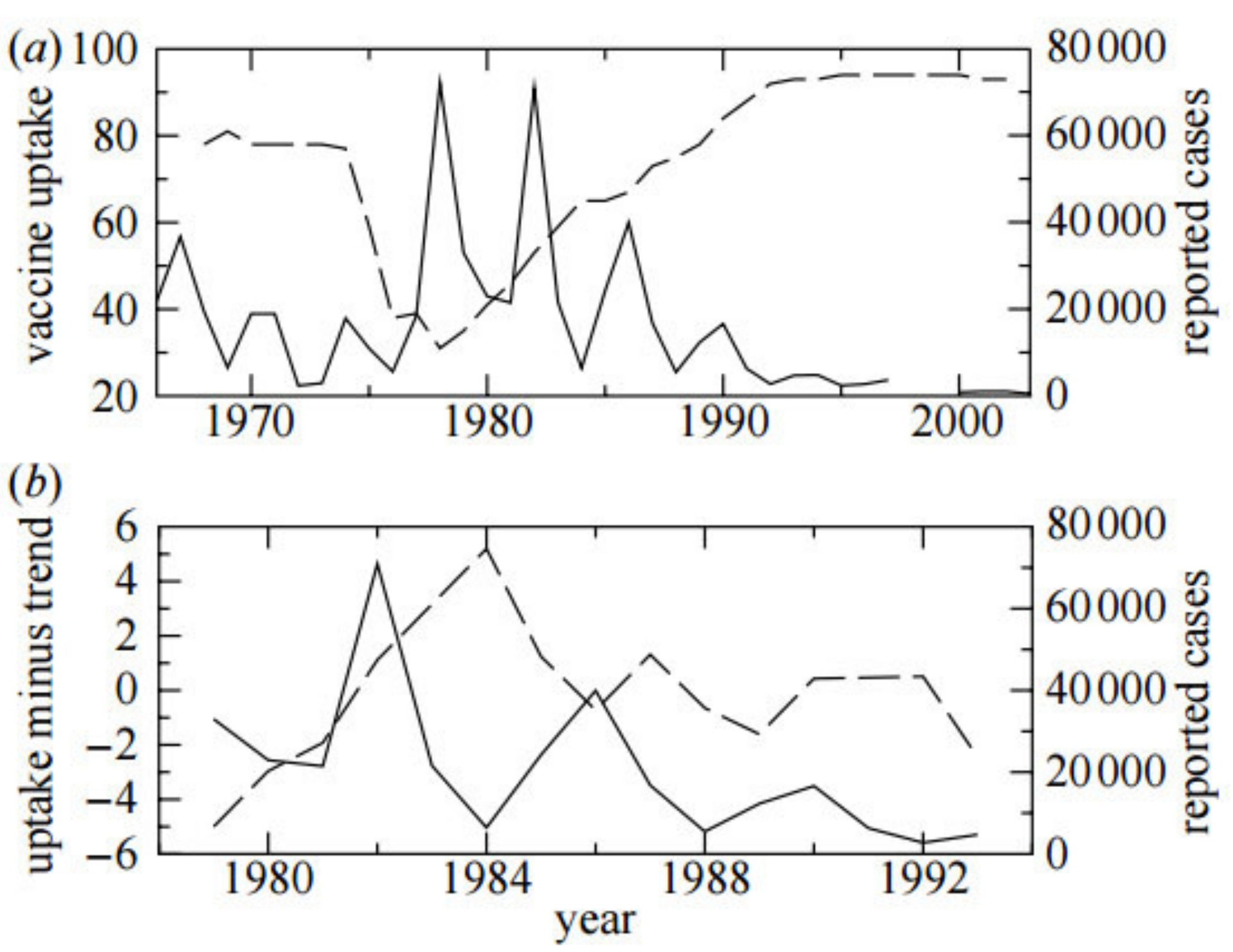}
\caption{Trends in pertussis vaccine uptake (dashed line) and case notifications (solid) in England and Wales 1967-2003 motivating the idea of prevalence-dependence in vaccine uptake: (a) absolute levels, (b) with vaccine uptake represented as deviations from a linear least-squares
fit. Note in particular that when coverage went abruptly down due to vaccine scare (mid seventies), large epidemics restarted, possibly uprising the perceived risk of infection, and eventually restoring  vaccine uptake at high levels. Nonetheless symptoms of prevalence-dependence persist, though at a lower scale, even after the overall relapse, as evidenced in (b). Moreover, the initial (pre-1975) persistently high coverage, bringing incidence to low and declining levels, might have sharply increased the number of VAE, eventually triggering the vaccine scare. Source: Reprinted figure from Ref.~\cite{Bauch2}.}
\label{Fig1_Bauch2005}   
\end{figure}

\subsubsection{An alternative model with myopic perception of the cost of vaccine adverse events}
\label{IGD_with_myopic_perception_of_VAE}
The hypothesis made in previous subsection that the perceived risk of vaccinating is constant, implies that the public correctly evaluates this risk, as the ratio between total reported VAEs for a given infection, and the total number of vaccinations for that infection per unit of time. However, due to the current high levels of herd immunity in modern industrialized countries, the perceived risk of serious disease from many infections is currently very low, while the large number of vaccinations administered unavoidably yields a steady flow of VAEs \cite{CDC_VAE}. The VAE burden might then make the perceived risk of suffering vaccine side effects (VSE) higher than risks perceived from infection.

Therefore, in \cite{noi5} it was assumed that the public \textit {myopically} evaluates the risk of VSEs by available information on the total number of VAE, which is proportional to actual vaccine uptake $x$. In particular, it was investigated the case $C_{V} = \alpha x$, defining the perceived cost of vaccination as the product of the (perceived) probability of being vaccinated, times a perceived probability of suffering VSE given vaccination. Similarly, $C_{N}$ is taken as an increasing function of infection prevalence: $C_{N}=h_{1}(I(t))$. Note that the case $h_{1}(0)>0$ is a realistic one as it might represent a scenario of local elimination where however infection re-emergence (e.g. by importation of cases) is feared.

The resulting dynamic equation for $x(t)$ is
\begin{equation}
x^{\prime}=\kappa\left(  h(I)-x\right)  x\left(  1-x\right),  \label{general}%
\end{equation}
where $\kappa =\kappa_{1} \alpha$, and $h_1(\cdot)=h(\cdot)/\alpha$.

Interestingly, if imitation dynamics is fast compared to infection dynamics, i.e. $\kappa>>(\mu+\gamma)$, then a quasi-stead state approximation yields
\begin{equation}
x(t)\approx\min\left(  h\left(  I(t)\right)  ,1\right),  \label{spert}%
\end{equation}
i.e. the information-dependent model with current information (i.e. Eqs~\ref{General_SI_S}-\ref{pM}) presented in the previous subsection is recovered. In other words, the phenomenological model \cite{noi1} is a special case of the imitation-based case when the imitation process is fast.

Comparing with the baseline model of previous section \ref{Bauch_2005}, we note that the alternative model of Eqs. \ref{sirvs}-\ref{general} also has the equilibria $A$ (unstable), $B$ (unstable) and $C$. Unlike section \ref{Bauch_2005} \cite{Bauch2}, the pre-vaccination endemic equilibrium $C$ is always unstable, reflecting the fact that introduction of a vaccine will always be successful, due to the low perceived cost of vaccination at low coverage levels.
Moreover, two further equilibria are induced by the inclusion of behavior i.e., a disease-free equilibrium $E$ with positive vaccine uptake (\textit{disease free partial vaccinator} equilibrium) $E=(1-h(0),0,h(0))$, and the partial vaccinator endemic state $D$
\begin{equation}
D=\left(  \mathcal{R}_{0}^{-1},I_{e},h\left(  I_{e}\right)  \right),
\end{equation}
where $\mathcal{R}_{0}=\beta/(\mu+\gamma)$ is the basic reproduction number of the infection, and $I_{e}$ is the unique solution of the equation
\begin{equation}
h(I)=  1-\mathcal{R}_{0}^{-1} -\frac{\mu+\gamma}{\mu}I . \label{equi}%
\end{equation}

Eq. \ref{spert}  alone can explain the behavior of the disease-free equilibrium $E$. Indeed, from $x^{\prime} \ge \kappa(h(0)-x)x(1-x)$ it follows that for large times $x(t)\ge min(h(0),1)$ and in turn: if $h(0)\ge x_{c}$ then $x(t)\rightarrow 1$, i.e. $E$ is globally attractive. Moreover, linearizing at $E$ it follows that if $h(0)<x_{c}$ then $E$ is unstable. Therefore, in case of myopic evaluation of VAE elimination turns out to be possible, but only if the perceived cost of disease associated with infection re-emergence is so large to bring the vaccine uptake in excess of the elimination threshold.

As all equilibria are independent of $\kappa$, it is natural to choose
$\kappa$ as a bifurcation parameter. Not surprisingly, the dynamics around the partial vaccinator endemic state $D$ depends on the steepness of function $h(I)$ at $D$. Namely, it exists a threshold:
$$ th = \frac{(\mu+\beta I_e)\left( (\mu+\beta I_e) + 2 \sqrt{\beta I_e(\mu+\gamma)} \right)}{\beta I_e  \mu} $$
such that i) if $h^{\prime}\left(I_e\right)<th$ then equilibrium $D$ is locally stable irrespective of the imitation speed $\kappa$; ii) if $h^{\prime}\left(I_e\right)>th$ then there is a range of values $(\kappa_1,\kappa_2)$ for the imitation speed $\kappa$ such that for $\kappa \in [\kappa_1,\kappa_2]$ and the system's orbits $x(t)=\left(S(t),I(t),x(t)\right)$ are oscillatory (i.e. recurrent epidemics are predicted) in the sense of Yabucovich \cite{yf3}, which onset via Hopf bifurcations at $\kappa_1$ and at $\kappa_2$.

Note that unlike the model reported in section \ref{Bauch_2005} where sustained oscillations around the endemic
state were triggered by large values of the imitation coefficient $\kappa$ \cite{Bauch2}, in the present model oscillations occur in an intermediate window of values of $\kappa$. This suggests that both slow and fast imitation might be stabilizing forces, quite similarly to what is reported in section \ref{delay_oscillations} for the phenomenological model. This is not surprising since $x(t)$ depends on $I(t)$ via a differential equation, i.e. $x(t)$ "follows" $h(I(t))$ with an appropriate delay.

\begin{figure}[t]
\centering \includegraphics[width=0.75\textwidth] {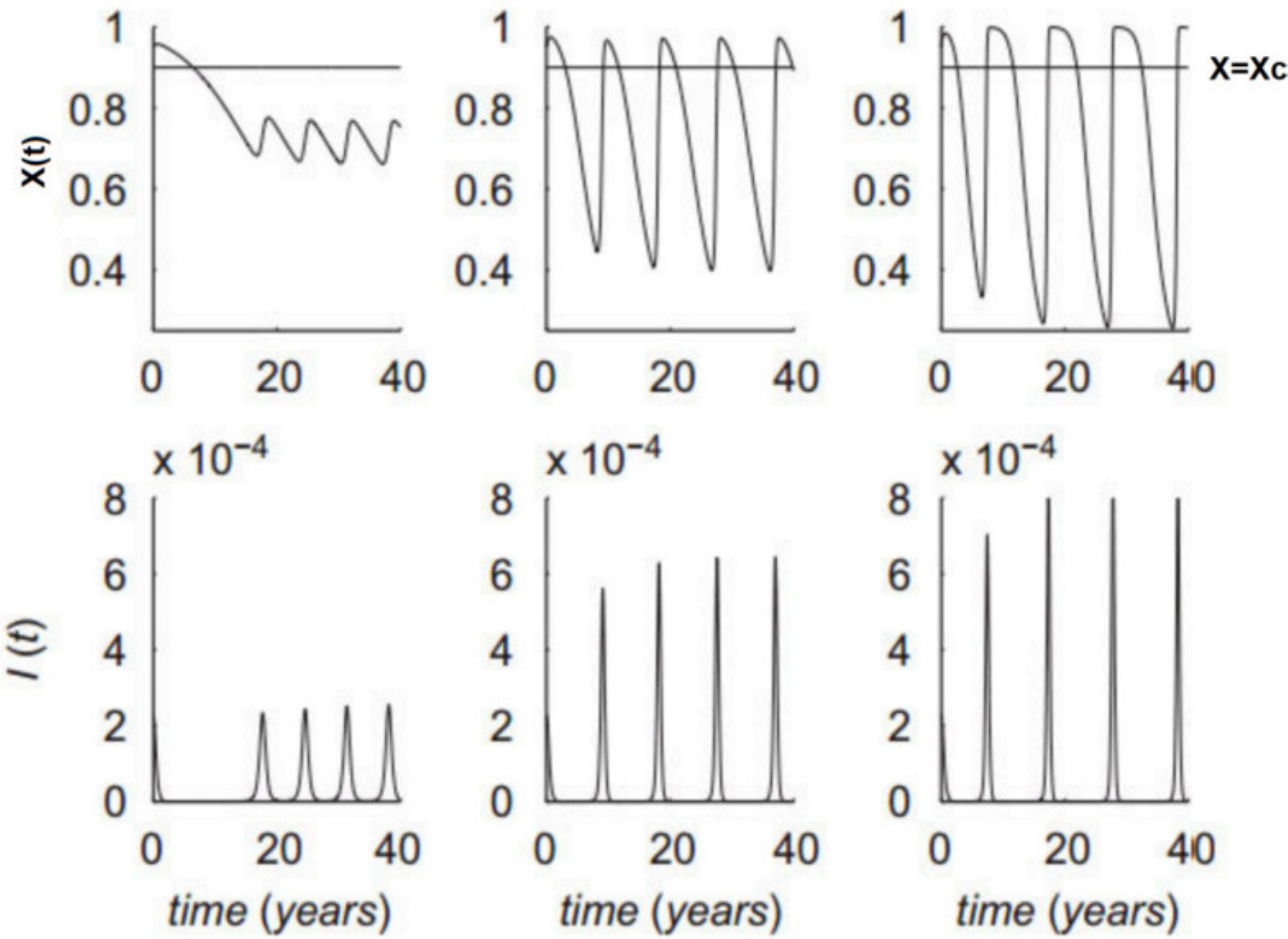} 
\caption{Oscillatory dynamics of the SIR model with vaccination decisions governed by Eq.~\ref{general} for a linear $h(I)$ of the form: $h(I)= \theta I, \theta>0$. The graphs report the dynamics of the vaccinating proportion $x(t)$ (top) and of infection prevalence $I(t)$ (bottom), for different values of the rescaled imitation coefficient $\kappa$: $\kappa=0.0005 day^{-1}$ (left), $\kappa=0.002 day^{-1}$ (centre), $\kappa=0.0035 day^{-1}$ (right). The system is initialized from the pre-vaccination endemic levels of variables $S,I$ and from $x(0)=0.95$. Demographic and epidemiological parameters are: $\mu = (1/75)year^{-1}$, $\gamma = 1/7 day^{-1}$, $\mathcal{R}_0 =10$,  $= 15,000$, $\kappa=0.002 day^{-1}$. Source: Reprinted figure from Ref.~\cite{noi5}. With permission from Elsevier.}
\label{Fig3_AlbertoPiero_JTB2011}       
\end{figure}

Possible dynamics of the system for some parametric constellations promoting sustained oscillations are illustrated in Fig.~\ref{Fig3_AlbertoPiero_JTB2011} for a measles-like infection. A noticeable fact is that large levels of the imitation coefficient $\kappa$, besides triggering very long inter-epidemic periods, can promote long-lasting epochs where the vaccine uptake $x$ lies above the elimination threshold $x_{c}$.

\paragraph{Modeling irrational behaviors}
A positive aspect of the imitation game-based seminal models \cite{Bauch2,noi5,noibeta} is that they parsimoniously describe, by means of a classical tool of game theory, the complex  social learning underlying the parental decision to vaccinate their children. Moreover, broadly speaking, the imitation process represent a first deviation from full rational behavior.  However, models
\cite{Bauch2,noi5,noibeta} are based on an important but implicit assumption, i.e. that the "force" driving the vaccine non-acceptance is para-rational. Indeed either it is constant (and fully dictated by a rational thought, as in \cite{Bauch2}) or it depends on the information on vaccine side effects (as in \cite{noi5,noibeta}). In reality in doing this the models \cite{Bauch2,noi5,noibeta} miss the fully irrational behavior underlying the periods of vaccine scares.

In \cite{BauchBhattaPLOSCB2012} Bauch and Bhattacharyya go an important step further by assuming that the "anti-vaccination force" (which depends on the vaccine risk perception) is an empirical function mimicking the rumours-driven sudden rise, stay and decay of vaccine scare. As far as the "force" favoring the vaccine uptake (i.e. the feedback from disease prevalence) they assume that it is proportional to the time-series of notified cases. The resulting dynamics for the fraction of vaccinated newborns $x(t)$ is thus of the type:
\begin{equation}\label{BM} x^{\prime}= \kappa x(1-x)( u L(t) -\omega(t) ), \end{equation}
where $L(t)$ is an interpolation of notified cases, $u$ is a proportionality constant and $\omega(t)$ is an appropriate unimodal function. Interestingly, the authors consider also other simplified models,\\
\noindent a) absence of social learning:
\begin{equation}
x(t) = u L(t) -\omega(t);
\end{equation}
b) absence of disease driven feedback:
\begin{equation}
x(t) = 1 -\omega(t);
\end{equation}
c) absence of pro-vaccination "force":
\begin{equation}
x^{\prime}= \kappa x(1-x)( -\omega(t) );
\end{equation}
d) explanatory model with force pro-vaccine depending on the state variable $I$ (as in \cite{Bauch2}):
\begin{equation}\label{BI}
x^{\prime}= \kappa x(1-x)( I(t) -\omega(t) ),
\end{equation}
complemented by an SIR model with vital dynamics, of course.

They applied the various models, and various models of the function $\omega(t)$, to available data concerning the whole cells pertussis vaccine scare and to data on the MMR vaccine scare. These multiple comparison not only were done in order to find the best fitting and predicting model, but also in order to disentangle the effect of each component: the social learning, the kind of curve $\omega(t)$ modelling the vaccine scare; the type of disease-related feedback. They obtained the following interesting results: i) the best fit to available data were obtained by the model of Eq.~\ref{BI} including both the disease-related feedback and the social learning mechanism; ii) this model also produced good predictions, up to 10 years in the case of pertussis data; iii) the model of Eq.~\ref{BM} performed better than the simplified models.

\subsubsection{Delaying vaccination as a further strategy}
A wide empirical evidence suggests that parents often adopt the strategy of delaying their children's \textit{age at vaccination} against common vaccine preventable infections (\cite{BhattaBauch2010} and references therein), mostly because they fear that the immune system of a newborn is still too weak to adequately tolerate immunization. Bhattacharyya and Bauch \cite{BhattaBauch2010} proposed a minimal 3-strategy model, with two possible vaccination rounds, the first at birth (age $0$), and the second one at a fixed subsequent age $a_1$ set to four years \cite{BhattaBauch2010}, always by a \textit{perfect} vaccine. This is remindful of the two-rounds MMR (measles-mumps-rubella) vaccination, with an early first dose and a second one offered during the childhood to reach unvaccinated or non-successfully vaccinated children. Parents can opt between the following strategies: i) \textit{timely vaccinator} ($V$) i.e. vaccinate at the first round; ii)  \textit{delayer} ($Del$) i.e. vaccinate at the second round; ii) \textit{Non vaccinators} ($N$). The three corresponding fractions of the population are denoted as $x_{V}$, $x_{Del}$ and $x_{N}=1-x_V-x_{Del}$. The adopted model is a two-age classes meta-population version of the SIR endemic model, with exponential transitions between age groups and inhomogeneous age-specific mixing (see section \ref{metapop}).

Individual learn and switch between strategies according to an IGD driven by the constant perceived risks of VAE from vaccination at ages $0$ and $a_1$ respectively, denoted by $r_{V,1}< r_{V,2}$, and by the direct prevalence-dependent perceived risk of disease in the two age groups \cite{Bauch2}. The latter depends on the corresponding risks $\eta_i$ ($i=1,2$) of acquiring infection in age groups $i=1,2$, that are modeled as follows  $\eta_i = \chi_i I_i/(\rho_1 +I_1)$, namely as increasing and saturating functions of the (perceived) group-specific infective prevalence $I_i$ ($i=1,2$) to mirror two stylized facts, namely that (i) social contacts relevant for transmission are largely assortative i.e., tend to mostly occur with individuals of the same age group \cite{Mossong}, and (ii) agents form their perceptions of risk according to simple heuristics. As for payoffs: (i) the timely vaccinator payoff only includes the vaccination cost at age zero, i.e. $C_V = -r_{V,1}$;  (ii) the delayer payoff includes the sum of the cost of acquiring disease in period 1, plus the cost of vaccination at age 2, which arises only if he/she escaped infection during period 1, i.e. $C_{Del}=-r_I \eta_1 - r_V (1-\eta_1)$, (iii) the non-vaccinator payoff is $C_{N}=-r_I \eta_1 - r_I \eta_2 (1-\eta_1)$.

The resulting dynamics of state variables $x_V$ and $x_{Del}$ (the dynamics of $x_N$ is given by $x_{N}=1-x_V-x_{Del}$) obey
\begin{align}
x_{V}^{\prime}  &  = \kappa_{v} (  \Delta E_{V,Del} x_{V}x_{Del} + x_{V} (1-x_{V}-x_{Del}) \Delta E_{V,N}), \label{p1_timely_vacc}\\
x_{Del}^{\prime}  &  = \kappa_{Del} ( x_{Del}(1-x_{V}-x_{Del}) \Delta E_{Del,N} - x_{Del} x_{V}  \Delta E_{Del,V} ), \label{p2_delayer}%
\end{align}
\noindent where (similarly to the previously illustrated IGD-based models) the quantities $\Delta E_{i,j}=$ $ e_{i}-e_{j}=$ $C_{j}-C_{i}$ $(i,j = V,Del,N)$ represent the pairwise perceived payoff gain resulting from social encounters between individuals belonging to group $i$ and group $j$, resulting from the difference of the related costs. Quite reasonably Eqs. \ref{p1_timely_vacc}-\ref{p2_delayer} postulate a simplified mixing pattern between players of various strategies (compared to the mixing relevant for infection transmission), as indeed parents of children eligible for vaccination represent a restricted subgroup of the population in age group 2. The full model is obtained by adding to Eqs. \ref{p1_timely_vacc}-\ref{p2_delayer} the equations for susceptible and infective individuals in age groups 1,2.

The qualitative behavior of the model is more complicated than the baseline model, although basic intuitions from section \ref{Bauch_2005} explain well the main results. For example, in addition to the unstable \textit{infection free-all vaccinators} equilibrium, there is an unstable endemic \textit{pure delayer} equilibrium with $I_2=0$. This further confirms the impossibility of disease elimination under prevalence-dependent cost of infection.

Several mutually exclusive other locally stable endemic states exist, including the pre-vaccination equilibrium and various \textit{mixed} states with sub-optimal vaccine uptake combining different proportions of individuals following the different strategies. Interesting oscillatory regimens were observed in the simulations: individuals switch between strategies depending on the pattern of prevalence, with frequently observed anti-phase oscillations between strategies $V$ and $Del$, e.g. epochs of high prevalence incentivate timely vaccination, which in turn reduce prevalence, thereby promoting the delayer attitude, which in turn favors a restart of prevalence in age group 1, and so on. However, inclusion of the third strategy has a stabilizing effect with respect to the basic ($V$-$N$) strategy of section \ref{Bauch_2005} \cite{Bauch2}. The model also offers some insight on the role of the delayer strategy during epochs of vaccine scare.

\subsubsection{Dynamical effects of social norms}
\label{social_norms}
Previous models assuming \textit{prevalence-dependent} vaccinating behavior predict that elimination is impossible since vaccine uptake eventually sets around some equilibrium levels below the critical threshold $x_{c}$. However, in western countries vaccination coverage for certain diseases such as poliomyelitis and dyptheria attained and maintained very high levels, ensuring persistent local elimination, even in countries with voluntary vaccination. This suggests that further behavioral mechanisms are at work. A number of psychological theories relevant for health decisions such as the theory of planned behavior \cite{Ajzen} and its extensions, have stressed the importance of social pressure as a determinant of individual behavior. In particular, our own perceptions of what persons relevant to us might think we ought to do are relevant - the so-called \textit{injunctive social norms} (ISNs) \cite{Cialdini_Trost}
might affect vaccinating behavior in a number of ways, i.e. by building self-legitimation for non-vaccinators, by enforcing their cohesion, or by increasing the \textit{conformism} of vaccinators as a majority.

Oraby et al. \cite{oraby2014influence} included ISNs into the IGD model of section \ref{Bauch_2005} by assuming, in line with the game-theoretical work by \cite{Helbing_Johansson_2010}, that ISNs add a new component to payoffs which is proportional to \textit{group pressure}, i.e. to how many other people in the population are also playing that strategy. The perceived payoffs for nonvaccinators and vaccinators become: $C_N=-r_i m I(t) + \delta_0 (1-x)$, and $C_V = -r_V + \delta_0 x$, where $\delta_0$ tunes the effects of social pressure on payoffs, taken as symmetric. Note that social pressure has the potential to mitigate, or even to sign-reverse, the costs of the two strategies.

After some rescaling, this yields the following equation for the vaccinators proportion $x$
\begin{equation}\label{sirvp_eq_social_norms}
x^{\prime}  =\kappa_{1} x(1-x) (I- \omega - \delta (1-2x) ),
\end{equation}
\noindent which, differently from section \ref{Bauch_2005}, emphasizes the relative cost of vaccination and the relative strength of social pressure with respect to the cost of disease, given by $\omega=r_V/(r_i m I)$ and $\delta=\delta_0/(r_i m I)$ respectively.

With respect to the model in section \ref{Bauch_2005} considering only the direct costs of vaccination or not, the inclusion of ISNs yields a far richer dynamics. First, besides the four basic equilibria $A,B,C,D$ observed in section \ref{Bauch_2005}, there is a fifth equilibrium $E$ induced by social pressure, where $S_E=1-x_E$, $I_E=0$, $x_E=2^{-1} (1+\omega/ \delta)$, with $x_E>1/2$. However, this  \textit{disease free partial vaccinator} equilibrium, which exists only if $\delta < \omega$ (i.e. if the strength of group pressure is smaller than the relative cost of vaccination) is always unstable.
Second, the pre-vaccination endemic state  $C=\left(S_{SIR}, I_{SIR},x_{SIR}=0\right)$ is locally stable if $I_{SIR} <\omega +\delta$, i.e. if the perceived cost of disease is small. Third, if $\delta > \omega$ ($\delta < \omega$), i.e. if the strength of group pressure is larger (smaller) than the cost of vaccination, then the pure vaccinator equilibrium $B$ is locally stable (unstable).
As a consequence of the above two stability conditions, both $B$ and $C$ can coexist in their stability ranges, i.e. the system exibits bistability and its mathematical behavior will depend on the initial conditions (see Fig. \ref{Fig2_Oraby_2014}). Bistability phenomena also involve the \textit{partial vaccinator} endemic state $D$ (see Fig. \ref{Fig2_Oraby_2014}). Finally, oscillations due to destabilization of $D$ are possible, as in section \ref{Bauch_2005}.

\begin{figure}
\centering \includegraphics[width=0.5\textwidth] {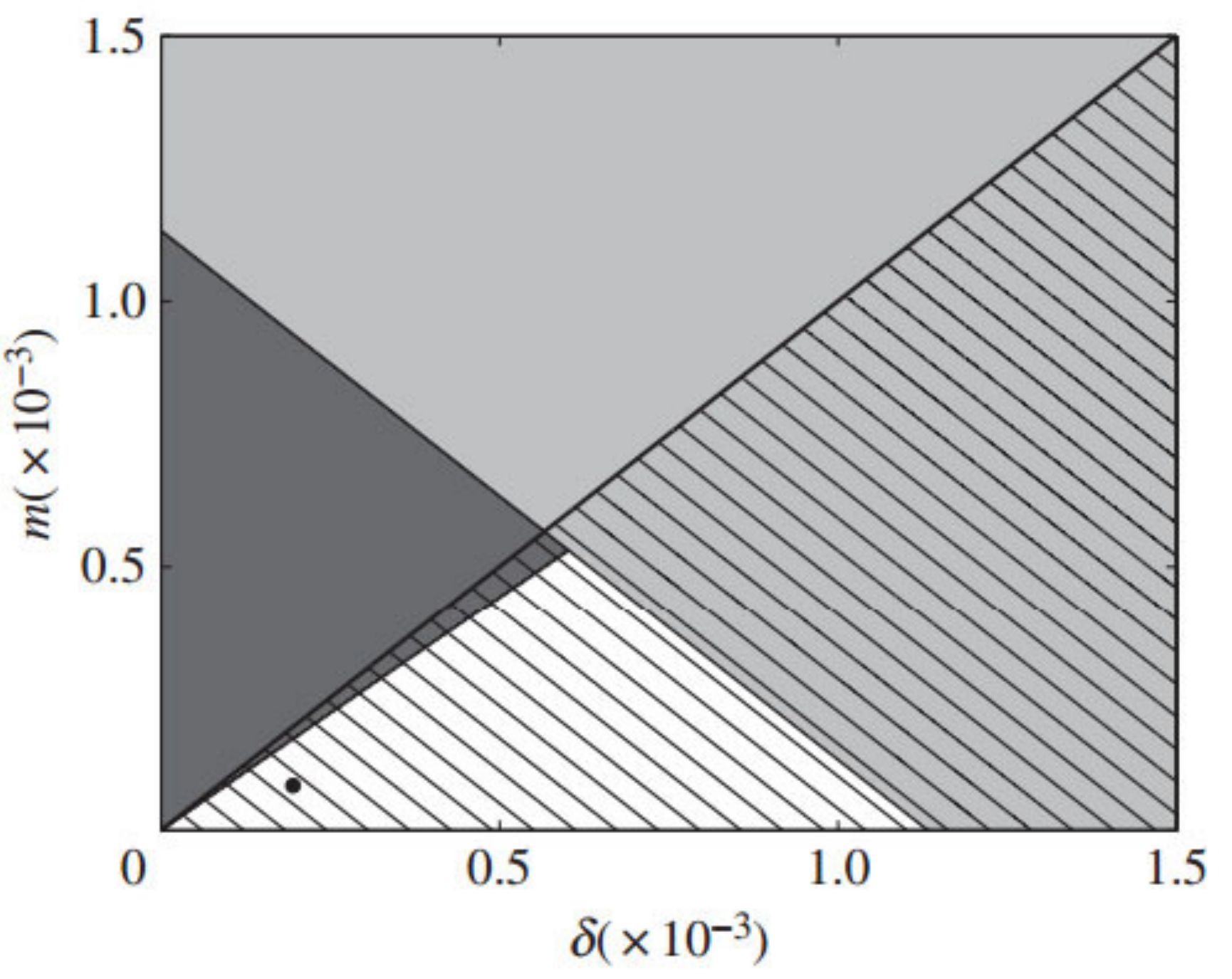}
\caption{Bistability of equilibria in the model with imitation dynamics and ISNs: regions of the $(\delta,m)$ parameters plane for the stability of the equilibria $B$ (disease-free full vaccinator equilibrium), $C$ (endemic pre-vaccination equilibrium), and $D$ (partial vaccinator endemic state). Equilibrium $B$ is stable in the hatched region below the thick black diagonal line, $C$ is stable in the light grey region, $D$ is stable in the dark grey region on the left. Details on other behavioral parameters and basic demographic and epidemiologic parameters are reported in \cite{oraby2014influence}. Source: Reprinted figure from Ref.~\cite{oraby2014influence}.}
\label{Fig2_Oraby_2014.pdf}       
\end{figure}

Overall, including ISNs widely expands the spectrum of qualitative mathematical behaviors w.r.t. the baseline case where payoffs only include direct costs of the alternative strategies. In particular, ISNs can mitigate the main perverse consequence of prevalence-dependence, namely the "elimination impossible" result, by allowing the pure vaccinator equilibrium to become stable at large levels of social pressure. Other effects are more controversial, as social pressures can induce bistability phenomena in a large portion of the parameter space, with the well-known amplification of uncertainity due to both the imprecise measurement of initial state and and the equilibrium switch that can be induced by extrinsic stochasticity. Thus, depending on the context, social norms can either support or hinder vaccination goals. Finally, inclusion of ISNs allows, under a few further assumptions, an excellent fit of prevalence and coverage data of pertussis in the UK since the seventies, including the epoch of the pertussis vaccine scare and the subsequent epoch of coverage relapse (see Fig. \ref{Fig1_Oraby_2014}).

\begin{figure}
\centering \includegraphics[width=0.7\textwidth] {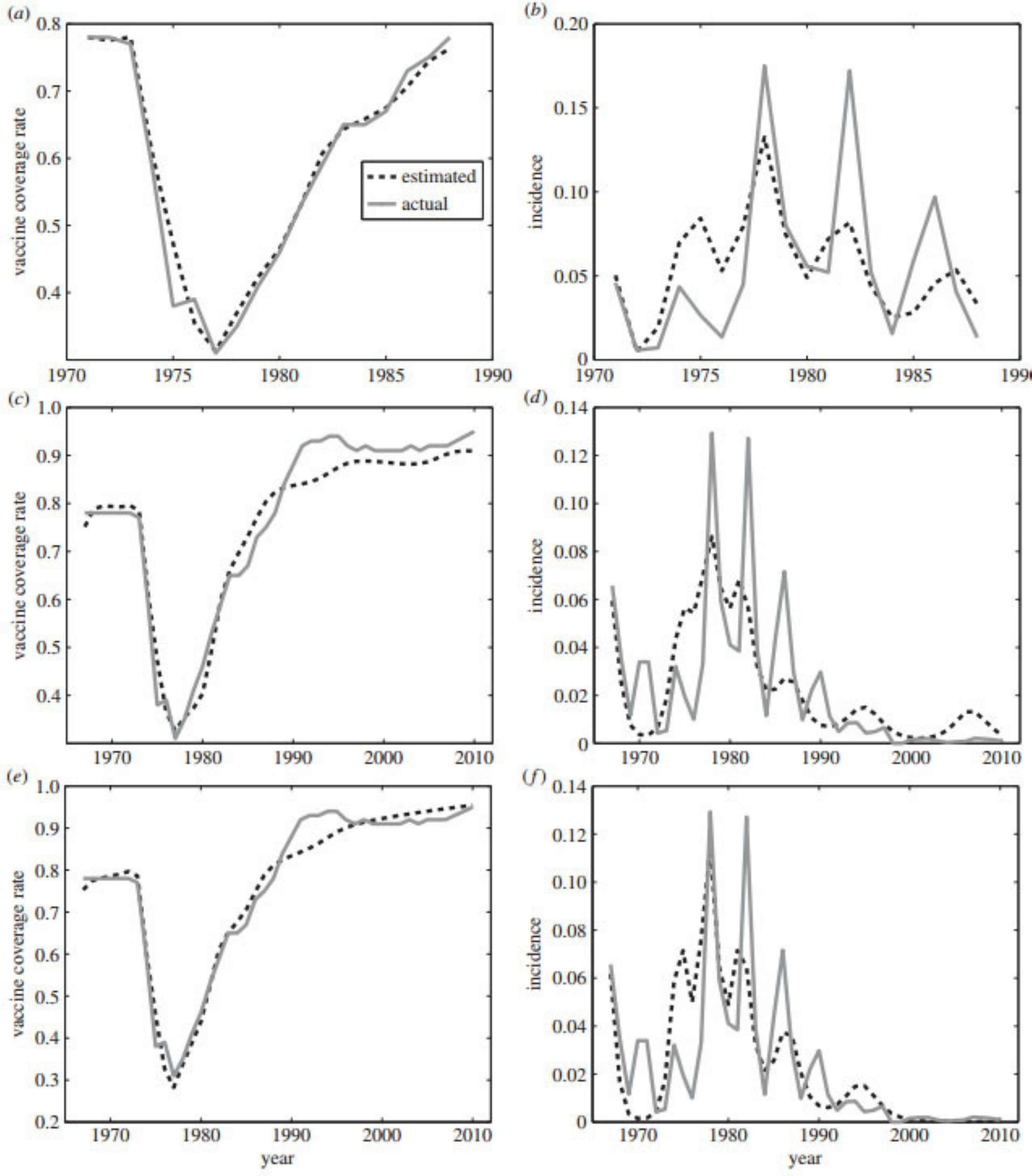}
\caption{Empirical and modelled (a,c,e) pertussis vaccine coverage and (b,d,f ) pertussis incidence in the UK, (a,b) from 1971 to 1988 without injunctive social
norms, (c,d) from 1967 to 2010 without injunctive social norms and (e,f ) from 1967 to 2010 with injunctive social norms. The solid line is the empirical data and the dashed line is the best-fitting model.  Source: Reprinted figure from Ref.~\cite{oraby2014influence} }
\label{Fig1_Oraby_2014.pdf}       
\end{figure}

\subsubsection{Implications of bounded rationality in vaccination decisions}
\label{sec:bounded_rational_vaccination}
Vaccinating (or not) is a typical example of \textit{decision under uncertainty} conditions, for which bounded rationality approaches, such as Kahneman and Tversky \textit{prospect theory} \cite{Kahneman_Tversky_1979} (see section \ref{PsychologicalModels}) typically work better than classical rational calculus. Oraby and Bauch \cite{OrabyBauch2014_bounded_rationality} have investigated the dynamic implications of bounded rationality in vaccinating behavior by an SIR model in which vaccination decisions are formulated according to the ideas of prospect theory.
The chosen coupled behavior-disease model is formulated as an IGD, i.e. parents still learn about other's strategies through imitation, also including the effects of ISNs (see the previous subsection) and a non-perfect vaccine. The $x$-equation for the proportion of parents choosing strategy $V$ is formulated as
\begin{equation}\label{sirvp_eq_bounded_rationality}
x^{\prime}  =\kappa_{1} x(1-x) F(\pi_{V|N}(I,x)) - F(\pi_{N|V}(I,x)),
\end{equation}
where $F(.)$ is a suitable probability distribution function, and $\pi_{V|N}(I,x)$ (alt. $\pi_{N|V}(I,x)$) is the net reward that a parent playing stategy $N$ (alt. $V$) perceives to gain from switching to strategy $V$ (alt. $N$), given current conditions on prevalence and vaccine uptake $I,x$. The map $F$ transforms net rewards to a probability scale, so that the term $\Delta_F = F(\pi_{V|N}(I,x)) - F(\pi_{N|V}(I,x))$, replacing the \textit{sure} perceived payoff gain of previous models, represents the \textit{probability gain} arising from the switch between strategies. The perceived net rewards $\pi_{V|N}(I,x)$, $\pi_{N|V}(I,x)$ are defined as
\begin{align}
\pi_{V|N}(I,x)  &  = \pi_{N}(V) - \pi_{N}(N) + (\delta_{V}x - \delta_{N} (1-x) ),  \label{pi_from_N_to_V}\\
\pi_{N|V}(I,x)  &  =  \pi_{V}(N) - \pi_{V}(V) + (\delta_{N}(1-x) - \delta_{V} x).  \label{pi_from_V_to_N}%
\end{align}

In Eqs. \ref{pi_from_N_to_V}-\ref{pi_from_V_to_N}, besides the social norm components (analogous to those described in the previous section but taken asymmetric for sake of generality), the true core is represented by the differences $\Delta \pi_{V|N}=\pi_{N}(V) - \pi_{N}(N)$, and
$\Delta \pi_{N|V} = \pi_{V}(N) - \pi_{V}(V)$, where the quantities $\pi_{N}(V), \pi_{N}(N), \pi_{V}(N), \pi_{V}(V)$ represent the expected utilities arising from each of the four possible \textit{prospects} faced by agents.  In \cite{OrabyBauch2014_bounded_rationality} it is hypothesized that agents classify the possible realizations of infection in a four item scale $r_{i,I}$ ($i=1,..,4$) i.e. (i) \textit{mild} infection, (ii) \textit{moderate}, (iii) \textit{morbid}, up to (iv) \textit{death} due to disease/vaccine adverse events (VAE), scaled to the base-case of absence of infection, with corresponding probabilities $p(r_{i,I})$. Analogous classification $r_{i,VAE}$ ($i=1,..,4$) is done for VAE. At each realization it is attached a utility evaluation $U_I (r_{i,I})$ as well as a weight $W(x(r_{i,I}))$ to the corresponding probability. The various prospects are defined consequently. For example, agents currently playing $N$ who continue to play $N$ can only experience the consequences of infection, so their prospect includes the utility outcomes $U_I (r_{i,I})$ with expected utility $\pi_{N}(N)=\sum U_I (r_{i,I})W(x(r_{i,I}))$. On the other hand agents currently playing $N$ who consider to switch to $V$ will include in their prospect both the utility outcomes from VAE, $U_{VAE} (r_{i,VAE})$, as well as those due to infection, as a consequence of vaccine imperfection, and so on.
The key step then lies in assigning the utility evaluations and corresponding weights, according to the principles of bounded rationality, i.e. accounting for the two different possible parents' attitudes, namely \textit {risk averting} vs \textit {risk seeking}, that classically cause departures from full rationality,
depending on a number of parameters reflecting \textit{boundedly rational} behavior. This is done by appropriate parametric specifications of $U$ and $W$ \cite{OrabyBauch2014_bounded_rationality} including \textit{perfect rationality } as a sub-case.

The resulting SIR vaccination model based on Eq. \ref{sirvp_eq_bounded_rationality} and Eqs. \ref{pi_from_N_to_V}-\ref{pi_from_V_to_N} is quite richer compared to section \ref{social_norms} as it includes, besides bounded rationality of vaccination decisions, also vaccine imperfection and asymmetric effects of social pressure. In particular, the underlying perfect rationality model is the extension of the model of previous section accounting for vaccine imperfection and asymmetric social pressure.

The ensuing equilibrium structure differs from that in \ref{social_norms} due to the appearance of a sixth endemic state $F$, which is a \textit{pure vaccinator} endemic state forced by vaccine imperfection. In particular (i) the \textit{disease free partial vaccinator} equilibrium $E$ remains always unstable, as in \ref{social_norms}, (ii) the pure vaccinator, disease-free equilibrium $B$ is stable for infections that are not not highly transmissible, if the
vaccinators social pressure is relatively strong compared to the vaccine cost, while the \textit{pure vaccinator endemic state} $F$ is stable if and only if vaccinator pressure is large and the vaccine efficacy is low.

To sum up, the effects of bounded rationality are mostly dynamical. Indeed, while the dynamics observed in the underlying rational model are much alike those in section \ref{social_norms}, with bistability regions having simple forms, the inclusion of bounded rationality yields a zoo of dramatic dynamical changes, which include complicate nonlinear shapes of the stability regions, and steady oscillations about the pure vaccinator disease-free equilibrium $B$. In addition, boundedly rational behavior makes it harder to eliminate infection, because it can offset the beneficial effects of the social pressure of the vaccinators group.

\subsubsection{Modeling public communication on disease and vaccine risks}
\label{public_communication_PONE2012 }
A parsimonious explanation of some observed features of vaccination systems can be obtained by removing an unrealistic feature of models based on IGD, namely that switching between strategies is entirely determined by learning from other agents (i.e., parents of children to be vaccinated) during social encounters. Indeed, the pure-imitation model poorly represents the process of information provision in modern countries because, even when a given vaccination is fully voluntary. Public health systems will nonetheless keep the role of main suppliers of the relevant information on risks associated to diseases and vaccines, as suggested by a body of evidence \cite{noibeta}. In this subsection we report an extension of the basic imitation-based framework of sections \ref{Bauch_2005}-\ref{IGD_with_myopic_perception_of_VAE}, which accounts also for perceptions based on information (about risks) supplied by the public health systems \cite{noibeta}. In this new framework, remindful of Bass classic model for information diffusion \cite{Bass}, switches between $N$ and $V$ strategies is determined by the balance between private information, exchanged through inter-personal communication between parents of children eligible for vaccination during their social contacts, and public information, communicated by the public health authorities through media and related channels. This allows to amend the equations for the dynamics of switch between the two strategies as follows
\begin{align}
y_V^{\prime} =& G(t)y_N- K_{VN}y_N y_V + k_{NV}y_N Y_V, \label{sVaw}\\
	y_N^{\prime} =& -G(t)y_N- k_{NV}y_N Y_V + K_{VN}y_N y_V, \label{sNaw}
\end{align}
where $G(t)$ represents the rate of transition from strategy $N$ to strategy $V$ induced by mediatic information (which is independent of the level of individuals' social activity), yielding
\begin{equation}\label{igeq}y_V^{\prime} = G(t)(1-y_V) + (K_{NV}- K_{VN})y_V(1-y_V),   \end{equation}
where $K_{NV} = \kappa \theta(I)$ and $K_{VN}=\kappa \alpha(x)$. In the simplest case one can assume that $G(t)$ is a strictly positive constant, i.e. $G(t)=k_{G}>0$. The underlying idea is that the information provided by public health systems aims to convey a different perception of risks related to disease and vaccination compared to what might happen in private social encounters. In particular, a "`wise"' public communication aims to convince that (i) vaccines are highly safe, with a very low, constant, risk of VAE, and that (ii) risks associated with the disease are prevalence-independent.

Setting $x=y_V$ and simplifying Eq. \ref{igeq} yields the following dynamic equation for the proportion favourable to strategy $V$
\begin{equation}\label{p_equation_G_model}
x^{\prime} = \kappa (1-x) \left( \left( \theta(I)-\alpha(x)\right) x + \psi \right),
\end{equation}
where parameter $\psi=G/\kappa$ summarizes the effectiveness of the public \textit{effort} (information, education, availability of vaccination infrastructures, subsides to vaccination staff, etc) in affecting perceptions on vaccines and disease.

The properties of the ensuing behavior-disease model arising by coupling Eq.  \ref{p_equation_G_model} with Eqs. \ref{sirvs}-\ref{sirvi} depend in a critical manner
on parameter $\psi$. Briefly, there exist appropriate thresholds $\psi_c < \psi_1$  such that
\begin{itemize}
\item For high levels of public effort ($\psi \ge\psi_1$) only the pure vaccinator equilibrium $B$ exists and it is globally asymptotically stable (GAS).
\item For intermediate levels of public effort (say $\psi_c <\psi < \psi_1$) the pure vaccinator equilibrium is unstable, but there is a \textit{disease free partial vaccinator} equilibrium $E$, with vaccine uptake above the elimination threshold, which is GAS.
\item For low levels of public effort ($\psi \le \psi_c$) the disease free partial vaccinator equilibrium $E$ exists but is unstable, and the \textit{partial vaccinator} endemic state $D$ appears ensuring endemicity, though it is not necessarily stable i.e. oscillations may occur.
\end{itemize}
In a control perspective, if the infection is endemic at $D$ in a scenario where vaccination is voluntary with public intervention absent, it is possible to increase the equilibrium coverage by increasing the public effort in providing information about benefits of vaccination. Suitable further increases in public effort can allow the equilibrium vaccine uptake to expand until the endemic state $D$ disappears by exchanging its stability with the disease free partial vaccinator $E$, thus achieving elimination. Further increases in $\psi$ can push $E$ to collapse on the pure vaccinator equilibrium $B$.

These results indicate that public intervention can completely mitigate the negative implications of prevalence-dependent, pure-imitation models (sections  \ref{Bauch_2005}-\ref{IGD_with_myopic_perception_of_VAE}), particularly removing the impossibility to eliminate the infection. A simple calibration to Italian data on measles coverage shows that this model explains the recent Italian story, characterized by a fast increase in measles vaccine uptake following substantial efforts by the public health system, largely better than models including imitation only \cite{noibeta}.

\subsection{\textit{Behavior-explicit} game-theoretic models of vaccinating behavior}
\label{game-theoretic models}
The first application of classical game theory and epidemiological modeling to investigate the implications of vaccination decisions dealt with the conflict between group interest and self-interest in relation to smallpox vaccination upon reintroduction following a terroristic attack \cite{Bauch0}. A companion paper \cite{Bauch1} applied the same approach to mass vaccination against common vaccine-preventable childhood infections under voluntary vaccination. The latter work gave much momentum to subsequent research in the field. In line with the previous subsections, this subsection reviews this seminal work and a few subsequent fundamental contributions focusing on childhood vaccination  \cite{Bauch1,Reluga1,RelugaGalvani,Shim_JTB_2012}. The flavor of the presentation is slightly different however, as the focus of these approaches, though framed within SIR models for vaccine-preventable infections, is more on classic game theoretic questions, namely existence of Nash equilibria for individuals' vaccine uptake and comparisons with the level which is optimum for the community as a whole, rather than on dynamic properties.

\subsubsection{The baseline equilibrium model}
\label{Earn_Bauch_2004}
The work in \cite{Bauch1} integrates ideas from the endemic SIR model with vaccination at birth (see section~\ref{SIR_vaccination}) into a game theoretical framework to investigate the interplay between individual vaccination decisions and the population-level dynamics of infection in a voluntary vaccination scenario. Game theory is the natural framework given that the vaccination strategy of each specific agent (i.e.  parents), whose sum defines the vaccine coverage at the population level, is  influenced by other parents' decisions in the sense that the corresponding payoff will depend on the strategies played by all other players.

Letting the individual strategy be represented by the parents' probability to vaccinate $x$, authors then seek a \textit{convergently stable Nash equilibrium} (CSNE) solution $x^*$, which is expected to be observed in a real population of players.

The key determinants of individual strategies are the perceived \textit {relative cost of vaccination} $r$, defined as the ratio between the perceived risk of VAE ($r_V$) and the perceived risk of disease given infection ($r_I$), and the probability $\pi_x$ of experiencing infection under a given vaccination coverage $x$. For the latter it is assumed that parents correctly know the underlying epidemic model and estimate $\pi_x$ by the corresponding lifetime probability of infection at endemic equilibrium for an unvaccinated individual, given by
\begin{equation}
\pi_x = 1-\frac{1}{(1-x)\mathcal{R}_0}.
\end{equation}
Note that the scenario of full absence of vaccinators ($x=0$) yields $\pi_0=x_{c}$.

\begin{figure}
\centering \includegraphics[width=0.9\textwidth] {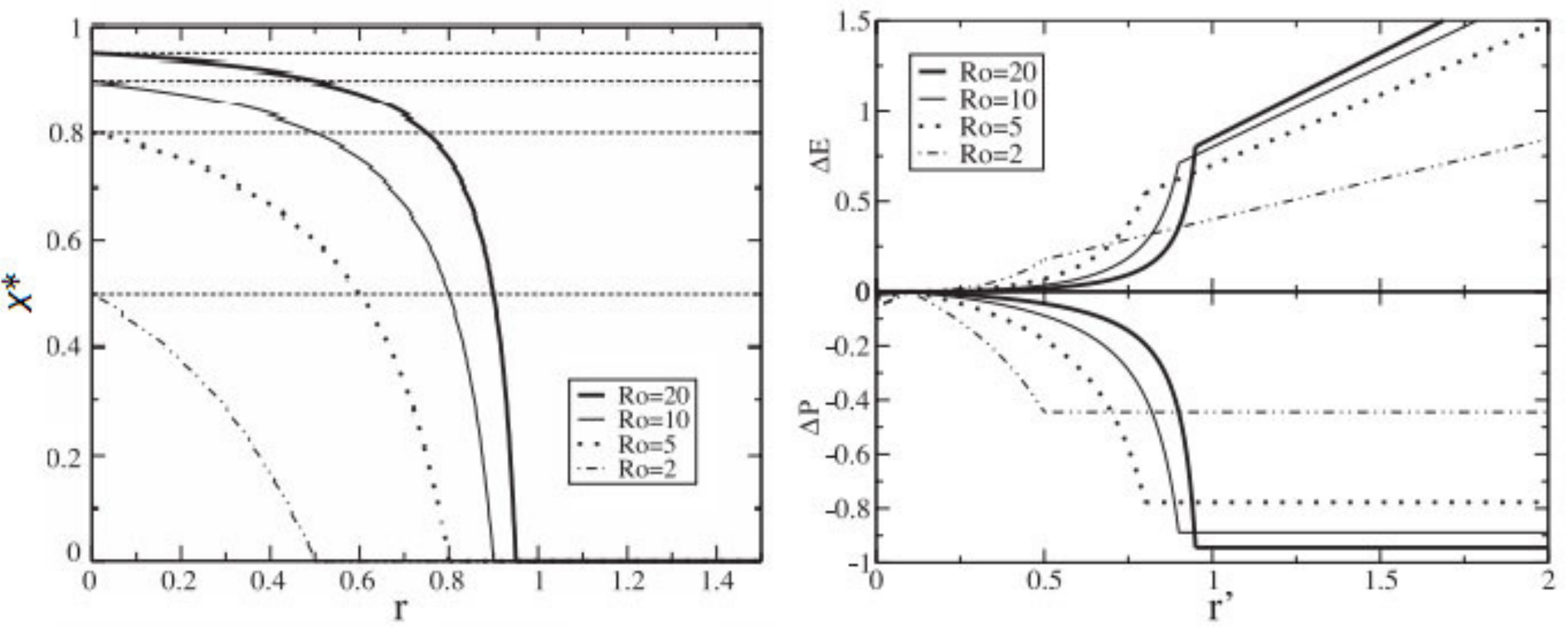}
\caption{The baseline equilibrium model of the vaccination game. Left panel: patterns of the vaccination coverage $x^*$ predicted by the CSNE as a function of the relative risk $r=r_V / r_I$ of vaccination for different values of the BRN $\mathcal{R}_{0}$. The dashed horizontal lines drawn at the level of the critical coverage $x_c=x^* (r=0)$ (for the different levels of $\mathcal{R}_{0}$) show the impossibility to eliminate the infection in a regime where individuals vaccinate according to self-interest driven by direct perceived risks. Right panel: impact of a vaccine scare as represented by a large increase of the relative risk of vaccination from a very low level $r^0<<x_{c}$ to a new level $r^{'}$, for different levels of the BRN $\mathcal{R}_{0}$. The graph reports the trend of the payoff gain $\Delta E$ and of the change $\Delta P$  in vaccine uptake for different levels of $r^{'}$. For moderate levels of $r^{'}$, namely $r^{'}<x_{c}$, the impact is moderate but becomes dramatic when $r^{'}$ exceeds $x_{c}$ causing the CSNE coverage dropping to zero. Source: Reprinted figure from Ref.~\cite{Bauch1}.}
\label{Figs1_2_BauchEarn2004}       
\end{figure}

The authors show that there is always a unique CSNE solution $x^*$. For high levels of the relative perceived risk of vaccination, namely for $r>\pi_0$, i.e. for $r>x_{c}$, the CNSE corresponds to the situation where no one vaccinates ($x^*=0$), i.e. the pure non-vaccinator strategy. On the other hand, if $r<\pi_0$, i.e. $r<x_{c}$, the CNSE solution predicts a strictly positive probability to vaccinate $x^*$. The condition $r<\pi_0$ implies
\begin{equation}
(1-r)\mathcal{R}_0 <1.
\end{equation}
Under the above constraint, the CSNE is
\begin{equation}
x^* = 1 - \frac{1}{(1-r)\mathcal{R}_0},
\end{equation}
which is such that
\begin{equation}
x^*(r)< x_{c}
\end{equation}
(unless $r=0$). In other words, under voluntary vaccination elimination turns out to be impossible if individual only take their self-interest into account (Fig. \ref{Figs1_2_BauchEarn2004}, left panel). Moreover, the case where $r$ exceeds the elimination threshold ($r>x_{c}$), a fact possibly observable during periods of vaccine scare, the CNSE predicts absence of any vaccinator at all. A key point stressed in \cite{Bauch1} is that when the perceived relative risk switches from an initial level $r^0<x_{c}$ to another one $r^{'}$ greater than $x_{c}$, then the resulting payoff gain  is increasing with $r^{'}$ and it has no plateau. In other words, the vaccine scare can go unlimited. In \cite{Bauch1} it is also considered the effect of measures induced by the public health systems to counteract the vaccine scare, i.e. a reduction of the relative risk from a value in excess of $x_{c}$ down to another value below $x_{c}$. The authors obtained that the incentive to vaccinate dimishes as the vaccine coverage approaches to its Nash equilibrium. Summarizing, a vaccine scare easily induces a substantial drop in vaccination, whereas restoring the pre-scare coverage levels is difficult. These findings are illustrated in the right panel of Fig.~\ref{Figs1_2_BauchEarn2004}.

\subsubsection{Best individual vs community vaccination response in a heterogeneous population}
\label{Reluga_Bauch_2006}
In \cite{Reluga1} a general game-theoretical approach was adopted, aimed to relate the population vaccine demand to individual decisions in a behaviorally heterogeneous population sharing a common information `signal' about the dynamic spread of an SIR-type infection. In their model the individuals choose between strategy $V$ (vaccinate their children) and $N$ (not to vaccinate) based on perceived benefits and costs of vaccinating, in turn based on the information signal $\sigma$ carrying information about risks of acquiring the infection. This information signal has the same meaning (and form) of the information index $M$ in the phenomenological models of section \ref{InfoDependentCoverageNoiTPB2007}. The population is assumed to be behaviorally heterogeneous, and its internal structure is described by a single real variable $y$. Individuals having type $y$ choose strategy $V$ (to vaccinate) with probability $\chi(y,\sigma)$ and strategy $N$ with probability $1-\chi(y,\sigma)$. Benefits of vaccination are summarized by the utility function of strategy $V$ which depends on $\sigma$ and $y$: $U_V(y,\sigma)$,  and is increasing in $\sigma$. On the other hand, the utility of vaccine refusal (strategy $N$) is assumed to also depend on the population average vaccine uptake
\begin{equation}
\bar{\chi}=\int_y \chi(y,\sigma)dy,
\end{equation}
i.e. $U_N(y,\sigma,\bar{\chi})$, where $U_N$ is decreasing in $\bar{\chi}$, mirroring a free-riding effect (compare this assumption with sections \ref{IGD_with_myopic_perception_of_VAE} - \ref{public_communication_PONE2012 } where the perceived cost of vaccination was increasing in the average vaccine uptake).

By taking the individual's expected utility
\begin{equation}
U(y,\sigma,\bar{\chi}) = \chi(y,\sigma )U_V(y,\sigma)+ (1-\chi(y,\sigma))U_N(y,\sigma,\bar{\chi}),  
\end{equation}\label{Utility_Reluga200}
one notes that the resulting individual's optimal strategy will necessarily depend on the strategies adopted by other individuals in the population i.e., the above problem defines a game.

The  search for the \textit{best response strategy} $\chi^*$ for the previous problem has to be framed into the theory of set-valued maps \cite{aubin}, yielding
\begin{equation}
\chi^*(y,\sigma,\bar{\chi})=  \begin{cases} 0 &\mbox{if } U_V(y,\sigma)<U_N(y,\sigma,\bar{\chi}), \\
1 &\mbox{if } U_V(y,\sigma)>U_N(y,\sigma,\bar{\chi}), \\
[0,1] &\mbox{if } U_V(y,\sigma)=U_N(y,\sigma,\bar{\chi}). \end{cases}
\end{equation}
In order to completely characterize the Nash equilibrium, it is needed to fulfill a self-consistency condition concerning the average vaccine uptake at equilibrium:
\begin{equation}
\bar{\chi^*} \in \int \chi^*(y,\sigma,\bar{\chi^*})dy,
\end{equation}
i.e.
\begin{equation}
\int \inf \chi^*(y,\sigma,\bar{\chi^*} ) dy< \bar{\chi^*}<\int \sup \chi^*(y,\sigma,\bar{\chi^*} )dy .
\end{equation}
This condition ensures that there exists a Nash equilibrium and that the population's average response is the same for all Nash equilibria.

On the other hand, at the population level, a benevolent social planner will seek the optimal community strategy by maximizing the \textit{community utility}
\begin{equation}
\widehat{U}(\chi(.)) = \int \left( U(y,\sigma,\bar{\chi})  \chi(y,\sigma )U_N(y,\sigma)+ (1-\chi(y;\sigma))U_V(y,\sigma,\bar{\chi}) \right) dy .
\end{equation}
By elementary functional analysis it is easy to verify that the above individual-wise optimum $\chi^*$ and the community-wise optimum $\chi^{C}$ are such that for all $y$ and given $\sigma$
\begin{equation}
\chi^*(y,\sigma)<\chi^{C}(y,\sigma).
\end{equation}
In other words, the Nash equilibrium for individuals will always be characterized by "less vaccination" than the corresponding community-based equilibrium, unless the Nash equilibrium is universal vaccination, confirming in a very general manner a long-standing result of the literature, namely that the pursuit of optimal self-interest leads to vaccine uptakes that are systematically lower than those that are optimal at community level.

Note that both the optimal $\chi^*$ and its average value $ \bar{\chi}^*$ depend implicitly on the information signal $\sigma$, which is a function of the state variables. For example in \cite{Reluga1} it is assumed that $\sigma$ is an increasing and saturating function of infection prevalence $I$: $\sigma(t)=\beta I / (\mu + \beta I)$. Note also that $\bar{\chi^*}$ results to be an increasing function of prevalence if the signal $\sigma$ has this property.

The implications of the above described approach for infection transmission and control are then investigated by setting the resulting vaccination strategies within the framework of an SIR model with vaccination. Three main cases are considered: (a) the vaccine coverage $x$ is equal to the average Nash equilibrium response $ x(t) = \bar{\chi^*}(I)$; (b) the vaccine coverage adapts dynamically to the differences between its departures from the average Nash response, according to the equation $ x^{\prime}(t) = \alpha(\widehat{\chi^*}(I) - x)$; (c) the vaccine coverage adapts dynamically according to an IGD where the switch rate from $N$ to $V$ strategy is determined by the payoff gain of vaccination (which, as it can be easily seen, is constant) and the inverse switch is determined by the perceived vaccine cost (which depends on $\sigma(I)$).
Unsurprisingly, model (a) (compare with the phenomenological model in section  \ref{InfoDependentCoverageNoiTPB2007}) leads to a unique locally stable endemic equilibrium with partial vaccination. Similarly in model (b), where $x$ 'follows' $\bar{\chi^*}$ with an exponential delay, the endemic equilibrium can be unstabilized by varying $\alpha$, yielding sustained oscillations. This confirms the already stated fact that phenomenological and game-theoretic coupled behavior-infection models are often indistinguishable when the game-theoretic solution is simply set, heuristically, within the framework of a dyamic epidemiological model.

\subsubsection{Markov decision processes show that the interaction between vaccine-believers and vaccine-skeptics yields sub-optimal measles vaccine uptake}
\label{Markov_decision_processes_measles}
Within game-theoretic approaches several efforts \cite{Reluga2,RelugaGalvani,Shim_JTB_2012} have been devoted to the investigation of the population level implications, of individual choices determined within the framework of Markov decision processes theory (MDPT). An attempt of systematization of the subject in reference to vaccination games was developed in  \cite{RelugaGalvani}. The framework rests on three interacting building blocks, namely a model for individual decisions, a model for changes at the individual scale
($\mathcal{I}$-scale), and a mean-field epidemiological model for describing the dynamics at the population scale ($\mathcal{P}$-scale). Individuals handle uncertainty by maximizing their (inter-temporal) expected utility, on the assumption that they are able to predict - in probabilistic terms - their future evolution in terms of both their future states $\mathcal{Y}(t)$ and the ensuing utility flow, that follow from taking a given strategy (i.e., vaccinating a child at a certain age, or not vaccinating him), or action, $A(t)$. Path $\mathcal{Y}(t)$ is a stochastic process which depends on the adopted strategy as well as on the strategies that have been adopted by other individuals, that is the average strategy at the population level. For example, for those not vaccinating the future utility path depends on the risk of catching infection which under homogeneous mixing depends on the overall coverage prevailing in the population. The $\mathcal{P}$-scale dynamics of infection is specified by an appropriate mean-field model, e.g. an SIR model, where the actual coverage is determined by the average of individuals' strategies. Finally, the $\mathcal{I}$-scale dynamics is specified by taking a suitable continuous-time Markov chain embedded in the $\mathcal{P}$-scale model. In \cite{Shim_JTB_2012} this framework was applied to investigate the consequences of a typical phenomenon in current vaccination systems, namely the coexistence of well-identified and stable groups valuing differently the costs arising from infection and vaccination.

\begin{figure}
\centering \includegraphics[width=0.5\textwidth] {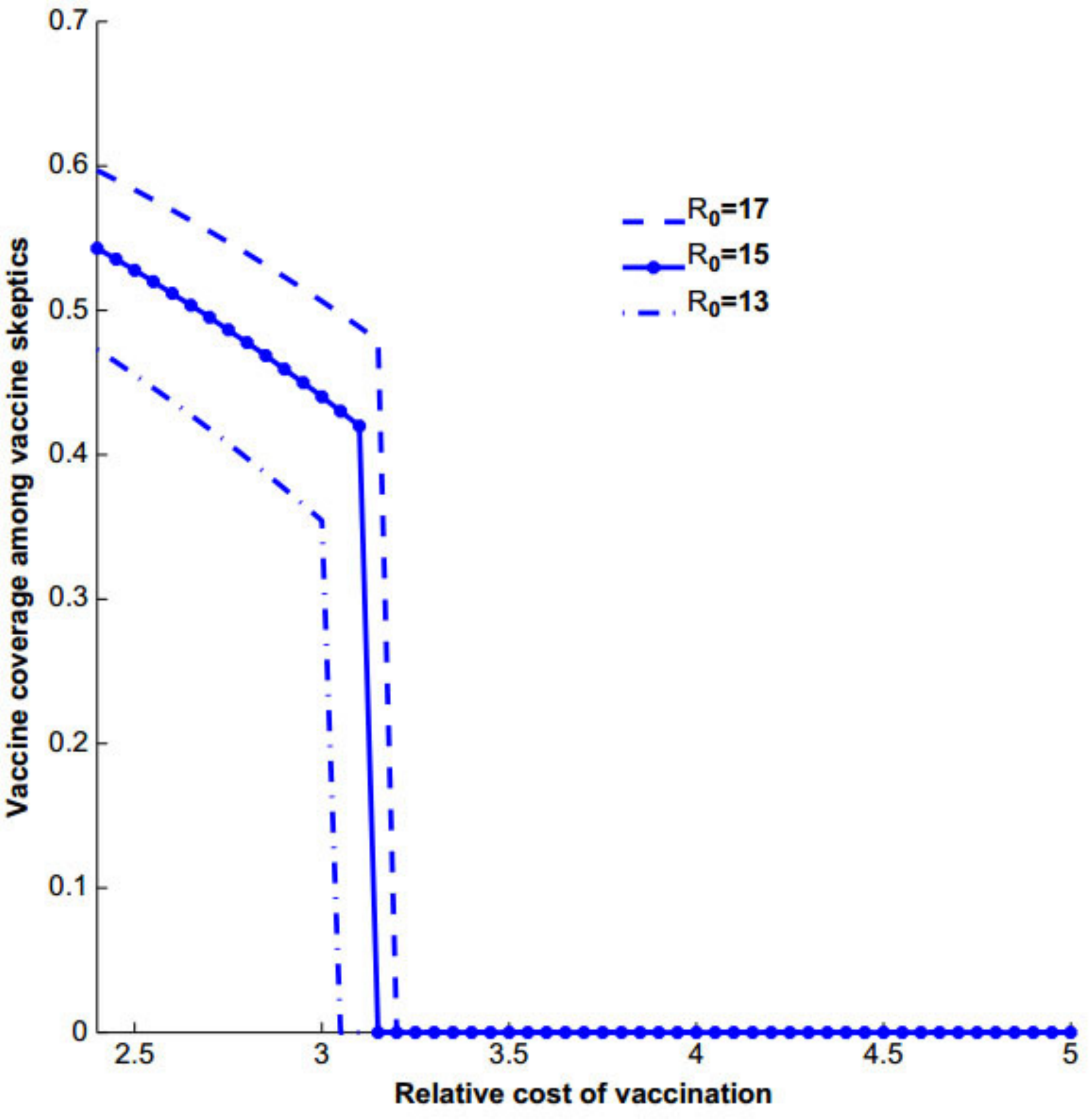}
\caption{Pattern of the best Nash response in terms of vaccine uptake among vaccine skeptics, drawn as a function of the cost of vaccination relative to the cost of infection ($C_{V,1}/C_{I,1}$) for different levels of the BRN $\mathcal{R}_{0}$ consistent with measles infection ($\mathcal{R}_{0}=13,15,17)$, and for a proportion of skeptics ($q_{1}=0.30$, showing the dramatic fall of the optimal vaccine uptake when the relative perceived risk exceeds a threshold level.  Source: Reprinted figure from Ref.~ \cite{Shim_JTB_2012}. With permission from Elsevier.}
\label{Fig3A_Shim_2012}       
\end{figure}

The simplest scenario, introduced in the static game-theory framework in \cite{Manfredi}, is represented by the conflict between two groups, namely vaccine vaccine \textit{skeptics} (group 1) and \textit{believers} (group 2), with different asymmetric perceived costs of infection and VAE: $C_{V,1}>C_{V,2}, C_{I,1}<C_{I,2}$. Let $q_{n}$ ($n=1,2, q_{1}+q_{2}=1$) denote the relative size of each group (assumed to be constant), $\phi_{n}$ the vaccination \textit{strategy} adopted by individuals in group $n$ and $\overline{\phi}_{n}$ the average coverage prevailing in group $n$. The $\mathcal{P}$-scale model is represented by the following simple two groups' SIRV model with a class vaccinated at birth by a perfect vaccine, in a constant population, where the two groups differ only in the average coverage:
\begin{align}
S_{n}^{\prime}&= \mu( 1- \overline{\phi}_{n})q_{n} -\beta S_{n} (I_{1}+_I{2}),   \label{SIRV_Shim2012_eqS_k}\\
I_{n}^{\prime}&= \beta S_{n} (I_{1} + I_{2}) -(\gamma +\mu)I_{n},   \label{SIRV_Shim2012_eqI_k}\\
V_{n}^{\prime}&= \mu \overline{\phi}_{n}q_{n} - \mu V_{n},     \label{SIRV_Shim2012_eqV_k}
\end{align}
\noindent where $S_{n},I_{n},V_{n}$ respectively represent the susceptible, infective, removed and vaccinated proportions in each group, with the removed compartment obeying $R_{n}=q_{n}-(S_{n} + I_{n} + V_{n})$. The rates determining the various state transitions in Eqs. \ref{SIRV_Shim2012_eqS_k}-\ref{SIRV_Shim2012_eqV_k} determine the associated Markov chain for state transition at the $\mathcal{I}$-scale, under the further assumption that agents estimate the FOI faced by their children by its endemic equilibrium value
$\lambda=\lambda_{e} = \beta (I_{1,e} + I_{2,e})$, which agents know from the knowledge of the $\mathcal{P}$-scale model. It holds
$\lambda=\mu (\mathcal{R}_{V}-1)$ where $\mathcal{R}_{V}=\mathcal{R}_{0}(1-\overline{\phi}_{1}q_{1}-\overline{\phi}_{2}q_{2} )$ is the vaccine reproduction number (section \ref{SIR_vaccination}), meaning that individuals correctly perceive the long-term risk as a function of the overall coverage prevailing in the community. The corresponding system for the individual's state probabilities has the form $\mathcal{Y}_{n}^{\prime} = \mathcal{Q}_{n} \mathcal{Y}_{n}$, driven by the $4 \times 4$ transition matrix $\mathcal{Q}_{n}$:
$$
\mathcal{Q}_{n} =\begin{pmatrix}
-(\lambda+\mu)&0&0&0\\
0&\lambda & -(\gamma+\mu) &0  \\
0& 0 & -\mu &0\\
0& 0 &0 &-\mu \\
\end{pmatrix}
$$
\noindent with initial probability distribution, actually representing the probabilities that the generic individual occupies at the moment of birth any of the four possible epidemiological states, given by the vector $\mathcal{Y}_{n}(0)=[1-\overline{\phi}_{n}, 0, \overline{\phi}_{n}, 0]$. Note that the transition matrix is the same for both groups, since perceptions about epidemiological and demographic hazards are assumed to be the same, so that difference in $\mathcal{I}$-scale dynamics is only due to the decision to vaccinate or not made at birth. Given the adopted individual strategy $\phi$, and the ensuing $\mathcal{I}$-scale path $\mathcal{Y}_{n} (t)$, individuals belonging to group $n$ have a (discounted) utility flow
\begin{equation}
U_{n} = F_{n} \mathcal{X}_{n}(0) + \int_{0}^{\infty} e^{-rt}f_{n}^{T} \mathcal{X}_{n}(t)dt, \label{Shim2012_Utility_flow_k} %
\end{equation}
\noindent where vector $f_{n}=[0, 0, -C_{V,n}, 0]^T, (n=1,2)$ includes the costs associated with occupancy of various states at any time, while
$F_{n}=[0, -C_{I,n}, 0, 0]^T$ includes the costs associated with transitions occurring at age $0$, which occur only in the event of vaccination. Integrating Eq. \ref{Shim2012_Utility_flow_k} yields $U_{n} = (F_{n}^{T} + f_{n}^{T}(rI-Q_{n})) \mathcal{X}_{n}(0) \label{Shim2012_Utility_k}$, which expands as

\begin{equation}
U_{n} = -\phi_{n}C_{V,n} - (1-\phi_{n})C_{I,n} (r+\mu+\gamma)^{-1}\frac{\lambda}{r+\mu+\lambda}. \label{Shim2012_Utility_k}%
\end{equation}
\\
The mathematical analysis shows that the Nash strategy, i.e. the \textit{best response curve} $\widehat{\phi}_{n}$ arising from the maximization of expected utility of Eq. \ref{Shim2012_Utility_k}, has the following form: (i) not to vaccinate ($\widehat{\phi}_{n}=0$) for levels of the perceived (equilibrium) FOI $\lambda$ strictly below a group-specific threshold $\lambda_{\mathcal{T},n}$, (ii) to always vaccinate ($\widehat{\phi}_{n}=1$) for values of of $\lambda$ strictly above $\lambda_{\mathcal{T},n}$, and (iii) to choose some (unspecified) intermediate level $\widetilde{\phi}_{n}, 0<\widetilde{\phi}_{n}<1$ for $\lambda=\lambda_{\mathcal{T},n}$. The threshold $\lambda_{\mathcal{T},n}$ depends on the group's perceived costs
$$
\lambda_{\mathcal{T},n}= \frac{C_{V,n}(r+\mu+\gamma)(r+\mu)}{C_{I,n}-C_{V,n}(r+\mu+\gamma)},
$$
\noindent and it is obviously larger for \textit{skeptics}.

Previous results allow a range of possibilities (ranging from both groups not vaccinating to partial coverage to both groups vaccinating) yielding a number of substantive implications. In particular, the community vaccine uptake determined by the Nash strategy, which is always lower than the community uptake based on the maximization of the (expected) average population utility (see also previous subsection), exhibits a sharp negative relationship with the proportion of vaccine skeptics in the population. Under a measles-like parametrization Nash strategies therefore imply that even moderate proportions of vaccine skeptics can bring the overall coverage sharply below the critical level. Moreover, the Nash vaccine uptake among vaccine skeptics shows a dramatic responsiveness to the relative cost of vaccination $C_{V,1}/C_{I,1}$ i.e. while for low relative vaccination cost the coverage among vaccine skeptics can be high, there is a level of vaccination cost in correspondence of which coverage abruptly falls to zero, as illustrated in Fig.  \ref{Fig3A_Shim_2012}.

\subsection{Other contributions to mean-field coupled disease-behavior models}
\label{Other_meanfield}

In the previous subsections an attempt was made to present some key modeling approaches, and related dynamic implications, from the recent literature on mean-field coupled behavior-disease models. No-doubt, deterministic mean-field models represented the vast majority of contributions in the first epoch of the behavioral epidemiology of infectious diseases. Nonetheless, in order to keep the presentation adequately self-contained we restricted the choice of the materials, focusing on SIR-like models for vaccine preventable infections, such as measles and pertussis. This was partly motivated by the fact that many seminal ideas arose from the conflict between public and self-interest in relation to common childhood vaccinations. Nonetheless it is to be acknowledged that this choice has left-out several excellent contributions to the field. We therefore devote this sub-section to depict some of the historical profile of behavioral epidemiology in order to recall the other many areas and contributions to coupled mean-field models for disease and vaccinating behavior.
After the seminal work by Capasso and Serio \cite{capasso1978generalization} and a few other phenomenological models mainly focusing on the effects of including nonlinear forces of infection in basic deterministic epidemiological model, behavioral epidemiology developed in the 1980-1990s along two main veins. The first one focused on the already mentioned conflict between public and self-interest in relation to common childhood vaccinations, and the free-rider problem \cite{fine1986individual, brito1991externalities, francis1997dynamic, Geoffard97}. These works prepared the humus for the outbreak of contributions presented in this subsection. The second vein yielded to the literature on behavioral change triggered by the HIV/AIDS threat since the 1980s. As noted in \cite{BDM} "the combination of a long incubation period, with difficult and costly treatment, and the lack of a vaccine, have made instilling preventive behavior through the dissemination of information on risky behavior with respect to sexual or intra-venous drug use the main control strategy, especially in poor resource settings." In this situation, where reliable data on individuals' responses to the spread of epidemics were mostly missing, mathematical modeling played a pioneering role in the understanding of the effects of behavior change on HIV dynamics, including the effect of prevalence-dependent switching to lower risk groups, reducing contact rates after screening or treatment, prevalence-dependent sexual mixing patterns, including the warning that availability of effective therapies and protective vaccine might increases disease severity by raising at-risk behavior \cite{scalia1991effect,li1992effects,stigum1997effect, velasco1994modelling, hadeler1995core,hsieh1996two,velasco1996effects}. Though most among the cited papers, which were pioneers in an endless list, considered phenomenological models, there have been also instances of more structured approaches to behavior, such as \cite{Kremer1996_HIV_behavior,kremer1998effect}.
This literature on behavior change in response to the HIV epidemics threat posed, among other things, the bricks for the current vast literature on behavioral responses to pandemics threat.

In particular, sticking on mean-field models of vaccinating behavior, the current phase has seen an explosion of contributions. A number of papers have applied the Markov decision process theory (DMPT) machinery presented in section~\ref{Markov_decision_processes_measles}
to formulate game-theoretic models of vaccinating behavior in relation to both seasonal \cite{galvani2007long,shim2012influence} and pandemic influenza vaccination~\cite{shim2011optimal}. Mean-field analyses were carried out for models investigating adaptive influenza vaccination decisions throughout several influenza seasons based on the inductive vaccination game \cite{breban2007mean,vardavas2007can}. Still, DMPT models for childhood immunization were considered for rubella \cite{shim2009insights} and varicella \cite{shim2009insights}. Mean field epidemiological models with vaccination choices derived from classical utility-based economic approaches were used in \cite{gersovitz2004economical}
to demonstrate the conflict between individual and social optimum, and the nature of externalities arising from vaccination. Deeper utility maximization frameworks were used to investigate a number of discrete-time coupled disease-behavior models considering both risky behavior and vaccination decisions in \cite{chen2006susceptible,chen2014economics}. In particular \cite{chen2006susceptible,chen2014economics} revisited Francis' work \cite{francis1997dynamic}, showing that (a) the free market solution always yields inefficiently low levels of vaccine uptake, and that (b) higher coverage can only result from frictions such as vaccine imperfection, agents heterogeneity, and in general from strategic interactions. In \cite{cojocaru2007dynamics} the implications of vaccination choices were investigated combining a game-theoretic approach with the framework of projected dynamical systems. A coupled behavior-disease model within a public choice economic framework was used in \cite{althouse2010public} to investigate the effects of the different types of externalities introduced by interventions against infections. An SIR model with bounded-rational vaccination decisions alternative to the model by Oraby and Bauch~\cite{OrabyBauch2014_bounded_rationality} presented in section~\ref{sec:bounded_rational_vaccination} has been developed and analysed in~\cite{voinson2015beyond}. This massive work has prepared the ground for the next step, i.e. moving towards more complicated and possibly more realistic, network models, as reported in the next sections.

\section{Behavior-vaccination dynamics in structured, networked populations}
\label{sec:beh-vac-net}

In Sections~\ref{sec:rationale}-\ref{sec:beh-vac-mean}, we have reviewed the progress made in the realm of behavioral vaccination, which is directly linked to public health measures aimed at eradicating vaccine-preventable infectious diseases. However, the majority of such research relies on the assumption of a homogeneous mixing population, where individuals interact randomly with one another, and thus a susceptible agent is equally likely to acquire the infection from any infected agent in the population. Despite this simplification, as reviewed above, research along these lines has provided important insights into the spreading properties of highly transmissible diseases, and has outlined effective ways for their containment. Nevertheless, the effectiveness and the predictive value of the theory relying on the assumption of a homogeneous mixing population falls sharply for close contact epidemics, which, for example, are brought about by sexually transmitted diseases~\cite{sturm2005parental,crepaz2004highly,gomez2008spreading,bauch2000moment} or the smallpox~\cite{del2005effects,kretzschmar2004ring,bauch2003ZhenPNAS}. To resolve this weakness, the assumption of a homogeneous mixing population needs to be replaced by an actual social network that describes the contacts among all involved. In this framework, the nodes of a network represent individuals, while the edges connecting them represent contacts that enable the disease to be transmitted. Actually, in Sections~\ref{concep-net}-\ref{sec:vac-net}, we have summarized the recent development of infectious diseases and vaccination on networks, where the dimension of individual behavior is neglected.  Here, from a new viewpoint,  we will systematically survey how social contact patterns influence the interplay between individual vaccination behavior and disease dynamics, i.e., we will review behavior-vaccination dynamics in structured, networked populations.

\subsection{Behavior-vaccination dynamics on static networks}
\label{traditional-net}

Here we review research concerning behavior-vaccination dynamics on classical static networks, including lattices, small-world networks (SW)~\cite{newman1999scaling,watts1998collective}, Erd{\"o}s-R{\'e}nyi random graphs (ER)~\cite{erd6s1960evolution}, and Barab{\'a}si-Albert scale-free networks (BA or SF)~\cite{barabasi1999emergence}. Under different microscopic rules, like payoff-driven imitation or opinion-associated dynamics, the vaccination decisions of individuals become dependent on the strategies (i.e., to vaccinate or not) of their neighbors, the perceived risk of infection in their neighborhood, the perceived safety and efficiency of the vaccine, as well as on the financial costs that are associated with vaccination and disease infection. This gives rise to collective phenomena that are akin to the universal behavior of particles near phase transitions points in classical models of statistical physics, and this in turn has far reaching implications for the successful mitigation of epidemic spreading.

\subsubsection{Impact of imitation dynamics: The universal ``double-edged sword'' effect}
\label{imitation-double}

\begin{figure}[t]
\centering \includegraphics[width=140mm,height=106mm] {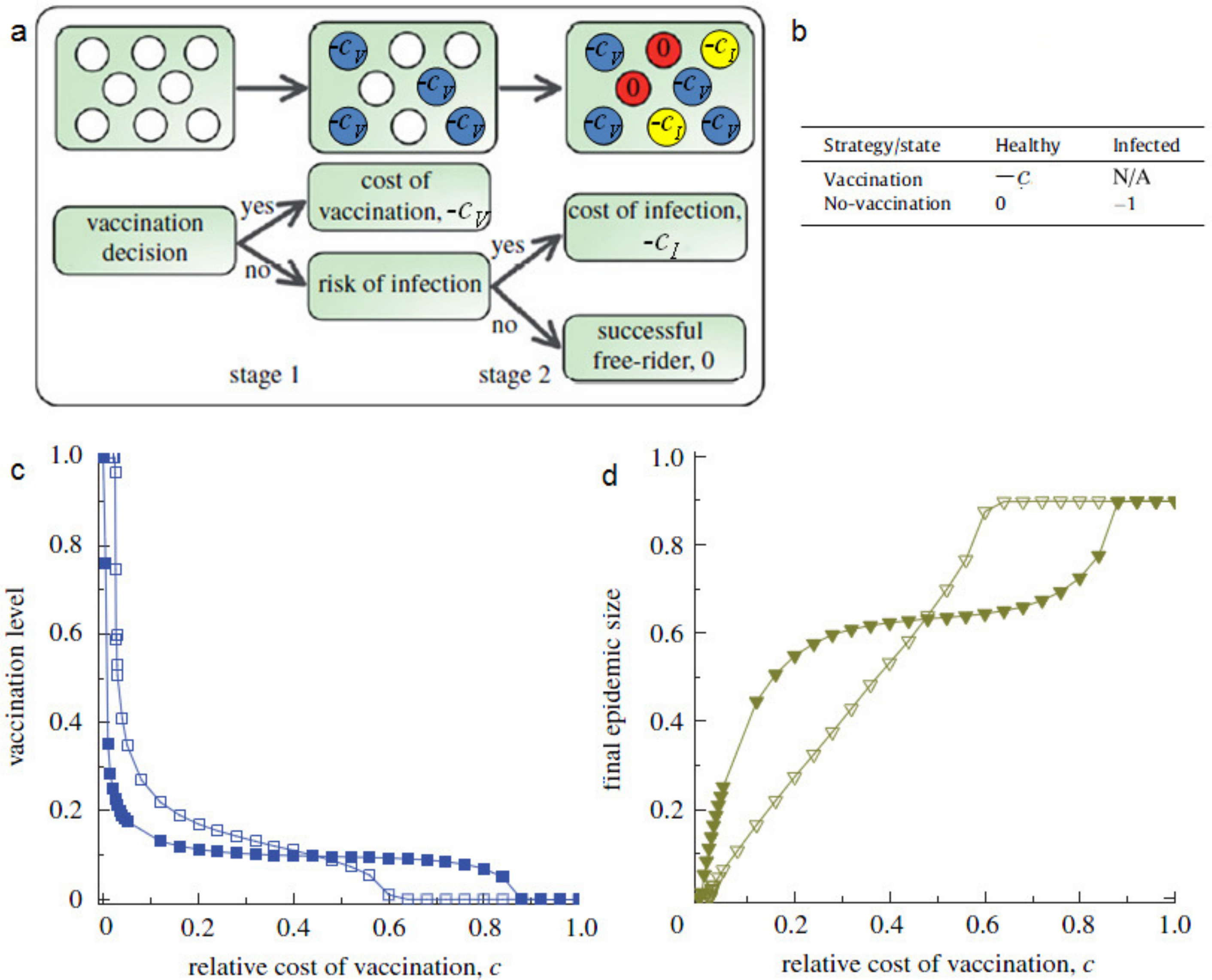}
\caption{(a) Schematic illustration of the voluntary vaccination dilemma game, which is further divided into two stages. The first stage is the public vaccination campaign, where vaccinated agents (blue) bear the vaccination cost $C_V$ in return for perfect immunity. During the second stage, the epidemic strain breaks out, and each unvaccinated individual faces the risk of infection. Infected agents (yellow) bear the cost of infection $C_I$. Unvaccinated individuals who remain healthy are free-riding on the vaccination efforts of others, and they are indirectly protected through ``herd immunity''. (b) The payoffs for different strategies (to vaccinate or not) and outcomes (healthy or infected). The fractions of  vaccinated (c) and infected (d) agents are shown as a function of the relative cost of vaccination $c$ for the selection strength $\kappa^{s} =1$ (open symbols) and $10$ (filled symbols), as obtained on a square lattice. Source: Reprinted figure from Ref.~\cite{fu2011ZhenPRSB}.}
\label{imitation-1}
\end{figure}

Like the public goods dilemma~\cite{rand2009positive,wang2015universal,perc2013evolutionary}, voluntary vaccination is often regarded as a problem of the commons~\cite{bauch2004ZhenPNAS}. In particular, if sufficiently few people vaccinate, the ``tragedy of the commons'' is likely \cite{hardin1968tragedy}. As the vaccination level increases and herd immunity is achieved, individual often forget the risk of some infectious disease and they lost the memory of their outbreaks. Increasingly often there is no memory at all of outbreaks of some infectious diseases. As a consequence, many decide to no longer vaccinate, and thus to effectively free-ride on the efforts of others who still vaccinate their offspring. With such behavior on the uprise, the herd immunity gets lost and the outbreak probability increases significantly. To mathematically describe such a dilemma, Fu \textit{et al}. have recently integrated the vaccination dynamics and epidemiological modeling into an agent-based evolutionary vaccination game, where vaccination and non-vaccination are mapped to strategies~\cite{fu2011ZhenPRSB}. More precisely, the vaccination dilemma is modeled as a two-stage game (see Fig.~\ref{imitation-1} (a) for a schematic illustration), with the first stage corresponding to a vaccination campaign, and the second stage corresponding to an outbreak of the virus. During the first stage, each vaccinated person acquires perfect immunity at a cost $C_V$, which covers the actual cost of the vaccine, the perceived risk, adverse long-term health impact (if any), and other intangibles. During the second stage, each infected individual bears the cost $C_I$, which covers all expenses related to the infection, including the cost of medication, missed work days, and so on. There are also those lucky individuals who remain uninfected during stage two despite being unvaccinated (marked by red circles in Fig.~\ref{imitation-1} (a)). These free-riders avoid all the costs and are thus better off as everybody else, benefiting of course from herd immunity. For simplicity but without loss of generality, to reduce the dimensionality of the parameter space, the aforementioned costs can be re-scaled by defining the relative cost of vaccination as $c=C_V/C_I$ ($0<c<1$ since $C_V<C_I$). The payoff function of an individual $i$ thus becomes
\begin{equation}
\Pi_i = \left\{ \begin{array}{ll}
-c, & \textrm{vaccinated;}\\
-1, & \textrm{infected;}\\
0, & \textrm{free-riding.}
\end{array} \right.
\label{payoff-1}
\end{equation}
It is worth pointing out that the payoff $\Pi_i$ is negative, simply because agents need to bear the cost rather than to enjoy a profit. Actually the most profitable outcome is a net zero payoff.

Once the epidemic ends, individuals update their strategies for the next season based on their payoffs. To accommodate imperfect information, errors in judgment, and irrational decision making in humans~\cite{pingle1995imitation}, the Fermi function
\begin{equation}
W_{i \to j}=f(\Pi_j -\Pi_i)=\frac{1}{1+\textrm{exp}[-\kappa^{s}(\Pi_j -\Pi_i)]},
\label{Fermi-1}
\end{equation}
can be used to determine whether an individual $i$ adopts the strategy of one randomly selected neighbor $j$ or not. Here $\Pi_i$ and $\Pi_j$ are their payoffs, while $\kappa^{s}$ ($0<\kappa^{s}<\infty$) denotes the strength of selection~\cite{blume1993statistical, szabo2007evolutionary, jrsiPGG} (note that $\kappa^{s}$ is different from the imitation coefficient $\kappa$ for well-mixed population in Section~\ref{sec:beh-vac-mean}). Small $\kappa^{s}$ values correspond to weak selection, which means that individuals are less sensitive to the payoff difference. Conversely, large $\kappa^{s}$ values correspond to strong selection, which means that the level of uncertainty during the strategy adoption process is small.

An important result is that, in comparison to the homogeneous mixing case~\cite{bauch2005ZhenPRSB, noi5}, networks act as a ``double-edged sword''. On the one hand, networks promote vaccination if the cost of vaccinating is low, but on the other hand, beyond a certain threshold value of the cost, they act as strong suppressors of vaccination. Also, the more heterogeneous the interaction network, the more pronounced this effect. Interestingly, a similar paradox exists also for different selection strengths, as shown in Fig.~\ref{imitation-1}(b) and (c). In particular, weak selection guarantees a larger low-cost threshold $c_L$, below which the epidemic is completely eradicated. As the selection strength increases, the number of free-riders starts going up, which in turn breaks apart the clusters of those who vaccinate. However, when the vaccination cost is sufficient large, the high-cost threshold $c_H$, above which nobody vaccinates, rises with an increasing selection strength. This means that the coexistence phase with vaccinators and non-vaccinators simultaneously present in the population widens with larger $\kappa^{s}$ values. Due to the special role of hubs, these phenomena become the more pronounced the more heterogeneous a network is. In the light of this research, this behavior corresponds to a scenario where voluntary vaccination provided herd immunity for some time but then it fails~\cite{vardavas2007can,colgrove2006state}.

Similarly to other binary-option games~\cite{breban2007mean,szabo2007evolutionary}, the above-reviewed vaccination dilemma game can also be studied with the mean-field approximation approach~\cite{Wu2013epl,liu2012impact,wang2015coupled}, as follows. On an uncorrelated and fully connected graph, we define $x$ as the fraction of vaccinated individuals and $r(x)$ as the probability that a susceptible individual gets infected in the population (in general, the whole population is initially divided into two classes: vaccinated agents with fraction $x$, and susceptible subjectives with fraction $1-x$. $r(x)$ depending on $x$ is equivalent to being a fraction of susceptible people). Each individual can take the states and payoffs: vaccinated ($V$, $\Pi_V=-c$), infected ($I$, $\Pi_{I}=-1$), non-vaccinated and healthy (or free-riding) ($H$, $\Pi_{H}=0$) (i.e., Eq.~\ref{payoff-1}). It is then possible to derive how the fraction of vaccinated individuals $x$ changes over time. Namely, when an individual with state $V$ turns to compartment $H$ or $I$, $x$ drops as
\begin{equation}
x^- = x(1-x)\{[(1-r(x)]W_{V \to H}+r(x)W_{V \to I}\},
\label{mean-drop}
\end{equation}
where $1-x$ is the fraction of neighbors holding the opposite strategy, $W_{V \to H}$ ($W_{V \to I}$) is the probability that individuals from the compartment $V$ change to the compartment $H$ ($I$). Accordingly, $x$ rises as
\begin{equation}
x^+ = x(1-x)\{[(1-r(x)]W_{H \to V}+r(x)W_{I \to V}\}.
\label{mean-gain}
\end{equation}
Taken together, the time evolution of $x$ is
\begin{equation}
\begin{split}
\frac{dx}{dt}
&=x^+ - x^-\\
&= x(1-x)\{[(1-r(x)]W_{H \to V}+r(x)W_{I \to V}\} -x(1-x)\{[(1-r(x)]W_{V \to H}+r(x)W_{V \to I}\}.\\
\end{split}
\label{mean-total}
\end{equation}
If the strategy transition probability is determined by Eq.~\ref{Fermi-1}, the above equation becomes
\begin{equation}
\begin{split}
\frac{dx}{dt}&= x(1-x)\{[(1-r(x)]\textrm{tanh}[\frac{\kappa^{s}}{2}(\Pi_V-\Pi_H)]+r(x)\textrm{tanh}[\frac{\kappa^{s}}{2}(\Pi_V-\Pi_I)]\} \\
&=x(1-x)\{[(1-r(x)]\textrm{tanh}[\frac{\kappa^{s}}{2}(-c)]+r(x)\textrm{tanh}[\frac{\kappa^{s}}{2}(1-c)]\}.\\
\end{split}
\label{mean-total12}
\end{equation}
Assuming the right-hand side of this equation equal to zero, we can get the equilibrium vaccination level. Then, the equilibrium fraction of infected individuals, i.e., the size of the epidemic, is expected to be $f_I=r(x)$, satisfying the self-consistent equation $f_I=(1-x)(1-e^{\mathcal{R}_0 f_I})$, where $\mathcal{R}_0$ is the basic reproductive number of the epidemic (it is usually related with degree variation or degree distribution of networks~\cite{lloyd2001viruses}, but here equal to the case of homogeneous mixing population owing to full-connections in network population).

\begin{figure}
\begin{center}
\includegraphics[width=1.0\textwidth]{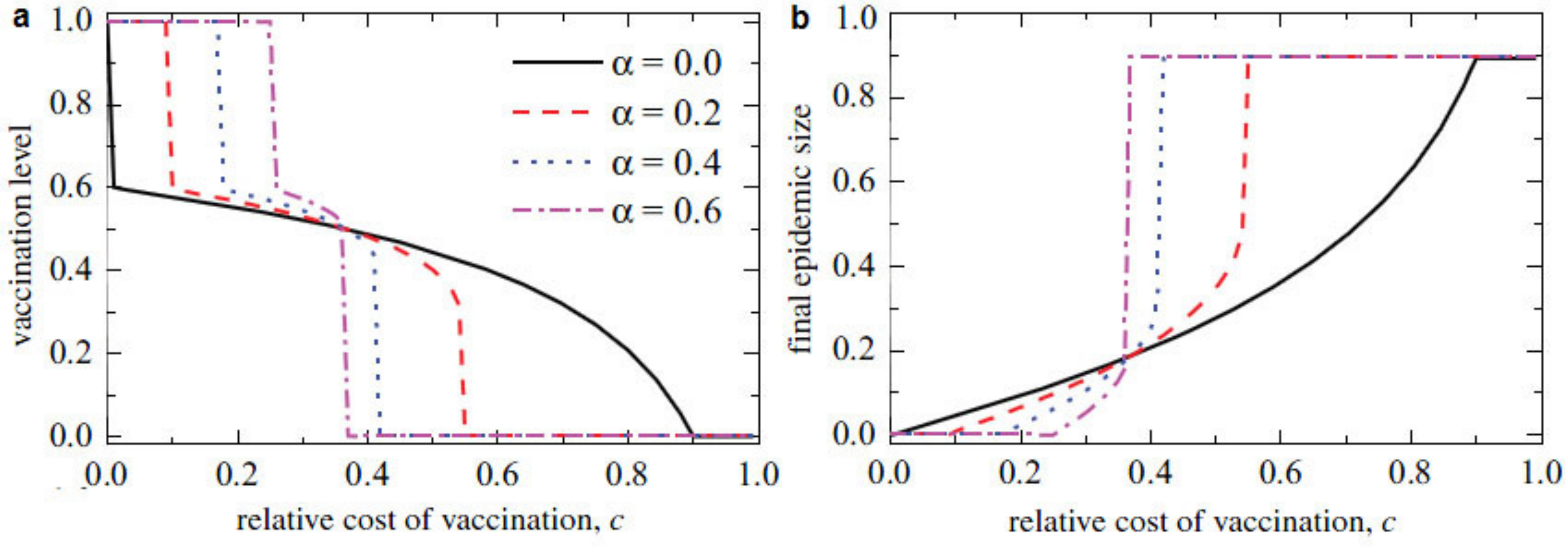}
\end{center}
\caption{The fraction of vaccinated (a) and infected (b) agents is shown as a function of the vaccination cost $c$, as obtained with the mean-field approximation of the vaccination dilemma game. The value of $\alpha$ defines peer pressure, where $\alpha=0$ recovers the setup used in~\cite{fu2011ZhenPRSB}, while positive $\alpha$ values introduce social influence in the imitation processes. Source: Reprinted figure with permission from Ref.~\cite{Wu2013epl}.}
\label{imitation-2}
\end{figure}

In a social dilemma game, whether or not to switch the vaccination strategy depends not only on the personal success of each individual, but also on the success of others~\cite{watts2007influentials,szolnoki2012wisdom}. Using this as motivation, one recent work extended the decision-making process on the basis of personal imitation and social influence (or peer pressure)~\cite{Wu2013epl}.  The pairwise comparison rule for the strategy-updating becomes
\begin{equation}
W_{i \to j}=f(\Pi_j -\Pi_i)= (k_o/k_i)^{\alpha} \frac{1}{1+\textrm{exp}[-\kappa^{s}(\Pi_j -\Pi_i)]},
\label{Fermi-2}
\end{equation}
where $k_o$ is the number of neighbors possessing the opposite strategy of individual $i$, $k_i$ is the number of all neighbors of $i$ (i.e., the degree of $i$), and $\alpha$ determines the so-called peer pressure. Here $\alpha=0$ recovers the setup used in~\cite{fu2011ZhenPRSB}, while positive $\alpha$ values make individuals more prone to sticking with the strategy that is representative for their neighbors. Accordingly, the mean-field equations for the vaccination level (i.e., Eq.~\ref{mean-total}) become
\begin{equation}
\begin{split}
\frac{dx}{dt}
&= x^+ - x^-\\
&= (1-x)x^{1+\alpha}\{[(1-r(x)]W_{H \to V}+r(x)W_{I \to V}\} -x(1-x)^{1+\alpha}\{[(1-r(x)]W_{V \to H}+r(x)W_{V \to I}\}.\\
\end{split}
\label{mean-total2}
\end{equation}

Interestingly, compared with the classical version of the model~\cite{fu2011ZhenPRSB}, the ``double-edged sword'' effect occurs again. Peer pressure (i.e., $\alpha>0$) promotes vaccination coverage as its cost is small, but plummets when costs exceed a critical value, irrespective of the structure of the interaction network. The larger the value of $\alpha$, the narrower the range of the coexistence phase for vaccination and non-vaccination. These findings can be quantitatively validated also by the mean-field approximation, as shown in Fig.~\ref{imitation-2}. If the vaccination is costly, peer pressure promotes the advantage of vaccinators, and thus forces their neighbors into following. This gives rise to larger clusters of vaccinators, which become unable to effectively prevent the spreading of the disease across the network. Only as the cost of vaccination decreases sufficiently is this cluster formation impaired, and then randomly distributed vaccinators are able to cover wider parts of the network to provide effective protection for their peers.

Along similar lines, the generality of the ``double-edged sword'' effect was further confirmed with a revised imitation dynamics. Recently, Zhang \textit{et al}. considered a more rational decision-making model from two aspects~\cite{zhang2012fu}. In the first place, the vaccination behavior of individuals was endowed with memory for their past experiences. The payoff function in Eq.~\ref{Fermi-1} was thus reformulated as
\begin{equation}
\Pi_i(t)=\Pi_i(t)+w\Pi_i(t-1), \quad   t\ge 1,
\end{equation}
where $w$ ($0 \le w \le 1$) is the memory weight. Secondly, a perceived risk threshold $\rho$ was introduced to avoid the blind imitation of individuals. That is, only when the infection risk surpasses this threshold, the willingness to vaccinate increases. Despite of these differences in the model, the results have shown that, as the cost of the vaccine is small, the coverage of the vaccination increases with increasing $w$ and $\rho$, leading to lower onsets of an epidemic. If the vaccination is costly, the results also remain qualitatively unchanged, and the main observations are attributed to the formation and decomposition of vaccination clusters, as described in~\cite{Wu2013epl}.

When modeling voluntary vaccination programs, we can choose different decision-making rules that govern the microscopic dynamics, and these different rules also frequently lead to different outcomes on the global scale. Social learning or imitation, as reviewed in this subsection, is a popular and realistic rule, which however is conducive to the emergence of clusters in structured populations. In vaccination dilemmas, this ultimately leads to a persistent ``double-edged sword'' effect, where below a certain cost vaccination is effective, while above this threshold it becomes ineffective. In what follows, we will review results obtained also with other decision-making rules at the microscopic level, and outline differences among them at larger scales.

\subsubsection{Impact of interaction network topology}

In network epidemiology research, it is thoroughly established that the structure of the interaction network plays an important role in the spread of infectious disease~\cite{Boccaletti2006, vespiromu, newman2002spread, moreno2002epidemic, kivela2014multilayer,li2014epidemic,li2011dimension,
boccaletti2014structure, pastor2015epidemic, salehi2015spreading,xia2013effects}. For example, epidemic spreading on scale-free networks is difficult to control, and with the lack of an outbreak threshold, the disease can propagate even with a low infection rate~\cite{vespiromu}. However, this holds if vaccination is absent. If an effective vaccination program is incorporated into the disease dynamics the outcome can be different. To account for this, a new modeling framework for evolutionary vaccination decision-making was proposed~\cite{zhang2010ZhenNJP}, where individuals resolved the vaccination dilemma in terms of the trade-offs between the infection risk, the side effects, and the cost of vaccination. More precisely, if a susceptible individual has $k_{inf}$ infected neighbors, the total infection probability will be
\begin{equation}
\lambda_I=1-(1-\beta_I)^{k_{inf}},
\end{equation}
where $\beta_I$ is the infection rate for a single contact between a susceptible and an infected individual.

\begin{figure}
\begin{center}
\includegraphics[width=1.0\textwidth]{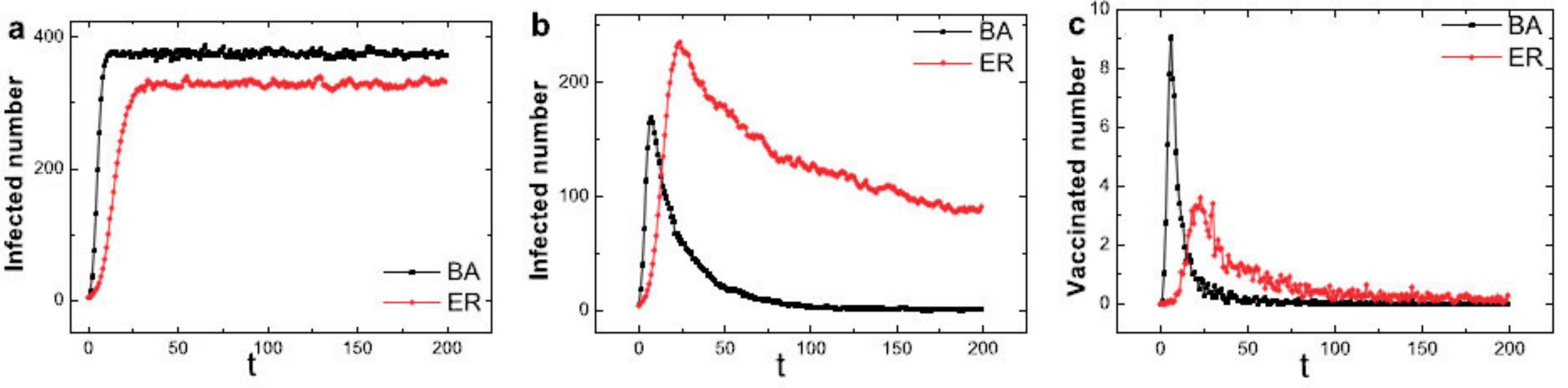}
\end{center}
\caption{Time evolution of the number of infected individuals without (a) and with (b) voluntary vaccination, and the number of vaccinated agents (c) on BA (or SF) and ER networks. The parameter values are $c_1=1$ and $c_2=0.7$. Source: Reprinted figure with permission from  Ref.~\cite{zhang2010ZhenNJP}. \textcircled{c} IOP Publishing \& Deutsche Physikalische Gesellschaft. CC BY-NC-SA. Reproduced by permission of IOP Publishing.}
\label{imit-topo-1}
\end{figure}

Similarly, the total perceived infection probability is
\begin{equation}
\lambda_{perc}=1-(1-\beta_{perc})^{k_{inf}},
\end{equation}
where if there is full knowledge for the infection risk $\beta_{perc}=\beta_I$ and $\lambda_{perc}=\lambda_I$.

If the susceptible agent does not choose to vaccinate, it bears the perceived cost
\begin{equation}
P_N=c_1 \lambda_{perc}, \quad or \quad P_N=c_1 \lambda_I.
\end{equation}
Conversely, missing out on vaccination yields
\begin{equation}
P_V=c_2 ,
\end{equation}
where $c_1$ and $c_2$ are related to the risk of infection, to potential side effects, and to the cost of vaccination.

Unlike with imitation considered in the previous subsection (see Eq.~\ref{Fermi-1}), here individuals adjust their strategies in agreement with another kind of decision-making rule that governs the microscopic dynamics, namely according to their self-interest or cost minimization. Thus, individuals try either to minimize their cost or to maximize their benefits (i.e.  self-interest rule). The strategy $s_i$ of individual $i$ will thus be
\begin{equation}
s_i =f(P_V-P_N)= \left\{ \begin{array}{ll}
\textrm{vaccination}, & \textrm{if $P_V-P_N<0$};\\
\textrm{non-vaccination}, & \textrm{otherwise}.\\
\end{array} \right.
\label{self-interested-1}
\end{equation}

Contrary to the result that the outbreak of disease on SF networks is more likely than on ER random graphs~\cite{Boccaletti2006, vespiromu, pastor2015epidemic}, in the realm of the model above it turns out that, if individuals are inclined to vaccinate at the initial stage of an outbreak, the epidemic is better constrained on the SF network (see Fig.~\ref{imit-topo-1}). Namely, on SF networks, the majority of middle- and large-degree nodes have a higher tendency to vaccinate. To verify this fact, the authors in~\cite{zhang2010ZhenNJP} provide a mathematical proof for the threshold degree of vaccination. Taking into account the vaccination condition given by Eq.~\ref{self-interested-1}, one obtains the threshold
\begin{equation}
P_N>P_V  \Rightarrow  \quad c_1\lambda_I > c_2 \Rightarrow \quad 1-(1-\beta_I)^{k_{inf}}>\frac{c_2}{c_1} \Rightarrow \quad k_{inf} \ge [log_{1-\beta_I}\frac{c_1-c_2}{c_1}]+1.
\end{equation}
Using $\beta_I=0.2$, $c_1=1$ and $c_2=0.7$ as an example, we have $ k_{inf} \ge 6$. For network of $<k>=6$, the vaccinated candidates are very limited on ER graphs, but more nodes prefer to vaccinate to reduce the loss because of the existence of middle- and large-degree nodes in SF networks. The vaccination program on SF networks, in some sense, is like the targeted vaccination in the mandatory model~\cite{pastor2002immunization,cohen2003efficient}, yet shows an adaptive character as well. In addition to permanently effective vaccination, temporarily effective vaccination (taking into account that the vaccine for some disease like influenza and hepatitis B will loose its efficiency after a given period of time) has also been considered. Though in this case there exist vaccination oscillations, the spread of disease ultimately can be controlled on complex networks. This further confirms that the larger inclination for hub nodes to get vaccinated plays an important role and can critically affect the success of vaccination campaigns, which ought to be considered in the design of incentives for vaccination.

Despite of the inspiring progress, there does not exist a direct comparison between a perfect and an imperfect vaccine in~\cite{zhang2010ZhenNJP}. Aiming to clarify this, Cardillo \textit{et al}. introduced a tunable parameter $\tau$ to modulate the quality of the vaccine in influenza-type disease, the vaccine being perfect at $\tau=0$ and useless at $\tau=1$~\cite{cardillo2013ZhenPRE}. When the vaccine is perfect, the research revealed that SF networks outperform ER graphs, since the overall vaccinated (infected) number is larger (smaller) on SF networks. This result is in agreement with the general conclusions reported in~\cite{zhang2010ZhenNJP}. However, if the vaccine is imperfect, a crossover effect occurs such that the ER graphs are more effective than SF networks in enhancing vaccination. In~\cite{cardillo2013ZhenPRE}, these results are mainly attributed to the competition of two processes taking place. One is the propagation of the diseases, and second is the uptake of vaccination behavior. If the vaccine works perfectly, or even if the vaccine is imperfect but the infection rate is small, the vaccination behavior spreads faster on SF networks than on ER graphs (i.e., the vaccination onset starts earlier on SF networks), as the natural epidemic threshold is smaller on SF networks~\cite{Boccaletti2006,vespiromu,pastor2015epidemic}. However, if the vaccine is imperfect and the infection rate is large, the outcome of the competition between disease and vaccination spreading reverses. Namely, under such conditions, the smaller benefits provided by the imperfect vaccine cause agents to forgo vaccination, which naturally promotes the propagation of the disease.

Taken together, these findings significantly enrich our understanding of behavior-vaccination dynamics on static, complex networks, and can relevantly inform on optimal vaccination strategies.

\subsubsection{External incentive programs}

\paragraph{The role of subsidy policy: a paradox}
Except for internal mechanisms, external incentive programs also play a vital role in the design of behavioral vaccination~\cite{jamison2004external,d2012interplay}. For example, historically, subsidy policy showed its particular effectiveness in terms of controlling epidemic spreading, which is of significance from the socioeconomic perspective~\cite{zhang2013impacts, geoffard1997ZhenPrinceton, culyer2000handbook, gersovitz2005tax, gersovitz2003infectious}. However, how different subsidy policies might work in structured populations is still an open issue. Motivated by this, Zhang \textit{et al}. recently considered two categories of subsidy policies and explored their influence on different networks, and how they affect vaccination decisions and disease dynamics~\cite{zhang2014effects}. The first class is the so-called free subsidy policy, where the total amount of subsidy $\Lambda$ is distributed to a certain fraction of individuals, who take vaccination without any personal cost. The (net) payoff function (i.e. Eq.~\ref{payoff-1}) thus becomes
\begin{equation}
\Pi_i = \left\{ \begin{array}{ll}
0, & \textrm{subsidized-vaccinating;}\\
-c, & \textrm{vaccinated;}\\
-1, & \textrm{infected;}\\
0, & \textrm{free-riding.}
\end{array} \right.
\label{payoff-2}
\end{equation}
where the number (fraction) of subsidized vaccinators or freely vaccinated individuals is $N_{SV}=\Lambda /c$ ($p=N_{SV}/N$, where $N$ is the size of the network). In this case, it is worth noting that, since the number of freely vaccinated individuals keeps constant in each vaccination stage, the selection process of them does not need to be repeated. If one individual is initially  chosen as the recipient (to receive the vaccine for free), it will remain a recipient forever. The second class is partial-offset subsidy policy, where each vaccinated individual is offset by a certain amount $\xi$, resulting in the actual vaccination cost $c(1-\xi)$. Correspondingly, the (net) payoff becomes
\begin{equation}
\Pi_i = \left\{ \begin{array}{ll}
(\xi-1)c, & \textrm{subsidized-vaccinating;}\\
-1, & \textrm{infected;}\\
0, & \textrm{free-riding.}
\end{array} \right.
\label{payoff-3}
\end{equation}
Similar to Eq.~\ref{mean-total}, the mean-field approximation for the fraction of vaccinated individuals $x$ for both the free subsidy policy and the partial-offset subsidy policy is given as
\begin{equation}
\begin{split}
\frac{dx}{dt}
&=x^+ - x^-\\
&= x(1-x)\{[(1-r(x)]W_{H \to V}+r(x)W_{I \to V}\}-(x-p)(1-x)\{[(1-r(x)]W_{V \to H}+r(x)W_{V \to I}\},\\
\end{split}
\label{mean-total3}
\end{equation}
and
\begin{equation}
\begin{split}
\frac{dx}{dt}
&=x^+ - x^-\\
&= x(1-x)\{[(1-r(x)]W_{H \to SV}+r(x)W_{I \to SV}\} -x(1-x)\{[(1-r(x)]W_{SV \to H}+r(x)W_{SV \to I}\},\\
\end{split}
\label{mean-total4}
\end{equation}
where $SV$ is the subsidized-vaccinating state and $x-p$ is the fraction of individuals who can change their vaccination strategies.

\begin{figure}
\begin{center}
\includegraphics[width=0.5\textwidth]{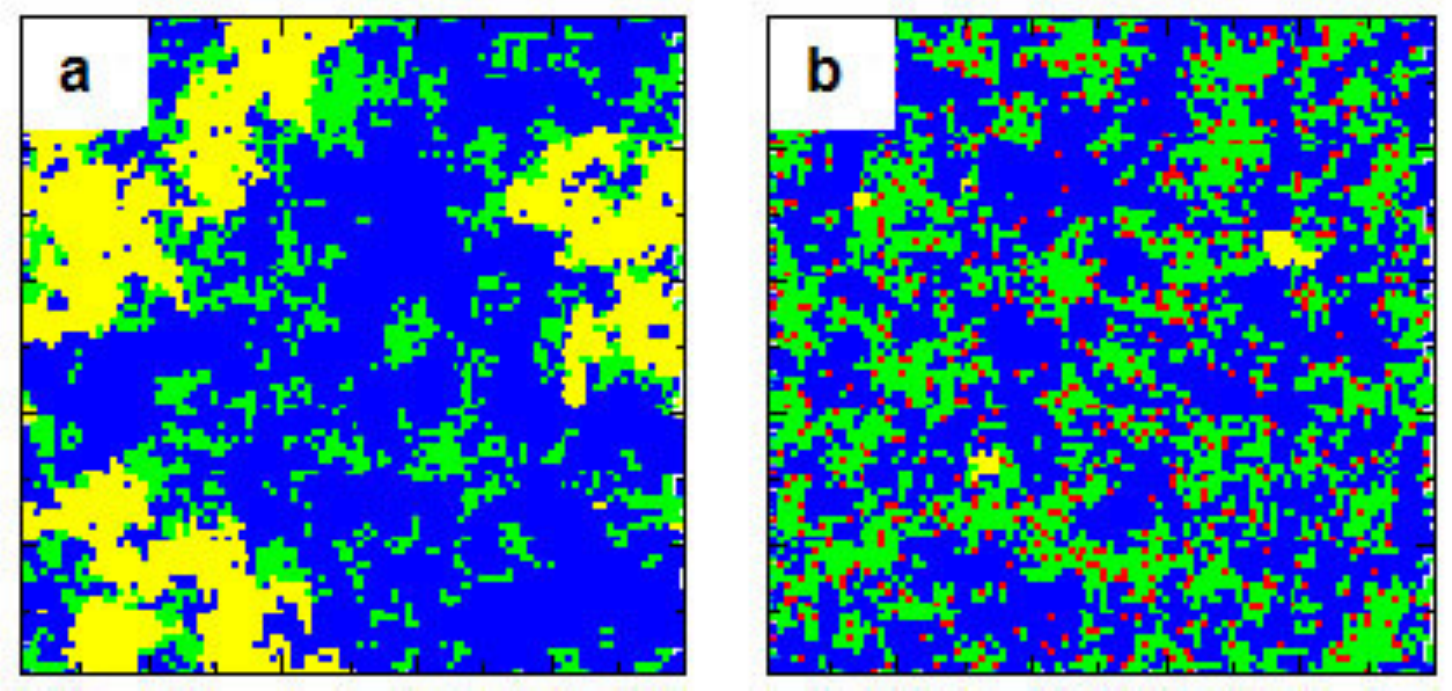}
\end{center}
\caption{Characteristic snapshots of the stationary state configuration for partial-offset (a) and free (b) subsidy policies on the square lattice. Blue, green, yellow, and red points represent free-riders (not vaccinated and healthy), vaccinators, infected individuals, and freely vaccinated recipients, respectively. Source: Reprinted figure from Ref.~\cite{zhang2014effects}.}
\label{subsidy}
\end{figure}

Based on the imitation dynamics (i.e. Eq.~\ref{Fermi-1}), it is found that the free subsidy policy in general leads to higher vaccination coverage than the partial-offset subsidy policy, indicating that the former case is more effective in controlling epidemic outbreaks. This phenomenon can also be verified by means of the mean-field approximation (i.e., Eqs.~\ref{mean-total3} and~\ref{mean-total4}). In the free subsidy case, the recipients act as role models, and are thus more appealing than the non-subsidized vaccinators in attracting their neighbors to take vaccination. This is expressed qualitatively in terms of the higher fraction of imitated vaccinators in the neighborhoods of recipients. Although freely vaccinated recipients (red) are randomly distributed on networks, it is quite remarkable how they are able to excite their neighbors and motivated them to imitate their behavior (see Fig.~\ref{subsidy}). This gives rise to vaccination clusters (green), which naturally results in a more effective control in comparison to the partial-offset subsidy policy. Snapshots in Fig.~\ref{subsidy} provide a compelling visual in favor of this argument.

As can be inferred from~\ref{imitation-double}, the updating dynamics based on individual behavior choices can directly decide the final vaccination coverage and the average epidemic size. When studying the impact of subsidies~\cite{zhang2014effects}, the authors considered also another Fermi-like updating rule, but one that does not involve imitation~\cite{zhang2013impacts}. Such updating is known as myopic~\cite{wang2012if}, because individuals update their vaccination preference by balancing their own advantages and disadvantages of vaccination (like the self-interest rule), rather than considering input from their neighbors. In particular, an individual first gets the perceived costs $P_N$ and $P_V$, and then opts for vaccination according to the probability
\begin{equation}
w_C=f(P_V-P_N)=\frac{1}{1+\textrm{exp}[-\kappa^{s}(P_V-P_N)]}.
\end{equation}
In other words, the strategy $s_i$ of individual $i$ is governed as
\begin{equation}
s_i =f(P_V-P_N)= \left\{ \begin{array}{ll}
\textrm{vaccination}, & \textrm{if $q<1/\{1+exp[-\kappa^{s}(P_V-P_N)]\}$};\\
\textrm{non-vaccination}, & \textrm{otherwise};\\
\end{array} \right.
\label{self-interested}
\end{equation}
where $q$ is a randomly distributed real number in $[0,1]$.

This myopic updating dynamics leads to completely different results in terms of the effect of different subsidy policies. Namely, in this case the partial-offset subsidy policy works better than the free subsidy policy. Under the latter, there is polarization between subsidized vaccinators and general agents, which is very non-beneficial for the emergence of herd immunity. Owing to the unequal distribution of subsidy, the perceived infection risk of those recipients greatly decreases, but non-recipients are more prone to taking a risk (i.e., the willingness of vaccination among non-recipients declines) after balancing the advantages and disadvantages of vaccination. In the partial-offset subsidy case, the cost of vaccination is evenly reduced for each vaccinated individual due to the endowment of the policy, which provides natural conditions for the formation of herd immunity. In addition to group-optimal vaccination, the study revealed that a moderate subsidy rate can generate the maximal socioeconomic benefic under the partial-offset subsidy policy, which is a particular meaningful message for the public health authorities.

Considering these results, it becomes clear that the optimization~\cite{gao2015selectively,li2014chaos,du2016heterogeneous,du2009asymmetric} of the subsidy policy is indeed a difficult yet very important issue, which can mean the difference between a successful and an unsuccessful vaccination campaign. For the optimal outcome, it is important to carefully consider the mechanisms of individual decision-making, as well as to actually test on a representative sample how people respond to different external incentives in realistic situations.

\paragraph{The impact of committed individuals, or zealots}
In real life, many people hold rigid viewpoints on certain aspects of reality, regardless of peer pressure, incentives, and media. Such people are known as committed individuals or, shorter, zealots. The impact of zealots on various aspects of social dynamics has been studied often before, for example in opinion dynamics~\cite{mobilia2003does, galam2005heterogeneous, xie2011social, singh2012accelerating} or in evolutionary social dilemmas~\cite{masuda2012evolution}. Of course, zealots are also found with regards to vaccination. Since some agents are much more sensitive to the infection risk than average, they consistently adopt the vaccinating strategy, i.e., the committed vaccinators do not engage in the decision-making process~\cite{liu2012ZhenPRE}. In essence, the setup of committed vaccinators is the same as the free subsidy policy (the mean-field analysis is the same as put forward forward with Eq.~\ref{mean-total3}), needing only one additional parameter $f_C$ to determine their fraction. Owing to the effect of ``steadfast role models'', committed vaccinators stimulate vaccination uptake and can thus contribute significantly to the prevention of infectious diseases. The larger the value of $f_C$, the more obvious this effect. As such, committed vaccinators persuade their neighbors to follow their behavior and form vaccinating clusters, which is, as highlighted often above, the typical effect of imitation~\cite{fu2011ZhenPRSB, Wu2013epl, zhang2014effects}.

Of course, zealots can also be committed non-vaccinators. It is thus necessary to look at the down side too, that is, considering committed vaccinators as good role models, but also considering committed non-vaccinators as bad role models. This has been done in~\cite{fukuda2016}, and the results were as follows. On lattices, the presence of committed vaccinators (non-vaccinators) exerts a positive (negative) effect on the epidemic spreading. Moreover, on SF networks the presence of committed vaccinators can not prevent an epidemic for small and intermediate vaccine cost, predominantly because the hub nodes consistently prefer to reject vaccination. In this respect, the effect of committed individuals depends not only on their fraction, but also on the network structure and the cost of vaccination.

\subsubsection{Multiple-strategy dilemma framework: A counter-intuitive observation}
\label{counter-intuitive}

\begin{center}
\begin{table*}
\centering
\caption{\label{tab-payoff} The payoffs for different strategies and states. Here $\Pi_V$, $\Pi_S$ and $\Pi_L$ are the (net) payoffs for vaccination, self-protection and laissez-faire, respectively. The superscripts $H$ and $I$ denote healthy and infected states, while $w$ is the probability of a susceptible to become infected. For more details, please refer to~\cite{zhang2013braess}.}
\begin{tabular}{p{5cm}p{5cm}p{2cm}ll|}\hline
Strategy \& State & Fraction &  Payoffs &\\
\hline
\rowcolor{mygray}
Vaccinated \& Healthy & $\rho_V$ &  $\Pi_V=-c$ &\\
Self-protected \& Healthy &  $\rho_S^H=\rho_S[\delta+(1-\delta)(1-w)]$  & $\Pi_S^H=-b$  &\\
\rowcolor{mygray}
Self-protected \& Infected &  $\rho_S^I=\rho_S(1-\delta)w$  & $\Pi_S^I=-b-1$  &\\
Laissez-faire \& Healthy &  $\rho_L^H=(1-\rho_V-\rho_S)(1-w)$  & $\Pi_L^H=0$  &\\
\rowcolor{mygray}
Laissez-faire \& Infected &  $\rho_L^I=(1-\rho_V-\rho_S)w$  & $\Pi_L^I=-1$  &\\
\hline
\end{tabular}
\end{table*}
\end{center}

When an infectious diseases threatens, vaccination is often the first and the most common line of defence to protect the public. However, due to high economic cost, religious beliefs, as well as possible side effect~\cite{schimit2011vaccination,parker2006implications,may2003clustering}, people sometimes prefer alternative, and frequently also cheaper self-protection measures~\cite{capasso1978generalization,noibeta}, such as avoiding contact with risk groups or wearing face masks to at least temporarily mitigate the risks. To take this into account, three different strategies, namely vaccination, self-protection, and laissez-faire (i.e., neither vaccination nor self-protection), were considered in the realm of an evolutionary epidemic game~\cite{zhang2013braess}. Assuming that the vaccine provides perfect protection at a cost $c$, self-protection is effective with probability $\delta$ at a lower cost $b$ ($b<c$). Moreover, with probability $1-\delta$ a self-protected individual turns to laissez-faire. That is, self-protection is an alternative in between traditional vaccination and laissez-faire. Analogous to the binary-choice vaccination game~\cite{fu2011ZhenPRSB}, the payoff of this new framework is portrayed in Table~\ref{tab-payoff}, where $\rho_V$, $\rho_S$ and $\rho_L$ are the factions of vaccinated, self-protected and laissez-faire individuals, respectively. According to the mean-field approximation framework, the time evolution of the fraction of strategies is governed by the following equations
\begin{equation}
\frac{d\rho_V}{dt}=(\rho_V \leftrightarrows \rho_S^H)+(\rho_V \leftrightarrows \rho_S^I)+(\rho_V \leftrightarrows \rho_L^H)+(\rho_V \leftrightarrows \rho_L^I),
\end{equation}
and
\begin{equation}
\frac{d\rho_S}{dt}=(\rho_S^H \leftrightarrows \rho_V)+(\rho_S^H\leftrightarrows \rho_L^H)+(\rho_S^H \leftrightarrows \rho_L^I)+(\rho_S^I \leftrightarrows \rho_V)+(\rho_S^I \leftrightarrows \rho_L^H)+(\rho_S^I \leftrightarrows \rho_L^I),
\end{equation}
where
\begin{equation}
\begin{split}
\rho_A \leftrightarrows \rho_B
&= (\rho_B \to \rho_A)-(\rho_A \to \rho_B)\\
&=\rho_A \rho_B (W_{B\to A}-W_{A\to B})\\
&=\rho_A\rho_B\{\frac{1}{1+\textrm{exp}[-\kappa^{s}(\Pi_A-\Pi_B)]}- \frac{1}{1+\textrm{exp}[-\kappa^{s}(\Pi_B-\Pi_A)]} \}\\
&=\rho_A\rho_B\textrm{tanh}[\frac{\kappa^{s}}{2}(\Pi_A-\Pi_B)].
\end{split}
\end{equation}
Here $A$ and $B$ represent the final states, and the expressions for the fractions and payoffs of each state are from Table~\ref{tab-payoff}. Since the preservation condition $\rho_V+\rho_S+\rho_L=1$ holds, it is straightforward to get the equilibrium fraction of all three strategies from the above equations.

\begin{figure}
\begin{center}
\includegraphics[width=0.65\textwidth]{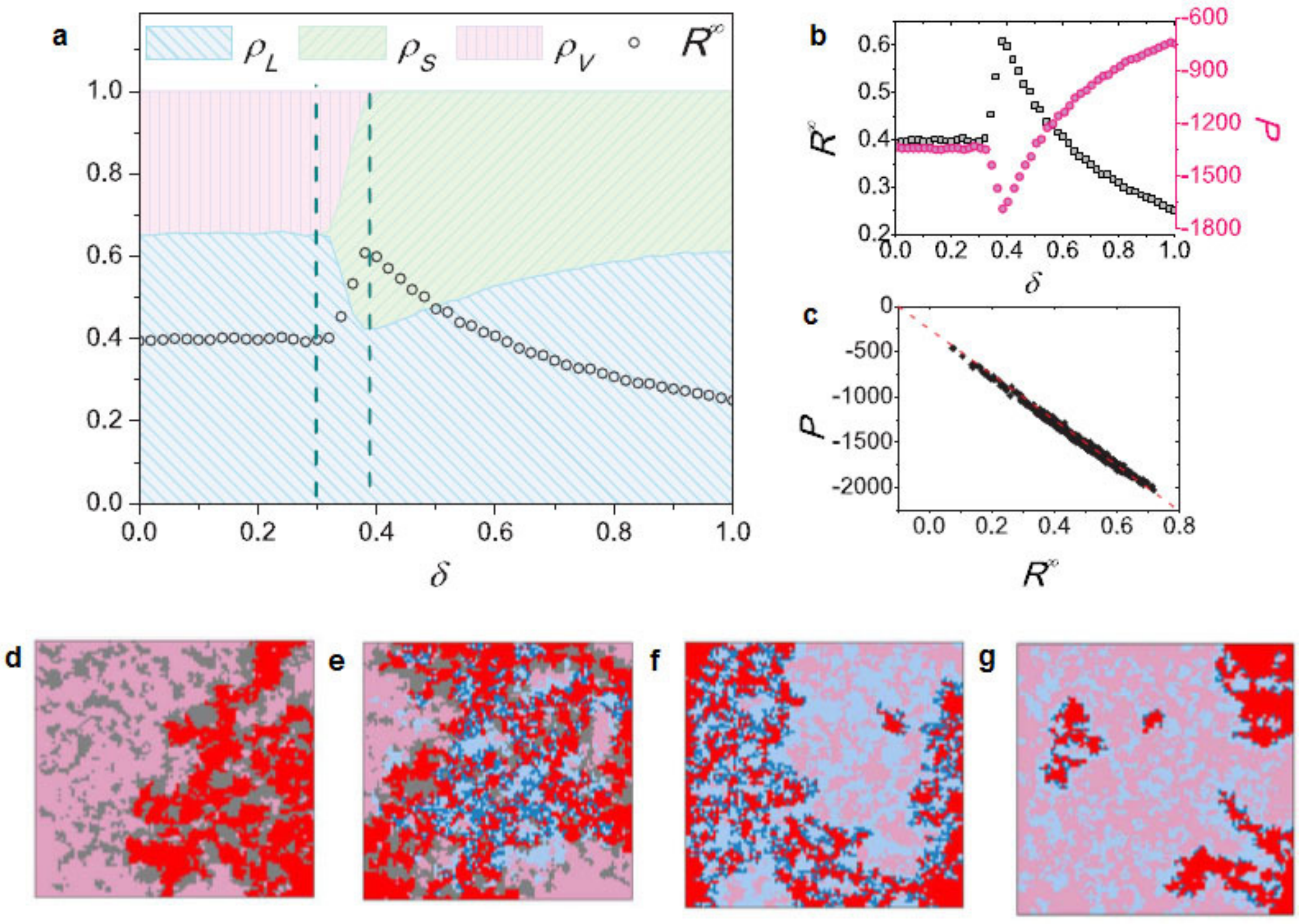}
\end{center}
\caption{(a) The fraction of strategies and the epidemic size $R^{\infty}$ are shown as a function of the successful rate of self-protection $\delta$. (b) The epidemic size $R^{\infty}$ and the total payoff $P$ are shown as a function of $\delta$. (c) Correlation between the population payoff $P$ and the epidemic size $R^{\infty}$. (d)-(g) Typical snapshots of the steady state for $\delta=0.2$, 0.35, 0.5 and 0.95, respectively. Grey, light red, dark red, light blue and dark blue denote vaccinated, uninfected laissez-faire, infected laissez-faire, uninfected self-protective, and infected self-protective individuals, respectively. Source: Reprinted figure from Ref.~\cite{zhang2013braess}.}
\label{paradox}
\end{figure}

With this setup, the authors in~\cite{zhang2013braess} discovered a counterintuitive phenomenon. Namely, intermediate values of $\delta$ lead to a larger epidemic size and a higher collective cost, which is similar to the well-known Braess's paradox~\cite{braess1968paradoxon, pala2012transport}. More precisely, the whole range of $\delta$ is divided into three regions, as shown in Fig.~\ref{paradox}(a). In the left region, owning to the low efficiency, no one self-protects, which returns the traditional binary selection model~\cite{fu2011ZhenPRSB}. Infected laissez-faire individuals (dark red) and uninfected laissez-faire individuals (i.e., free-riders, light red) are strictly separated by vaccinators and form percolating clusters (see Fig.~\ref{paradox}(d)). With middle region, for intermediate values of $\delta$ (i.e., the region between both dash lines), the payoff advantage of self-protection starts to play a noticeable role, so that vaccination and partial laissez-faire are subverted. However, this leaves the heard with insufficient protection, ultimately resulting in a larger epidemic and the fragmentation of percolating free-rider clusters. Another counterintuitive fact is that more self-protected individuals (whose individual payoff is higher) generate a lower (rather than higher) total payoff of the population. Thus, the collective payoff and the epidemic size are negatively correlated with the spread of self-protected individuals, as shown in Figs.~\ref{paradox}(b) and (c). In the third region, when $\delta$ is sufficiently large, self-protection is sufficiently effective. Epidemics decline and vaccination is not a viable option. Under such circumstances, the percolating cluster of free-riders (light red) recovers due to herd immunity.

The universality of this interesting phenomenon can be further extended to other networks, and it can be qualitatively validated by means of the mean-field approximation. In heterogeneous networks, due primarily to the hubs, those that become infected and laissez-faire individuals fare worse still than they do on more homogeneous networks where the differences in degree are smaller. Similarly counterintuitive phenomena can also be observed with myopic updating (self-interest rule)~\cite{wang2014multiple}, although the differences with the more commonly considered imitation rule~\cite{fu2011ZhenPRSB,Wu2013epl,zhang2010ZhenNJP,zhang2014effects} are significant. This should be taken into account, specifically when considering competing strategies in vaccination programs.

\subsection{Behavior-vaccination dynamics on empirical networks}

Different from the network models reviewed above, realistic contact networks usually exhibit assortative and disassortative mixing patterns~\cite{Newman2010, newman2003structure, Newman_mixing} and community structure~\cite{Newman2, Fortunatoregino2010Physrep}, both of which significantly impacts the coupled dynamics of behavior and vaccination taking place on them. In this subsection we review recent research devoted to behavior-vaccination dynamics on empirical networks.

\subsubsection{How does feedback influence well-known passive and pro-active policies?}

\paragraph{The voluntary ring vaccination strategy}

\begin{figure}
\begin{center}
\includegraphics[width=0.9\textwidth]{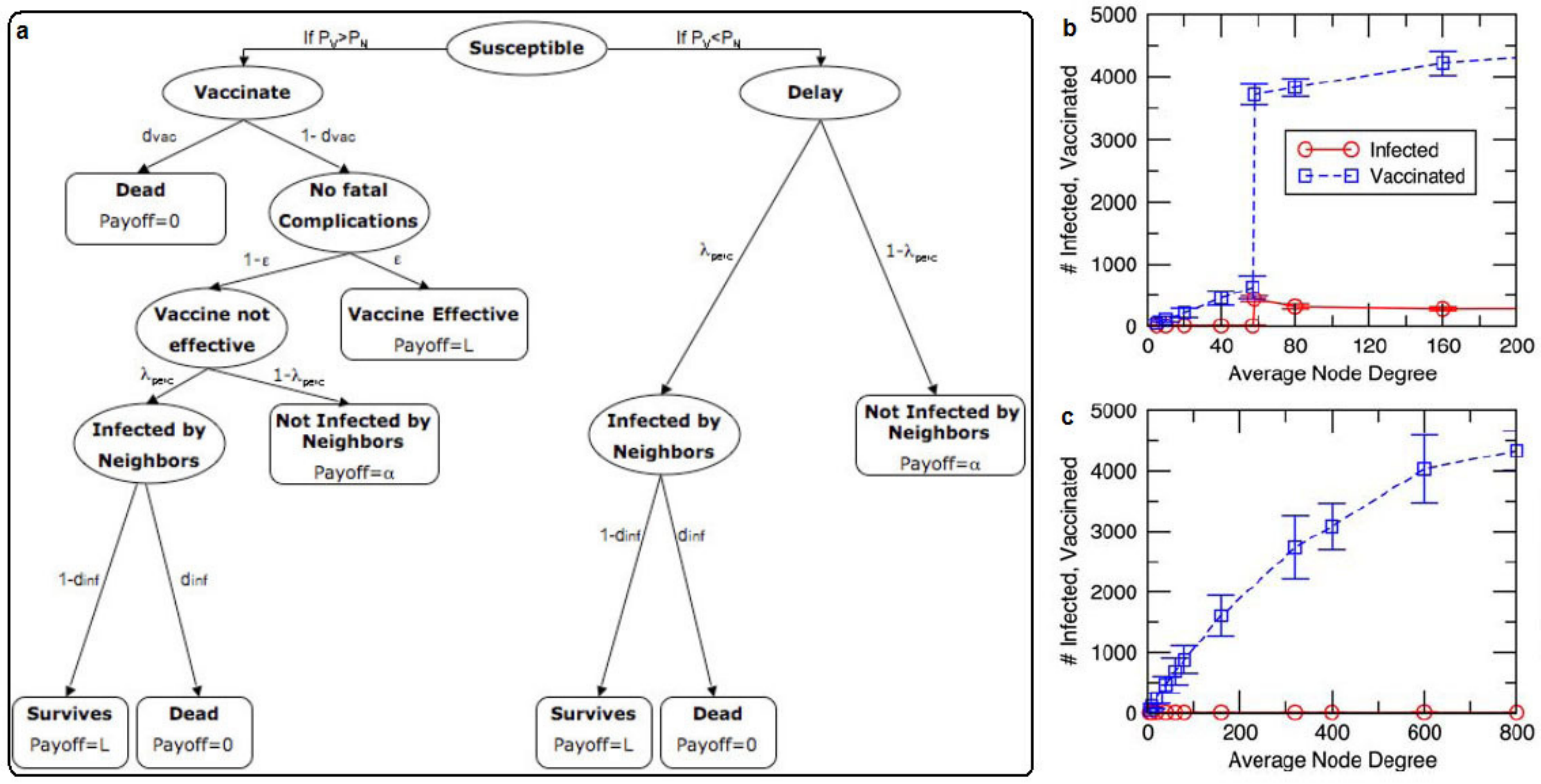}
\end{center}
\caption{(a) Schematic illustrations of the perceived costs of individuals. Ellipses represent transition states while boxes represent formal states. Parameters on arrows denote transition probabilities and expressions in boxes denote costs for entering that state. For the meaning of used parameters we refer to~\cite{perisic2009simulation,perisic2009social}. The numbers of infected and vaccinated agents are shown as a function of the average neighborhood size in the theoretical model (b) and the smallpox model (c). The error bars represent two standard deviations from the mean across 20 simulations per datum. Source: Reprinted figure from Refs.~\cite{perisic2009simulation,perisic2009social}.}
\label{ring-vaccinate1}
\end{figure}

In a homogeneous mixing population, it is a well-recognized fact that voluntary vaccination (without any economic incentives) is difficult or impossible to control, and thus can not be used effectively to eradicate vaccine-preventable disease~\cite{fine1986ZhenOUP, geoffard1997ZhenPrinceton, bauch2003group, bauch2004ZhenPNAS, galvani2007long, noi1, vardavas2007can, breban2007mean, barrett2007smallpox}. As vaccination coverage levels increase, herd immunity can provide greater protection. However, soon the low infection risk reduces the enthusiasm of individual vaccination and leads to the reoccurrence of high disease prevalence. That is to say, voluntary policy generates oscillations in the uptake of vaccination~\cite{bauch2005ZhenPRSB}, which is particular true for highly transmissible diseases. However, the history of smallpox teaches us that complete eradication by means of voluntary vaccination is possible~\cite{fenner1988smallpox,cochi2014global,brilliant1985management}. Aiming to elucidate this observation, Perisic \textit{et al.} have integrated disease dynamics with a voluntary ring vaccination policy in a social network~\cite{perisic2009social}.

Based on the perceived risk of infection and benefit of vaccination, they obtained the perceived cost functions of non-vaccination $P_N$ and vaccination $P_V$ (see Fig.~\ref{ring-vaccinate1}(a) for the transition probability of individual states):
\begin{equation}
P_N=(1-\lambda_{perc})a+\lambda_{perc}[(1-d_{inf})L],
\end{equation}
\begin{equation}
P_V=\varsigma [(1-d_{vac})L]+(1-\varsigma)\{( 1-d_{vac})[(1-\lambda_{perc})a+\lambda_{perc}(1-d_{inf})L]\},
\end{equation}
where $a$ ($L$) is the payoff for individuals with continued susceptibility (lifelong immunity), $\varsigma$ is vaccine efficacy and $d_{inf}$ ($d_{vac}$) represents the probability of death due to infection (vaccine-related complications). In particular, $P_V$ consists of two terms: the first term denotes the total payoff if vaccine is efficacious; while the second is the total payoff once vaccine loses efficacy. The decision-making process was in accordance with rational self-interest rule (i.e., Eq.~\ref{self-interested-1}). If the cost to vaccinate is smaller than the cost not to vaccinate on a given day (i.e., $P_V<P_N$), individuals would turn to vaccination. Interestingly, there is a threshold in the neighborhood size, which guaranteed ``first-order'' phase transition of vaccination behavior, like the explosive phenomena of collective dynamics in statistical physics~\cite{gomez2011explosive, achlioptas2009explosive}. From Fig.~\ref{ring-vaccinate1}(b), it is clear that voluntary ring vaccination quickly contains the outbreak until the average neighborhood size is larger than $57$. Due to effective control, the total number of infected and vaccinated agents remains very small below this threshold. Interestingly, as the neighborhood size increases (while the average force of infection is held constant), the model recovers the homogeneous mixing scenario~\cite{fine1986ZhenOUP, bauch2003group, noi1, bauch2004ZhenPNAS, galvani2007long}, where each infected agent could influence more neighbors. As a result, infection breaks through the imperfect rings of vaccinated individuals, percolates through the whole network and leads to a considerable final epidemic size and numerous vaccinated. To further validate the accuracy of these theoretical predictions, the authors in~\cite{perisic2009social} also carried out the investigation for smallpox-specific disease history and vaccine properties, with similar results (see Fig.~\ref{ring-vaccinate1}(c)). Their efforts thus illustrate the significant difference between continuous population and spatial structure population in coupled behavior-vaccination dynamics. On the other hand, it suggests that the design of voluntary vaccination, as a primary control measure, to eradicate outbreak of some close contact diseases such as smallpox should rely on the number of social contacts and on the local structure of the population.

\begin{figure}
\begin{center}
\includegraphics[width=1.0\textwidth]{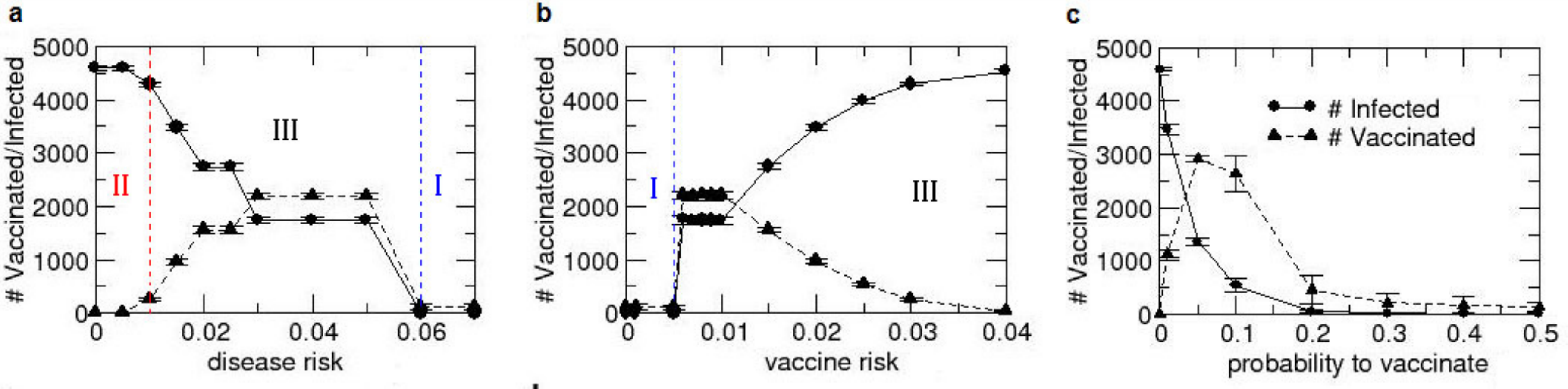}
\end{center}
\caption{The numbers of infected and vaccinated individuals are shown as a function of disease risk (a, i.e., the probability of death due to disease), vaccine risk (b, i.e., the probability of death due to vaccine) and probability to vaccinate (c). Note that (a)-(b) are further divided into three phases. The error bars represent two standard deviations from the mean across $20$ simulations per datum. Source: Reprinted figure from Ref.~\cite{perisic2009simulation}.}
\label{ring-vaccinate2}
\end{figure}

Though voluntary vaccination policy enables eradication of the disease in contact-structured networks, the foregoing model made a couple of simplifying assumptions regarding the natural history of the disease, pre-existing immunity, and human behavior. To relax some of these simplifications, the authors further investigated the dynamics via weighing infection risk (i.e., the probability of death due to disease) versus vaccine risk (i.e., the probability of death due to vaccine)~\cite{perisic2009simulation}. Of significance, they have shown a counterintuitive phenomenon for the coupled dynamics, which was then divided into three phases (see Fig.~\ref{ring-vaccinate2}). Phase \uppercase\expandafter{\romannumeral1} exhibits a negligible final size of the epidemic and a small number of vaccinated individuals. In this phase, any node with an infected neighbor would vaccinate immediately, preventing disease transmission. Thus, the infection risk will be fast, and effectively controlled under a voluntary ring vaccination. Then, in phase \uppercase\expandafter{\romannumeral2}, nobody is likely to vaccinate due to low disease risk or high vaccine risk (this phase is not shown in Fig.~\ref{ring-vaccinate2}(b)). This finally results in a large final epidemic size. For intermediate epidemic risk and vaccine risk (i.e., phase \uppercase\expandafter{\romannumeral3}, the area between two dash lines), though a great number of agents prefer to vaccinate, the disease still percolates through the whole population due to imperfect protection. Except for a deterministic decision process, the authors also tested the effect of a probabilistic setup. Accordingly, even if $P_V<P_N$, individuals may still opt to vaccinate with a given probability. If this probability is low, then very few people are allowed to vaccinate and infection takes over the population (see Fig.~\ref{ring-vaccinate2}(c)). As the probability increases, the outbreak is effectively contained by means of the voluntary ring vaccination. Interestingly, the continuous enhancement of vaccination probability quickly reduces the essential number of those that need to be vaccinated for better controlling the spread of the disease (namely, there is a vaccination peak). Altogether, the study~\cite{perisic2009simulation} revealed that the final outcome of rational self-interest dynamics under a voluntary vaccination policy is closely related with contact patterns, with the perceived vaccine risk, as well as with the risk of contracting the disease.

\paragraph{The superspreader vaccination strategy}

In the context of network vaccination (i.e. Section~\ref{sec:vac-net}), it has been thoroughly established that targeted vaccination towards superspreaders can significantly reduce the spread of disease~\cite{wang2015immunity, pastor2002immunization, lloyd2001viruses, zhao2014immunization,
buono2015immunization}. However, the assumption that superspreaders (who typically have many connections and are responsible for the majority of secondary infections~\cite{kemper1980identification, woolhouse1997heterogeneities, galvani2005epidemiology,
lloyd2005superspreading, stein2009lessons}) always accept vaccination does not capture the realistic transmission dynamics with behavioral response, especially for seasonal influenza-like vaccination. In this regard, Wells \textit{et al}. analyzed how behavioral feedback affects the effectiveness of vaccination strategies~\cite{wells2013policy}. In addition to the passive vaccination as the baseline strategy (whose update is governed by a function of payoff difference $f(P_V-P_N)$, individuals can also implement one of four pro-active strategies: 1) random vaccination (RV) targeting a randomly chosen individual; 2) nearest neighbor vaccination (NN) aiming to a randomly chosen individual and one of their neighbors; 3) chain vaccination (CV) targeting either a randomly chosen individual or a neighbor of an individual targeted the previous day; 4) improved nearest neighbor vaccination (INN) targeting a randomly chosen individual and one of their most connected neighbors.

Surprisingly, it was revealed that the proposed superspreader vaccination strategies provide little or no improvements, compared to traditional ad-hoc delivery of vaccination~\cite{cohen2003ZhenPRL, holme2004efficient}. The behavioral feedback facilitates policy resistance, which nearly completely undermines the effectiveness of superspreader strategies. Actually, superspreader strategies enhance the vaccination level among superspreaders. However, this effort is offset by the low coverage among non-superspreaders (representing the majority of nodes in a heterogeneous network), which naturally leads to a low overall level of success. These observations indicate that selecting connection heterogeneity as an important criterion of implementing superspreader strategies is not sufficiently accurate. For example, individuals with higher infectiousness or a longer period of infectiousness may not coincide with highly connected individuals~\cite{miller2007effective,lloyd2005superspreading}. Moreover, to improve the readiness to vaccinate in the population, an economic incentive to the targeted individual has also been proposed. As expected, the undermined benefit of pro-vaccination strategies can be somewhat recovered under a substantial economic stimulus. Thus, the design of vaccination strategies targeting superspreaders must consider the possible policy resistance caused by strong effects of behavioral feedback, and mitigate it whenever possible.

\paragraph{The mixed vaccination strategy}

Up to now, we have acquired a thorough understanding for the interrelationship between vaccination decision and disease dynamics: via discrete, memoryless strategies, human decision-making can profoundly influence epidemic incidence as well as vaccination coverage~\cite{funk2010modelling, wang2015coupled, wang2014spatial, bauch2004ZhenPNAS}. However, as the cooperation strategy in evolutionary games~\cite{szolnoki2007cooperation,kokubo2015spatial}, vaccination sometimes can be regarded as one mixed strategy, i.e., choosing vaccination with a given probability and taking the non-vaccination stance with the remaining probability. It is interesting to consider how does this setup affect the vaccination program and the spread of disease.

To address this, Cornforth \textit{et al.} defined the vaccination strategy $v_k$ of an individual with degree $k$ as the probability that it will opt for vaccination~\cite{cornforth2011erratic}. The total perceived payoff of this individual is thus given by
\begin{equation}
\pi_k(v_k)=u-v_k(C_V+o_k C_R)-(1-v_k)\alpha_k C_I,
\label{mix-strategy}
\end{equation}
where $u$ is the baseline payoff corresponding to the value of a healthy life over the course of an epidemic, while $C_V$ ($C_I$) represents the monetary cost of vaccination (infection). Because the influenza vaccine is only partially effective, some vaccinated agents remain susceptible to the infection. Here, $C_R$ denotes the cost of infection to a partially resistant (immune) agent, while $o_k$ ($\alpha_k$) is the perceived probability of becoming infected if vaccinated (non-vaccinated). Thus, different from respective expressions of discrete strategies~\cite{fu2011ZhenPRSB, zhang2010ZhenNJP, zhang2013braess}, the second term in the right hand side of Eq.~\ref{mix-strategy} is the perceived payoff of vaccination selection, and the third accounts for the perceived payoff of non-vaccinated behavior. It is also worth mentioning that the estimation of perceived risk depends on the information of past epidemic seasons.

\begin{figure}
\begin{center}
\includegraphics[width=0.8\textwidth]{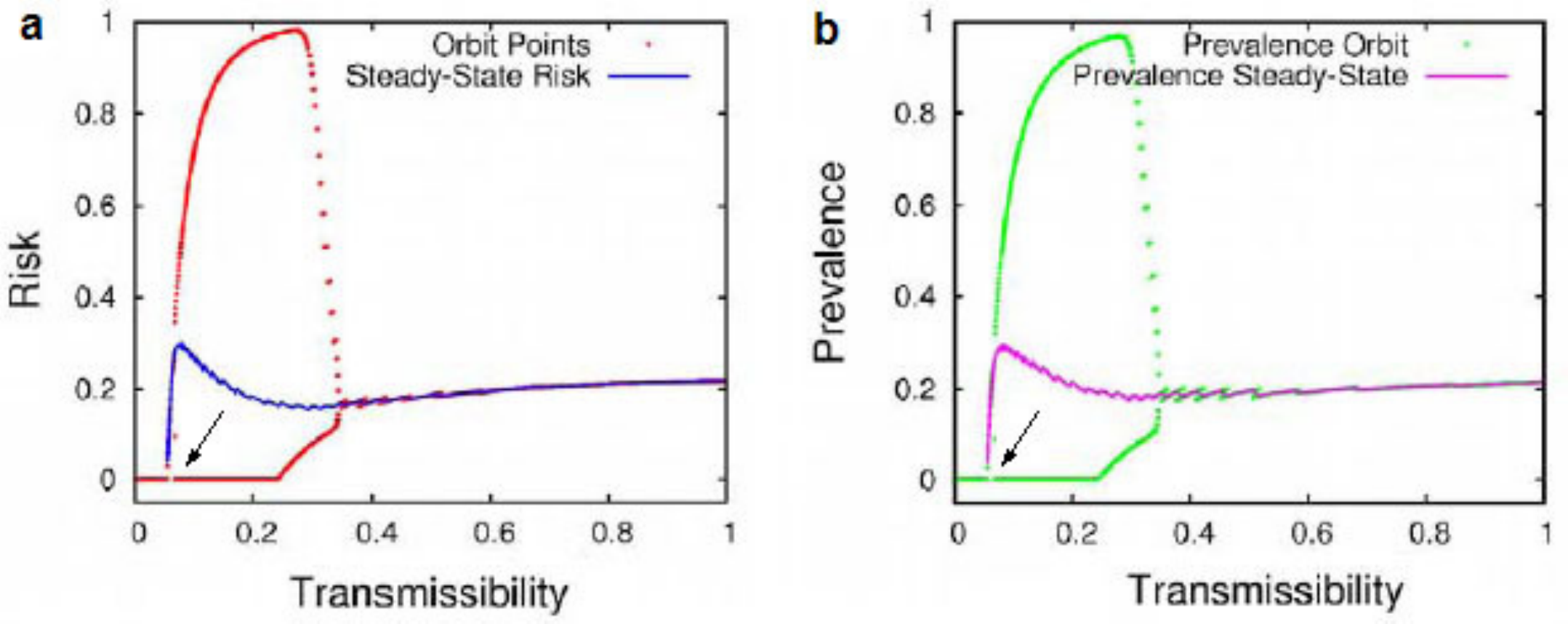}
\end{center}
\caption{The impact of transmissibility on risk (a) and disease prevalence (b) (steady-states and unstable limit cycles) in urban networks. Before the threshold (indicated by black arrows), there is no epidemic risk and vaccination. With transmissibility increasing, these two first diverge and then end at the same state again. Source: Reprinted figure from Ref.~\cite{cornforth2011erratic}.}
\label{mixed-vaccinate}
\end{figure}

Regardless of the properties of the interaction networks, it was shown in~\cite{cornforth2011erratic} that the likelihood of vaccination increases with degree (because the infection risk also increases with degree), which is the same with prediction of voluntary ring vaccination~\cite{perisic2009social}. Nevertheless, the degree threshold for vaccination is closely related with the overall structure of the network, and increases with the heterogeneity of the degree distribution (though the perceived fraction of vaccination does not change monotonically with network heterogeneity). Interestingly, it was also found that vaccination relies critically on the transmissibility of disease~\cite{cornforth2011erratic}. Results presented in Fig.~\ref{mixed-vaccinate} show how the steady-state risk/prevalence (corresponding to Nash Equilibria) and unstable limit cycles vary as a function of transmissibility. For small transmissibility, there is no epidemic risk and thus no vaccination under the threshold (indicated by black arrows). As transmissibility increases, epidemic risk/incidence is predicted to converge to a stable Nash Equilibrium. However, in reality there are considerable oscillations from one season to the next. This is caused by the polarization of vaccination behavior: non-vaccination is always followed by wide-spread disease and consequently high vaccination coverage in the next season. At a high level of transmissibility, the oscillatory pattern in vaccination behavior goes back to the steady state. These oscillations seem empirically inconsistent with the relatively stable vaccination pattern of influenza. To elucidate this discrepancy, it is further unveiled that increasing the number of prior seasons that individuals recall during the decision-making process enables these oscillations collapsing onto a stable vaccination strategy.

\subsubsection{Social clustering of behavior and opinions -- assortativity of (non-)vaccination}
\label{clusters}

\paragraph{Empirical evidences -- clusters driven by strategy and opinion updating}

\begin{figure}
\begin{center}
\includegraphics[width=1.0\textwidth]{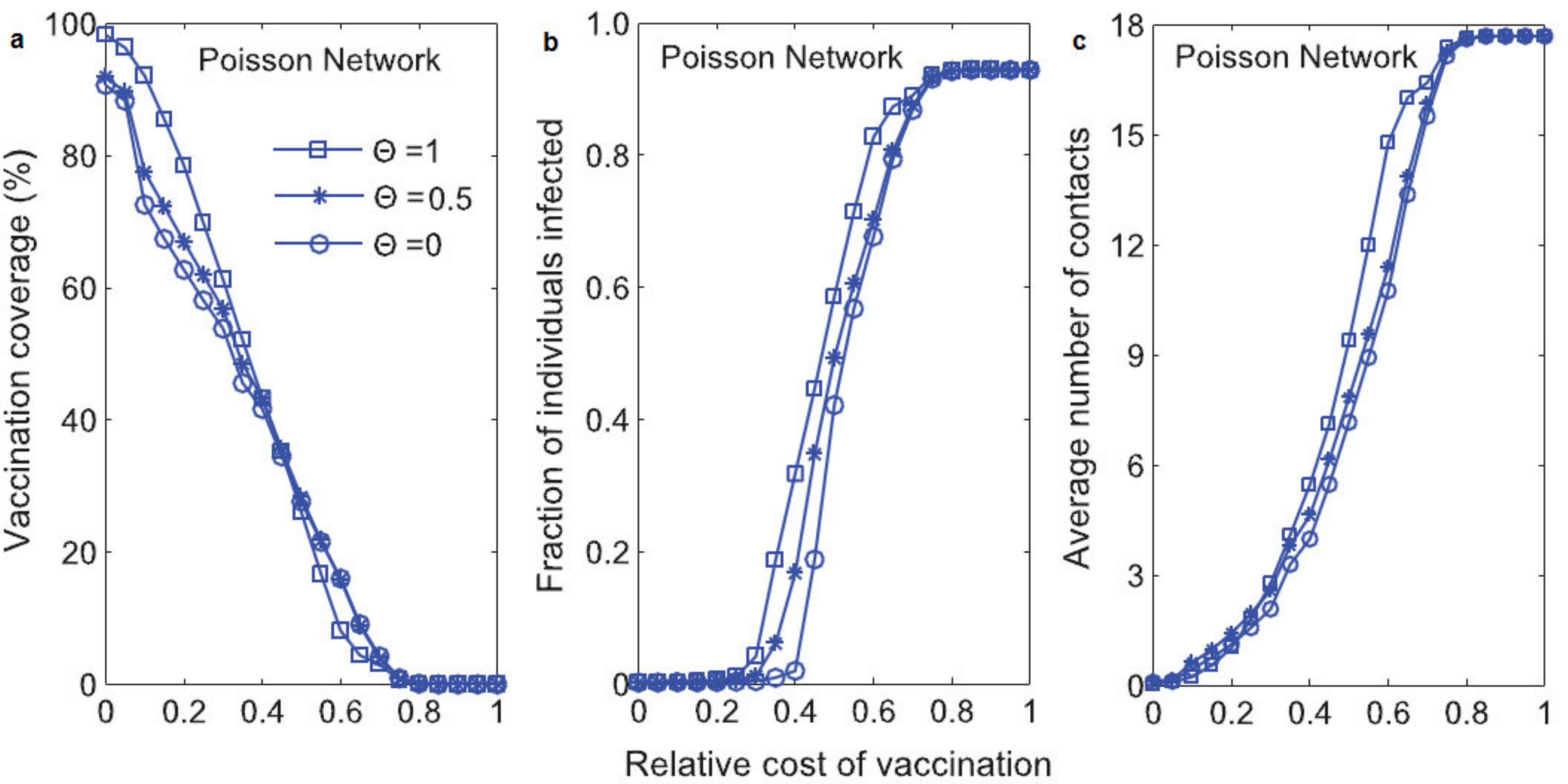}
\end{center}
\caption{Vaccination coverage (a), epidemic size (b), and average number of contacts between non-vaccinators are shown as a function of the relative cost of vaccination in passion (or homogeneous) networks. The fraction of individuals with imitation behavior is $\Phi$, while the remaining $1-\Phi$ follow a payoff maximization strategy. Source: Reprinted figure from Ref.~\cite{mbah2012impact}.}
\label{cluster-a}
\end{figure}

When reviewing results presented in~\ref{traditional-net}, we have emphasized that the updating dynamics plays a crucial role in the final size of epidemic and vaccination coverage. Notably, during the decision-making process, there exist different criteria and factors that affect each particular choice, such as payoff maximization~\cite{zhang2010ZhenNJP}, stochastic imitation~\cite{fu2011ZhenPRSB}, and social impact theory~\cite{xia2013computational}. While it is clear that taking into account one or the other criterion leads to different evolutionary outcomes, few works have focused on the impact of joint action. In this regard, Mbah \textit{et al}. incorporated imitation dynamics (i.e., Eq.~\ref{Fermi-1}) and self-interest dynamics (i.e., Eq.~\ref{self-interested-1}) into a coupled epidemiological model on different contact networks~\cite{mbah2012impact}. In particular, individuals could make their vaccination decision either based on imitation or based on payoff maximization, with probability $\Theta$ and $1-\Theta$, respectively.

Surprisingly, in this framework the ``double-edge sword'' effect appears again. Namely, imitation behavior (i.e., $\Theta=1$) increases vaccination coverage for relatively low costs of vaccination, yet impedes the incentive to vaccinate at high costs (see Fig.~\ref{cluster-a}(a)). While in terms of completely eradicating the epidemic, imitation behavior fails and even enlarges the ultimate epidemic size, irrespective of vaccination costs (Fig.~\ref{cluster-a}(b)). This unexpected outcome is related to behavioral clusters resulting from imitation dynamics, where vaccinators tend to contact vaccinators, and non-vaccinators tend to contact non-vaccinators (Fig.~\ref{cluster-a}(c)). Although vaccination levels increase in some regions, clusters of non-vaccinators (susceptible individuals) exacerbate transmission and elevate the probability of an outbreak. In this regard, imitation behavior may shed new light into the high incidence of measles in many countries with high vaccine uptake~\cite{salathe2008effect, schmid2008ongoing, richard2008measles,wang2011effects,zhang2012evaluating}. Realistic policies should thus consider the impact of non-vaccinated clusters, and empirical data on vaccination behavior should be explored also in terms of geolocating potential hotspots of non-vaccinated individuals by public health authorities.

Besides payoff driven updating, opinion dynamics is another possible explanation for the large outbreaks of some vaccine-preventable diseases such as measles in many high-vaccination countries~\cite{Maybook, orenstein2004summary}. Recently, Salath{\'e} \textit{et al}. assigned either a positive or a negative opinion about vaccination to each node~\cite{salathe2008effect}, where positive opinion means that the agent will take up vaccination immediately, while individual with negative opinion reject vaccination due to personal beliefs, such as the safety and efficiency of vaccine, religious beliefs, and related considerations. Subsequently, an opinion formation process occurs based on the dissimilarity index and the strength of opinion formation. Interestingly, such a process leads to clusters of non-vaccinated individuals, since agents with negative opinion are more likely to contact same-minded counterparts. The invasion of a single infected agent into such a susceptible clusters may have dire consequences and dramatically decrease herd immunity. Moreover, the research also found that, when the vaccination coverage is close to or at the level required to provide herd immunity, this effect of opinion formation (or clustering) facilitating outbreaks is the strongest. This in turn indicates that the estimated coverage level to avert an outbreak of measles may be much higher than realistic demands~\cite{parker2006implications, richard2008measles}.

\paragraph{Quantitative characterization}

\begin{figure}
\begin{center}
\includegraphics[width=0.8\textwidth]{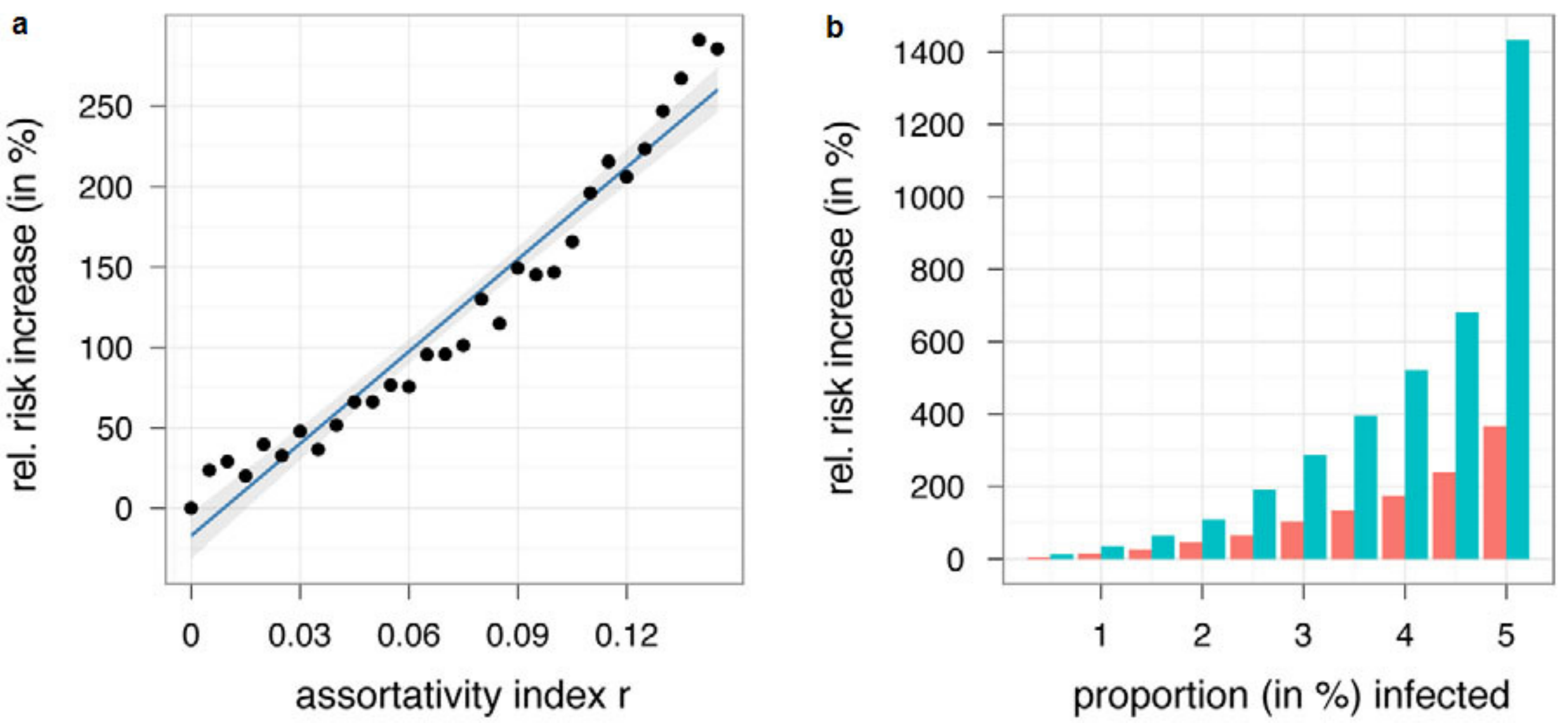}
\end{center}
\caption{(a) The impact of the assortativity index $r$ on the relative risk increase (compared to risk at $r\sim 0$) of disease outbreaks. Blue line shows the best fit of linear regression (confidence interval based on standard error). (b) Relative risk increase (compared to risk at $r\sim 0$) of disease outbreaks of a given fraction of the population for two values of the assortativity index $r=0.075$ (red) and $r=0.145$ (green). Source: Reprinted figure from Ref.~\cite{salathe2011assessing}.}
\label{cluster1}
\end{figure}

Relative to the above primarily qualitative endeavors, Salath{\'e} \textit{et al}. also defined a quantitative framework for the mixing patterns of vaccination behavior and opinions related to vaccinating~\cite{salathe2011assessing}. Based on publicly available data from Twitter, they firstly constructed a directed social network of information flow, where sentiment towards new influenza A (H1N1)-like vaccination was classified into four types: positive, negative, neutral, and irrelevant. Then, building on the Pearson coefficient~\cite{Boccaletti2006, newman2002assortative, wang2014emergence}, the assortativity coefficient $r$ is given as (relative to Eq.~\ref{degree-correlation-eq}, the following formula is actually another expression of correlation coefficient, and their detailed comparison has been proved in~\cite{qu2015effects})
\begin{equation}
r=\frac{\sum_i e_{ii}-\sum_i a_i b_i}{1-\sum_i a_i b_i},
\label{correlation}
\end{equation}
where
\begin{equation}
a_i=\sum_j e_{ij}, \quad b_j=\sum_i e_{ij}
\label{correlation2}
\end{equation}
and $e_{ij}$ is the fraction of edges connecting nodes of types $i$ and $j$. Moreover, here $r=0$ means random mixing of opinions, while positive (negative) $r$ values introduce assortative (disassortative) patterns, where nodes are more likely to connect with nodes of the same (different) type. Interestingly, the research found that vaccination is positively assortative with the sentiment score, which means that agents expressing positive (negative) sentiment are more often connected and take (reject) vaccination. In this sense, opinion distributions in different communities (each of which usually shares common opinion) profoundly affect the ultimate size of an epidemic. Results presented in Fig.~\ref{cluster1} show that the likelihood of an outbreak is enlarged, if unprotected agents are positively assorted and form susceptible clusters (due to the clusters of negative sentiment). For further details on the role of vaccination sentiment in realistic networks, we also refer to Sections~\ref{sec:survey_data} and~\ref{section:digital_epidemiology} for more details.

The aforementioned studies have proved that positively assortative non-vaccinated individuals in contact networks can lead to an enlarged likelihood of large disease outbreaks~\cite{mbah2012impact, salathe2008effect, salathe2011assessing}. To extend the universality of this finding, Barclay \textit{et al}. investigated whether such a pattern existed in influenza vaccination of contact networks among high schoolers~\cite{barclay2014positive}.  Contact network were found to be positively assortative (measured by Eq.~\ref{correlation}) with respect to influenza vaccination in that vaccinated (non-vaccinated) individuals are more inclined to link with their like. Although the overall coverage level may be temporarily high, clusters of non-vaccinated individuals will destroy the herd immunity and may take over the largest component of the network in the long term. Relative to the states of randomly assigned vaccination, positive assortativity thus increases the likelihood of large disease outbreaks. Of particular importance, non-vaccinated males contribute more than non-vaccinated females towards positive mixing patterns. Another interesting point is that the effect of non-vaccination clusters (or positive assortativity) becomes more visible if contact durations are longer or combined.

\subsubsection{Belief-decision models and dual-perspective systems}

\begin{figure}[t]
\centering \includegraphics[width=150mm,height=110mm] {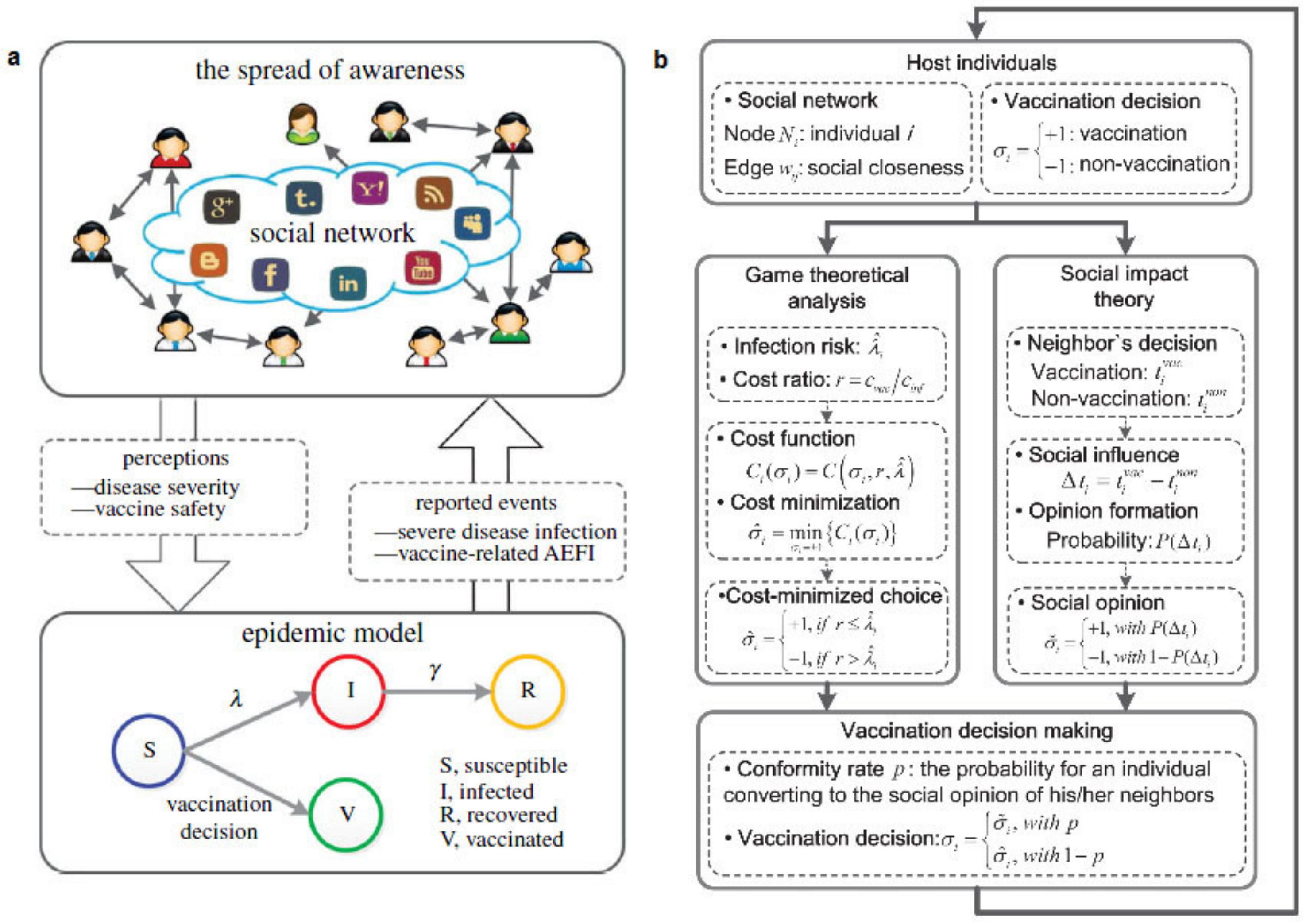}
\caption{(a) Schematic illustration of a belief-decision model for a vaccination campaign. A social network is used to describe the structure of interactions among individuals. There exists a feedback between the awareness and the vaccination choices of individuals.  Namely, vaccination decisions depend on the perceptions about disease severity and vaccine safety. (b) Schematic illustration of a dual-perspective view on modeling vaccination decision making. Here, vaccination decisions (i.e., to vaccinate $+1$, or not $-1$) are in accordance with either cost minimization or the impact of social influence. The conformity rate $p$ is the probability that an individual finally converts to the formalized social opinion, or otherwise follows a cost-minimization choice. Source: Reprinted figure from Refs.~\cite{xia2014belief, xia2013computational}.}
\label{illustrate-belief}
\end{figure}

As surveyed in \ref{clusters}, except for payoff-driven vaccination decision-making (i.e., the game-theoretical approach)~\cite{fu2011ZhenPRSB, zhang2010ZhenNJP, zhang2013impacts}, vaccination choices can also be driven by the awareness concerning the disease, and the perception one has about the vaccine and its safety. Actually, awareness has been widely identified as a significant factor in other disease prevention measures such as social distancing~\cite{funk2009ZhenPNAS, wu2012impact, ruan2012epidemic, valdez2012intermittent,
zhang2014suppression}. Motivated by this fact, Xia \textit{et al}. designed a belief-decision framework (see Fig.~\ref{illustrate-belief}(a) for a schematic illustration), where three possible vaccination decisions, i.e., to accept, to reject, and yet to decide (or no decision), are assorted to a set of belief values~\cite{xia2014belief}. The belief function for vaccination is given as follows
\begin{equation}
\left\{ \begin{array}{ll}
m(\textrm{Yes}) \in [0,1], \\
m(\textrm{No}) \in [0,1], \\
m(\Theta)=1-m(\textrm{Yes})- m(\textrm{No}),
\end{array} \right.
\end{equation}
where $m(\textrm{Yes})$ ($m(\textrm{No})$) means the probability of taking (rejecting) vaccination immediately, while  $m(\Theta)$ represents no firm decision and postponing the decision to the next time step. Owning to the spread of awareness of disease severity and vaccine safety, individuals update their belief values based on the extended Dempster-Shafer theory (a theory of belief functions or a mathematical theory of evidence, it is a general framework for reasoning with uncertainty, with understood connections to other frameworks such as probability, possibility and imprecise probability theories. In details, such a theory allows one to combine evidence from different sources and arrive at a degree of belief that takes into account all the available evidence)~\cite{shafer1976mathematical}, which incorporates information of connected neighbors. Interestingly, the authors found that the reporting rates of disease- and vaccine-related events mainly affect the vaccination coverage. While another key factor, the awareness fading coefficient, decides the time of individuals' vaccine administration.

After the above finding, another question naturally poses itself, namely, if both the game-theoretical approach and opinion dynamics are put together in a unified framework, how does this dual-perspective system influence the vaccination uptake and potential epidemic outbreaks? A computational model to evaluate the impact of vaccination associated cost and social opinions on vaccination decision-making (i.e., the interplay between payoff optimization and social influence) has therefore been proposed and studied by Xia \textit{et al.}~\cite{xia2013computational}. In more detail, the cost minimization scenario assumed that individuals (with mixed strategy) take up vaccination if the relative cost $c$ of the vaccine is larger than the perceived infection risk $\lambda_{perc}$ (i.e. $c>\lambda_{perc}$), where
\begin{equation}
\lambda_{perc}=\beta_I .(\frac{n_{non}}{ n_{vac}+ n_{non}})
\end{equation}
and $n_{vac}$ ($n_{non}$) denotes the number of neighbors being vaccinated (not vaccinated). In the opposite case, individuals give up vaccination. Moreover, according to the social impact theory~\cite{nowak1990ZhenPsR, lyst2002social}, to either vaccinate or not can be regarded as two types of opinions, which are acceptance and rejection, respectively. Individuals then update their opinions in terms of the Fermi-like function, where a normalized discrepancy between viewpoints replaces the previously used payoff difference~\cite{fu2011ZhenPRSB}. Finally, the conformity rate $p$ is introduced to control the tendency of individuals towards social opinions (see Fig.~\ref{illustrate-belief}(b) for an illustration). When the conformity rate is relatively small (i.e., decision are mainly cost-based), the effectiveness of vaccination declines with larger costs, which is as reported before in the realm of a purely game-theoretical model~\cite{fu2011ZhenPRSB}. However, when individuals follow social opinion (i.e., $p=1$), the resulting vaccination converges to a certain level, depending on individuals' initial level of willingness rather than the associated cost. For intermediate conformity rates (i.e., the joint impact of cost-minimization and opinion formation), the ultimate vaccination coverage and epidemic incidence are completely determined by the competition between vaccine cost and initial vaccination willingness. Thus, in addition to the determinants associated with the costs and benefits of vaccination, a voluntary vaccination program should carefully consider also the social aspects and opinion formation in contact networks.

\subsubsection{Competing controlling measures (going beyond traditional binary selection)}

As mentioned in section~\ref{counter-intuitive}, apart from vaccination, there might also exist non-pharmaceutical interventions that may help mitigate the risks and seem of particular importance before effective vaccines are widely available~\cite{del2005effects, kelso2009simulation, reluga2010game, funk2009ZhenPNAS, gross2006epidemic, zanette2008infection, glass2006targeted}. However, the majority of existing research considers either vaccinating behavior or non-pharmaceutical interventions separately. The effectiveness of the combination of both types of interventions on social or empirical networks may thus, at least to some extent, be beyond our knowledge (although both are used in some disease such as influenza). Aiming for a better understanding of this issue, Andrews \textit{et al}. recently proposed a model of competing interventions, which assesses the decision on the basis of subjective expected utility theory~\cite{andrews2015disease}. It is worth mentioning that, unlike by vaccination, susceptible/infectious individuals will implement non-pharmaceutical interventions on different neighbors.

Of note, there exists a tradeoff between two measures. If the vaccination coverage increases, the decay of infections and the resulting changes in transmission patterns enable a decrease in the practices of non-pharmaceutical interventions. Similarly, if non-pharmaceutical interventions become more common, vaccination coverage will decrease by roughly a similar extent. Even if changing the utility or the perceived risk, it seems impossible to simultaneously enhance the effectiveness of both measures. Moreover, it was also revealed that efforts to reduce spreading by expanding non-pharmaceutical interventions are almost completely mitigated by the resulting drop in vaccine coverage, whereas efforts to expand vaccine uptake are only weakly affected by the response of non-pharmaceutical interventions. Thus, (lack of) vaccination interferes more strongly than non-pharmaceutical interventions as the primary determinant of influenza incidence. Along this line, when considering multiple intervention programs, officials should pay particular attention to the potential interference among them, in particular looking out for (dis)assortativity between vaccination and non-pharmaceutical interventions in real-life situations. It is important to communicate to the public that these are complementary rather than exclusive measures to mitigate an outbreak.

\subsection{Behavior-vaccination dynamics on temporal networks}

In the above subsections, we have reviewed the remarkable advances in understanding vaccination and the spread of epidemics in structured populations, which were described by different complex, yet static networks. While many empirical observations could be explained in this theoretical framework~\cite{fu2011ZhenPRSB, zhang2010ZhenNJP, perisic2009social, salathe2011assessing}, it is nevertheless a fact that most social interactions among people change over time. Accordingly, a more apt description involves using time-varying or temporal networks~\cite{karsai2014time, holme2012temporal, gross2006epidemic, gross2008adaptive,liu2014controlling,starnini2013immunization}. Such adjustments may of course be unrelated to vaccination and epidemics, or they can happen precisely because of this. Either way, changes in the interaction structure of social networks affect how we vaccinate and how effective our measures are aimed at mitigating epidemics. In this subsection, we therefore cover an important aspect of behavior-vaccination dynamics in structured, networked populations, namely that on temporal networks. To that effect, we consider two representative examples.

\subsubsection{Outcome inelasticity and outcome variability on adaptive networks}
Early on in epidemiology, it has been pointed out that adaptations of various kinds can lead to a series of interesting phenomena, such as degree mixing patterns, oscillations, hysteresis, and phase transition between disease-free and endemic states~\cite{gross2006epidemic}. Along this line one might wonder, if a vaccination program is placed onto an adaptive network, how do the epidemic incidence and vaccination uptake co-evolve with time? To address this question, Morsky \textit{et al}. have recently proposed and studied an edge turnover effect on networks~\cite{morsky2012outcome}. Naturally, if an individual in a network is vaccinated (not vaccinated) and dies from adverse side effects (an infection), any links from and to it are removed from the network (technically, since the links are undirected, simply an edge is removed). To maintain the balance in the system, i.e, the average number of edges in the network, the same number of edges are re-introduced randomly among pairs of individuals, whereby not allowing multiple connections. Interestingly, such a relatively simple and straightforward setup gives rise to three vaccination regimes. These are successful ring vaccination, widespread vaccination, and no vaccination, all owing to a different frequency of edge turnovers that are related to the death of individuals in the network. Moreover, it was shown that stochastic effects can lead to a large variability in outcomes. For the same parameter values, some realizations guarantee fast control, while others can not prevent a fast-spreading epidemics. This indicates that the outcome of voluntary vaccination on adaptive networks may be highly unpredictable, precisely because the interaction structure changes over time. Thus, many arguments put forward above for static networks simply do not hold. There exist inherently nonlinear feedback mechanisms between behavior and disease dynamics, which can result in policy resistance or outcome inelasticity, neither of which would be observed on static networks in the traditional models that did not take strategic interactions or behavioral responses to disease dynamics into account~\cite{pastor2015epidemic, Boccaletti2006}. This gives rise to extreme events that we are only beginning to uncover and study, and which certainly shed new light into just how complex successful vaccination policies might turn out to be.

\subsubsection{The impact of closeness driven time-varying topology}
Along similar lines, in~\cite{perra2012activity} authors proposed a closeness-driven time-varying framework. In an instantaneous network, every individual connects to an acquaintance with probability $p$, or with probability $1-p$ to a random node~\cite{han2016evolutionary}. After an instantaneous network is formed in this way, the vaccination game proceeds with an imitation rule at the heart of the microscopic dynamics~\cite{fu2011ZhenPRSB}. These two steps, i.e., the creation of the instantaneous network and the evolutionary vaccination game proceed repeatedly until a steady state is reached. In the absence of vaccination there is monotonous increase in the size of the epidemics. Conversely, if vaccination is introduced, the epidemic incidence reaches a peak and then drops due to the enhanced crisis awareness. More importantly, the increase of $p$ greatly weakens the vaccination enthusiasm of individuals, since the links among acquaintances provide convenient conditions for herd immunity. These results indicate that closeness likely plays an important role in decreasing both the epidemic propagation and vaccination.

Although existing research already conveys clearly that temporal networks bring about novel phenomena in the mitigation of epidemics and the promotion of vaccination, there is still ample room for further research, for example related to the importance of different time scales that describe changes in the network, in the epidemic transmission, and in the vaccination uptake, where differences can markedly determine the fate of the population~\cite{paulsen2008detection, guerra2012adaptive}. Further research on behavior-vaccination dynamics on temporal networks also requires better integration with empirical data, which we cover in more detail in Section~\ref{section:digital_epidemiology}~\cite{brockmann2013hidden,balcan2009multiscale}.

\subsection{Behavior-vaccination dynamics on multilayer networks}

In addition to static and temporal networks, the most recent frontier in network science are the so-called interdependent or multilayer networks~\cite{kivela2014multilayer, poggio1990regularization, boccaletti2014structure, hornik1989multilayer}. The main premise behind multilayer networks is that a single layer of interactions is likely insufficiently accurate to describe the complexity of our social interactions. Arguably, each individual is likely a member in many social networks, including online social networks, and all these interactions shape markedly our perception of vaccination in the broadest possible sense. Although one may argue that all these interactions could be placed under the umbrella of a single network, the multilayer network perspective nevertheless creates better conditions for properly studying such social systems.

Not surprisingly, multilayer networks have recently been the subject of many novel frameworks to study various dynamical processes, such as epidemic spreading~\cite{buono2015immunization, salehi2015spreading, wang2012epidemics,sun2016pattern}, diffusion~\cite{gomez2013diffusion}, voting~\cite{halu2013connect}, synchronization~\cite{zhang2015explosive}, and the evolution of cooperation in game-theoretical models~\cite{wang2015evolutionary,wang2014degree} (see~\cite{kivela2014multilayer,boccaletti2014structure} for systematic reviews). In this subsection, we therefore cover the still very young field of behavior-vaccination dynamics on multilayer networks.

\subsubsection{The general framework and an example}

\begin{figure}
\begin{center}
\includegraphics[width=0.55\textwidth]{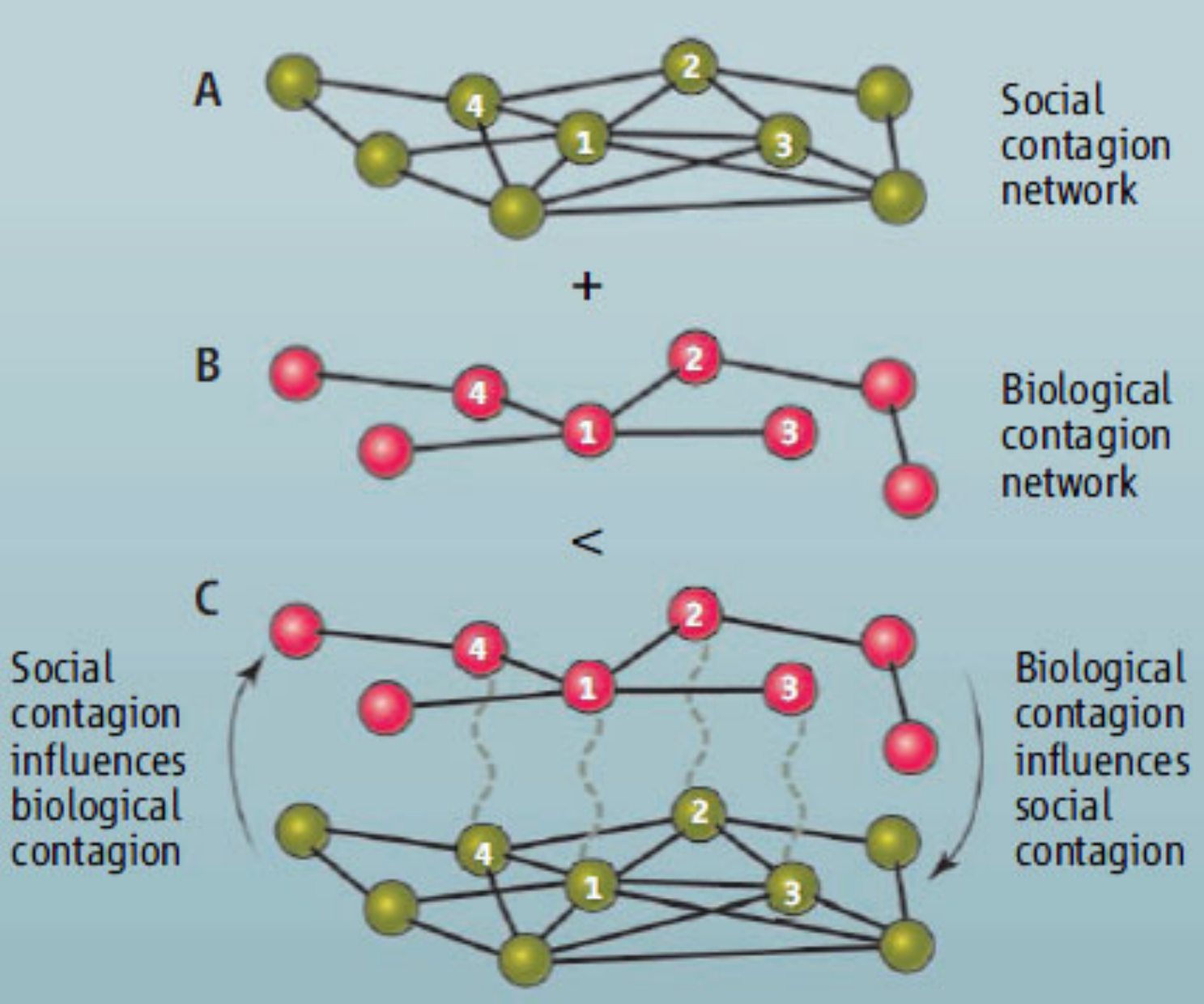}
\end{center}
\caption{Schematic illustration of behavior-vaccination dynamics on multilayer networks, which consist of social and biological contagion networks. The former layer supports the propagation of information related to the disease, such as opinions and sentiments, while the latter layer allows the actual spreading of the virus. More importantly, the resulting interplay on the new architecture leads to complex dynamics that is not present in either network layer in isolation, such as the rise and fall of vaccine coverage during a scare scenario. Source: From Ref.~\cite{bauch2013ZhenScience}. Reprinted with permission from AAAS.}
\label{multi-illu-be}
\end{figure}

With respect to coupled disease-behavior dynamics on multilayer networks, one recent work firstly provided a brief, conceptual framework, where the system is composed of two networks: the social contagion network~\cite{wang2016dynamics,wang2015dynamics} and the biological contagion network~\cite{bauch2013ZhenScience}. The former is assumed to support the spreading of opinions and sentiments related with vaccination, while the latter allows the propagation of the virus through local intra-layer connections. When both networks are coupled together, a feedback loop between layers spontaneously emerges, which generates more complex dynamics beyond either isolated network layer (see Fig.~\ref{multi-illu-be} for a schematic illustration). For instance, the vaccination opinions on the social contagion network may play a positive or a negative role in the epidemic trajectory on the biological contagion network and vice versa.

Along this main idea, Fukuda \textit{et al}. recently proposed a multilayer infrastructure, which is composed of a disease transmission network and an information propagation network through which individuals update their vaccination strategies and health information~\cite{fukuda2015influence}. Compared with  the evolutionary outcomes in the traditional single-layer network~\cite{fu2011ZhenPRSB}, it has been discovered that multilayer infrastructures typically suppress the vaccination coverage, resulting in relatively larger and more common epidemics. Although middle-degree and large-degree nodes show higher vaccination enthusiasm, numerous small-degree nodes make the propagation of vaccination very difficult, which is further exacerbated by the propagation of negative opinions in the other layer. Overall, successful vaccination policies in multilayer networks require even more care and consideration to account for multiple channels along which potentially discouraging information about vaccinating may spread, as we review also in the next subsection.

\subsubsection{Parallel, interacting dynamics on different networks -- The impact of opinion clusters}

\begin{figure}
\begin{center}
\includegraphics[width=1.0\textwidth]{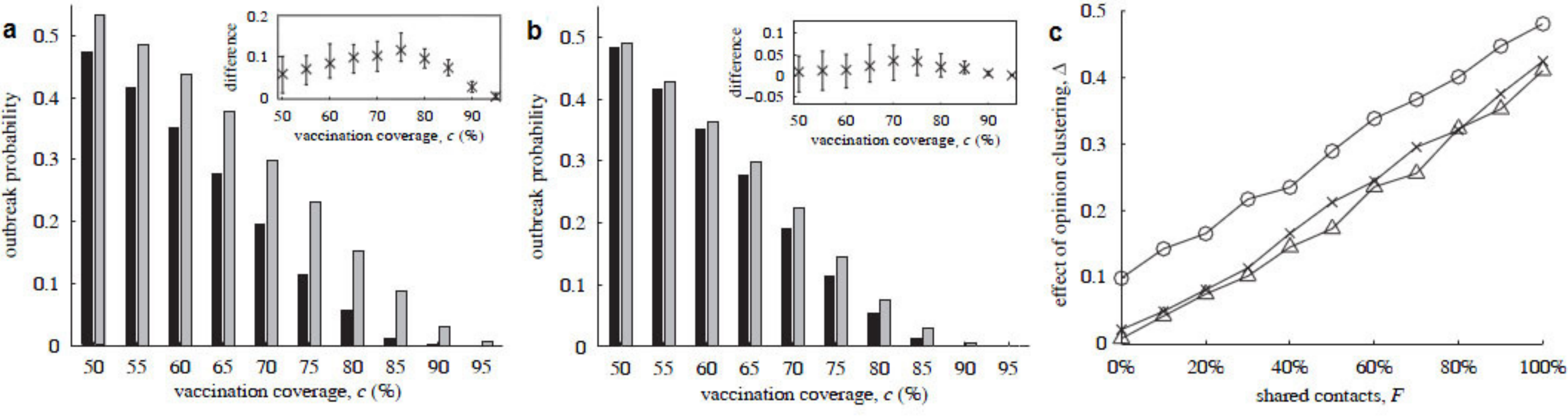}
\end{center}
\caption{Outbreak probability with (grey) and without (black) opinion formation in dependence on the vaccination coverage. The insets show the difference of outbreak probability between grey and black vertical lines. The parent and offspring networks share $F=1$ (a) and $F=0.2$ (b) of the intra-layer connections. (c) The impact of shared connections $F$ on the effect of opinion clustering $\Delta$. Circles, crosses and triangles represent the probability of forming a intra-layer connection equal to 0.1, 0.2 and unlimited, respectively. Source: Reprinted figure from Ref.~\cite{eames2009networks}.}
\label{multi-opinion}
\end{figure}

As in the case of~\cite{mbah2012impact, salathe2008effect, salathe2011assessing, barclay2014positive}, when opinions of individuals towards vaccination are influenced by their neighbors or social groups on networks, the susceptible/unvaccinated clusters can readily arise, which directly affects the likelihood of disease outbreak. Even with very high vaccination coverage, outbreaks can still occur via these clusters~\cite{mbah2012impact, salathe2008effect}. However, empirical evidence suggests that the transmission of infections is sometimes separated from decision-making. One typical example for this reasoning are childhood diseases, for which the primary risk group are young children, while the decisions about vaccination are of course taken by their parents~\cite{may2003clustering, leask2006maintains, abbasi2008mmr, heathcock2008measles}. Motivated by this fact, already in 2009 Eames used a multilayer network setup, where one network supports the opinion formation among parents (i.e. parent network), while the network of children allows infection transmission among offspring (i.e. offspring  network)~\cite{eames2009networks}. When parents and children are distributed on each network layer, offspring share their parents' locations (denoted by the inter-layer links) and intra-layer connections, whose fraction is controlled by the parameter $F$. $F=1$ means that each intra-layer link in the offspring network also appears in the parent network, i.e. both network layers are identical; while $F=0$ indicates that parent and offspring networks have no shared intra-layer connection. Using an identical opinion updating dynamics as in~\cite{salathe2008effect}, it was reported that even if the vaccination level is extremely high, the opinion formation process greatly enhances the outbreak likelihood, relative to the case without opinion dynamics (see Fig.~\ref{multi-opinion}(a)). Because the opinion of not to vaccinate is inclined to clustering among parents, this naturally gives rise to susceptible offspring clusters and thus creates significant infection risks. Interestingly, the above observation still remains intact, though less dramatic, when the overlap fraction is low, as shown in Fig.~\ref{multi-opinion}(b).

To quantify the difference between outbreak probability with and without opinion formation, in \cite{eames2009networks} the author further defined the effect of opinion clustering
\begin{equation}
\vartriangle =\frac{\sum_{\textrm{vaccination levels}, c}\{\textrm{Prob (Outbreak with opinion) - Prob. (Outbreak  without opinion)}\}}{ \sum_{\textrm{vaccination levels}, c}\textrm{Prob. (Outbreak without opinion)}}.
\end{equation}
As shown in Fig.~\ref{multi-opinion}(c), the effect of social influence monotonously increases with the amount of overlap between the two networks. When the two networks share many contacts, clusters of parents who choose not to vaccine are more likely to have interacting children, generating clusters of unvaccinated offspring and thus ultimately enhancing the risk of an infection, which is then likely followed by an outbreak. In this sense, non-overlapping intra-layer connections between the networks of parents and their offspring seem to be an important factor for determining the desired vaccination coverage for a successful mitigation of disease.

While the above-reviewed research already delivered inspiring results, just as with temporal networks, the study of behavior-vaccination dynamics on multilayer networks is still very much a developing field with vast possibilities for innovative future research. The evolution of links between different network layers, as well as their weight in determining the course of vaccination  in comparison to within-layer links are still very much under-explored subjects, just as potential extension of the basic theoretical framework to more than three layers.

So far, we have surveyed most achievements about behavior-vaccination dynamics in networked populations, which, as a whole, provides novel insight into real-world disease prevention under vaccination policy. To obtain a clearer understanding, Table~\ref{sum-behavior-vaccination} provides a summary of these new characteristics. We also notice because temporal networks and multilayer networks are two novel frameworks in network science, more efforts from statistical physics and mathematics are still needed to enrich the content of behavior-vaccination dynamics in future. Moreover, another fantastic point is that, except for infectious diseases, complex contagion~\cite{centola2010spread} and corresponding behavioral vaccination strategy~\cite{campbell2013complex} deserve great attention as well, especialy combining with lots of emerging data in this realm.

\begin{sidewaystable}
\newcommand{\tabincell}[2]{\begin{tabular}{@{}#1@{}}#2\end{tabular}}
\centering
\caption{Summary of behavior-vaccination in network populations.} \label{sum-behavior-vaccination}
\begin{tabular}{|m{3.5cm}|m{3.5cm}|m{7cm}|m{6cm}|}
\hline
Used networks & Employed methods & Novel characteristics \& effects of behavior-vaccination & Observed physics phenomena\\\hline
\tabincell{l}{Lattice \\ ER graphs \\ SW networks\\ SF networks} & \tabincell{l}{Monte Carlo simulation\\ mean-field theory\\ stochastic processes} & \textcircled{1}updating dynamics and network topology profoundly affect vaccination level and epidemic incidence; \textcircled{2}external incentives have the dual role in vaccination uptake; \textcircled{3}there may exist tradeoff in the competition of multiple prevention strategies. & \textcircled{1}second-order phase transition of vaccination/infection; \textcircled{2}self-organization and percolating of non-/vaccination clusters; \textcircled{3}positive feedback between vaccination and epidemic \\\hline
\tabincell{l}{social contact networks \\ urban networks \\ technological networks} &  \tabincell{l}{Monte Carlo simulation\\ data analysis \\ network creation \\ social influence theory}  & \textcircled{1}efficiency of behavior-vaccination is closely related to contact patterns, vaccine/disease risk and external incentives; \textcircled{2}the emergence of non-vaccination clusters could lead to outbreaks in spite of high vaccination level; \textcircled{3}opinion/awareness plays a significant role in vaccination campaign and disease prevention.  & \textcircled{1}'first-order' phase transition of vaccination/infection; \textcircled{2}assortative pattern of strategies/opinions; \textcircled{3}positive feedback between opinion/awareness and vaccination\\\hline
\tabincell{l}{temporal networks \\adaptive networks }& \tabincell{l}{Monte Carlo simulation \\  network creation} & outcome of voluntary vaccination becomes highly unpredictable & nonlinear feedback between vaccination and disease\\\hline
multilayer networks & \tabincell{l}{Monte Carlo simulation \\ social influence theory} & opinion/awareness on one layer may impede/accelerate disease spreading on other layers & positive/negative feedback among dynamics of different layers\\
\hline
\end{tabular}
\end{sidewaystable}

\section{Measuring and assessing vaccination levels: traditional approaches}
\label{sec:survey_data}

Due to the overwhelming public health benefits of vaccination, governments and public health institutions around the globe have invested substantial efforts to immunize their populations against vaccine-preventable diseases. The population dynamics of vaccine-preventable diseases depend strongly on the magnitude and distribution as well as the efficacy of vaccines in the population, but measuring these indicators is not always easy. First, measuring the vaccination coverage - the percentage of the population that has been vaccinated - is subject to numerous difficulties and biases. Second, even when vaccination coverages can be measured accurately, measuring the distribution of vaccination in the population - i.e. which parts of the population have been vaccinated - remains challenging. Third, the effect of vaccination on population immunity will ultimately depend crucially on the effectiveness of the vaccine, a hard-to-measure but crucial indicator for many public health policies. In this section, we review the challenges associated with assessing the vaccination status of a population. These challenges are not only relevant from a data perspective, but they are also crucial to inform modeling efforts.

\subsection{Measuring vaccination coverage}
In an ideal world, vaccination coverage would be easy to assess. Every individual in a population who receives a vaccine would be centrally registered, allowing for an always accurate calculation of the percentage of the population vaccinated. In practice, numerous problems stand in the way. On the numerator side, the number of people vaccinated is rendered inaccurate by the non-continuity of registration systems, by migration flows, and record duplications, as well as other challenges. On the denominator side, estimates can be equally inaccurate for a number of reasons. The various challenges are particularly difficult in low- and middle-income countries, where registry systems are often manual, patchy, or entirely absent.

\begin{figure}
\begin{center}
\includegraphics[width=1.0\textwidth]{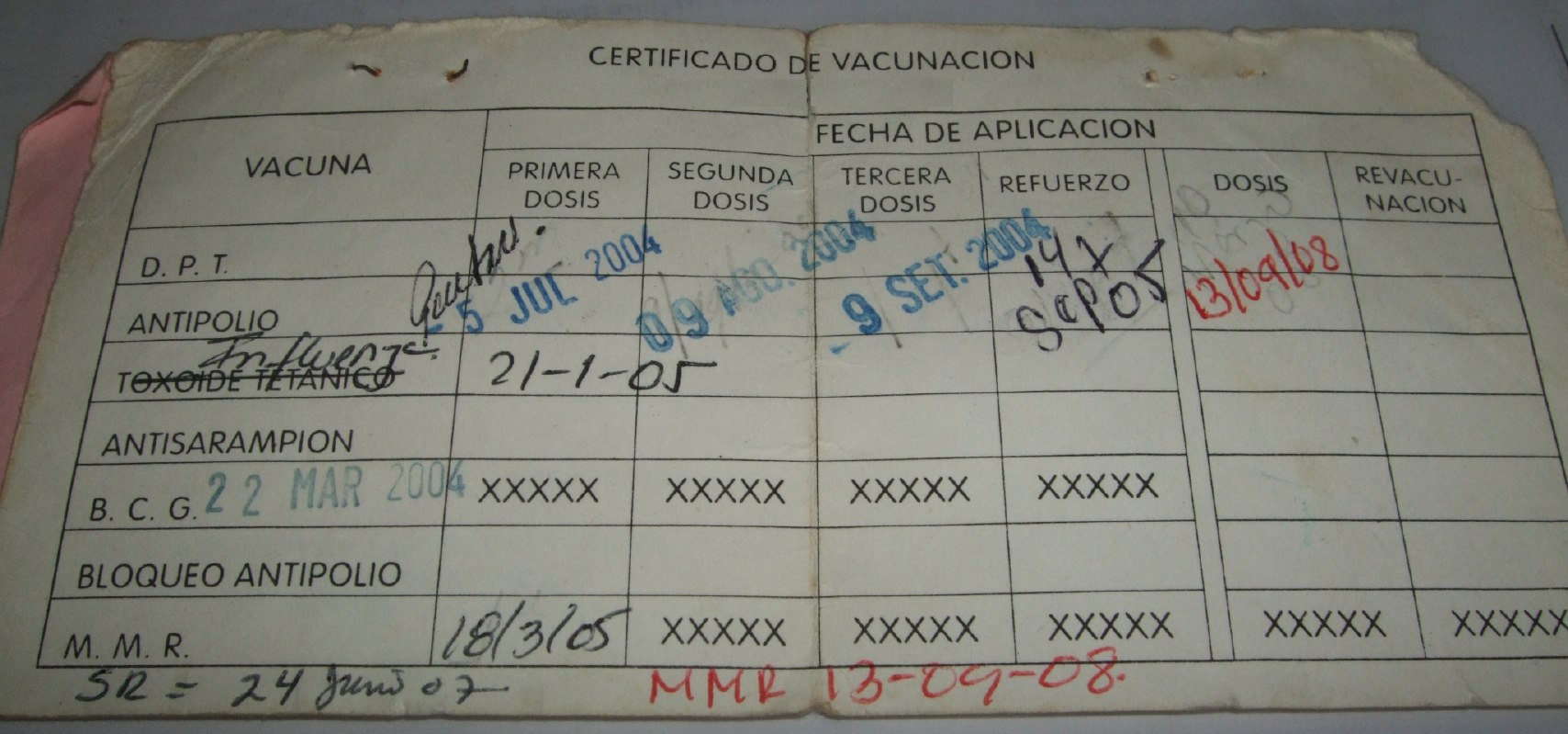}
\end{center}
\caption{A vaccination paper card, showing the various improvisations made over time. (Photo courtesy of Carolina Danovaro, Pan American Health Organization.) Source: Reprinted figure from Ref.~\cite{cutts2013measuring}.}
\label{papercard}
\end{figure}

The potential inaccuracy in estimates at the administrative levels means that measures from alternative data sources need to complement the numbers. Surveys in particular are a well-established source for alternative vaccination coverage estimates. Globally, there are four types of surveys that are commonly employed: Demographic and Health Surveys (DHS), Multiple Indicator Cluster Surveys (MICS), the Expanded Programme on Immunization cluster survey (EPI), and surveys using the Lot Quality Assurance Sampling (LQAS) method. These common surveys have different characteristics with respect to the primary objectives, the sampling scheme, household selection, total sample size, questionnaire length, duration, and so on ~\cite{cutts2013measuring}. They all suffer from a number of biases, of which two are typically highlighted: i) a selection bias due to sampling methods that are non-representative or reflective of poor field practices, and ii) an information bias generally - but not exclusively - due to inaccurate data recording manual records involving a paper card retained in the household (Fig.~\ref{papercard}). Many of these problems are expected to worsen in the future because of the increased complexity of immunization schedules.

In recent times, biomarker surveys have been added to the methods to measure vaccination coverage. While the presence of antibodies is a good marker of immunity, it's typically impossible to distinguish whether their presence is due to vaccination or to natural infection. Furthermore, while natural infection often causes life-long immunity, most current vaccines require booster shots because of the waning levels of antibodies following vaccination~\cite{slifka2014advances}. On the other hand, the case is reversed for some pathogens such as tetanus, where natural infection causes a short-lived immunity while vaccination can provide life-long immunity~\cite{wagner2012immunity}. Some vaccines, such as the Hepatitis B vaccine, cause the presence of antibodies that are not created following natural infection~\cite{cutts2004vaccines}, providing a clear measurement of vaccination, but this is the exception rather than the rule. And because of waning antibody levels, the number of doses can still not be reliably estimated~\cite{tapia2006measurement}. To complicate matters further, re-exposure to pathogens can trigger an increase of antibody levels ("immune boosting"), often without developing symptoms, as appears to be the case in whooping cough~\cite{pebody2005seroepidemiology,saemann2001stability,cattaneo1996seroepidemiology,teunis2002kinetics}. Indeed, it has been suggested that immune boosting must be more easily triggered than primary infection to account for age-incidence data found in the recent increase in whooping cough cases in teenagers and adults~\cite{lavine2011natural}.

\subsection{Vaccine efficacy and effectiveness}

Many studies looking at the effects of vaccination on disease dynamics at the population level assume that vaccines have 100\% efficacy. In other words, they assume that in the vaccinated group, the disease incidence following vaccination is zero. This assumption is generally wrong. At its most basic level, vaccine efficacy can be defined as the opposite of the relative risk ($RR$) among vaccinated individuals~\cite{weinberg2010vaccine}. It corresponds to the relative reduction of disease attack rate in the unvaccinated group ($ARu$), compared to that of a vaccinated group ($ARv$), all else being equal:

\begin{equation}
\frac{ARu - ARv}{ARu} = 1 - RR.
\end{equation}

In order to comply with the "all else being equal" requirement, vaccine efficacy is usually assessed through double-blind, randomized, clinical controlled trials, representing best case scenarios. The vaccine efficacy so established is usually a requirement for a new vaccine to be licensed by the relevant governmental authorities.

Vaccine efficacies vary widely. In influenza, where a new vaccine is developed each year to prevent seasonal influenza, vaccine efficacy can vary dramatically depending on the match of the vaccine with the dominant circulating flu strains. A recent meta-analysis~\cite{osterholm2012efficacy} has shown that influenza vaccine efficacy has been measured to be between 8\% and 93\%, depending on the vaccine type, the year, and the study population. The pooled efficacy of the trivalent inactivated vaccine in adults of age 18-65 was 59\%. For the vaccine preventing Human Papillomavirus (HPV) infections, meta-analyses have shown a vaccine efficacy of 78\% - 90\%~\cite{la2007hpv}. The efficacy of an acellular pertussis vaccine among adolescents and adults was shown to be 92\%~\cite{ward2005efficacy}.

Vaccine effectiveness needs to be distinguished from vaccine efficacy. While vaccine effectiveness measures the same outcome of interest - relative reduction of disease attack rate - it does so in an observational way under field conditions, rather than in clinically controlled experiments. Vaccine effectiveness is therefore influenced by many factors, including, but not limited to, vaccine efficacy. These factors include vaccine characteristics as well as host factors such as age, previous vaccinations and infections, etc. It's easy to see that all the challenges relate to vaccine coverage surveillance outlined above apply also to studies of vaccine effectiveness. Because vaccine effectiveness studies do not require clinically controlled trials, there are more data available on vaccine effectiveness than on vaccine efficacy. Vaccine efficacy is often assessed during or after outbreaks. For example, recent data obtained from a large-scale outbreak of measles in a German public school was analyzed resulting in an estimate of the measles vaccine effectiveness of over 99\% (following two doses of vaccination)~\cite{van2010estimation}.

Vaccine effectiveness and vaccination coverage combine to create the dynamics of infectious diseases observed in populations. In all populations, there will be non-vaccinated individuals, and there will be individuals for which the vaccine does not prevent disease. During a disease outbreak, we can therefore expect both unvaccinated and vaccinated individuals to fall ill. Under the assumption of random mixing, the fraction of unvaccinated individuals among individuals with disease can be easily calculated. The susceptible population is the sum of the number of unvaccinated individuals $U$, and the vaccinated individuals who remain susceptible despite vaccination, $V$, i.e. $S = U+V$. If these two types of susceptible individuals get infected with the same probability, then the fraction of unvaccinated individuals among diseased individuals will be $U / (U+V)$. Expressed in terms of vaccination coverage $x$ and vaccine effectiveness $\tau$, individuals can be susceptible because they did not get vaccinated $(1 - x)$ or because they did get vaccinated but the vaccination did not render the individual immune $(x(1 - \tau))$. Thus, the total susceptible fraction of the population is ${1 - x + x(1 - \tau) = 1 - \tau x}$, and the fraction of unvaccinated individuals among diseased individuals is
\begin{equation}
\frac{x(1 - \tau)}{1 - \tau x}.
\end{equation}

This relationship can be leveraged for multiple purposes. First, it can be used to estimate vaccine effectiveness. If the proportion of cases occurring in vaccinated and unvaccinated as well as the vaccination coverage $x$ is known, vaccine effectiveness $\tau$ can be easily calculated~\cite{orenstein1985field}. Secondly, it can be used to estimate vaccine coverage, if the proportion of cases occurring in vaccinated and unvaccinated as well as the vaccination effectiveness $\tau$ is known. This can be particularly helpful in situations where the vaccine effectiveness is well known, such as in the case of measles~\cite{althaus2015measles}. An increase in the fraction of cases among unvaccinated individuals may seem counter-intuitive and worrisome, but indeed is most likely an indicator of increasing vaccination coverages. The opposite case is just as true, and particularly relevant in situations of increasing vaccination hesitancy.

Overall, the ability to measure, or at least accurately estimate vaccination and immunization levels in a population are of crucial importance to public health, and the modeling efforts. As outlined above, the practical challenges have historically been quite substantial, and continue to pose problems. At the same time, the advance and widespread adoption of technology, and in particular also consumer technology such as the internet and mobile computing devices, can help supplement these efforts. In the next chapter, we will discuss the approaches based on a new set of methods broadly encapsulated under the term digital epidemiology.

\section{Digital epidemiology}\label{section:digital_epidemiology}

As discussed in many of the previous sections, the ability to timely monitor  the unfolding of outbreaks, as well as the capacity to capture our contacts, movements and communications are keys to characterize the spreading of infectious diseases.  This observation is quite intuitive and clear; perhaps since the very first epidemiological studies. However, major challenges in the data collection process have impeded developments for many years. Let us think first about surveillance systems. In many countries, health departments have the mandate to collect epidemiological data. In the USA for example, the surveillance for influenza like illnesses (ILI) is conducted by a network of about $3,000$ healthcare providers~\cite{CDC}. As we noted in Section~\ref{sec:survey_data} traditional surveillance approaches are affected by many limitations~\cite{chunara2013we}. Data is typically affected by large delays and backlogs. The geographical granularity of the sample is heterogeneous across the country. Furthermore, selection biases might play a major role. Indeed, only a fraction of infected individuals visit a doctor. These cases are typically associated with complications or legal requirements (leave from work). Data collections challenges have been even more pronounced in the observation of human dynamics relevant for diseases spreading. Indeed, the characterization of the way we interact, move and communicate has been relegated to ad-hoc small-scale studies for many years. The inherent complexity of such processes strongly limited the attempts to study their properties in these settings~\cite{vespignani2009predicting,gonccalves2015social}.

The digital revolution helped, and is helping, to lift such limitations~\cite{lazer2009computational}.  Nowadays more than $3$ billion of people had access to the Internet.  About $7$ billion phone subscriptions have been activated, and more than $20\%$ of them are associated to smartphones. As result a large fraction of our activities is digital, often online. Communication using social networks, emails, blog posts, and other forms of digital media are now commonplace.  Countless online resources are now used to access information that researchers can efficiently mine using search engines. Furthermore, the miniaturization of devices created a wide range of wearable able to measure our interactions, movements, as well as vital signs~\cite{piwek2015rise}.

It did not take long before researchers saw the potential of these new technologies for epidemiology~\cite{salathe12-1,chunara2012new,hartley2013overview}. Indeed, the large amount of data we generate, or that we can collect, contains crucial epidemiological indicators. Just to mention few examples, think people speaking about their health on social networks, or searching for diseases' symptoms on search engines. Furthermore, think about the uses of GPS data to characterize human mobility at different geographical scales, of RFID (radio frequency identification devices) tags to identify the features of human face-to-face contacts, or of mobile apps to gather health information from users that sign-up. This unprecedented wealth of data is at the basis for the development, test, and refinement of many modeling approaches discussed in the previous sections.

In the following, we will review recent developments in this arena (also see Table~\ref{dig_epi_tab} for a systematic summary in advance). In doing so, we will consider three main approaches according to the data collection process. We will first consider participatory methodologies that rely on individuals voluntary contribution to the data collection. We will then present passive approaches that identify and extract epidemiological relevant indicators from the digital traces of our daily activities. We will then conclude presenting mixed approaches of the previous two.

\begin{sidewaystable}
\footnotesize
\newcommand{\tabincell}[2]{\begin{tabular}{@{}#1@{}}#2\end{tabular}}
\centering
\caption{Summary of approaches in digital epidemiology} \label{dig_epi_tab}
\begin{tabular}{|m{3cm}|m{3cm}|m{4cm}|m{4cm}|m{4cm}|}
\hline
Method    & Data collection process &Relevance for epidemiology  &  Main advantages    & Main disadvantages    \\\hline
Participatory digital epidemiology & Surveillance platforms and wearable devices & Estimation of disease indicators and characterization of face-to-face interactions & Near real-time estimations, access to different geographical granularities, detailed characterization of social interactions responsible for the spreading of human to human diseases & Sample size is typically small, reliance on users' participation \\\hline
Non participatory digital epidemiology & Social media and mobile phones              & Estimation of disease indicators, characterization of social interactions and human mobility & Large sample size, cheap data collection process, near real-time analysis, access to different geographical granularities, characterization of human mobility and its impact in disease spreading & Correlations might be spurious, the interactions observed might not be relevant for disease spreading, keyword selection not trivial, changes in platform design might influence users' behaviors \\\hline
Mixed methods  & Social media and users participation  & Estimation of disease indicators and discovery of new outbreaks & Near real-time estimations, data classification improved by users' participation                                                                                                                  & Performance tied to users participation \\\hline
\end{tabular}
\end{sidewaystable}

\subsection{Participatory digital epidemiology }
The widespread diffusion of Internet connections and the miniaturization of devises is providing unprecedented opportunities to collect epidemiological relevant information with the direct participation of large number of individuals.

\subsubsection{Participatory platforms for digital surveillance}

Mobile and web-based applications can be effectively used to monitor the emergence and unfolding of outbreaks in near real-time~\cite{wojcik2014public,kass2013social}. FluNearYou~\cite{flunearyou,smolinski2015flu}, and InfluenzaNet~\cite{flunet,paolotti14-1} are two examples. The first platform engages individuals living in USA or Canada. The second instead is based in Europe and includes $11$ different countries. The main idea behind both initiatives is to complement the classic ILI surveillance systems crowdsourcing the data collection. Indeed, users that sign-up become sentinels. Every week, during the influenza season, they receive a request to answers few questions covering geographical, behavioral and medical information. In other words, the two platforms are systems for digital surveillance. Remarkably, several studies have shown that the ILI incidence estimated using these platforms is highly correlated with the classic ILI surveillance data~\cite{smolinski2015flu,paolotti14-1,chunara2015estimating}. With minimal participation from the subscribed users they allow to overcome several limitations of the classic systems. The data collection is inexpensive, and scalable. The data can be visualized in near real-time, providing interactive maps describing the current unfolding of ILIs that are readily available for modeling efforts at the sub-population level as those described in Sections~\ref{sec:compart-model} and \ref{sec:beh-vac-mean}. Another crucial advantage of these approaches is the reduction of back-logs and delays in cases reporting. This issues is particular relevant during the initial phases of the flu season when its severity and general characteristics are mostly unknown. The delays and back-logs of the traditional surveillance introduce a posteriori change in the projections making forecasts a difficult task. In other words, the "ground truth" used to inform models is far to be stable and  reliable. Instead, data from digital surveillance can be used to inform realistic and geographically resolved epidemiological models with timely and high resolution intelligence that can be adopted to define precise initial conditions necessary for the simulations~\cite{zhang2015social}.

Another important aspect of the data collection is the ability to access behavioral information as for example the vaccination status of each user. On one side this allows geographically resolved real-time evaluation of vaccination rates complementing those gathered via traditional approaches (i.e. Section~\ref{sec:survey_data}) and might be used to refine the input for the modeling efforts described in Sections~\ref{sec:compart-model} and \ref{sec:beh-vac-mean}. On the other side, it allows to estimate the vaccination efficacy overcoming the issues of classic observational studies~\cite{edmunds2012using}. These are often cohort studies performed considering, retrospectively, health records of vaccinated and not vaccinated patients. Clearly, the same selection biases, black-log and delays we saw afflict clinical surveillance are a problem also in this case. Digital surveillance approaches, although not perfect, clearly mitigate these issues. For example, while selection biases can be still a problem, well crafted questionnaires can alleviate them. Again, the main advantage is the scalability of such methods, and the possibility to perform near real-time analysis.

When considering participatory platform for digital surveillance, it is important to stress that these application have been developed to complement,  not  substitute, classic surveillance systems. Indeed, the detailed characterization of circulating strains is crucial and can be achieved, at the moment, just in clinical settings. However, the penetration of mobile and connected devices in developing countries is offering unprecedented opportunities to gather richer and timely epidemiological data in cases where classic surveillance methods are typically limited.

\subsubsection{Measuring and understanding close proximity interactions}
The data and collection methods we just introduced, provide critical information to estimate diseases prevalence and vaccination rates. However, they do not provide any explicit information about the spreading pathways. From a modeling standpoint, such data can inform just mean-field modeling efforts (see Sections~\ref{sec:compart-model} and \ref{sec:beh-vac-mean}). Our close proximity contacts are responsible for the spreading of infectious diseases. As discussed in Section~\ref{concep-net} social networks are far from homogeneous and their non trivial properties have critical effects on the spreading of infectious diseases. For many years, empirical studies devoted to characterize their features have been centered on surveys and diaries~\cite{barrat2015face}. Potentially, these methods allow to gather information regarding the number, location (work, home, or other places), duration, and frequency of contacts as well as the health status of the respondent~\cite{danon2013social,van2013impact}. However, they are affected by several limitations as the scalability in terms of costs and individuals engagement~\cite{danon2013social}, the introduction of biases by the self-reporting of events/contacts~\cite{danon2013social,smieszek2012collecting,smieszek2014should,read2008dynamic}, and the introduction of biases by the survey design~\cite{danon2013social}.

The incredible advances in technology as the reduction of costs and sizes of devices allowed the development of a wide new spectrum of methods to measure close proximity contacts. In the following, we will focus on those that are participatory in nature: individuals are asked to voluntarily participate to a data collection process where their close interactions are measured via some device as smartphones or RFID tags~\cite{hui2005pocket,o2006instrumenting,eagle2006reality,pentland2008honest,raento2009smartphones,salathe2010high,hashemian2010flunet,cattuto2010dynamics,kiukkonen2010towards,liu2011accurate,olguin2011mobile,aharony2011social,hornbeck2012using,striegel2013lessons,stopczynski2014measuring}. According to the technology used, the data collection can span few hours, days, or months. Smartphones allow longitudinal data collections. However, they can record proximity interactions (via bluetooth or WiFi) that are not necessarily relevant for the spreading of infectious diseases~\cite{barrat2015face,eagle2006reality,pentland2008honest,raento2009smartphones,kiukkonen2010towards,olguin2011mobile,aharony2011social,stopczynski2014measuring}. Furthermore, as we will see later, smartphones allow to access/collect a rich set of data that can be used to better understand different aspects of human dynamics ranging from digital interactions to mobility patterns~\cite{stopczynski2014measuring}.  RFID tags instead can be tuned to detect actual face-to-face interactions. However, they require i) participants to wear specifically designed devices ii) the presence of stationary nodes deployed to collect information about the location of each interaction~\cite{kazandjieva2010experiences,barrat2015face}. For these reasons, the data collection with this technology is limited to short time periods and controlled settings.

In general, all these tools have major advantages respect to traditional methods. First, technology makes their deployment cheap, simple and scalable. Second, they provide an objective measurement of close proximity contacts. Few recent comparative studies have shed some light in the difference between contact networks estimated by traditional approaches (surveys and diaries) and those gathered via wearables~\cite{mastrandrea2015contact,smieszek2014should}. The findings show that traditional methods miss the large majority of short contacts, and overestimate the duration of all interactions. Furthermore, the engagement of individuals with traditional methods is reduced respect to those based on wearables. Notwithstanding these limitations, traditional methods reproduce with good accuracy strong ties that are likely to be responsible to sustain the local unfolding of infectious diseases. However, they provide an incomplete picture. Indeed, they miss weak ties that have been identified as crucial connections to bridge separated clusters of individuals, thus fundamental for large scale outbreaks~\cite{mastrandrea2015contact,onnela2007structure,jensen2015detecting}.

\begin{figure}
\begin{center}
\includegraphics[width=0.55\columnwidth,angle=0]{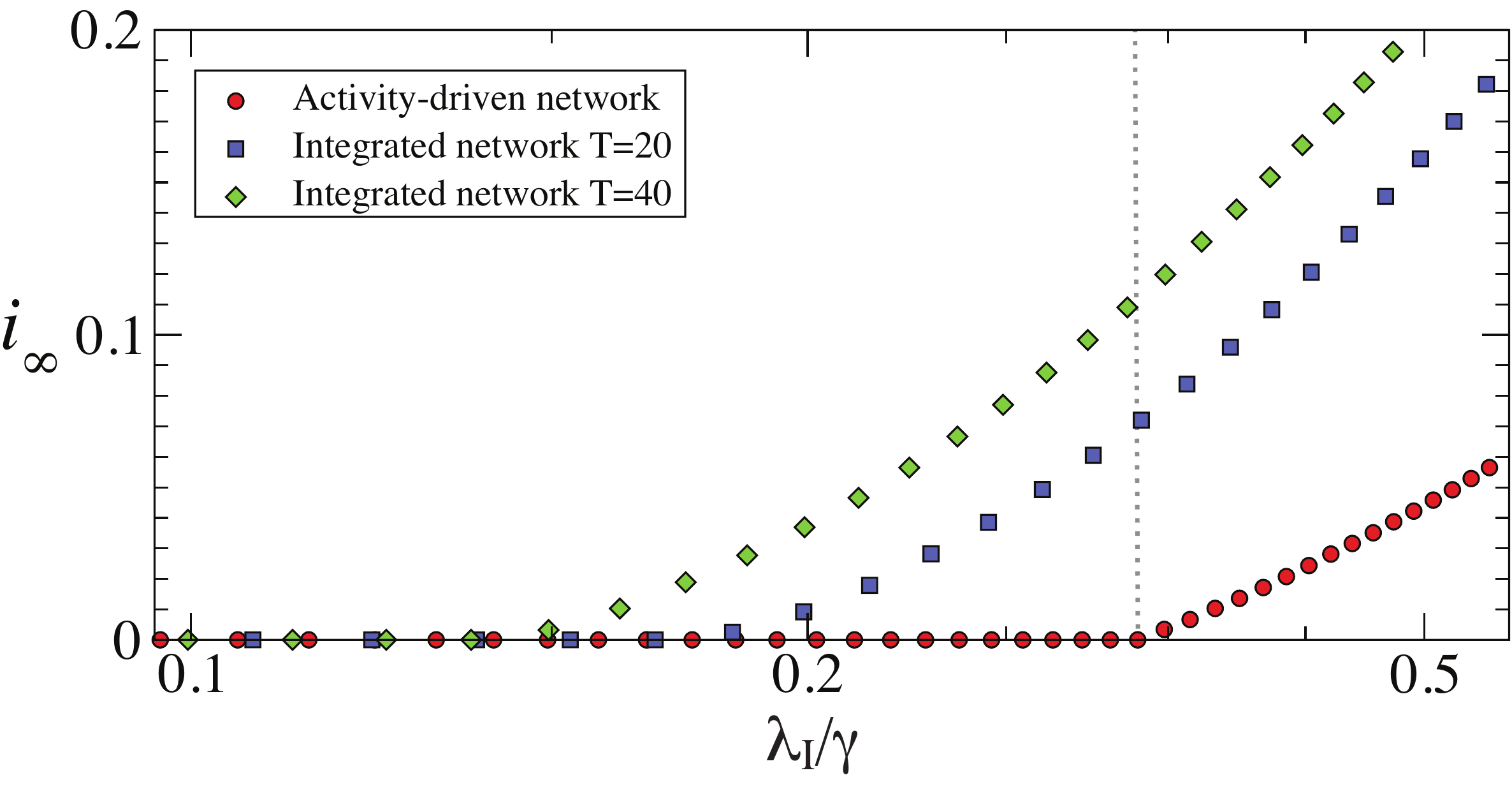}
\end{center}
\caption{Density of infected nodes, $i_\infty$, in the stationary state, obtained from numerical simulations of the SIS model on activity driven networks and two others resulting from an integration of the model over $20$ and $40$ time steps, respectively. $N=10^6$, $m=5$, $F(a) \propto a^{-\zeta}$ with $\zeta =2.1$ and $\epsilon \le a \le 1$ with $\epsilon =10^{-2}$. Each point represents an average over $10^2$ independent simulations. The dashed lines marks the analytical epidemic threshold in Eq.~\ref{thre}.}
\label{fig_10_1}
\end{figure}

The analysis of the data collected with wearable devices allows to characterize with precision the features of close proximity contacts. The first important observation is that contact networks are time-varying~\cite{holme2011temporal,barrat2015face,sekara2015fundamental,holme2015modern}. The distribution of individuals activity, the duration and time intervals of contacts are typically heterogeneous implying the absence of characteristic scales, i.e. averages are not good descriptors. Consequently, the role played by each individual in the spreading is heterogeneous. Furthermore, for diseases in which the probability of transmission is function of the time spent in contact, different interactions might be associated with very different transmission potentials. In particular, the large majority of contacts are short and thus associated with small transmission probabilities. However, some contacts have duration that is much larger than the average and thus can be extremely important for the spreading. The heterogeneity in time intervals between contacts, burstiness~\cite{barabasi2005origin}, has been associated with a rich, case-dependent, phenomenology that can either slow down~\cite{min2011spreading,stehle2011high,karsai2011small,miritello2011dynamical,vazquez2007impact,karrer2010message} or speed up~\cite{rocha2011simulated,takaguchi2013bursty} the spreading. Interestingly, the dynamics ruling the evolution of contacts change depending on the time-scale considered~\cite{barrat2015face,sekara2015fundamental}. In other words, the mechanisms driving the formation of contacts at the time-scale of minutes are different than those at the time scale of months~\cite{sekara2015fundamental}. This is of particular importance when studying the spreading of infectious diseases~\cite{morris1995concurrent,perra2012activity,rocha2011simulated,stehle2011simulation,masuda2013predicting,moody2002importance,liu2013contagion,riolo2001methods,fefferman2007disease,zhu2014effect,horvath2014spreading,machens2013infectious,liu2014controlling,starnini2014temporal,sun2015contrasting,han2015epidemic,sunny2015dynamics,holme2015basic,liljeros2007contact,holme2014birth,toth2015role,pastor2015epidemic,ribeiro2013quantifying}. Indeed, contacts dynamics that affect the spreading of transmissible illnesses are those unfolding at comparable time-scale respect to the disease~\cite{holme2011temporal,barrat2015face,holme2015modern,morris1995concurrent}. For example, the characteristics of contacts networks at the time-scale of days are relevant for the spreading of ILIs.
In order to show more clearly this crucial point we will now consider a simple, yet not trivial, model of temporal human interactions. This modeling framework is based on the empirical observation that the propensity of individuals of being engaged in social acts, \emph{activity}, is heterogeneous~\cite{perra2012activity}. Interestingly, such observations have been performed in a wide range of real datasets describing different types of human interactions or human dynamics ranging from conversations on Twitter to R\&D alliances between firms~\cite{perra2012activity,tomasello2014role}. For a given time interval $\Delta t$ the activity of each node $i$ can be measured as the fraction between the number of interactions made by $i$, $n_i$, divided by the total number of interactions made by all the nodes:
\begin{equation}
a_i=\frac{n_i}{\sum_l n_l}.
\end{equation}
While the value of $a_i$ might change in time~\cite{moinet2015burstiness}, observation in real data shown that the distribution of activity is virtually independent of the choice of $\Delta t$~\cite{perra2012activity,ribeiro2013quantifying,tomasello2014role}. This candidates the activity as good variable to describe some important aspects of time-varying networks. The modeling framework we are describing starts from such evidence and generates so called activity-driven networks~\cite{perra2012activity}. Each node is assigned to an activity $a$ extracted from a distribution $F(a)$. At any time step $t$ the network $G_t$ is build starting from $N$
disconnected vertices. In their simplest form the creation of activity networks is as follows:
\begin{itemize}
\item With probability $a_i \Delta t$ the node $i$ becomes active
  and generates $m$ links connected to $m$
   randomly selected nodes
\item With probability $1-a_i\Delta t$ the node will not be active. However, it will still be able to
  receive connections from other active vertices.
\end{itemize}
To reproduce empirical observations, activities are extracted from heavy-tailed distribution as for example $F(a)\sim a^{-\zeta}$ with $a \in [\epsilon; 1]$. The average number of active nodes at each time step is $N<a>$ that is in general much smaller than $N$. Consequently, the large number of nodes will be not active and the topology of each $G_t$ will be based on set of mostly disconnected stars centered around active nodes. On the other end, it is possible to prove that integrating links over sufficient $T$ time-steps such that $k/N \ll 1 $ and $T/N \ll 1$ the resulting network will have a degree distribution that follows the activity distribution~\cite{perra2012activity,starnini2014temporal}. In other words, the heterogeneity in the number of contacts integrated over time is driven by the heterogeneity in the propensity of nodes to be engaged in social acts. In this model, \emph{hubs} emerge in time due to their constant engagement rather than due to some first mover advantages as in classic preferential attachment models. In general, the full dynamics of the network and its ensuing structure is completely encoded in the
activity distribution. Interestingly, it is possible to derive the epidemic
evolution equation in which a spreading process and the network
dynamics are coupled together. We will consider an SIS model, but the same results holds for an SIR. As done throughout the review, let us define $\beta_I$ the infection probability per contact, and $\gamma$ the recovery rate. At a mean-field level, we can assume that nodes with the same activity $a$ are statistically equivalent. In these settings, we can describe the epidemic process considering the number of infected individuals in the class of activity rate $a$, at time $t$, as $I^{t}_{a}$.  The number of
infected individuals of class $a$ at time $t+\Delta t$ is given by
\begin{equation}
  I^{t+\Delta t}_{a}= I^{t}_{a}-\gamma \Delta t I^{t}_{a} +\beta_I m (N_{a}^{t}-I_{a}^{t})a\Delta t \int d a' \frac{I_{a'}^{t}}{N}+\beta_I m (N_{a}^{t}-I_{a}^{t})\int d a' \frac{I_{a'}^{t}a' \Delta t}{N},
\label{pp1}
\end{equation}
where $N_a$ is the total number of individuals with activity $a$~\cite{perra2012activity}. In Eq.~\ref{pp1}, the second term on the right hand side refers to infected nodes that recover, the  third term takes into account
the probability that a susceptible node of class $a$ is active and acquires
the infection getting a connection from any other infected individual, while the last term takes into
account the probability that a susceptible node, independently of his
activity, gets a connection from any infected active individual~\cite{perra2012activity}.  In order to solve Eq.~\ref{pp1} we can consider the total number of infectious nodes in the system by summing over all the activity classes:
\begin{equation}
\label{pp}
\int da I_{a}^{t+\Delta t}= I^{t+\Delta t}=I^{t}-\gamma \Delta t I^{t} + \beta_I m \langle a
\rangle \Delta t I^{t}+ \beta_I m \theta^{t} \Delta t,
\end{equation}
where $\theta^{t}=\int da' I^{t}_{a'}a' $ and we have dropped all
second order terms in the activity rate $I_a^{t}$. Indeed, we are interested in the early stages of the spreading where the number of infected individuals is much smaller than the number of susceptible nodes. In this regime the equation can be linearized. In order to obtain an
closed expression for $\theta$ we multiply both sides of
Eq.~\ref{pp1} by $a$ and integrate over all activities, obtaining the auxiliary equation
\begin{equation}
\theta^{t+\Delta t}=\theta^{t}-\gamma \Delta t \theta^{t}+\beta_I m \langle a^2 \rangle
I^{t} \Delta t+\beta_I m \langle a \rangle\theta^{t}\Delta t.
\label{othereq}
\end{equation}
In the continuous time limit we obtain the following closed system of differential equations
\begin{eqnarray}
  \partial_{t} I &=&-\gamma  I + \beta_I m \langle a \rangle I + \beta_I m \theta,\\
  \partial_{t} \theta &=& -\gamma  \theta +\beta_I m  \langle a^2\rangle
  I+\beta_I m \langle a \rangle\theta ,
\end{eqnarray}
whose Jacobian matrix  has eigenvalues
\begin{equation}
\Lambda_{(1,2)} = \langle a \rangle\beta_I m  - \gamma  \pm \beta_I m
\sqrt{\langle a^2\rangle }.
\end{equation}
The epidemic threshold for the system is obtained requiring the
largest eigenvalue to be larger than $0$, which leads to the condition
for the presence of an endemic state~\cite{perra2012activity}:
\begin{equation}
\label{t_m1}
\frac{\beta_I}{\gamma}> \frac{1}{m}\frac{1}{\langle a
  \rangle+\sqrt{\langle a^2 \rangle}}.
\end{equation}
The epidemic threshold can be derived considering that the per capita infection rate is given by the infection probability per contact, $\beta_I$, times the contact rate, $\langle k \rangle$:
$\lambda_I= \beta_I \langle k \rangle$. It is easy to prove~\cite{perra2012activity} that $\langle k
\rangle = 2 m \langle a \rangle $, and consequently
\begin{equation}
\label{thre}
\frac{\lambda_I}{\gamma}> \frac{2 \langle a \rangle}{\langle a
  \rangle+\sqrt{\langle a^2 \rangle}}.
\end{equation}
The threshold is function of the first and second moments of the activity distribution thus it takes into account the dynamics of interactions. Furthermore, the epidemic threshold is not function of the time-aggregated network presentation. It depends just on
the interaction rate of nodes. This results show the importance of time-scales. Indeed the spreading condition is dependent on the interplay between the time-scales of the network and spreading process. It is important to remember how in the case of uncorrelated annealed networks the threshold is function of the first and second moment of the degree distribution (see Section~\ref{concep-net} for the detailed derivation). In Fig.~\ref{fig_10_1}, we show the density of infected individuals in the endemic state as a function of ${\lambda_I}/\gamma$ for three different networks~\cite{perra2012activity}. In green diamonds and blue squares we show the results obtained running an SIS model in static networks generated aggregating links of an activity-driven network over $T=40$ and $T=20$ time-steps respectively. In red circles instead, we show the results obtained running an SIS model on activity-driven networks in which both links and disease evolve in time. Clearly, the results are quite different and show the effects of dynamical connectivity patterns on disease spreading. In particular, the threshold in static networks is significantly smaller than in their time-varying networks counterparts. Indeed, in static networks the disease can spread using all links activated in a certain time period independently of their activation time or order. In time-varying graphs instead, the disease spreads over a different, more homogeneous, topology in which at each time-step just few links are available for the contagion. Consequently, the disease will need a larger per capita infection rate or smaller recovery rate in order to percolate in the system.\\
In their simplest form\footnote{The models has been extended to include more realistic link's formation mechanisms~\cite{karsai2014time,ubaldi2015asymptotic}}  activity-driven networks consider a very simple and unrealistic link formation mechanism. Indeed, empirical contact networks are characterized by several types of correlations as moderate modularity, transitivity and different types of assortativity~\cite{salathe2010high,fournet2014contact,stopczynski2014measuring,stehle2011high,barclay2014positive}. In order words, individuals don't form random connections but tend to connect in groups/communities~\cite{fortunato2010community}.
For example, a study conducted in a high-school showed significant levels of clustering and different types of assortativity. In particular, the project collected data for three days in a high-school with a population of about $1000$ individuals (students, teacher, and staff)~\cite{barclay2014positive}. Each day about one third of them volunteered to wear a tag to collect their close proximity contacts and shared their flu vaccination status via an online survey.  The assortativity measured considering the vaccination was significantly more positive than what expected by chance~\cite{barclay2014positive}. Furthermore, the levels of assortativity increased for longer contact duration and combining all data collected across the three days. The mechanisms driving such features could be associated to differences in vaccination coverage and assortativity within sub-networks such as age, gender, role, and ethnicity~\cite{barclay2014positive}. Some hints supporting these hypothesis are found in the coverage respect to gender and role. Around $48\%$ of female reported to be vaccinated respect to $33\%$ of male. Around $52\%$ of staff reported to be vaccinated respect to $39\%$ of students. Remarkably, epidemics spreading simulations suggest that the observed features of contacts networks might induce larger average outbreak size and increase the probability of such outbreaks ~\cite{barclay2014positive}. More in general, the presence of correlations in links' creation might have contrasting effects on disease propagation. Locally the presence of dense connections might promote the propagation (especially in SIS models), the low connectivity between communities might hamper and slows down the global spreading~\cite{keeling2005implications,newman2003properties,miller2009spread,sun2015contrasting,karsai2014time}. 

Summarizing, empirical contact networks are characterized by heterogeneous activity, duration and time intervals between interactions as well as different types of correlations shaping the creations of links. Interestingly, a recent and simple model able to reproduce many of these properties has been developed~\cite{starnini2013modeling,starnini2014model}. The basic ingredient of the model is the idea that each node $i$ is characterized by an \emph{attractiveness} $b_i$. This is a proxy for social appeal/status. In the model, attractiveness are extracted from a uniform distribution with $b \in (0,1]$. In order to mimic close proximity interactions, nodes move in a box of fixed size performing a biased random walk. More precisely, at each time step $t$, each node $i$ moves, with probability $p_i(t)$, a step of length $r$ along a direction given by a randomly chosen angle. Instead, with probability $1-p_i(t)$, the node does not move. The probability $p_i(t)$ is defined as
\begin{equation}
p_i(t)= 1- \max_{j \in E_j(t)}{b_j},
\end{equation}
where $E_j$ are the nodes within a distance $d$ from $i$ at time $t$. In these settings, the node $i$ will create a connection with all the nodes in its proximity, and it will be driven to stop and interact with a probability that is proportional to the attractiveness of the most attractive node in the circle of radius $d$. The model considers also that nodes might not all be present, and willing to make connections, at the same time. To include this aspect, nodes might be active or inactive. At each time step, an inactive node $i$ become active with probability $t_i$, and an active node becomes inactive with probability  $1-t_i$. These probabilities are extracted from an uniform distribution with $t_i \in [0,1]$. Although, nodes do not keep memory of past interactions and despite its extreme simplicity, the model is able to reproduce contact duration, time interval between contacts, and weight distributions of real contact networks. Also, it captures the evolution of the average degree as function of time and the distribution of groups sizes. It is of particular interest consider that a simple biased random walk in 2D driven by a random distribution of attractiveness is able to capture highly not trivial properties of empirical close proximity networks.

\begin{figure}
\begin{center}
\includegraphics[width=0.4\columnwidth,angle=0]{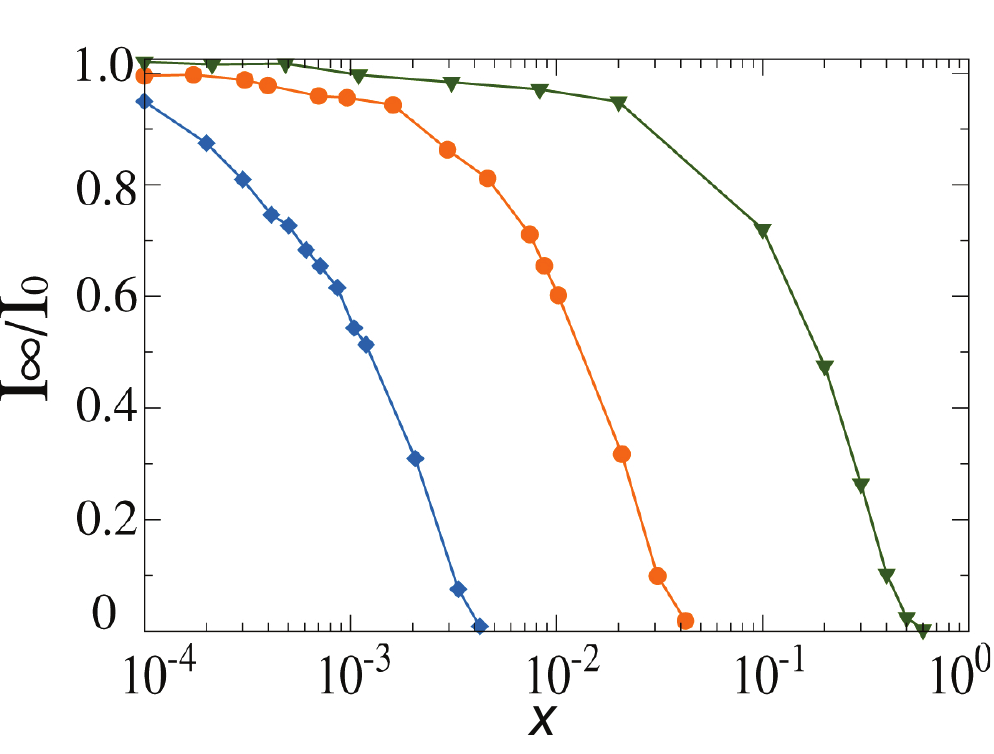}
\end{center}
\caption{Comparison of the stationary state of an SIS model with and without control strategy, $I_\infty/I_0$, as a function of $x$ when $\lambda_I/\gamma=0.81$. In green triangles, we consider the random strategy, in blue diamonds the targeted strategy, and in orange circles the egocentric strategy. Each plot is made averaging $10^2$ independent simulations started with $1 \%$ of random seeds. $N=10^4$, $m=3$, $\epsilon=10^{-3}$, activity distributed as $F(a)\sim a^{-2.2}$. Adapted with permission from Ref.~\cite{liu2014controlling}. Copyrighted by the American Physical Society.}
\label{fig_10_2}
\end{figure}

The study of close proximity contacts shows robust statistical patterns from one day to the next~\cite{fournet2014contact}. Interestingly, the same holds considering different years (for different students)~\cite{fournet2014contact}. However, the interactions of each individuals are not always the same. In fact, the study of the renewal of contacts in different days shows significant variations in the ego-net of each individuals~\cite{fournet2014contact}.
At this time-scale the distribution of number of contacts does not show large heterogeneities~\cite{salathe2010high,fournet2014contact,stopczynski2014measuring,stehle2011high,newman2010networks,gonccalves2015social}. As we saw above in the analysis of the epidemic threshold in activity-driven networks,  this observation is of particular importance. Indeed, as we discussed in details in Section~\ref{concep-net} heterogeneities in degree distributions make systems extremely fragile to the spreading of infectious diseases (lowering the epidemic threshold) and at the same time make targeted vaccination strategies extremely efficient~\cite{newman2010networks,barrat2008dynamical}. Consequently, contacts networks at the time-scale of the day are more resilient to the spreading of infectious diseases, but also less susceptible to targeted vaccination strategies than networks whose contacts are integrated over longer time-scales, i.e. months or years. These observations clearly point out the importance of better understanding the features of close proximity contacts in order to devise efficient vaccination strategies~\cite{salathe2010high,sun2015targeted,genois2014data,liu2014controlling,starnini2013immunization}. To this end, let us extend what discussed in Section~\ref{sec:vac-net} for annealed and static networks by considering the efficiency of different vaccination protocols on time-varying networks. We will focus on the case of activity-driven networks~\cite{liu2014controlling}. First, let us consider a random strategy (RS) in which a fraction of nodes $x$ is vaccinated. In this case, it is easy to modify the Eq.~\ref{pp} and obtain~\cite{liu2014controlling}
\begin{equation}
\label{thre_RI}
\frac{\lambda_I}{\gamma} \ge \xi^{RS} \equiv \frac{1}{1-x}\frac{2 \langle a \rangle}{\langle a \rangle +\sqrt{ \langle a^2 \rangle}} =\frac{\xi}{1-x}.
\end{equation}
In words, when a fraction $x$ of nodes is randomly vaccinated, the epidemic threshold can be written as the threshold with no intervention, $\xi$, rescaled by the number of nodes still available to the spreading process. Indeed, vaccinating random nodes is equivalent to rescale the per capita spreading rate by the fraction of available nodes $\lambda_I \rightarrow \lambda_I(1-x)$. As observed in Section~\ref{concep-net} this is the same in static and annealed networks.  Fig.~\ref{fig_10_2} shows the ratio between the number of infected nodes in the endemic state with random and no interventions, $I_\infty/I_0$, as function of $x$. As clear from the plot, the random strategy (green triangle) requires more than $50\%$ of vaccinated to reduce the endemic state to less than $20\%$ respect to the scenario without interventions. Let us now consider a targeted strategy (TS) in which $xN$ nodes are vaccinated in decreasing order of activity.  This method is equivalent to fix a value $a_c$ so that any node with activity $a\ge a_c$ is immune to the contagion process~\footnote{The value of $x$ and $a_c$ are linked by the relation $x=\int_{a_c}^{1}F(a)da$}. Also, for this scheme, it is possible to derive the analytic expression for the epidemic threshold~\cite{liu2014controlling}
\be
\label{thre_targ}
\frac{\lambda_I}{\gamma}\ge \xi^{TS}\equiv\frac{2 \langle a \rangle}{\langle a \rangle^{c}+\sqrt{(1-x) \langle a^2 \rangle^{c} }},
\ee
where $\langle a^n \rangle^c=\int_{\epsilon}^{a_c}a^{n}F(a)da $ describes the moments of the activity distribution discounting the vaccinated nodes. Eq.~\ref{thre_targ} is not a simple rescaling of the original threshold expression and implies a drastic change in the behavior of the contagion process. Interestingly, vaccinating a very small fraction of the most active nodes is enough to stop the contagion process. Indeed, as clear from Fig.~\ref{fig_10_2} vaccinating just the top 0.04\% of nodes is enough to halt the disease~\cite{liu2014controlling}.
As discussed in Sections~\ref{concep-net}-\ref{sec:vac-net}, the network-wide knowledge required to implement targeted control strategies is generally not available~\cite{cohen03-1}. In the case of time-varying networks, this issue is even more pronounced as the characterization of nodes depends on how long it is possible to observe the network dynamics~\cite{starnini2013immunization}. Understanding the efficiency of strategies that use partial information about the network is crucial. A possible implementation of this is based on egocentric samples (ES) on the networks~\cite{liu2014controlling}. In particular, a fraction $w$ of randomly selected nodes might act as ``probes". During an observation time $T$, their egonet generated by their interactions in the network can be monitored. After the observation window, each probe is asked to randomly select a node in the egonet that is then vaccinated. In this strategy, the probability of vaccination for one node with activity $a$ after a single time step is:
\be
\label{p_a}
P_a=a  w\int d a'  \frac{m N_{a'}}{N} + w\int d a' a'\frac{m N_{a'}}{N}\frac{1}{m}.
\ee
The first term on the right considers the probability that a node of class $a$ is active and reaches one of the probes; the second term, instead, takes into account the probability that one node of class $a$ is connected by an active probe. Solving the integrals in Eq.~\ref{p_a}, we can write $P_a= w \left ( am+\langle a \rangle \right)$.  Starting from this consideration, the probability of vaccination after $t$ time steps is $P_a^t=1-(1-P_a)^t$. The total number of nodes protected from the disease can be obtained summing over all the activity classes: $R^T = \sum_{a}N_aP_a^T=\sum_{a}N_a \left [ 1-(1-P_a)^T \right ]$.\footnote{The equation for $P_a$ does not consider the depletion of nodes in each class due to the vaccination process. The formulation is then a good approximation for small $w$ and $T$, when the probability that a probe is selected more than once is very small.} Using this expression, it is possible to obtain the epidemic threshold~\cite{liu2014controlling}:
\be
\label{thre_ES}
\frac{\lambda_I}{\gamma} \ge \xi^{ES} \equiv \frac{2 \langle a \rangle}{\Psi_1^T +\sqrt{ \Psi_0^T \Psi_2^T}},
\ee
where $\Psi_{n}^T=\int da \,\ a^{n}(1-P_a)^TF(a)$. This last integral is a function of the observing time window $T$. As clear from Fig~\ref{fig_10_2} this strategy is much more efficient than the random one, although not as performant as the targeted one. It is important to stress how the efficiency of this strategy is due to the ability to reach active nodes by a local exploration done observing the systems for few time steps.

Another particularly important application of understanding close proximity contacts is the characterization of pathogen transmissions in hospitals settings~\cite{voirin2015combining,vanhems2013estimating,obadia2015detailed,mastrandrea2015enhancing,hornbeck2012using}. Indeed, nosocomial infections are a major problem for the safety and health of patients and induce high economical costs~\cite{cosgrove2006relationship}. However,  due to challenges in the data collection, very little was understood until very recently. Wearables provide an unique opportunity in this context. To the present day, RFID tags have been used to identify or estimate pathways of transmission~\cite{voirin2015combining,vanhems2013estimating,obadia2015detailed} and to help the test and develop prevention as well as containment measures~\cite{mastrandrea2015enhancing,hornbeck2012using}.

\subsection{Non participatory digital epidemiology}

Our daily interactions with technology generate an unprecedented amount of data that can be harvested for digital epidemiology applications. Contrary to what we discussed above, here we will focus on methods that use data not generated, nor stored, explicitly for epidemiological studies. In other words, we will review recent non participatory approaches in which data is mined in search for proxies of epidemiological relevant behaviors/indicators.

\subsubsection{Social media based methods}

Social media provides unique opportunities to communicate, gather and spread information. The incredible progresses of technology and the consequent diffusion of mobile smart devices boosts the penetration and relevance of such rich medium. It did not take much time before our queries on search engines~\cite{ginsberg2008detecting}, communication on online social networks~\cite{signorini2011use,paul2011you,bodnar2013validating,broniatowski2013national,culotta2010towards}, visits to Wikipedia pages~\cite{generous2014detecting,mciver2014wikipedia,hickmann2015forecasting},  cancellations of online reservations~\cite{nsoesie2014guess}, satellite images of parking lots of hospitals~\cite{butler2014satellite}, or online sources~\cite{collier2008biocaster,torii2011exploratory} started to be adopted for epidemiological applications. Indeed, all these behaviors and interactions between us or with technology can be extremely relevant to model, characterize, anticipate, and forecast the unfolding of infectious diseases~\cite{salathe12-1}.

While the details behind each methodology are case specific and clearly beyond the scope of this review, the basic underlying ideas, as well as the associated pitfalls, are a common feature. In order to better understand them, we will describe the prototypical example of Google Flu Trends (GFT)~\cite{gft}. The service premiered in late $2008$ and used our queries on Google to predict the unfolding of the seasonal influenza (and in some countries also dengue) in $29$ countries. As we will describe later in more details, despite the initial glory, the service has been now discontinued and it has been reshaped in the form of a data sharing collaboration with few institutes~\cite{gft_blog}. The basic intuition behind GFT is extremely simple and powerful: traditional surveillance data (as for example the percentage of ILI visits) can be estimated and predicted by mining search engine queries. In particular, the epidemiological real data describing the progress of ILIs in a specific country, $P$, is estimated using proxy data, $Q$, describing the fraction of ILI related queries. More formally, the value of $P$ at a given week is obtained via a liner model:
\begin{equation}
\tilde{P}=\vartheta_0+\vartheta_1 \tilde{Q}+\varrho,
\end{equation}
where the tilde refers to the logit of each quantity ($\tilde{y}=\log\left(\frac{y}{1-y}\right)$), the $\vartheta$s are fitting parameters, and $\varrho$ is an error term. In other words, GFT obtains an estimation of $P$ considering the fraction of ILI relevant queries, $Q$. Interestingly, this quantity is evaluated mining about hundred billions of queries on Google's search engine. The methodology focuses just aggregated signals, no information/features about users are considered. Furthermore, the classification of queries as ILI relevant is performed measuring, independently, the correlation between $50$ millions queries and historic surveillance data. Remarkably, this unsupervised method, returns as most relevant queries, searches associated to influenza, influenza complications, and remedies~\cite{ginsberg2008detecting}. It is important to mention that the expression reported above could be generalized to include a set of $Q$ variables:
\begin{equation}
\tilde{P}=\vartheta_0+\sum_{i=1,n}\vartheta_i \tilde{Q_i}+\varrho,
\end{equation}
each one associated to a different query~\cite{bodnar2013validating}. However, as noted in the original paper, the performance of the method was higher adding up all the signal in a signal explanatory variable~\cite{ginsberg2008detecting}.  The initial results reported were extremely impressive: GFT had a mean correlation with surveillance data of about $0.97$ and was able to predict it one/two weeks ahead. It is worth thinking about this a bit more. A simple linear regression model fed with an unparalleled dataset of queries to a search engine is able to estimate with great accuracy information about the unfolding of ILI.

Clearly the methodology described is quite simple. The strength of the approach is the unique and unprecedented dataset used. For many different reasons this data cannot be easily shared. Consequently, researchers applied, extended, and improved the GFT archetype to other data streams that are instead accessible. Examples are Twitter and Wikipedia. Indeed, in both cases large samples and/or relevant complete information can been gathered via their standardized APIs. Methods using Twitter data rely on the assumption that people post tweet, information about their health status. In particular, tweets are mined in search of ILI related terms that are fed to regression models~\cite{signorini2010using,paul2011you,bodnar2013validating,broniatowski2013national,culotta2010towards}. The relevant terms are typically provided a priory or learned via machine learning techniques. Indeed, researchers do not have, typically, the computational resources of Google to test independently each $n$-gram used on Twitter. An increasing fraction of tweets contains geographical information that can be adopted to create models at different granularity going from cities~\cite{broniatowski2013national} to countries~\cite{signorini2010using,paul2011you,bodnar2013validating,culotta2010towards}. Methods based on Wikipedia consider pages view of particular entries as data proxies~\cite{generous2014detecting,mciver2014wikipedia,hickmann2015forecasting}. The underlying assumption is the same of GFT: users visit specific pages to gather information about diseases, their symptoms, and possible medications. The approaches presented so far have been not focused just on ILI, but they considered a wide range of transmissible diseases going from dengue to ebola~\cite{generous2014detecting}.
As we observed in the case of participatory digital surveillance data, the results of GFT or similar tools based on different data proxies can serve as input for modeling efforts based on unstructured (sub)populations (see Section~\ref{sec:compart-model}). A nice summary the results of such approaches in providing actual predictions for the 2013-2014 flu season in the USA is detailed in Ref.~\cite{biggerstaff2016results}.

Social media offers unprecedented opportunities to access information and communicate. Unfortunately, they can also be used effectively to spread misinformation, myths, or conspiracy theories~\cite{del2016spreading}. In the realm of public health this phenomenon has been extremely clear. Indeed, as discussed in Section~\ref{sec:behaviormodeling}-\ref{sec:beh-vac-net} vaccination coverage has been affected by the spreading of myths (as the link between vaccines and autism or sterilization) or miscommunication and poor policy making from public officials (as in the case of Nigeria boycott of polio vaccination)~\cite{larson2013measuring,kata2012anti}. Social media has changed significantly the way people acquire health information. We transitioned from classic doctor-patient to more patient-patient or patient-social media interactions~\cite{larson2011addressing}. On one side, this allows digital epidemiology based on social media to work. On the other, it makes us susceptible to the spreading of unfounded and unscientific information.

Many experts are now referring to the decrease of vaccination rates as a \emph{confidence crisis} linked to the spreading of unscientific myths on social media, but also to socio-economical, socio-cultural, psychological, and political factors~\cite{larson2011addressing,ibuka2014free,shim2012influence,hershey1994roles,menzies2006vaccine,chapman2012using}. The features of online discussions about these and other topics can be mapped and as discussed in Section~\ref{sec:beh-vac-net} used to model behavior-vaccination dynamics~\cite{salathe11-1,del2016spreading,larson2013measuring,conover2011political,quattrociocchi2014opinion,salathe2013dynamics}. Interestingly, different opinions (pro and anti vaccination) induce effects of polarization in which individuals of different ideas do not interact much. As consequence, communication networks cluster in separated communities, \emph{eco-chambers}, that exacerbates the challenges in spreading factual and expert-based knowledge. Furthermore, a recent study on the diffusion of sentiments regarding vaccines on Twitter suggests other peculiar dynamics~\cite{salathe2013dynamics}. First, individuals that are connected with large number of opinionated users are inhibited  from expressing their own opinions. Second, exposure to negative sentiments is contagious while exposure to positive sentiments is not. All together these results indicate that peer influence and social contagion are content dependent and far from being trivial. It is important to mention how social media and more in general technology are also used to improve vaccine communication and coverage via texts, emails, calls, apps, and web services~\cite{stockwell2013utilizing,witteman2012defining,huston2015searching}.

Despite good performances, all digital epidemiological methods based on social media are affected by several limitations (see Table~\ref{social_media_table}). Also in this case, we will consider GFT as example to better understand the shortcomings and potential pitfalls of these approaches. GFT has been released in November $2008$ showing great results. However, the tool was not able, despite being updated, to predict the unfolding of the pandemic in $2009$~\cite{lazer14-1,olson2013reassessing}. Furthermore, GFT overestimated i) $92\%$ of the weeks in the seasons $2011$ and $2012$, ii) the peak in $2013$ (the prediction was more than double the real value). As we wrote above the service has been now discontinued~\cite{gft_blog} and it is limited to data sharing with research partners. The reasons behind these bad performances are several. Firstly, GFT adopted simple correlations between online queries and ILI surveillance data. In doing so, it captured other phenomena correlated with ILI as for example winter patterns. The pandemic in $2009$ peaked much earlier (late October early November) than the classic flu. Probably, this is what impeded GFT to capture its unfolding. Secondly, Google search engine is highly dynamical. Changes are constantly implemented. Some of them might modify users' searching behaviors, as for example the auto-complete suggestions. It is important to consider all of these points as they are likely to affect all digital tools for epidemiology. Indeed, they are typical shortcomings and issues associated to all non participatory data collections. In particular, simple correlations might not be enough to describe complex phenomena unfolding at different scales in time and spaces respect to those used for training purposes. More in general, correlation does not guarantee causation. A recent study on Twitter shows that the time series of the word zombie is a very good predictor of the seasonal flu~\cite{bodnar2013validating}. Furthermore, matching searches or keywords to context (in this case disease) without taking in account that the same word can be used in discussions completely unrelated to illness might introduce spurious correlations or decrease the signal to noise ratio~\cite{weeg2015using}.  Notwithstanding, it is fundamental to acknowledge the potential of these tools in the fight against infectious diseases. They offer unprecedented opportunities to estimate epidemic proxies at high geographical resolution or in locations where surveillance data is not available. Furthermore, these data can be used to inform realistic epidemic models ~\cite{Shaman11122012,zhang2015social}  as well as a large number of more theoretical models as those described in Sections~\ref{concep-net}-\ref{sec:beh-vac-net}.

\begin{center}
\begin{table*}
\centering
\caption{\label{social_media_table} Summary of the main advantages and disadvantages of social media based methods in digital epidemiology.}
\begin{tabular}{|p{5cm}|p{5cm}|}\hline
\textbf{Main Advantages} & \textbf{Main Disadvantages} \\
\hline
Sample size & Possible biases in sample \\
\hline
Access to different geographical granularities & Correlations might be spurious \\
\hline
Near real-time analysis & Simple word matching might not sufficient \\
\hline
Cheaper data collection & $n$-gram selection computationally expensive \\
\hline
Rich set of information available & Changes in platform design and functionalities might influence users behaviors \\
\hline
\end{tabular}
\end{table*}
\end{center}

\subsubsection{Mobile phone based methods}

Mobile phones are one of the most diffuse technological device. About $96\%$ of the worldwide population has one~\cite{ict}. For billing purposes phone carriers collect call data records (CDRs) that can be mined and used to study our communications, social networks, and mobility (see Ref.~\cite{blondel2015survey} for a recent review). The last two phenomena are of epidemiological relevance. In particular, contact networks estimated via CDRs can be used to characterize human interactions at large and long-term scales well beyond those accessible via participatory experiments mentioned above. Although, not all connections mediated by mobile phones are relevant for disease transmission, recent studies clearly show the possibility of inferring close proximity networks from CDRs~\cite{eagle2009inferring,calabrese2011interplay}. The temporal features describing mobile phone networks are similar to those of actual closed proximity interactions~\cite{karsai2012universal,karsai2011small,miritello2011dynamical,karsai2014time}. However, CDRs can provide just contact networks integrated in time. Indeed, the empirical timestamp of each interaction is not per se relevant for disease spreading. Interestingly, also these, time-aggregated and static, networks share several statistical features with close proximity contacts. In particular, high clustering, transitivity, and assortativity have been reported~\cite{blondel2015survey}. Furthermore, integrated mobile phone datasets are typically characterized by heavy-tailed distributions of contacts and weights. However, their heterogeneity are, also in this case, function of the time window used to aggregate data~\cite{krings2012effects}. From a modeling stand point, CDRs perfectly lend themselves as input and test bed for the different modeling efforts beyond the homogeneous mixing approximation discussed in Section~\ref{concep-net}.
Human mobility is another crucial ingredient for spreading of infectious disease. CDRs provide unprecedented opportunity to study this human phenomena. The large body of literature on the subject identifies three main characteristics of our movements~\cite{blondel2015survey,toole2015modeling}. Firstly, the exploration of new places is very slow. Secondly, our movement is relatively predictable. Thirdly, the way each one of us moves is rather unique. The understanding of such features has been crucial for modeling and understanding of the spreading of infectious diseases. For example approaches informed with data from CDRs have been used to map and control the spreading of Malaria~\cite{wesolowski2012quantifying,pindolia2012human}, Ebola~\cite{wesolowski2014commentary}, and ILIs~\cite{tizzoni2014use,eubank2004modelling}. In order to better understand the potential and importance of this data we will briefly summarize the methods and results recently reported in Ref.~\cite{wesolowski2012quantifying}. The diffusion of Malaria is the results of a complex process that involves the interaction between its parasite and two hosts: mosquitoes and humans~\cite{wesolowski2012quantifying}. In particular, the movement of people increases the dispersal of the parasite beyond the reach of mosquitoes. CDRs in combination with spatially resolved information about Malaria prevalence have been used to define how human movements drive parasite importation between different regions of Kenya. In particular, CDRs were used to quantify the travel patterns of about $15$ millions of people (around $34\%$ of the total population) over one year. The calls of such individuals have been mapped to one of the $12,000$ cells towers spread across $692$ settlements~\cite{wesolowski2012quantifying}. In order to define the primary settlement for each person, the researchers considered a majority rules according to the time spent in each location. Furthermore, they collected information, such as destination and time, about the movements of individuals made out of their primary settlement. This data has been combined with an simple epidemic transmission model based on high resolution maps of Malaria prevalence. The fusion of these two data sources allowed to define the networks of parasite movements across the country as well as source (net emitters of people and parasites) and sink (net receivers of people and parasites) regions~\cite{wesolowski2012quantifying}. These results provided unprecedented lens to characterize Malaria dynamics and spreading. More importantly, they suggested a novel set of strategies for containment that targets large flows of human movements associated with high risk of importations. These encode many aspects of the complex dynamics driving the spreading and go well beyond the typical based on local transmission hot spots or routine surveillance in high-risk areas~\cite{wesolowski2012quantifying}.

\subsection{Mixed approaches}

We distinguished among the various methods and tools of digital epidemiology considering the strategy used for data collection. So far we presented active (participatory) or passive (non participatory) approaches. This division is quite clear and includes most of the notable tools. However, there are some that are a mix between the two. In the case of \emph{Health Map}~\cite{healthmap} the mix is quite asymmetric and it could have been classified as non participatory tool. This project is a data grinder. It considers a wide range of data sources such as different online data streams, eyewitness reports, expert-curated discussion, and official reports~\cite{healthmap,freifeld2008healthmap,brownstein2007healthmap}. It has been lunched in $2006$ with the goal of monitor, nowcast, and predict the spreading of infectious disease. The data is fetched, classified, and visualized 24/7/365. The system is automatized and allows to process and digest incredible amount of relevant information providing comprehensive reports that are used to facilitate the detection or monitoring of public health threats across the world. The Health Map infrastructure is behind an impressive number of projects and specific applications~\cite{healthmap_projects,healthmap_pubs} . Examples include Health Map Flu Trends in which all the different data streams are combined to provide forecasts of flu activity in the USA~\cite{healthmap_flu,santillana2015combining}, Dengue Map that provides real-time alerts of dengue outbreaks~\cite{healthmap_dengue}, or Health Map Vaccine which connects the population with vaccines~\cite{healthmap_vaccine,huston2015searching}. As for the partecipatory survelliance systems described above, the data provided by Health Map does not provided any information about the contacts of individuals. It provided instead, data at the (sub)population level that can be used to develop unstructured compartmental models (see Sections~\ref{sec:compart-model} and \ref{sec:beh-vac-mean})

Another application that mixes data from social media to active participation of users is Crowdbreaks~\cite{crowdbreaks}. This tool mines Twitter in search of tweets of epidemiological relevance. Crowdbreaks is a system for digital surveillance able to provide geographically resolved information about diseases in the USA. The selection of relevant tweets process is what makes it unique. Indeed, the task is crowdsourced to users visiting the webpage. In particular, it adopts the wisdom of the crowd to classify tweets as relevant, not relevant, or potentially relevant to a particular disease. With the collaboration and direct engagement of people the performance of machine learning methods used to automatically detect relevant tweets increases in time. The system, is currently in a beta version during which the contribution of individuals is crucial to help make it operational.

\section{Conclusion, discussion and future research}\label{conclusion}

This review has summarized the rich opportunities and challenges for studying theoretical and empirical aspects of the complex interactions between vaccinating behavior and disease dynamics, as well as the many opportunities to apply methods and concepts from statistical physics to these interactions.  This was first seen in the contrast between mean-field models that do not incorporate human behavior versus mean-field models that do incorporate them (Sections \ref{sec:compart-model} and \ref{sec:beh-vac-mean}).  Mean-field models continue to play an invaluable role in the development and application of the behavioral epidemiology of infectious diseases in at least two respects. Firstly, they have shown the intrinsic complexity of coupled behavior-disease dynamics: coupled behavior-disease models generate emergent dynamics that are not present in models where vaccinating behavior is ignored, and coupled models also exhibit more dynamical outcomes in general. For instance, these models can exhibit oscillations in disease outbreaks and vaccine coverage even when the corresponding compartmental model does not exhibit any such dynamics. By incorporating human behavior (either explicitly or implicitly) we can ask questions that are not possible with models lacking behavior, such as how do different modes of communication (e.g. word-of-mouth versus mass media) influence the probability of eradication of a vaccine-preventable infection? Secondly, coupled behavior-disease models have demonstrated the ability to explain observed data better compared to simple behavior-free epidemiological models, in cases where data availability permits such a comparison.  As a result, mean-field behavior-disease models show  potential for informing public health policy makers of modern industrialized countries planning the introduction of a new vaccine; planning the response to a decline in vaccination coverage following a rumor; or more generally to strategically act to increase the degree of resilience of public vaccination  systems in face of the threat posed by human behavior in a world where risks from most common infections are typically perceived as ``minor".

Future research on mean-field behavior-disease models should aim to further exploit benefits and potentialities of simple models, such as representing simple and parsimonious environments for coupling behavior-disease dynamics and allowing predictions at a high level of abstraction (e.g. population-level prevalence and coverage). As such, a first line of inquiry should use mean-field models for investigating the implications of further social, economic, or psychological theories of behavior relevant for vaccination in order to achieve a more detailed understanding of the ``typically observed" behaviour (see e.g. \cite{oraby2014influence}). A second line should aim to more extensively develop the--just initiated--empirical use of mean-field models by applying them over different datasets (e.g. different countries) and different infections (e.g. measles vs pertussis). This would allow to test the role of different behavioral factors and improve our understanding of why such factors promoted different behavioral responses to the same rumour, such as observed in the MMR vaccine-autism rumor. A third line of inquiry should aim to systematically include stochasticity in simple behavior-disease models, with the aim to preserve parsimony but at the same time to better account for the fact that most challenges due to vaccine refusal will appear under circumstances of disease rarity, i.e. when infections become sub-critical due to efficient control, thereby lowering the perceived payoff of vaccination. These situations are of special relevance for large scale elimination plans, as the World Health Organization plans for measles elimination, where local infection rarity is expected to become the rule of infection dynamics and chance effects become fundamental. Last, straightforwardly incorporating behavioral responses in the realistic age-structured models typically used to evaluate the impact of intervention programs would help such models account for possibly the most important source of uncertainty about the future success of vaccination programs--human behavior.  We emphasize that these suggestions do not exhaust the possibilities for further development of mean-field behavior-disease models.

Mean-field models assume that the population can be approximated as a continuum, and they are often deterministic in nature.  However, for many applications, it may not be possible to ignore the discrete, individual-based nature of real populations.  Section \ref{sec:vac-net}--on models for disease dynamics and vaccinating on networks that do not explicitly represent behavior dynamics--showed that even without taking individual behavior into account, understanding networks of individual contacts in a population has a critical role to play in shaping our understanding of epidemiological processes. It is vitally important that network structure is taken into account if we attempt to predict population-level dynamics from individual-level observations. In addition, many intervention strategies such as contact tracing or ring vaccination can only be accurately captured and modeled using network-based approaches. The emergence of network modeling tools enables different strategies to be tested in an artificial environment. Finally, vaccination on time-varying networks  is another interesting research topic in this area, requiring high-resolution temporal data to track and understand the control of disease and the dynamic evolution of networks.

Given that network models lend themselves so easily to representation of individuals, it is perhaps no surprise that the recent growth in network-based coupled behavior-disease models has been so rapid, as described in Section \ref{sec:beh-vac-net}.  As with the mean-field models, many behaviors were noted in network behavior-disease models that were not possible in network models lacking behavior, such as the difficulty of eradicating infection.  Moreover, the importance of local structure and stochasticity take on new meaning, in networks where nodes actively adapt to their local epidemiological environment. In Section \ref{sec:beh-vac-net}, we provided an overview of current results about behavioral vaccination in networked populations. Starting from the simple single-layer networks, we  first reported the influence of benefit-inspired imitation dynamics, network topology, external incentives as well as additional protection measures on individual vaccination willingness and total vaccination level. We  then demonstrated how social clusters expand and induce a large outbreak probability even if there exists extremely high vaccination coverage on empirical networks. It is worth mentioning that opinions about vaccination have been well identified as another type of method for protection measure updating. Finally, we briefly discussed the recent progress of behavioral vaccination on the novel framework of temporal and multilayer networks, where the links among nodes are time-varying or multiple. Along this line, behavioral vaccination on metapopulation networks deserves our considerable attention in future, because this topology supports the human mobility and information exchange among different patches, which will directly affect how human update their measures and seems closer to realistic situations. This framework, on the one hand, will further enrich content of behavioral vaccination (especially the feedback loop between human behavior and disease) with mobile fashion, similar to reaction-diffusion processes in physical chemistry. On the other hand, such a framework can be well-integrated with empirical data, such as the control of an infectious disease pandemic driven by air travel \cite{colizza2006role}. As mentioned above, (benefit-inspired or opinion-based) updating dynamics plays a crucial role in the final vaccination level and epidemic incidence. Most existing literature assumes that everyone has the same updating criterion, namely, identical cognition process for infection risk, vaccine risk and benefit brought by vaccination, which in fact greatly differs among individuals due to social norms, education background, age structure, etc. This point may attract great interest in future research. Risk perception on networks is another unexplored field of voluntary vaccination. The vast majority of models assume that information transmission of infection risk is completely local (i.e. coming from the contacting neighbors) and neglects information dissemination to the whole population or network, which is ubiquitous in reality, due to public media. Thus, how to incorporate local and global information into one modeling framework will be of particular significance in future.

Other unexplored problems concerning behavior-disease interactions on vaccination networks also merit great attention in the future. For instance, these approaches could be extended to other types of vaccination on multilayer networks, like local vaccination, or targeted vaccination based on various centrality indices. However, the development of a theoretical framework for multilayer networks still lies at the beginning stages: there are no uniform definitions of these centrality indices, which indicates that such a task needs in-depth interdisciplinary collaborations, especially with network scientists.  Another scenario that must attract more attention is when multilayer networks support different dynamical processes on each layer. An infectious disease can spread simultaneously on different layers, or  disease and awareness can diffuse in their respective layers.  In such cases, how to implement vaccination becomes a novel challenge.  Additionally, one of the most sophisticated and commonly used models for non-behavioral infectious disease modeling is the meta-population network, where individuals can move between different patches due to migration or commuting, as in many real-world populations. Vaccination on these and other social structures using endogenized behavioral dynamics remains little explored.  Finally, one of main  needs in theoretical epidemiology of infectious diseases is the development of new analytical or approximated theoretical physics methodologies in order to make simulation-independent inferences concerning individual-based models. Indeed, inference uniquely based on simulations can sometime hide the underlying big picture of a complex system.  We suggest that Ising models, percolation theory, and related  approaches from statistical physics may be useful for studying coupled behavior-disease dynamics on networks, just as these approaches has proven useful in the study of (non-behavioral) epidemiological processes on lattices \cite{durrett1995spatial,newman2002spread,meyers2005network}.

The ultimate standard in any scientific discipline should be testing hypotheses against data, and the standard for models of coupled behavior-disease dynamics should be no different.  The last few sections of our review explored existing data sources like surveys and vaccine coverage records, which could be used to parameterize or validate such models. Measuring and assessing immunization levels remains a difficult problem. As more health systems globally are moving into the digital area, we can expect an improvement in the quality of the data. Furthermore, novel biomedical approaches may make it easier to assess whether individuals could mount an immune response, the expected waning time, and how immunity came about (vaccination vs. infection).  Such individual-level characteristics are highly relevant to population-level infectious disease models.  Nevertheless, as problems of vaccinations are particularly pronounced in low-income countries, the general challenges will remain in the foreseeable future. Surprisingly, the problems described here--uncertainty in vaccination coverage, and variability in vaccine efficacy and effectiveness--are only rarely taken into account in theoretical studies of infectious disease dynamics. We hope that in the future, these real world issues will get more attention in the modeling literature.

The last few sections of our review also explored new digital data sources that could facilitate a new era in the theoretical and empirical development of coupled behavior-disease models.  Digital epidemiology is a very new and promising area of research: a natural evolution of epidemiology in the digital era.  Such data sources lend themselves very naturally to network coupled behavior-disease models.  Recent advances in mobile phone technology, information from twitter, Facebook, and GPS location enables us to determine individual connections, or even track the movement of people in real time. This would help us to build full and comprehensive networks for many airborne diseases such as influenza. It will also track how these network topologies change in the face of a severe epidemic, or even how individual behaves when an intervention measure is implemented.  The vibrant body of literature on digital epidemiology shows incredible potential as well as future directions and challenges. The size and heterogeneity of the datasets we will face is destined to increase dramatically. This will require the development of new algorithms and techniques to effectively collect, store and analyze large-scale epidemiological data. In particular, new approaches, specifically designed for epidemiology, will be needed to filter, classify, and model massive data streams. Cheaper and more scalable devices will be required to improve our understanding of face-to-face interactions and thus of spreading pathways. From a theoretical standpoint, new frameworks able to further characterize the complexity of human contacts and their effects on epidemic spreading will be needed. In particular, it will be crucial to deepen our understanding of the temporal, and high-order structural features of social interactions. Furthermore, a better understanding of the dynamics of social contagion will be necessary to characterize behavioral changes induced by spreading of information/misinformation that affect vaccination rates and more in general disease spreading. The future of digital epidemiology is also tied to the development of new approaches for data access, sharing and privacy. Indeed, the ethical and legal challenges that arise from mining human digital traces are in general far from be understood or well regulated. Digital epidemiology is a truly interdisciplinary endeavor that currently connects a wide range of disciplines ranging from data science and mathematics to network science and medicine. A new breed of practitioners able to bridge all of these disciplines are necessary to face the challenges ahead.  Furthermore, the prospect of being able to relate individual-level behavioral dynamics to population-level behavioral dynamics with unprecedented spatial and temporal resolution raises the possibility of attempting to develop a statistical physics of behavior.  The extensive literature on non-equilibrium thermodynamic theory may prove useful in this endeavor \cite{kreuzer1981nonequilibrium}.

Clearly, the opportunities for new theory and new applications, as well as the potential awards of success, are enormous in the field of coupled behavior-disease modeling.  But how can the field move forward on these opportunities?  Review papers often include a discussion paragraph on how more interdisciplinary interactions are needed.  We fall to the temptation of including such a paragraph here, but we will explain exactly why interdisciplinary interactions are valuable in the context of this particular field.  The field itself emerged from interdisciplinary interactions when researchers--usually trained in other areas like physics or mathematics--took an interest in disease systems.  Despite this, there remains a need for more interdisciplinary interactions than currently take place.  For instance, physicists, mathematicians, and economists often develop models of behavior-disease interactions in parallel with one another and apparently without interfacing with the knowledge of the other fields.  This not only risks duplication, but also means that research fails to benefit from cross-fertilization of ideas.  Interactions with epidemiologists and public health researchers could also be more pervasive than they currently are, in order to ensure that these models are ultimately applied to improving population health outcomes in the real world.   In a similar vein, the scope of interdisciplinary interactions needs to expand to include new fields.  For instance, digital data and machine learning hold significant promise for helping us better understand behavior-disease interactions and parameterize coupled behavior-disease models.  However, properly harnessing these new sources of data will require help from statisticians and computer science experts.

Interdisciplinary interactions are stymied by different terminology used by differently trained researchers.  For instance, the theoretical study of interactions between disease transmission and behavior has been variously referred to using terms such as ``behavioral epidemiology", ``economic epidemiology", ``coupled disease-behavior systems" and other terms although in some cases, it should be noted that these terms do not completely overlap because they are used differently to marry different sets of concepts.   Similarly, economists use the word ``rational" to describe selfish, socially blind behavior, although other fields would not describe such behavior as rational in its usual sense.  The problem of inconsistency in language could be addressed through workshops devoted to bringing together different fields, which would also help the more general lack of interaction between different disciplines.  A degree of coordination as exhibited by the nomenclature committee of the International Astronomical Union for studying interactions between behavior and disease is probably impossible due to the nature of the subject, but efforts to bring together different types of experts could at least accomplish some of the benefits of greater coordination.

Shakespeare famously ended his play ``A Midsummer Night's Dream" with an apology from the spirit Puck to the audience: ``If we shadows have offended, Think but this, and all is mended---That you have but slumbered here While these visions did appear."  We also wish to apologize to the readers in advance, for having inadvertently missed some excellent papers in the literature, having unintentionally misrepresented other papers that we did include, or being completely wrong about where the field should go next.  However, there is no need for the reader to pretend our review was something they dreamed while slumbering. Instead, please write us if you feel our review was lacking in some respect, or better yet, publish some research that shows us the value of other perspectives on the field.  If nothing else, we hope that our review has convinced the readers that coupled behavior-disease modeling is a rapidly growing area that promises exciting theoretical developments and empirical applications in the coming decade, especially if we draw on the accumulated knowledge of many decades of statistical physics.

\section*{Acknowledgements}

We would like to acknowledge gratefully Lin Wang, Irene Sendi{\~n}a-Nadal, Shlomo Havlin, Mark Newmann, Alessandro Vespignani, Yamir Moreno, Jesus G{\'o}mez-Garde{\~n}es, Gao-xi Xiao, Ke-ke Huang, Zhi-xi Wu, Cheng-yi Xia, Gui-quan Sun, Jin Zhen, Hai-feng Zhang, Ming Tang, Da-qing Li, Wen-bo Du, Xue-long Li, Xin-yang Deng, Chao Gao, Yong Deng, Han-xin Yang, Marko Jusup, Yoh Iwasa, Shingo Iwami, Boris Podobnik, Lidia Braunstein, and Bruno Goncalves for their constructive help and discussions. We also appreciate all our other friends with whom we maintained (and are currently maintaining) interactions and discussions on the topic covered in our review.

This work was supported by the National Natural Science Foundation of China (Grant No. 61374169), Natural Science Foundation of Anhui Province (Grant No. 1508085MA04), and Project of Natural Science in Anhui Provincial Colleges and Universities (Grant No. KJ2015ZD33) to ZW, an NSERC Individual Discovery Grant (RGPIN-04210-2014) to CTB, Shandong Province Outstanding Young Scientists Research Award Fund Project (Grant No. BS2015DX006), Shandong Academy of Sciences Youth Fund Project (Grant No. 2016QN003), and National Natural Science Foundation of China (Grant No. 61572297) to DWZ, and the Slovenian Research Agency (Grant Nos. J1-7009 and P5-0027) to MP.

\clearpage

\section*{Appendix}
Here, we provide a summary of abbreviations of the used terminologies in the main text.

\newcommand{\tabincell}[2]{\begin{tabular}{@{}#1@{}}#2\end{tabular}}
\begin{longtable}{|c|c|c|}
\caption*{Table A1: \bf Abbreviations of the used phrases in our report.}\\
\hline\textbf{\large Abbreviations}    &  \textbf{\large Full names}    &  \textbf{\large First appeared}         \\
\hline
\tabincell{c}{SIR\\ SIS\\ SEIR\\ SEIRS\\ ODE\\ GAS\\ AIDS\\ FOI\\ BRN\\ ERN\\ VRN\\ NGO\\ AR\\ DFE\\ EE\\ WAIFW matrices\\ ME\\ CLE\\ FPE\\ SDE\\ CCS} & \tabincell{c}{Susceptible-Infectious-removed\\ Susceptible-Infectious-Susceptible\\ Susceptible-Exposed-Infective-Removed\\ Susceptible-Exposed-Infective-Removed-Susceptible\\ ordinary differential equation\\ globally asymptotically stable\\ acquired immune deficiency syndrome\\ force of infection\\ basic reproduction number\\ effective reproduction number\\ vaccine reproduction number\\ next generation operator\\ attack rate\\ disease-free equilibrium\\ endemic equilibrium\\ Who Acquires Infection From Whom matrices\\ Master equation\\ Chemical Langevin equation\\ Fokker-Plank equation\\ Stochastic Differential equation\\ critical community size}& Section 3\\
   \hline
   \tabincell{c}{ER graph\\ SW network\\ WS model\\ SF network\\ BA network\\ DMS model\\ DBMF\\ IBMF\\ ICM\\ GEMF framework\\ STD}& \tabincell{c}{Erd{\"o}s-R{\'e}nyi graph\\ small-world network\\ Watts and Strogatz model\\ scale-free network\\ Barab{\'a}si-Albert network\\ Dorogovtsev-Mendes-Samukhin model\\ Degree-based mean-field theory\\ Individual-based mean-field theory\\ Independent cascade model\\ generalized epidemic mean-field framework\\ sexually transmitted diseases}& Section 4\\
   \hline
   \tabincell{c}{CBF\\ BHD}& \tabincell{c}{community bridge finder\\ bridge-hub detector}& Section 5\\
   \hline
   \tabincell{c}{SIAs\\ MMR\\ AIC}& \tabincell{c}{special immunization activities\\ measles-mumps-rubella\\ Aikaike Information Criterion}& Section 6\\
   \hline
   \tabincell{c}{SEU\\ ESS}& \tabincell{c}{subjective expected utility\\ evolutonarily  stable state}& Section 7\\
   \hline
   \tabincell{c}{VAE\\ IGD\\ EGDT\\ VSE\\ ISNs\\ CSNE\\ MDPT}& \tabincell{c}{vaccine adverse effects\\  Imitation Game Dynamics\\ Evolutionary Game Dynamics Theory\\ vaccine side effects\\ injunctive social norms\\ convergently stable Nash equilibrium\\ Markov decision processes theory}& Section 8\\
   \hline
   \tabincell{c}{RV\\ NN\\ CV\\ INN}& \tabincell{c}{random vaccination\\ nearest neighbor vaccination\\ chain vaccination\\ improved nearest neighbor vaccination}& Section 9\\
   \hline
\tabincell{c}{DHS\\ MICS\\ EPI\\ LQAS\\ HPV} & \tabincell{c}{Demographic and Health Surveys\\ Multiple Indicator Cluster Surveys\\ Expanded Programme on Immunization\\ Lot Quality Assurance Sampling\\ Human Papillomavirus}& Section 10\\
   \hline
   \tabincell{c}{ILI\\ RFID\\ ES\\ GFT\\ CDRs \\RS\\ TS} & \tabincell{c}{influenza like illnesses\\ radio frequency identification devices\\ egocentric samples\\ Google Flu Trends\\ call data records\\ random strategy\\ targeted  strategy} & Section 11\\
   \hline
\end{longtable}

\clearpage

\bibliography{references}
\label{conclusion}

\end{document}